\documentclass[article, twocolumn, 
   aps, pra,
  amsmath,amssymb,
  longbibliography,
  ]{revtex4-1}
\usepackage{graphicx,color}
\usepackage{amsmath}
\usepackage{natbib}
\usepackage{epsfig}
\begin{document}

\title{\color{blue} Elementary vibrational model for transport properties of dense fluids}

\author{S. A. Khrapak}\email{Sergey.Khrapak@gmx.de}
\affiliation{Joint Institute for High Temperatures, Russian Academy of Sciences, 125412 Moscow, Russia}

\begin{abstract}
A vibrational model of transport properties of dense fluids assumes that solid-like  oscillations of atoms around their temporary equilibrium positions dominate the dynamical picture. The temporary equilibrium positions of atoms do not form any regular structure and are not fixed, unlike in solids. Instead, they are allowed to diffuse and this is why liquids can flow. However, this diffusive motion is characterized by much longer time scales compared to those of solid-like oscillations. Although this general picture is not particularly new, only in a recent series of works it has been possible to construct a coherent and internally consistent {\it quantitative} description of transport properties such as self-diffusion, shear viscosity, and thermal conductivity. Moreover, the magnitudes of these transport coefficients have been related to  the properties of collective excitations in dense fluids. Importantly, the model is simple and no free parameters are involved. Recent achievements are summarized in this overview. 
Application of the vibrational model to various single-component model systems such as plasma-related Coulomb and screened Coulomb (Yukawa) fluids, the Lennard-Jones fluid, and the hard-sphere fluid is considered in detail. Applications to real liquids are also briefly discussed.  Overall, good to excellent agreement with available numerical and experimental data is demonstrated. Conditions of applicability of the vibrational model and a related question concerning the location of the gas-liquid crossover are discussed.             
\end{abstract}

\date{\today}

\maketitle

\section{Introduction}

Our understanding of transport processes in fluids (throughout this paper we use the term ``fluid'' to denote subcritical liquid and supercritical dense fluid regions in the phase diagram of a substance; the crossover between the gas-like and liquid-like behaviours of supercritical fluids will be also discussed) 
remains incomplete and fragmented as compared
to gases and solids, although certain progress has been achieved over decades~\cite{FrenkelBook,GrootBook,BalucaniBook,MarchBook,HansenBook}. 
Moreover, it is very unlikely that a general theory of transport processes in fluids can be developed. The main problem is the absence of a small parameter. In solids the small parameter is provided by the ratio of the vibrational amplitude of an atom around its equilibrium position (lattice site) to the distance between neighboring lattice sites. In gases the small parameter is the ratio of the characteristic radius of interatomic forces to the mean interatomic separation~\cite{LifshitzKinetics}. No such parameter exists in fluids.  

Difficulties with theoretical description of liquid state dynamics in comparison with solids and gases have been very well formulated by Brazhkin~\cite{BrazhkinUFN2017}. Solids and gases can be considered in some (dynamical) sense as ``pure'' aggregate states. In solids the motion of atoms is purely vibrational. Hence diffusion is greatly suppressed, shear viscosity reaches extremely high values, and the thermal conductivity is well described by phonon theory~\cite{ZimanBook,Klemens1993}. In dilute gases atoms move freely along straight trajectories between pair collisions, and this can be considered as a random walk process. Kinetic theory with the Boltzmann equation and the Chapman-Enskog approach lead to accurate expressions for the transport coefficients~\cite{LifshitzKinetics,ChapmanBook}. From this perspective liquids constitute a ``mixed'' aggregate state. Both vibration and random walk mechanisms are present. Their relative importance depends on the exact location in the phase diagram. Near the liquid-solid phase transition vibrational motion dominates and solid-like approaches to transport properties are more relevant. At lower densities and higher temperatures ballistic motion is more important and transport is similar to that in gases.      

Thus, although it is unreasonable to expect a unified theory of transport properties applicable to the entire fluid regime, models that focus on a particular regime might be more successful.   
In a series of recent papers a consistent view on transport properties of sufficiently dense fluids, where dynamics is dominated by solid-like vibrational properties, has been put forward. The purpose of this paper is to provide an overview of this approach and to demonstrate how it applies to various simple fluids. Its predictions concerning the self-diffusion, viscosity, and thermal conductivity coefficients will be compared with those from extensive numerical simulations. The applicability regime will be identified. Some interesting consequences will be discussed.

This paper deals exclusively with classical fluids. Main attention is also given to dielectric fluids.  Specifically, presentation is merely focused on one-component
simple fluids and one component plasma-related systems. Therefore, the electron contribution to the thermal conductivity (like e.g. in liquid metals and plasmas) is not considered.  

The paper is organized as follows. In Section~\ref{Norm} we introduce normalization used for the transport coefficients and the concept of excess entropy scaling. Qualitative behaviour of the transport coefficients as the density changes from dilute gaseous values to dense liquid values close to the freezing point is discussed in Section~\ref{qualitative}. In Section~\ref{picture} a simple physical picture of liquid dynamics within the vibrational model paradigm is provided. In Section~\ref{SErel} self-diffusion in liquids is described as random walk process, which leads to the Stokes-Einstein relation between the self-diffusion and viscosity coefficients. Relation between different relaxation times is briefly discussed in Section~\ref{relaxation}. Vibrational model of thermal conductivity is introduced in Section~\ref{ThCo}. In Section~\ref{CollProp} interrelations between the transport and collective mode properties are discussed. Section~\ref{OCP} provides a detailed illustration of the vibrational model performance using a special case of one-component plasma. Further examples, including the screened Coulomb (Yukawa), Lennard-Jones, and hard-sphere fluids are provided in Section~\ref{Examples}. A link between the vibrational model of thermal conductivity and the excess entropy scaling is sketched in Section~\ref{Link}. The location of the crossover between gas-like and liquid-like dynamics is discussed in Section~\ref{Crossover}. In Section~\ref{Realliq} relevance of the discussed results to the transport properties of real liquids is briefly discussed. The paper ends by a brief discussion, conclusion and outlook in Section~\ref{Concl}.

\section{Normalization and excess entropy scaling}\label{Norm}

Numerical values of transport coefficients for different fluids and different conditions can differ by orders of magnitude~\cite{Refprop}. To introduce some systematics, it makes sense to use a rational normalization. It has been proven particularly useful to employ a system-independent normalization,  which is sometimes referred to as Rosenfeld's normalization~\cite{RosenfeldPRA1977,RosenfeldJPCM1999}, although it can be traced to much earlier works (for instance by Andrade~\cite{Andrade1931,Andrade1952}). The normalized self-diffusion ($D$), shear viscosity ($\eta$), and thermal conductivity ($\lambda$) coefficients are 
\begin{equation}\label{Rosenfeld}
D_{\rm R}  =  D\frac{\rho^{1/3}}{v_{\rm T}} , \quad\quad
\eta_{\rm R}  =  \eta \frac{\rho^{-2/3}}{m v_{\rm T}}, \quad\quad \lambda_{\rm R}=\lambda\frac{\rho^{-2/3}}{v_{\rm T}},
\end{equation}
where the subscript ${\rm R}$ emphasizes that Rosenfeld's normalization is used. Here $\rho$ is the atomic density so that $\rho^{-1/3}=\Delta$ corresponds to the mean interatomic separation, $v_{\rm T}=\sqrt{T/m}$ is the thermal velocity, $T$ is temperature in energy units ($\equiv k_{\rm B}T$), and $m$ is the atomic mass. This normalization will be employed throughout this paper.   

Additionally, Rosenfeld suggested useful scaling relationships for transport coefficients of dense simple fluids -- their excess entropy scaling~\cite{RosenfeldPRA1977}. According to this scaling the reduced self-diffusion,viscosity and thermal conductivity coefficients of simple fluids can be expressed as exponential functions of the reduced excess entropy~\cite{RosenfeldJPCM1999}
\begin{equation}\label{Rosenfeld1}
D_{\rm R}  \simeq  0.6{\rm e}^{0.8 s_{\rm ex}} , \quad
\eta_{\rm R}  \simeq  0.2{\rm e}^{-0.8 s_{\rm ex}}, \quad \lambda_{\rm R} \simeq 1.5 {\rm e}^{-0.5 s_{\rm ex}}.
\end{equation}
Here the reduced excess entropy per particle, expressed in units of $k_{\rm B}$, is defined as $s_{\rm ex}=s-s_{\rm id}$, where $s_{\rm id}$ is the reduced entropy of an ideal gas at the same temperature and density determined by Sackur-Tetrode equation
\begin{equation}
 s_{\rm id}=\frac{5}{2}+\ln\left[\frac{1}{\rho}\left(\frac{mT}{2\pi\hbar^2}\right)^{3/2}\right].   
\end{equation}
The excess entropy $s_{\rm ex}$ is negative because interactions enhance the structural order compared to that in a fully disordered ideal gas. This implies that the reduced diffusion coefficient decreases towards the freezing point, while the viscosity and thermal conductivity coefficients increase. In the ideal gas limit ($s_{\rm ex}=0$) Eqs.~(\ref{Rosenfeld1}) predict finite reduced transport coefficients, while in reality $D_{\rm R}$, $\eta_{\rm R}$ and $\lambda_{\rm R}$ all diverge as density goes to zero.   

By now it is well recognized that many simple and not so simple systems conform to the approximate excess entropy scaling. A couple of recent examples include comprehensive study based upon viscosities obtained from experimental measurements of molecular fluids as well as molecular simulations of model potentials~\cite{BellPNAS2019}. A very useful modified excess entropy scaling of the transport properties of the
Lennard-Jones fluid has been discussed in detail in Ref.~\cite{BellJPCB2019}. There are also counterexamples, where the original excess entropy scaling is not applicable. Typical situations when this happens include anisotropic interactions, potentials with negative curvature, bounded potentials, and systems with weak energy-virial correlations. Among known examples are water models, the Gaussian core model, the Hertzian model, the soft repulsive-shoulder-potential model, models with flexible molecules, etc. ~\cite{KrekelbergPRE03_2009,KrekelbergPRE12_2009,
FominPRE2010,DyreJPCB2014}. For a recent review of excess entropy scaling and related topics see Ref.~\cite{DyreJCP2018}.

The excess entropy scaling can be rationalized in terms of isomorph theory~\cite{DyreJPCB2014,GnanJCP2009}. Isomorphs
are defined as the lines of constant excess entropy in the
thermodynamic phase diagram. Certain structural and dynamical properties in properly reduced units are invariant along isomorphs to a good approximation. This includes traditional measures such as radial distribution function, the instantaneous shear modulus, normalized time-autocorrelation functions~\cite{DyreJPCB2014}, the bridge function~\cite{CastelloJCP2021}, and even some higher-order structural correlations~\cite{RahmanMolecules2021}. Macroscopically reduced transport coefficients (diffusion, viscosity, and thermal conductivity) are among isomorph invariants. The concept of isomorphs implies correlations between isomorph-invariant quantities and thus it explains scaling of reduced transport coefficients with excess entropy in the Rosenfeld approach. Note that within the isomorph theory concept the excess entropy plays no special role, because any
other {\it isomorph-invariant} quantity can be used in place of excess entropy~\cite{DyreJPCB2014}. This explains for instance pronounced correlations between the viscosity and thermal conductivity coefficients of dense fluids discussed recently~\cite{KhrapakJETPLett2021}. 

It should be noted that a somewhat different variant of entropy scaling of atomic diffusion in condensed matter was also proposed by Dzugutov~\cite{DzugutovNature1996}, who used the excess entropy in the pair approximation, $s_2$, instead of the full excess entropy (note that the total excess entropy can be approximated by the pair contribution only in some vicinity of the freezing point~\cite{LairdPRA1992,GiaquintaPhysA1992,GiaquintaPRA1992,
SaijaJCP2006,FominJCP2014,KlumovResPhys2020,KhrapakJCP2021}).
This relation is consistent with the isomorph concept, because $s_2$ is also isomorph invariant~\cite{GnanJCP2009}.

Neither the excess entropy scaling approach nor the isomorph concept can explain the exact form of scaling and physical mechanisms behind the transport processes. They should be considered as merely heuristic approaches. Rosenfeld originally referred to the entropy scaling as a semi-empirical ``universal'' corresponding-states relationship~\cite{RosenfeldJPCM1999}. Later in this paper we will illustrate how an approximate excess entropy scaling of the thermal conductivity coefficient naturally arises within the vibrational model of dense liquid dynamics.   

\section{Qualitative behaviour of reduced transport coefficients}\label{qualitative}

As we have seen, Rosenfeld's excess entropy scaling describes a decrease of the diffusion coefficient and an increase of the viscosity and thermal conductivity coefficients when approaching freezing point. This holds in the dense fluid regime, not too far from the fluid-solid coexistence. It is important to understand the qualitative behaviour of transport coefficients in the entire domain corresponding to the gaseous and fluid regimes. In this Section an illustrative example is provided.  

\begin{figure}
\includegraphics[width=8.5cm]{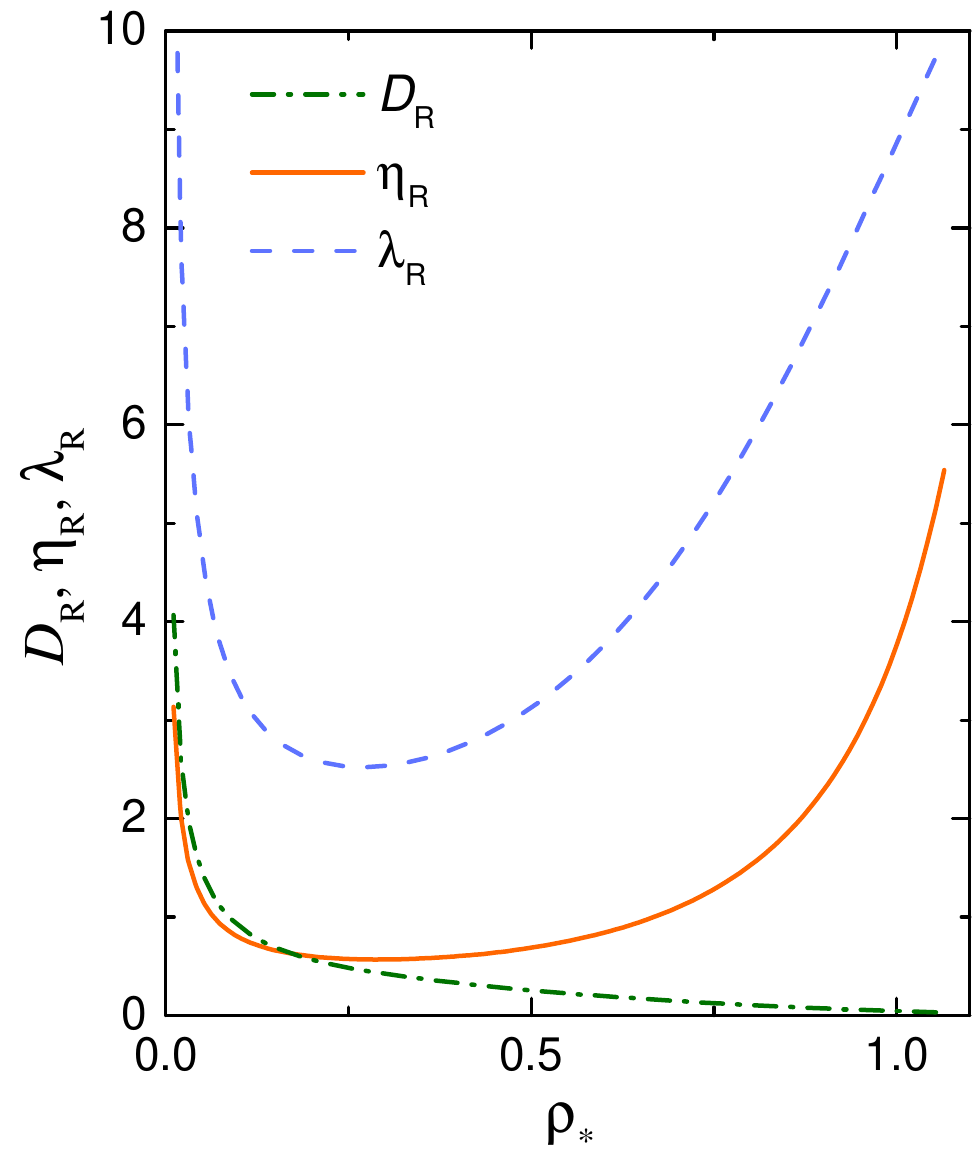}
\caption{(Color online) Reduced self-diffusion ($D_{\rm R}$, dash-dotted curve), viscosity ($\eta_{\rm R}$, solid curve), and thermal conductivity ($\lambda_{\rm R}$, dashed curve) coefficients versus reduced density $\rho_*$ of the LJ fluid along a supercritical isotherm $T_*=2$. Liquid boundary of the solid-fluid coexistence corresponds to $\rho_*\simeq 1.065$.  }
\label{FigLJ_example}
\end{figure}  

Such an example is presented in Figure~\ref{FigLJ_example}. Here the reduced diffusion, viscosity and thermal conductivity coefficients of the Lennard-Jones (LJ) fluid are plotted versus the reduced density $\rho_*$ along a supercritical isotherm $T_*=2$ (the definition of LJ units and other details regarding LJ fluids are given in Section~\ref{LJliq}). Transport coefficients are calculated using the approach from Ref.~\cite{BellJPCB2019}. The qualitative behaviour of the reduced transport coefficients is quite general and relatively universal for a broad range of simple fluids.  The specific shape of the LJ interaction potential plays only a minor role in this respect. At small densities $D_{\rm R}$, $\nu_{\rm R}$ and $\lambda_{\rm R}$ are all decreasing with the density. This can be understand as follows. In dilute gases the transport properties are determined by rare events of pairwise collisions between the constituent atoms (atoms move along straight trajectories between collisions most of the time). Consider elementary kinetic formulas for the transport coefficients in dilute gases~\cite{LifshitzKinetics}:
\begin{equation}\label{elkinform}
D\sim \ell v_{\rm T}, \quad \eta \sim mv_{\rm T} \rho\ell, \quad \lambda\sim c_{\rm p}v_{\rm T}\rho\ell,
\end{equation}   
where $\ell$ is the mean free path between collisions and $c_{\rm p}$ is the reduced specific heat at constant pressure. The mean free path can be expressed via the effective momentum transfer cross section $\Sigma$ as $\ell\sim 1/\rho\Sigma$ ($\Sigma$ is udsed instead of conventional $\sigma$ to avoid confusion with the hard-sphere diameter and LJ length scale, which will appear below). In dilute gases we have $\Sigma\ll\rho^{-2/3}$, which automatically implies $\ell\gg \rho^{-1/3}$. In Eqs.~(\ref{elkinform}) numerical coefficients of order unity are omitted. Note that for monatomic dilute gases there exists exact relation between the viscosity and thermal conductivity, $\eta = 4m\lambda/15$, which does not depend on the exact mechanisms of interatomic interactions~\cite{LifshitzKinetics}. If we now apply Rosenfeld's normalization to Eq.~(\ref{elkinform}) we get the transport coefficients in the dilute gaseous regime
\begin{equation}\label{Rgas}
D_{\rm R}\sim \frac{1}{\Sigma\rho^{2/3}}, \quad \eta_{\rm R} \sim  \frac{1}{\Sigma\rho^{2/3}}, \quad \lambda_{\rm R}\sim \frac{c_{\rm p}}{\Sigma\rho^{2/3}}.
\end{equation}   
All the reduced transport coefficients are of the same order of magnitude and they all decrease as $\rho^{-2/3}$ in this regime. This can be very clearly seen in Fig.~\ref{FigLJ_example}: The diffusion and shear viscosity coefficients are rather close in dilute gaseous regime, $D_{\rm R}\simeq \eta_{\rm R}$; the thermal conductivity coefficient is about $2.5$ times larger (for monatomic gases $c_{\rm p} = 2.5$). 

As density increases, the diffusion coefficient continues to decrease. A cage of nearest neighbours (potential well) develops around each atom. This suppresses atomic diffusion dramatically. At freezing point a quasi-universal value of $D_{\rm R}\simeq 0.03$ is reached~\cite{PondSM2011,KhrapakAIPAdv2018}. The situation is different for the shear viscosity and thermal conductivity coefficients. They achieve minima at some intermediate density and then {\it increase} with the density as the freezing point is approached (the values of $\eta_{\rm R}$ and $\lambda_{\rm R}$ at freezing are also to some extent quasi-universal~\cite{KhrapakAIPAdv2018,KhrapakPoF2022}). This implies that the change in mechanisms of momentum and energy transfer occurs. Namely, the momentum and energy in dense fluids are transferred collectively due to strong interatomic interactions. We will see below that atomic vibrations around their temporary equilibrium positions play a decisive role in the transport properties of dense fluids. Minima in the reduced shear viscosity and thermal conductivity coefficients indicate at the crossover between gas-like and liquid-like mechanisms of momentum and energy transport and can serve as indicators of gas-to-liquid dynamical crossover. This topic received considerable attention in recent years and it will be further discussed in Section~\ref{Crossover}.            
Recently, the existence and the magnitudes of the minima of shear viscosity and thermal conductivity coefficients have been discussed from an interesting perspective.    
It has been suggested that the {{\it kinematic viscosity} and {\it thermal diffusivity}} of liquids and supercritical fluids have lower bounds determined by fundamental physical constants~\cite{TrachenkoSciAdv2020,TrachenkoPRB2021},
\begin{equation}\label{bonds}
\nu_{\rm min}=\alpha_{\rm min}=\frac{1}{4\pi}\frac{\hbar}{\sqrt{m_e m}}, 
\end{equation}
where $\nu$ is the kinematic viscosity, $\alpha$ is the thermal diffusivity, $\hbar$ is the Planck's constant, $m_e$ is the electron mass and $m$ is the atom or molecule mass. The very existence of such universal bounds and their closeness is an intriguing result. On the other hand, in a later paper
purely classical arguments were shown to be sufficient to provide an adequate estimate of the transport coefficients at their minima~\cite{KhrapakPoF2022}. These were simply estimated by extrapolating the gas-like and liquid-like asymptotes for the shear viscosity and thermal conductivity coefficients into the crossover regime. The typical values at the minima are $\eta_{\rm R}^{\rm min}\simeq 0.6$ and $\lambda_{\rm R}^{\rm min}\simeq 3$ for several important real liquids (see Table I from Ref.~\cite{KhrapakPoF2022}). The minimal values are somewhat lower for soft interactions, which are relevant in the plasma-related context (Coulomb and weakly screened Coulomb potentials).

\section{Physical picture of dense liquid dynamics}\label{picture}

The main assumptions adopted in the vibrational model of atomic transport have been discussed by many authors over decades, see e.g. Refs.~\cite{FrenkelBook,Hubbard1969,Stillinger1982,ZwanzigJCP1983,GoldenPoP2000}.  
Namely, it is assumed that atoms in liquids exhibit solid-like oscillations about temporary equilibrium positions corresponding to a local minimum on the system's potential energy surface~\cite{FrenkelBook,Stillinger1982}. These positions do not form a regular lattice like in crystalline solids, but correspond to a liquid-like structure~\cite{Hubbard1969}. They are also not fixed, and change (diffuse or randomly drift) with time (this is why liquids can flow), but on much longer time scales (separation of time scales corresponding to fast solid-like atomic oscillations and their slow drift plays a very important role in the approach). Effectively, one can assume that local configurations of atoms are preserved for some time until a fluctuation in the kinetic energy allows rearranging the positions of some of these atoms towards a new local minimum in the multidimensional potential energy surface. The waiting time distribution of the rearrangements scales with time as $\exp(-t/\tau)/\tau$,  where $\tau$ is a lifetime. Atomic motions after the rearrangements are uncorrelated with motions before rearrangements~\cite{ZwanzigJCP1983}. This picture allows to make important approximations about the properties of atomic motion and mechanisms of momentum and energy transport in the liquid state. With appropriate elaborations, it will allow us to come up with quantitative expressions for the transport coefficients.  
 
\section{Self-diffusion as random walk process (Stokes-Einstein relation)} \label{SErel}

Self-diffusion usually describes the displacement of a test particle immersed in a medium with no external gradients. If this motion can be considered a random walk process, then the diffusion coefficient in three spatial dimensions can be defined as~\cite{FrenkelBook}
\begin{equation}\label{RW}
D=\frac{1}{6}\frac{\langle r^2 \rangle}{\tau},
\end{equation}  
where $r$ is an actual (variable) length of the random walk, $\tau$ is the time scale, and we focus on sufficiently long times ($t\gg\tau$). Consider first an ideal gas as an appropriate example. The atoms move freely between pairwise collisions. If the distribution of free paths between collisions follows the $e^{-r/\ell}/\ell$ scaling, then $\langle r\rangle=\ell$ and  $\langle r^2\rangle=2\ell^2$, where $\ell$ is the mean free path~\cite{FrenkelBook}. Combining this with the relation for the average atom velocity $\langle v\rangle=\ell/\tau$, we recover the elementary kinetic formula for the diffusion coefficient of a dilute gas 
\begin{equation}
D=\frac{1}{3}\langle v\rangle \ell.
\end{equation}    

The dynamical picture is quite different in liquids, but the concept of random walk remains relevant~\cite{KhrapakMolecules12_2021}. 

Based largely on the physical picture drawn in Sec.~\ref{picture}, Zwanzig suggested the following simple approach to estimate the self-diffusion coefficient~\cite{ZwanzigJCP1983}. He approximated the velocity autocorrelation function of an atom $j$ by  
\begin{equation}\label{Z}
Z_j(t)\simeq \left(\frac{T}{m}\right)\cos (\omega_j t)e^{-t/\tau},
\end{equation}
which corresponds to the time dependence of a damped harmonic oscillator. This is apparently a simplest approximation corresponding to the vibrational picture of atomic dynamics. Here $\omega_j$ is an effective vibrational frequency of an atom $j$. The self-diffusion coefficient $D$ is given by the Green--Kubo formula
\begin{equation}\label{D}
D=\frac{1}{N}\int_0^{\infty}\sum_j Z_j(t)dt.
\end{equation}
Zwanzig then assumed that vibrational frequencies $\omega_j$ are related to the collective mode spectrum and performed averaging over collective modes. After the evaluation of the time integral, this yields  
\begin{equation}\label{D1}
D =\frac{T}{3 mN}\sum_{\bf k}\frac{\tau}{1+\omega_{\bf k}^2\tau^2},
\end{equation} 
where the summation runs over $3N$ normal mode frequencies (clearly, $N\gg 1$ is assumed). Central to the dynamical picture sketched in Sec.~\ref{picture} is the separation of time scales corresponding to fast solid-like atomic oscillations and their much slower drift. This picture makes sense only if the waiting time $\tau$ is much longer than the inverse characteristic frequency of the solid-like oscillations. In this case, we can neglect unity in the denominator of Eq.~(\ref{D1}) and rewrite it as
\begin{equation}\label{D2}
D=\frac{T}{m\tau}\left\langle\frac{1}{\omega^2}\right\rangle, 
\end{equation}
where the conventional definition of averaging, 
\begin{equation}
\left\langle \frac{1}{\omega^{2}}\right\rangle = \frac{1}{3N}\sum_{\rm k}\omega_{\bf k}^{-2},
\end{equation}
has been used. 

Equation (\ref{D2}) allows for a simple physical interpretation. It represents a diffusion coefficient of a random walk process, Eq.~(\ref{RW}). The length scale of this process is identified as
\begin{equation}
\langle r^2\rangle=\frac{6T}{m}\left\langle\frac{1}{\omega^2}\right\rangle,
\end{equation}
which is twice the mean-square displacement of an atom from its local equilibrium position due to solid-like vibrations~\cite{BuchenauPRE2014,KhrapakPRR2020}. The coefficient of two appears, because the initial atomic position is not at the local equilibrium, but randomly distributed with the same properties as the final one (after the waiting time $\tau$). The characteristic time scale of the random walk process is just the waiting time $\tau$, which has not yet been specified. It appears that the appropriate scale for the waiting time is given by the Maxwellian shear relaxation time~\cite{FrenkelBook,KhrapakMolPhys2019}    
\begin{equation}\label{tau}
\tau_{\rm M}=\frac{\eta}{G_{\infty}}=\frac{\eta}{m\rho c_t^2},
\end{equation}
where $G_{\infty}$ is the infinite frequency (instantaneous) shear modulus, and $c_t$ is the transverse sound velocity. 

Substituting Equation~(\ref{tau}) into Equation~(\ref{D2}), we obtain a relation between the self-diffusion and viscosity coefficients in the form of the Stokes--Einstein (SE) relation 
\begin{equation}\label{SE}
D\eta\left(\frac{\Delta}{T}\right)=\frac{c_t^2}{\Delta^2}\left\langle\frac{1}{\omega^2}\right\rangle=\alpha_{\rm SE},
\end{equation}
where $\alpha_{\rm SE}$ is the SE coefficient (this relation is also sometimes referred to as Stokes--Einstein--Sutherland relation). Eq.~(\ref{SE}) essentially coincides with Zwanzig's original result~\cite{ZwanzigJCP1983}. Having no better option, Zwanzig used a Debye approximation, characterized by one longitudinal and two transverse modes with acoustic dispersion, which allowed him to estimate the constant $\alpha_{\rm SE}$. We will discuss this approximation in more details below.
It makes sense to discuss first the main qualitative implications of the vibrational approach. 

First of all, it appears that self-diffusion in the liquid state can be viewed as a random walk due to atomic vibrations around temporary equilibrium positions over time scales associated with rearrangements of these equilibrium positions. In this paradigm, consecutive changes of temporary equilibrium positions (jumps of liquid configurations between two neighboring local minima of the multidimensional potential energy surface in Zwanzig's terminology) are relatively small, much smaller than the vibrational amplitude. Hopping events with displacement amplitudes of the order of interatomic separation may be present, but they have to be relatively rare so that they do not contribute to the diffusion process. This picture is different from the widely accepted hopping mechanism of self-diffusion in liquids. Previously, the concept of random walk was suggested in the context of molecular and atomic motion in water and liquid argon~\cite{BerezhkovskiiPRE2002}. Vibrational model provides a more quantitative basis for this treatment.   

Second, the vibrational approach does not allow us to derive separately the coefficients of self-diffusion and viscosity, but only their product in the form of SE relation (\ref{SE}). The latter is also sometimes referred to as SE relation without the hydrodynamic diameter~\cite{CostigliolaJCP2019}. Other derivations of the SE relation on the atomic scale have been also proposed~\cite{BalucaniBook,Balucani1990}. From the excess entropy scaling perspective, Eqs.~(\ref{Rosenfeld}) and (\ref{Rosenfeld1}) tell us that $D_{\rm R}\eta_{\rm R}\equiv \alpha_{\rm SE}\simeq 0.12$, which is not too far from the actual range $\alpha_{\rm SE}\simeq 0.13 - 0.17$ for simple fluids (see below).   

Third, formula (\ref{SE}) particularly emphasizes the relation between the liquid transport and collective mode properties. Since the exact distribution of frequencies is generally not available, some approximations have to be employed at this point. We will discuss these approximations later in Sec.~\ref{CollProp}. 

Finally, it should be mentioned that very important questions related to the breakdown of the SE relation in supercooled and glass forming liquids~\cite{HodgdonPRE1993,TarjusJCP1995,BordatJPCM2003,ChenPNAS2006,
PuosiJCP2018} are completely beyond the scope of this paper.

\section{Relaxation time scales}\label{relaxation}

An important time scale of a liquid state is a structure relaxation time. This can be defined as an average time it takes an atom to move the average interatomic distance $\Delta$ (sometimes it is referred to as the Frenkel relaxation time~\cite{BrazhkinPRE2012,BrazhkinUFN2012,BrykJPCB2018}). It is interesting to compare it with the waiting time scale (Maxwellian relaxation time) introduced above. Taking into account the diffusive character of atomic motion, we can write $\tau_{\rm R}=\Delta^2/6D$. From Equation~(\ref{RW}), we immediately~get
\begin{equation}
\tau_{\rm R}=\frac{\Delta^2}{\langle r^2 \rangle}\tau_{\rm M}.
\end{equation} 

This implies that quite generally $\tau_{\rm R}/\tau_{\rm M}\gg 1$. The time scale ratio $\tau_{\rm R}/\tau_{\rm M}$ has a maximum at near-freezing conditions, where, according to the Lindemann melting criterion $\Delta^2/\langle r^2 \rangle\sim 100$~\cite{Lindemann,KhrapakPRR2020}. 
This picture correlates well with the results from numerical simulations (see, e.g., Fig.~3 from Ref.~\cite{BrykJPCB2018}). Thus, there is a huge separation between the structure relaxation and individual atom dynamical relaxation time scales, justifying the main assumption behind the vibrational model.

\section{Thermal conductivity}\label{ThCo}

\begin{figure}
\includegraphics[width=8.5cm]{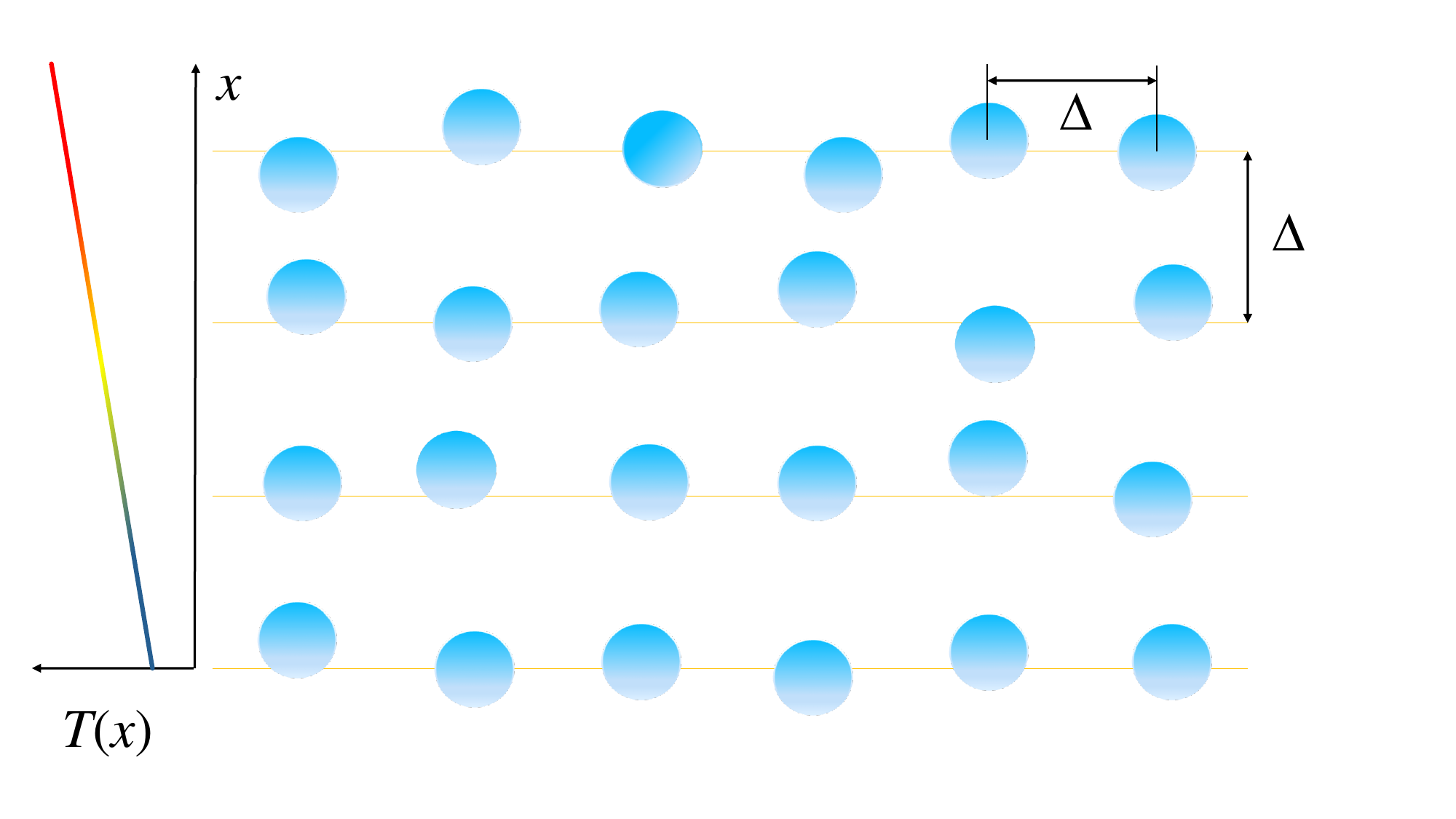}
\caption{(Color online) Two-dimensional slice of a quasi-layered three-dimensional fluid model structure. The average inter-particle separation within one quasi-layer and between quasi-layers is $\Delta = \rho^{-1/3}$. The temperature increases from bottom to top. Reproduced from S. Khrapak, Phys. Rev. E {\bf 103}, 013207 (2021).}
\label{Fig1}
\end{figure}

In this Section a vibrational model of thermal conductivity is outlined following Ref.~\cite{KhrapakPRE01_2021}. The picture drawn in Sec.~\ref{picture} remains valid. In addition, liquid is approximated by a layered structure with layers perpendicular to the temperature gradient and separated by the distance $\Delta = \rho^{-1/3}$. The particle density in each such quasi-layer is $\Delta^{-2}$. A sketch of the considered idealization is shown in Fig.~\ref{Fig1}.   

Now, if a temperature gradient is applied, the average difference in energy between the atoms of adjacent layers is $\Delta (dU/dx)$, where $U$ is the internal energy (per atom). In the considered model, the energy between successive layers is transferred when two vibrating atoms from adjacent layers ``collide'' (this should not be a physical collision; the atoms just need to approach by a distance that is somewhat shorter than the average interatomic separation). The characteristic vibrational frequency of the liquid's quasi-lattice is $\nu$ and it defines the characteristic energy transfer frequency. The energy flux per unit area is
\begin{equation}\label{flux}
\frac{dQ}{dt}=-\frac{\nu}{\Delta} \frac{dU}{dx},
\end{equation}  
where the minus sign indicates that the heat flow is down the temperature gradient. On the other hand, Fourier's law for the heat flow reads
\begin{equation}\label{Fourier}
\frac{dQ}{dt}=-\lambda\frac{dT}{dx},
\end{equation}
where $\lambda$ is the thermal conductivity coefficient, which is a scalar in isotropic liquids. Combining Eqs.~(\ref{flux}) and (\ref{Fourier}) we immediately get
\begin{equation}\label{Cond0}
\lambda=\frac{dU}{dT}\frac{\nu}{\Delta}=\frac{dU}{dT}\frac{\langle \omega \rangle}{2\pi \Delta}.
\end{equation}
In the last step it has been implicitly assumed that the characteristic frequency of energy exchange is equal to the average vibrational frequency, $\nu=\langle \omega \rangle/2\pi$. This appears adequate in dense liquids characterized by soft interatomic interactions~\cite{KhrapakPRE01_2021}. The derivative of internal energy with respect to temperature should be taken at constant pressure since pressure should be constant in equilibrium.
Thermodynamic identities yield
\begin{equation}
 \left(\frac{\partial U}{\partial T}\right)_P=c_{\rm p}-\frac{P}{N}\left(\frac{\partial V}{\partial T}\right)_P = c_{\rm v}+ \left(\frac{\partial U}{\partial V}\right)_T\left(\frac{\partial V}{\partial T}\right)_P.  
\end{equation}
Dense fluids close to the freezing point can be considered as essentially incompressible (in particular, when interactions are sufficiently soft), hence $(\partial U/\partial T)_P\simeq c_{\rm p}\simeq c_{\rm v}$, i.e. the difference between specific heats $c_{\rm p}$ and $c_{\rm v}$ is usually insignificant. Since normally $c_{\rm v}$ is easier to evaluate, it is more appropriate to use $c_{\rm v}$ for practical estimates. We get therefore
\begin{equation}\label{Cond1}
\lambda\simeq c_{\rm v}\frac{\langle \omega \rangle}{2\pi \Delta}. \end{equation}

Equation (\ref{Cond1}) emphasises again relations between the liquid transport and collective mode properties. In the case of thermal conductivity, we need to evaluate the average vibrational frequency $\langle \omega \rangle$, instead of $\langle \omega^{-2}\rangle$ in the case of SE relation. As already pointed out, since the actual frequency distribution can be quite complex in liquids and can vary considerably from one type of liquid to another, some simplifying assumptions are necessary at this step. In the next Section several possible practical simplifications are discussed. 

\section{Relation to collective properties}\label{CollProp}

Accurate information about liquid collective properties is often unavailable. Moreover, collective properties can differ from one liquid system to another and depend considerably on the state point in the phase diagram. Therefore, some simplifications are usually involved. The question may arise how sensitive are the considered transport models to the assumptions about the collective mode properties? This Section provides some examples whose appropriateness will then be checked against existing results from numerical simulations for various simple fluids.   

We start with a simplest approximation in which all atoms are oscillating with the same Einstein frequency $\Omega_{\rm E}$ (known as Einstein model in the solid state physics).
For a pairwise interaction potential $\phi(r)$, the Einstein frequency is defined as~\cite{BalucaniBook}
\begin{equation}
\Omega_{\rm E}=\frac{\rho}{3m}\int d{\bf r}\nabla^2\phi(r)g(r),
\end{equation}
where $g(r)$ is the radial distribution function (RDF). If all atoms are oscillating with the same frequency, averaging is trivial: 
\begin{equation}
\langle \omega^{-2}\rangle = \Omega_{\rm E}^{-2}, \quad\quad \langle \omega\rangle = \Omega_{\rm E}.
\end{equation}

For the SE coefficient the Einstein model then yields
\begin{equation}
\alpha_{\rm SE}\simeq \frac{c_t^2}{\Delta^2\Omega_{\rm E}^2}. 
\end{equation} 
For sufficiently steep interactions, the transverse instantaneous sound velocity can be estimated as $c_t^2\simeq \Omega_{\rm E}^2\Delta^2/10$, see e.g. Eq.~(6.102) from Ref.~\cite{BalucaniBook}. We arrive at $\alpha_{\rm SE}\simeq 0.1$, which provides correct order of magnitude estimate. The obtained numerical value is nevertheless smaller than the actual values (see below). This is because the low frequencies, which are not included in the Einstein model, contribute considerably to the magnitude of $\langle \omega^{-2}\rangle$. This leads to some underestimation of $\alpha_{\rm SE}$. 
        
Within the vibrational model of thermal conductivity we obtain
\begin{equation}\label{Horrocks}
\lambda \simeq c_{\rm v} \frac{\Omega_{\rm E}}{2\pi \Delta},
\end{equation}         
where $c_p$ has been substituted by $c_{\rm v}$, as discussed above. We recover immediately the expression proposed by Horrocks and McLaughlin~\cite{Horrocks1960}. It is also similar to the result obtained earlier by Rao~\cite{Rao1941}. Neglecting the low frequency domain does not produce large errors in this case and Eq.~(\ref{Horrocks}) represents an appropriate estimate in this case.   
     
In his derivation of SE relation, Zwanzig originally used a Debye approximation, characterized by one longitudinal and two transverse modes with acoustic dispersion. The sum over frequencies can be converted to an integral over $k$ using the standard procedure
\begin{equation}
\sum_{\bf k}(...)\rightarrow \frac{V}{(2\pi)^3}\int (...) d{\bf k},
\end{equation}
where $V$ is the volume. For $\langle \omega^{-2}\rangle$  this yields
\begin{equation}\label{kaverage}
\left\langle\frac{1}{\omega^2}\right\rangle = \frac{1}{6\pi^2 \rho}\int_0^{k_{\rm max}}k^2dk\left(\frac{1}{\omega_l^2}+\frac{2}{\omega_t^2}\right),
\end{equation} 
where the cutoff $k_{\rm max}= (6\pi^2 \rho)^{1/3}$ is chosen to provide $\rho$ modes in each branch of the spectrum. This ensures that the averaging procedure applied to a quantity that does not depend on $k$ will not change its value. 
Substituting $\omega_l=c_lk$ and $\omega_t=c_tk$ into \mbox{Eq.~(\ref{kaverage})} we can evaluate the SE constant as 
\begin{equation}\label{alpha}
\alpha_{\rm SE}=\frac{2}{(6\pi^2)^{2/3}}\left(1+\frac{c_t^2}{2c_l^2}\right)\simeq  0.13\left(1+\frac{c_t^2}{2c_l^2}\right).
\end{equation}
This is essentially Zwanzig's original result, except he expressed the SE coefficient in terms of the longitudinal and shear viscosity as $\alpha_{\rm SE}\simeq 0.13 (1+\eta/2\eta_{l})$. The equivalence was pointed out later~\cite{KhrapakMolPhys2019}. Note that since the sound velocity ratio $c_t/c_l$ is confined in the range from $0$ to $\sqrt{3}/2$, the coefficient $\alpha_{\rm SE}$ can vary only between $\simeq$0.13 and $\simeq$0.18~\cite{ZwanzigJCP1983,KhrapakMolPhys2019}. This is very close to actual values of the SE coefficient in simple fluids as will be demonstrated below.     

For the problem of thermal conductivity, averaging is performed similarly:
\begin{equation}\label{kaverage1}
\langle \omega \rangle = \frac{1}{6\pi^2 \rho}\int_0^{\rm k_{\rm max}}k^2dk\left[\omega_l(k)+2\omega_t(k)\right].
\end{equation}   
If acoustic dispersion relations are used, the thermal conductivity coefficient becomes~\cite{KhrapakPRE01_2021}
\begin{equation}\label{Cahill1}
\lambda \simeq \frac{1}{4}\left(\frac{3}{4\pi}\right)^{1/3} c_{\rm v} \frac{c_l+2c_t}{\Delta^2}. 
\end{equation}
If we substitute $c_{\rm v}\simeq 3$ near freezing (according to Dulong-Petit law), we get a formula similar to that of minimal thermal conductivity model proposed by Cahill and Pohl~\cite{Cahill1989,Cahill1992} for amorphous solids. Their expression is in good agreement with the measured thermal conductivities of many amorphous inorganic solids, highly disordered crystals, and amorphous macromolecules~\cite{XiePRB2017}. Its modification (mainly concerning the numerical front factor), obtained using the vibrational model, works rather well in the liquid regime.

The most accurate theoretical estimate of the thermal conductivity coefficient would be obtained if the exact vibrational density of states (VDOS) were known. Although for a given liquid and for a particular state point on the phase diagram, VDOS can be computed by molecular dynamics~\cite{Berens1983,BalucaniBook,OhtaPoP2000}, no general approaches to VDOS across the liquid regime exist. Some important progress in this direction has recently been reported~\cite{ZacconePNAS2021,StamperJPCL2022}. Namely, Zaccone and Baggioli have developed an analytical model for VDOS, based on overdamped Langevin liquid dynamics~\cite{ZacconePNAS2021}. Distinct from the Debye approximation, $g(\omega) \propto \omega^2$, for solids, the universal law for liquids reveals a linear scaling, $g(\omega)\propto \omega$, in the low-energy region. Stamper {\it et al.} have confirmed this universal law with experimental VDOS measured by inelastic neutron scattering on real liquid systems~\cite{StamperJPCL2022}. Nevertheless, the applicability regime and accuracy level of this model need to be clarified. Reasonable simplifications and approximations are therefore still in use. In this respect, substituting accurate dispersion relations for $\omega_l(k)$ and $\omega_t(k)$ in Eqs. (\ref{kaverage})  and (\ref{kaverage1}) instead of their acoustic asymptotes can improve the accuracy compared to the Debye model. This is particularly relevant to fluids with non-acoustic dispersion relations (see below).

To conclude this section, let us estimate the thermal conductivity coefficient of simple fluids at freezing conditions, demonstrating that a quasi-universality can be expected. We can use Eq.~(\ref{Horrocks}) and take $c_{\rm v}\simeq 3$ according to the Dulong-Petit law, which is adequate near the fluid-solid phase transition. According to the vibrational paradigm, atomic dynamics in fluids near the freezing point is dominated by solid-like oscillations around their temporary equilibrium positions. The properties of these oscillations are close to those in solids near melting. The Einstein frequency can therefore be estimated as follows. According to the Lindemann’s melting criterion~\cite{Lindemann}, melting of a solid occurs when the vibrational amplitude reaches a threshold value, which is approximately $\sim 0.1$ of the interatomic spacing. We can write this condition as
\begin{equation}\label{Lindemann}
\langle \xi^2\rangle \sim 0.01\Delta^2,
\end{equation}
where $\langle \xi^2\rangle $ is the average vibrational amplitude. It can be expressed as~\cite{KhrapakPRR2020,BuchenauPRE2014} 
\begin{equation}\label{xi1}
\langle \xi^2\rangle = \frac{3T}{m}\left\langle\frac{1}{\omega^2}\right\rangle,
\end{equation}      
where the averaging is performed over 3$N$ normal modes. In the simplest Einstein approximation this leads to 
\begin{equation}\label{xi2}
\langle \xi^2\rangle = \frac{3T}{m\Omega_{\rm E}^2}.
\end{equation} 
The same result could be obtained by equating the harmonic potential energy $\tfrac{1}{2}m\Omega_{\rm E}^2\langle \xi^2\rangle$ to the average kinetic energy $\tfrac{3}{2}T$  (i.e. from energy equipartition). Combining these approximations we easily arrive at   
\begin{equation}\label{lambda8}
\lambda_{\rm R}\sim 8 
\end{equation}
at freezing. This rough estimate is in fact quite successful. For model systems considered later in this paper an approximate relation $\lambda_{\rm R}\sim 10$ works quite well on the freezing line (except hard sphere fluids where $\lambda_{\rm R}\simeq 14$ at the freezing point~\cite{Pieprzyk2020,KhrapakApplSci2022}). This quasi-universality also holds for many real atomic and molecular liquids, such as Ne, Ar, Kr, Xe, N$_2$, O$_2$, CO$_2$, and CH$_4$~\cite{KhrapakPoF2022,KhrapakJCP2022_1,KhrapakJETPLett2021}. 
  
\section{Special case: One-component plasma}\label{OCP}

Let us now proceed to the application of the vibrational model to some selected fluids. In this Section we start with analysing in detail its application to a strongly coupled one-component plasma (OCP) fluid. The OCP fluid is chosen for the following three main reasons: (i) vibrational (caging) motion is most pronounced due to extremely soft and long-ranged character of the interaction potential~\cite{DonkoPRL2002,Daligault2020}; (ii) Zwanzig's assumption about acoustic spectra is not directly applicable to the OCP case, because the longitudinal mode is not acoustic (but plasmon), and thus it is a good opportunity to demonstrate how the model should be modified for non-acoustic spectra; (iii) collective modes in the OCP system are very well studied and understood and simple analytical expressions for the long-wavelength dispersion relations are available.   

The OCP model is an idealized system of mobile point charges immersed in a neutralizing 
fixed background of opposite charge (e.g. ions in the immobile background of electrons or vice versa)~\cite{BrushJCP1966,HansenPRA1973,deWitt1978,BausPR1980,IchimaruRMP1982,
StringfellowPRA1990}. From the fundamental point of view OCP is characterized by a very soft and long-ranged Coulomb interaction potential, $\phi(r)= q^2/r$, where $q$ is the electric charge. The particle-particle correlations and thermodynamics of the OCP are characterized by a single dimensionless coupling parameter $\Gamma=q^2/aT$, where $a=(4\pi \rho /3)^{-1/3}$ is the Wigner-Seitz radius. At $\Gamma\gtrsim 1$ the OCP is called strongly coupled and this is where it exhibits properties characteristic of a fluid phase (a body centered cubic phase becomes thermodynamically stable at $\Gamma\gtrsim 174$, as the comparison of fluid and solid Helmholtz free energies demonstrates~\cite{IchimaruRMP1982,DubinRMP1999,KhrapakCPP2016}).
Dynamical scales of the OCP are usually characterized by the plasma frequency $\omega_{\rm p}=\sqrt{4\pi q^2 \rho/m}$. In the OCP, the Einstein frequency is directly related to the plasma frequency, $\Omega_{\rm E}=\omega_{\rm p}/\sqrt{3}$. The transverse sound velocity can be expressed in terms of the plasma frequency and mean interparticle separation as $c_t^2 \simeq 0.015 \omega_{\rm p}^2\Delta^2$~\cite{KhrapakPoP2016,KhrapakMolecules12_2021}. 

\begin{figure}
\includegraphics[width=8.5cm]{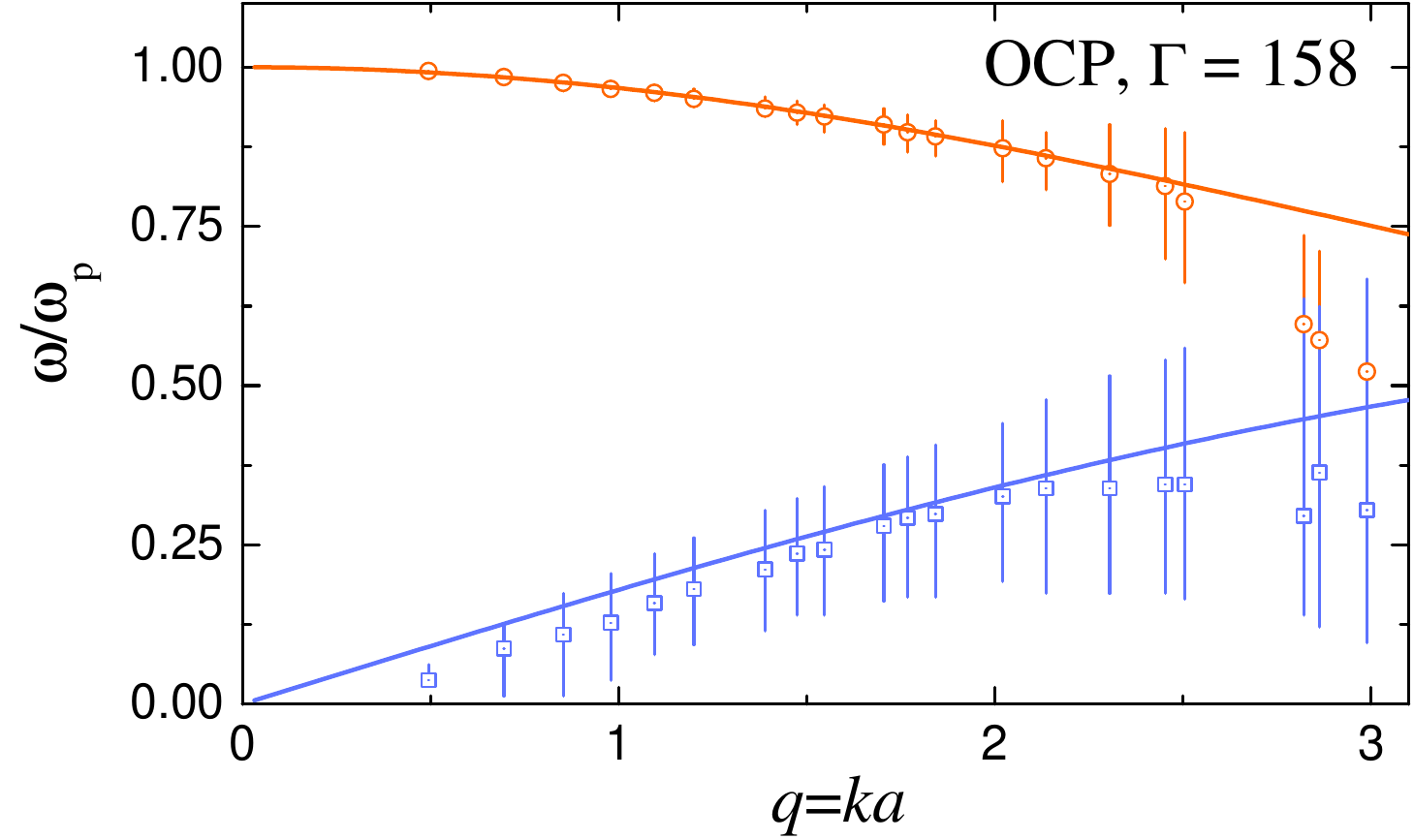}
\caption{(Color online) Dispersion of the longitudinal (upper curve and symbols) and transverse (lower curve and symbols) modes of the
strongly coupled OCP with $\Gamma = 158$. Symbols correspond to the results from MD simulations by Schmidt et al.~\cite{SchmidtPRE1997}. The solid curves are plotted using Eqs.~(\ref{L2}) and (\ref{T2}). For further details see e.g. Ref.~\cite{KhrapakIEEE2018}.
}
\label{FigOCP}
\end{figure}

To estimate the SE coefficient and the coefficient of thermal conductivity from the vibrational model we make use of dispersion relations based on the quasi-crystalline approximation (QCA)~\cite{Hubbard1969,SingwiPRA1970,Takeno1971,KhrapakSciRep2017} also known as the quasi-localized charge approximation (QLCA)~\cite{GoldenPRA1992,RosenbergPRE1997,GoldenPoP2000,KalmanPRL2000,
SchmidtPRE1997,KhrapakPoP02_2017,KhrapakIEEE2018} in the plasma related context. When combined with the excluded cavity model for the radial distribution function, particularly simple and fully analytical expressions for $\omega_l(k)$ and $\omega_t(k)$ can be derived~\cite{KhrapakPoP2016}:
\begin{equation}\label{L2}
\omega_l^2=\omega_{\rm p}^2\left(\frac{1}{3}-\frac{2\cos Rq}{R^2q^2}+\frac{2\sin Rq}{R^3q^3} \right)
\end{equation}
and
\begin{equation}\label{T2}
\omega_t^2=\omega_{\rm p}^2\left(\frac{1}{3}+\frac{\cos Rq}{R^2q^2}-\frac{\sin Rq}{R^3q^3} \right),
\end{equation}   
where $q=ka$ is the reduced wave-number and $R$ is the reduced excluded cavity radius. In the strongly coupled OCP regime we have $R= \sqrt{6/5}\simeq 1.09545$~\cite{KhrapakPoP2016}. Expressions (\ref{L2}) and (\ref{T2}) are rather accurate in the long-wavelength regime~\cite{KhrapakPoP10_2016,KhrapakAIPAdv2017,
KhrapakIEEE2018,FairushinResPhys2020}. This is illustrated in Fig.~\ref{FigOCP}, where comparison with the existing numerical results~\cite{SchmidtPRE1997} is presented. The agreement is rather good, except the existence of $k$-gap (zero-frequency non-propagating domain at $k<k_{\rm gap}$) in the transverse mode is not accounted for~\cite{KhrapakJCP2019}. Expressions (\ref{L2}) and (\ref{T2}) can therefore be used to perform averaging over collective modes frequencies. The results are summarized in Tab.~\ref{Tab1}. 

The result for $\langle\omega^2/\omega_{\rm p}^2\rangle\equiv 1/3$ is exact by virtue of Eqs.~(\ref{L2}) and (\ref{T2}). The quantity $\langle\omega_{\rm p}^2/\omega^2\rangle$ is used to estimate the SE coefficient. The result is $\alpha_{\rm SE}\simeq 0.146$.  The quantity $\langle\omega/\omega_{\rm p}\rangle$ allows us to write the thermal conductivity coefficient in the form
\begin{equation}\label{TC_OCP}
\lambda_{\rm R}\simeq 0.2284 c_{\rm v}\sqrt{\Gamma}.
\end{equation} 
These results can be now checked against the results from existing MD simulations. The quantity $\langle\ln\omega/\omega_{\rm p}\rangle$ provided for completeness in Tab.~\ref{Tab1} emerges in a variant of cell theory of liquid excess entropy~\cite{KhrapakJCP2021}.  

\begin{table}
\caption{\label{Tab1} Averaged frequencies of a strongly coupled OCP fluid obtained with the help of dispersion relations~(\ref{L2}) and (\ref{T2}).}
\begin{ruledtabular}
\begin{tabular}{cccc}
$\langle\omega^2/\omega_{\rm p}^2\rangle$ & $\langle\omega_{\rm p}^2/\omega^2\rangle$ & $\langle\omega/\omega_{\rm p}\rangle$ & $\langle\ \ln\omega/\omega_{\rm p}\rangle$    \\ \hline
$\tfrac{1}{3}$ &  9.7623  & 0.514 & -0.8023  
\end{tabular}
\end{ruledtabular}
\end{table}    

\begin{figure}
\includegraphics[width=8.5cm]{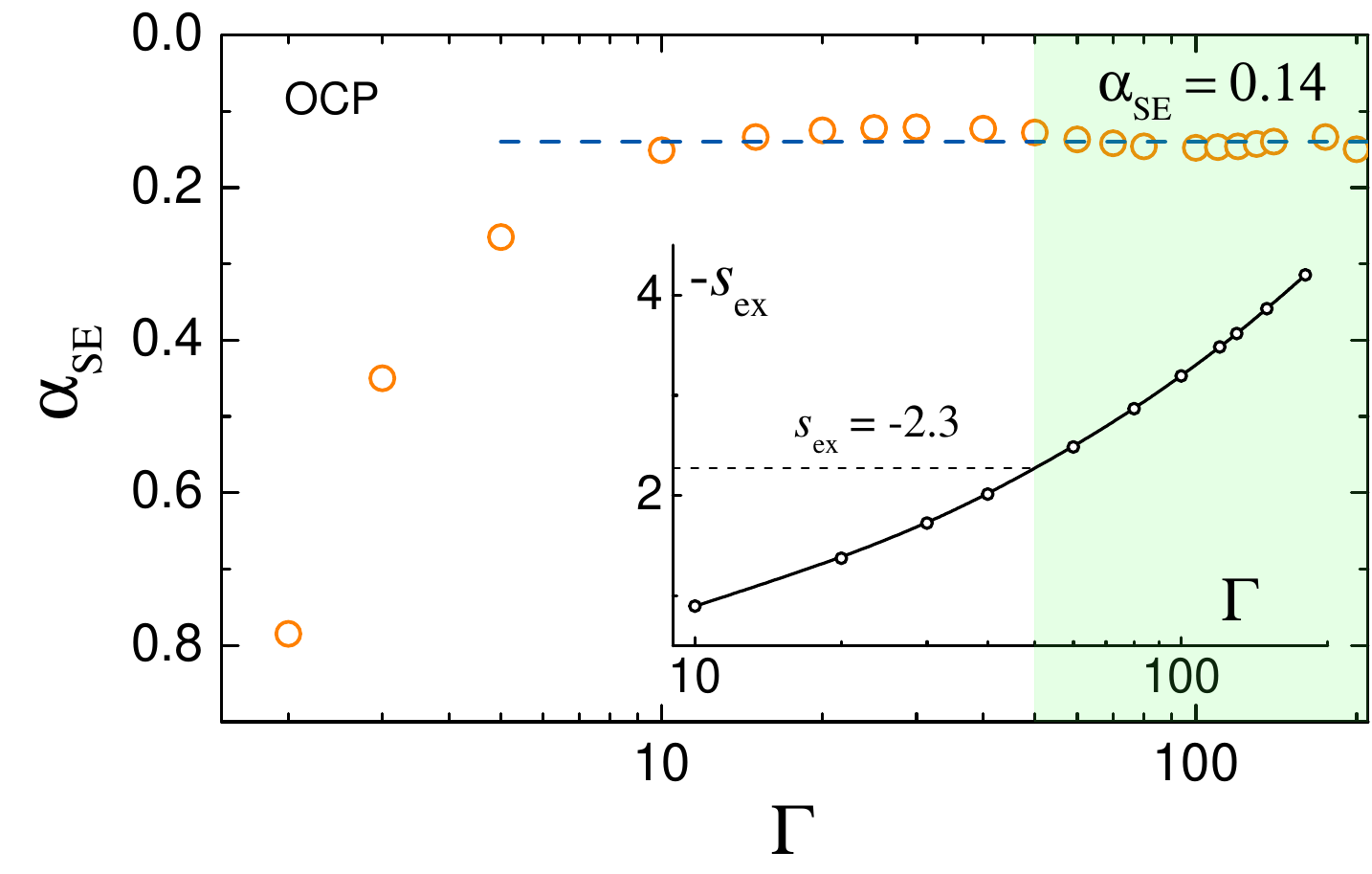}
\caption{(Color online) Stokes-Einstein coefficient $\alpha_{\rm SE}$ versus the coupling parameter $\Gamma$ for the OCP fluid. The symbols correspond to MD simulation results from Refs.~\cite{DaligaultPRE2012,DaligaultPRE2014}. The dashed line is a strong coupling asymptote $\alpha_{\rm SE}\simeq 0.14$. In the shaded area at $\Gamma\gtrsim 50$ the SE coefficient is practically constant (lies in a narrow range $\alpha_{\rm SE}\simeq 0.14\pm 0.01$). The inset shows the dependence of the negative excess entropy on the coupling parameter. Onset of the validity of the SE relation corresponds to $s_{\rm ex}\lesssim -2.3$. Reproduced from S. Khrapak and A. Khrapak, Phys. Rev. E {\bf 104}, 044110 (2021).}
\label{Fig2}
\end{figure}

Transport properties of the OCP and related system are very well investigated in classical MD simulations. Extensive datasets and fits are available in the literature~\cite{HansenPRA1975,DonkoPRL1998,DonkoPoP2000,SalinPRL2002,VaulinaPRE2002,BasteaPRE2005,DaligaultPRL2006,
DaligaultPRL2012,DaligaultPRE2012,KhrapakPoP2013,DaligaultPRE2014,
KhrapakAIPAdv2018,ScheinerPRE2019}. To evaluate the SE coefficient emerging in numerical simulations we have used an accurate fitting formula for the self-diffusion coefficient proposed in Ref.~\cite{DaligaultPRE2012} along with the MD data on the shear viscosity coefficient tabulated in Ref.~\cite{DaligaultPRE2014}. The resulting dependence of the SE coefficient $\alpha_{\rm SE}$ on the coupling parameter $\Gamma$ is shown by circles in Fig.~\ref{Fig2}. Note the reversed vertical axis in the figure to highlight the level of accuracy of SE relation.

The strong coupling asymptote, $\alpha_{\rm SE}\simeq 0.14$, is approached at $\Gamma\simeq 10$. The numerical value of this asymptote is very close to that the theory predicts. At $\Gamma\gtrsim 50$ the SE coefficient lies in a narrow range $\alpha_{\rm SE}\simeq 0.14\pm 0.01$.  We use this as a pragmatical (although to some extent arbitrary) definition of the region of validity of SE relation~\cite{KhrapakPRE10_2021} (shaded area in Fig.~\ref{Fig2}). Note that already starting from $\Gamma\simeq 10$ the deviations from the strong coupling asymptote are relatively small.  

The inset in Fig.~\ref{Fig2} shows the dependence of the minus reduced excess entropy $-s_{\rm ex}$ on the coupling parameter $\Gamma$ as tabulated in Ref.~\cite{LairdPRA1992}. The change in the slopes of asymtotes at $\Gamma\simeq 10$ corresponds to $s_{\rm ex}\simeq -0.9$. The onset of validity of the SE relation at $\Gamma\simeq 50$ corresponds to $s_{\rm ex}\lesssim -2.3$. Excess entropy appears as a very good indicator of the validity of the SE relation and of vibrational model of transport processes, as well as the gas-like to liquid-like dynamical crossover. We will elaborate on this further below. 

\begin{figure}
\includegraphics[width=8.5cm]{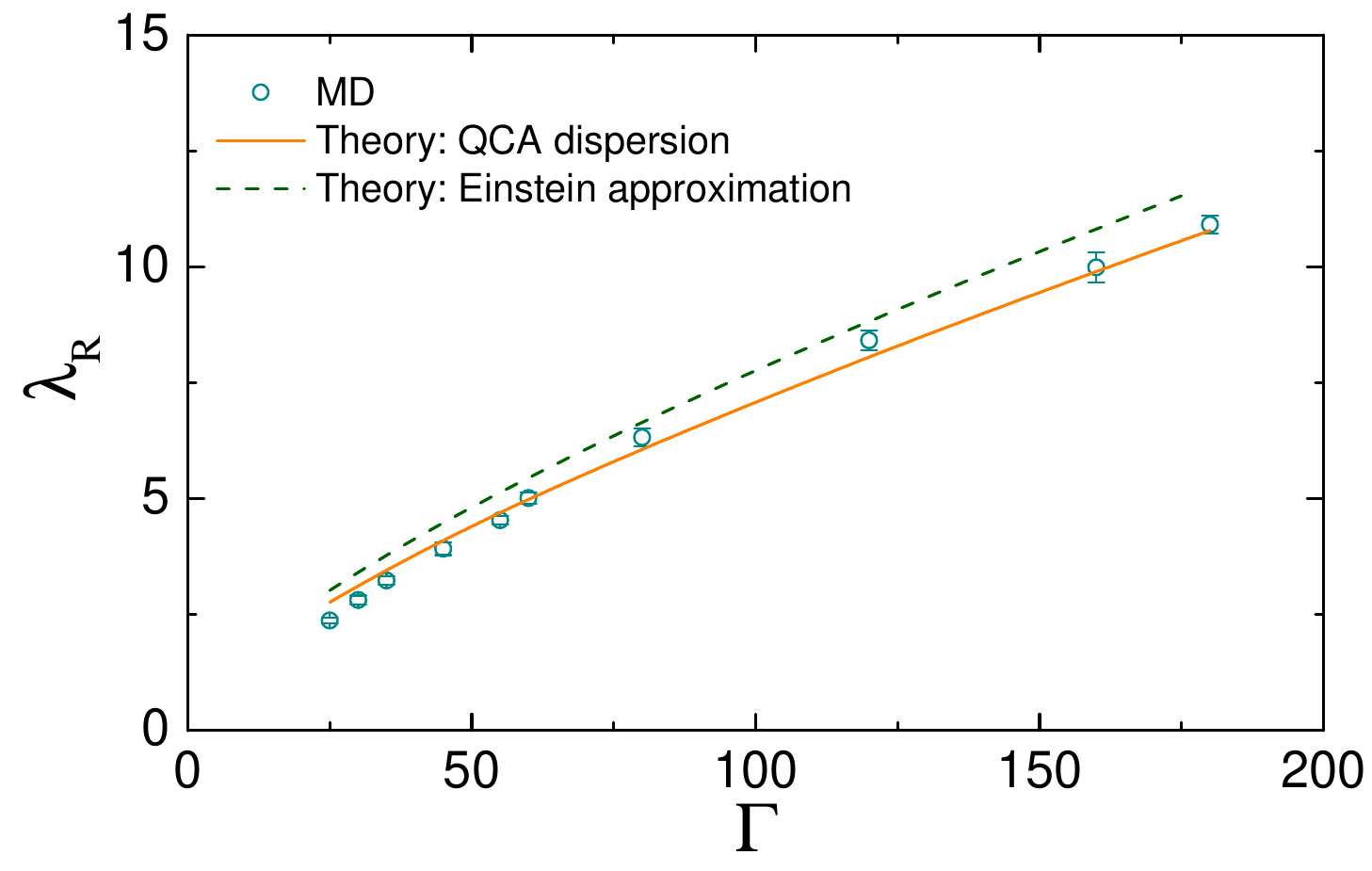}
\caption{(Color online) Reduced thermal conductivity coefficient 
$\lambda_{\rm R}$ of a strongly coupled OCP vs the coupling parameter $\Gamma$. Symbols correspond to numerical results~\cite{ScheinerPRE2019}. Curves are calculated using the
vibrational model: Solid curve using Eq.~(\ref{TC_OCP}); dashed curve using Eq.~(\ref{Horrocks}).
}
\label{Fig3}
\end{figure}

The vibrational model of thermal conductivity has been applied to the strongly coupled OCP fluid in Ref.~\cite{KhrapakPRE01_2021}. The specific heat was estimated from a simple three-term equation of state proposed in Ref.~\cite{KhrapakPoP10_2014}, based on extensive Monte Carlo simulation data from Ref.~\cite{CaillolJCP1999}.
In Figure~\ref{Fig3} we compare theoretical results from Eqs.~(\ref{Horrocks}) and (\ref{TC_OCP}) with MD simulation data from Ref.~\cite{ScheinerPRE2019}. Overall, the agreement between the theory and simulation in the strongly coupled regime $\Gamma\gtrsim 50$ is remarkably good, especially taking into account the absence of free parameters in the theory. In the OCP case we have $\langle \omega/\omega_{\rm p}\rangle\simeq 0.514$ and $\Omega_{\rm E}/\omega_{\rm p}\simeq 0.577$ (see Tab.~\ref{Tab1}). Thus, quite expectedly, the Einstein approximation somewhat overestimates the accurate results. The calculation based on the QCA dispersion relations is in very good agreement with numerical data. For weaker coupling the vibrational model overestimates the thermal conductivity coefficient considerably, but it should not be applied in this regime, because vibrational motion (caging) does not dominate particle dynamics in this regime.

\section{More examples}\label{Examples}

\subsection{Screened Coulomb (Yukawa) systems}

Screened Coulomb or Yukawa system represents a collection of point-like charges immersed into a neutralizing polarizable medium (usually conventional electron-ion plasma), which provides screening. The pairwise screened Coulomb repulsive interaction  potential (also referred to as Debye-H\"uckel potential) is
\begin{equation}\label{Yukawa}
\phi(r)=(q^2/r)\exp(-\kappa r/a),
\end{equation}
where $\kappa = a/\lambda$ is the dimensionless screening parameter, which is the ratio of the Wigner-Seitz radius to the plasma screening length. The screening length is normally related to the Debye radius of a screening medium, $\lambda_{\rm D}=\sqrt{T/4\pi e^2 n_s}$, where $n_s$ is the number density of the screening particles, which are assumed singly charged for simplicity. In the simplest case $\lambda=\lambda_{\rm D}$, but more complicated scenarios can be also realized, in particular in complex (dusty) plasmas~\cite{KhrapakCPP2009,SemenovPoP2015}.     Yukawa potential is widely used as a reasonable first approximation for actual  interactions in three-dimensional isotropic complex plasmas and colloidal suspensions~\cite{TsytovichUFN1997,FortovUFN,FortovPR,
IvlevBook,KhrapakCPP2009,KhrapakPRL2008,KlumovUFN2010,
ChaudhuriSM2011,LampePoP2015,SemenovPoP2015}. 

The dynamics and thermodynamics of Yukawa systems are characterized by the two dimensionless parameters, $\Gamma$ and $\kappa$. Detailed phase diagrams of Yukawa systems are available in the literature~\cite{RobbinsJCP1988,HamaguchiJCP1996,HamaguchiPRE1997,
VaulinaJETP2000,VaulinaPRE2002,KhrapakEPL2010,YazdiPRE2014}. 
Note that the screening parameter $\kappa$ determines the softness of the interparticle repulsion. It varies from the very soft and long-ranged Coulomb potential at $\kappa\rightarrow 0$ (corresponding to the OCP limit considered above) to the hard-sphere-like interaction limit at $\kappa\rightarrow \infty$ (this limit is considered below). In the context of complex plasmas and colloidal suspensions the relatively ``soft'' regime, $\kappa\sim {\mathcal O}(1)$, is of particular interest. 

Transport phenomena in three-dimensional Yukawa fluids have been relatively well investigated~\cite{RobbinsJCP1988,OhtaPoP2000,SanbonmatsuPRL2001,
SaigoPoP2002,VaulinaPRE2002,SalinPRL2002,SalinPoP2003,
FaussurierPRE2003,DonkoPRE2004,
DonkoPRE2008,DaligaultPRE2012,KhrapakPoP2012,KhrapakPoP2013,
DaligaultPRE2014,
KhrapakJPCO2018,KhrapakAIPAdv2018,
KahlertPPR2020}. Recent approaches to thermodynamics of Yukawa systems are discussed in Refs.~\cite{ToliasPRE2014,ToliasPoP2015,KhrapakPRE02_2015,KhrapakJCP2015,
KhrapakPPCF2015,VeldhorstPoP2015,CastelloCPP2020}. There has been also considerable interest in two-dimensional Yukawa systems, which are related to laboratory realizations of dusty plasmas, but these are beyond the scope of this paper and are not considered. 

\begin{figure}
\includegraphics[width=8.5cm]{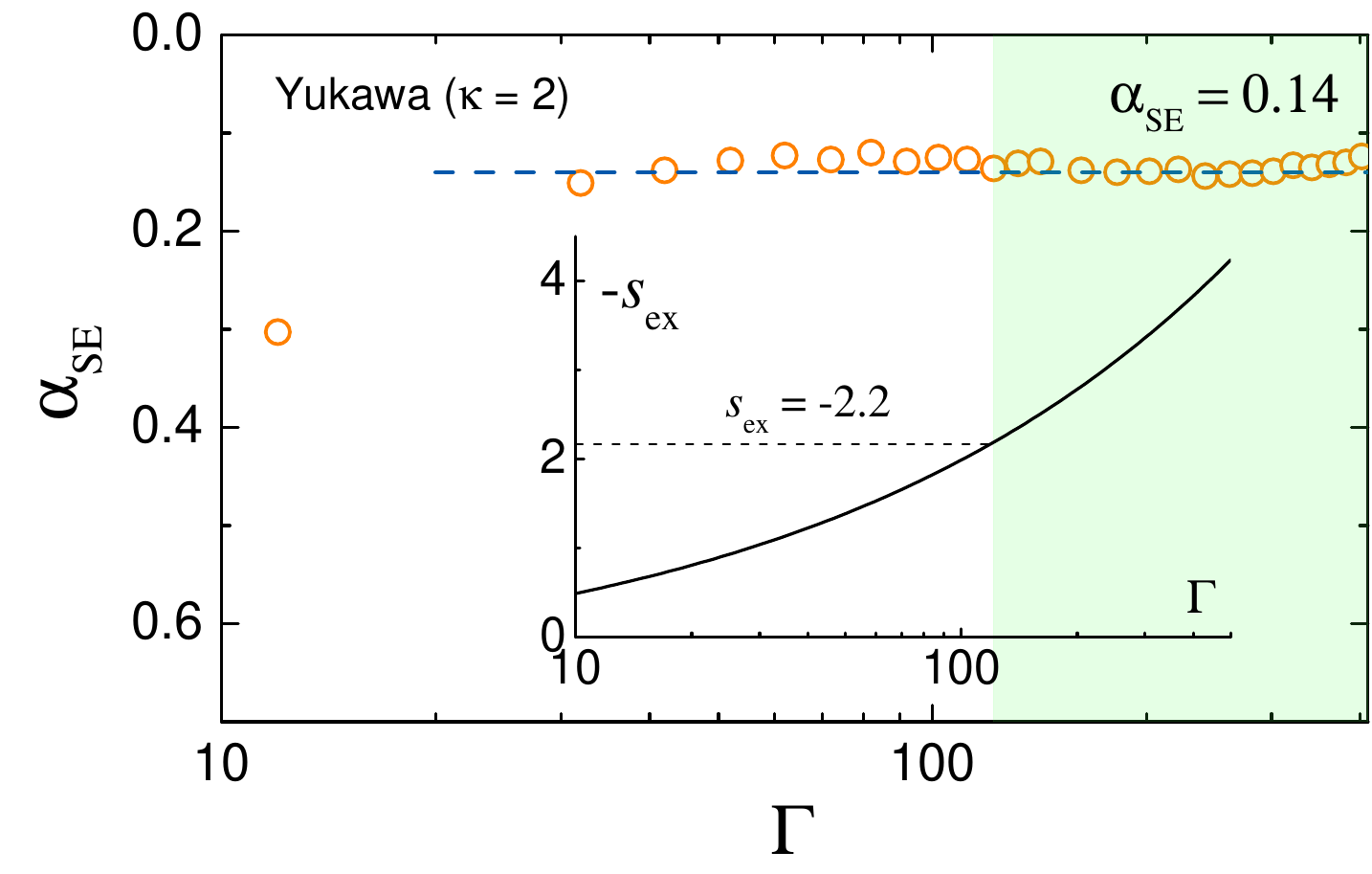}
\caption{(Color online) Stokes-Einstein parameter $\alpha_{\rm SE}$ versus the coupling parameter $\Gamma$ for a Yukawa fluid with $\kappa=2$. The symbols correspond to MD simulation results from Refs.~\cite{DaligaultPRE2012,DaligaultPRE2014}. The dashed line is the dense fluid asymptote $\alpha_{\rm SE}\simeq 0.14$. In the shaded area at $\Gamma\gtrsim 120$ the SE coefficient is practically constant (lies in a narrow range $\alpha_{\rm SE}\simeq 0.14\pm 0.01$). The inset shows the dependence of the negative excess entropy on the coupling parameter. Onset of the validity of the SE relation corresponds to $s_{\rm ex}\lesssim -2.2$.  Reproduced from S. Khrapak and A. Khrapak, Phys. Rev. E {\bf 104}, 044110 (2021). }
\label{Fig4}
\end{figure}

In the context of the SE relation we combine the accurate fits for the self-diffusion coefficient in Yukawa fluids~\cite{DaligaultPRE2012} with the tabulated numerical data on shear viscosity coefficient for $\kappa = 2$~\cite{DaligaultPRE2014}. The results are plotted in Fig.~\ref{Fig4}. The emerging picture is similar to that in the OCP case, except higher $\Gamma$ values are involved. The slope of the dependence $\alpha_{\rm SE}$ on $\Gamma$ changes at approximately $\Gamma\simeq 30$. The strong coupling asymptote is $\alpha_{\rm SE}\simeq 0.14$, the same as in the OCP case. At $\Gamma\gtrsim 120$ the SE coefficient lies in a narrow range $\alpha_{\rm SE}\simeq 0.14\pm 0.01$ and this is identified as the region of validity of the SE relation (shaded area in Fig.~\ref{Fig4}). As previously observed, the deviations from the asymptotic strong coupling value are already relatively small at $\Gamma\gtrsim 30$. 

The inset in Fig.~\ref{Fig4} shows the dependence of the minus reduced excess entropy $-s_{\rm ex}$ on the coupling parameter $\Gamma$. The curve is calculated using the Rosenfeld-Tarazona (RT) scaling of the thermal component of the excess internal energy~\cite{RosenfeldMolPhys1998,RosenfeldPRE2000}. This approach combines relative good accuracy with simplicity and relatively wide applicability and has been extensively used to construct simple practical approximations for thermodynamic properties of Yukawa and other related fluids~\cite{IngebrigtsenJCP2013,KhrapakPRE02_2015,KhrapakPPCF2015,
KhrapakJCP2015,ToliasPoP2019,CastelloPoP2019}.
The particular form used here is taken from Ref.~\cite{RosenfeldPRE2000}. The change in the slopes of asymptotes at $\Gamma\simeq 30$ corresponds to $s_{\rm ex}\simeq -1.0$. The onset of validity of the SE relation at $\Gamma\simeq 120$ corresponds to $s_{\rm ex}\lesssim -2.2$.

\begin{figure}
\includegraphics[width=8.cm]{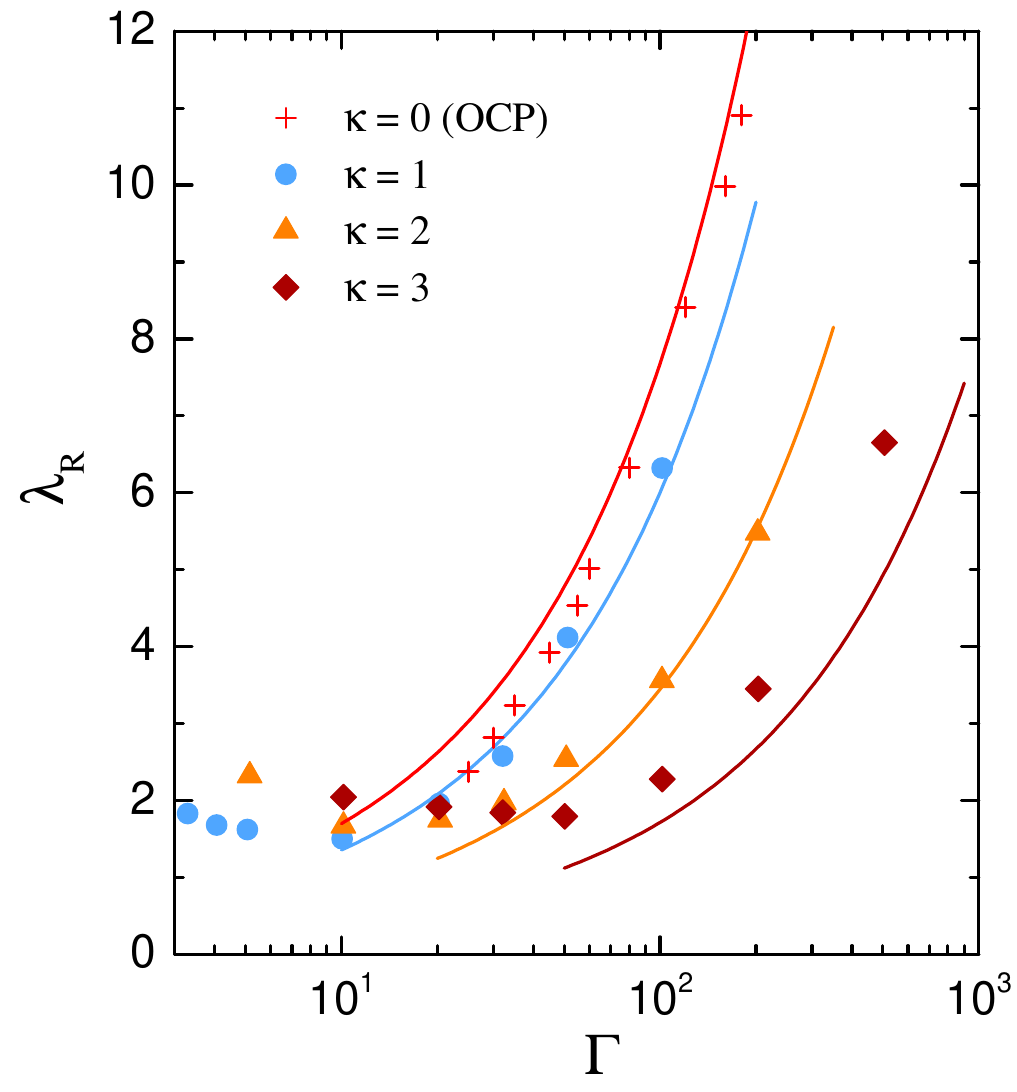}
\caption{(Color online) Reduced thermal conductivity coefficient $\lambda_{\rm R}$ vs the coupling parameter $\Gamma$ of Yukawa fluids with different screening parameters.
Circles, triangles, and rhombs correspond to molecular dynamics results from Ref.~\cite{DonkoPRE2004} for $\kappa=1$, $\kappa=2$, and $\kappa=3$, respectively. Crosses are the OCP numerical
results from Ref.~\cite{ScheinerPRE2019}. The solid curves are the theoretical calculations using the Einstein approximation of Eq.~(\ref{Horrocks}). Reproduced from S. Khrapak, Plasma Phys. Rep. {\bf 49}, 15 (2023).
}
\label{Fig5}
\end{figure}

The vibrational model of heat conduction was applied to estimate the thermal conductivity coefficient of strongly coupled Yukawa fluids in Ref.~\cite{KhrapakPoP08_2021}. The specific heat was estimated using the RT scaling~\cite{RosenfeldMolPhys1998,RosenfeldPRE2000}. In Figure~\ref{Fig5} we compare results from MD simulations in Ref.~\cite{DonkoPRE2004} with a theoretical calculation using the Einstein approximation of Eq.~(\ref{Horrocks}). As could be expected, the agreement is relatively good in the soft interaction limit (small $\kappa$), but worsens as $\kappa$ increases. For example, for $\kappa=3$, Eq.~(\ref{Cahill1}) provides a considerably better agreement with numerical results (see Fig. 1 from Ref.~\cite{KhrapakPoP08_2021}). Overall, we observe that the vibrational model describes relatively well the thermal conduction in strongly coupled Yukawa fluids.  

There exists a useful quasi-universal scaling approach to transport coefficients of strongly coupled Yukawa fluids. This scaling was in particular elaborated by Rosenfeld in connection with the diffusion and viscosity coefficients of Yukwa fluids~\cite{RosenfeldPRE2000,RosenfeldJPCM2001}. He demonstrated that the reduced transport coefficients are quasi-universal functions of the reduced coupling parameter $\Gamma/\Gamma_{\rm fr}$, where $\Gamma_{\rm fr}$ is the value of $\Gamma$ at the fluid-solid phase transition (at freezing). The values of $\Gamma_{\rm fr}$ for different $\kappa$ were tabulated in Refs.~\cite{HamaguchiJCP1996,HamaguchiPRE1997}; accurate practical expressions for $\Gamma_{\rm fr}(\kappa)$ can be found in Refs.~\cite{VaulinaJETP2000,VaulinaPRE2002}. Rosenfeld originally arrived at this scaling by combining the excess entropy scaling with the Rosenfeld-Tarazona~\cite{RosenfeldMolPhys1998} scaling of the excess entropy with $\Gamma/\Gamma_{\rm fr}$~\cite{RosenfeldJPCM2001}. Similar conclusion could be reached by combining the isomorph theory with excess entropy scaling~\cite{VeldhorstPoP2015}. Neither approach specifies explicitly the functional form of the dependence of $D_{\rm R}$ and $\eta_{\rm R}$ on $\Gamma/\Gamma_{\rm fr}$ (the same is true for the vibrational paradigm). A simple practical {\it ad hoc} formula of the form 
\begin{equation}\label{etascaling}
\eta_{\rm R}\simeq 0.13\exp\left(3.64\sqrt{\Gamma/\Gamma_{\rm fr}}\right)
\end{equation}
demonstrates reasonable agreement with available results from molecular dynamics simulations in the regime $0.1\lesssim \Gamma\lesssim \Gamma_{\rm fr}$ ~\cite{KhrapakAIPAdv2018}. This functional form was originally proposed by Costigliola {\it et al}. as a general temperature dependence of viscosity of dense fluids~\cite{CostigliolaJCP2018}. To be consistent with the SE relation the self-diffusion coefficient should scale as  
\begin{equation}\label{Dscaling}
D_{\rm R}\simeq \exp\left(-3.64\sqrt{\Gamma/\Gamma_{\rm fr}}\right).
\end{equation}
Note that the combination of Eqs.~(\ref{etascaling}) and (\ref{Dscaling}) leads to $\alpha_{\rm SE}\simeq 0.13$, which is slightly smaller than the value $\simeq 0.14$ evidenced in Fig.~\ref{Fig4}. Nevertheless, they agree relatively well with individual datasets on viscosity and diffusion, as documented in Ref.~\cite{KhrapakAIPAdv2018}. 

A quasi-universal scaling of the thermal conductivity coefficient with $\Gamma/\Gamma_{\rm fr}$ appears to be a natural consequence of the vibrational model of fluid transport properties. The details of the derivation can be found in Ref.~\cite{KhrapakPPR2023}. Here only the final result for the reduced thermal conductivity coefficient is provided:
\begin{equation}\label{lambdascaling}
\lambda_{\rm R}\simeq 3.40\left[1.5+1.86\left(\frac{\Gamma}{\Gamma_{\rm fr}}\right)^{2/5}\right]\sqrt{\frac{\Gamma}{\Gamma_{\rm fr}}}.
\end{equation}      

\begin{figure}
\includegraphics[width=8.5cm]{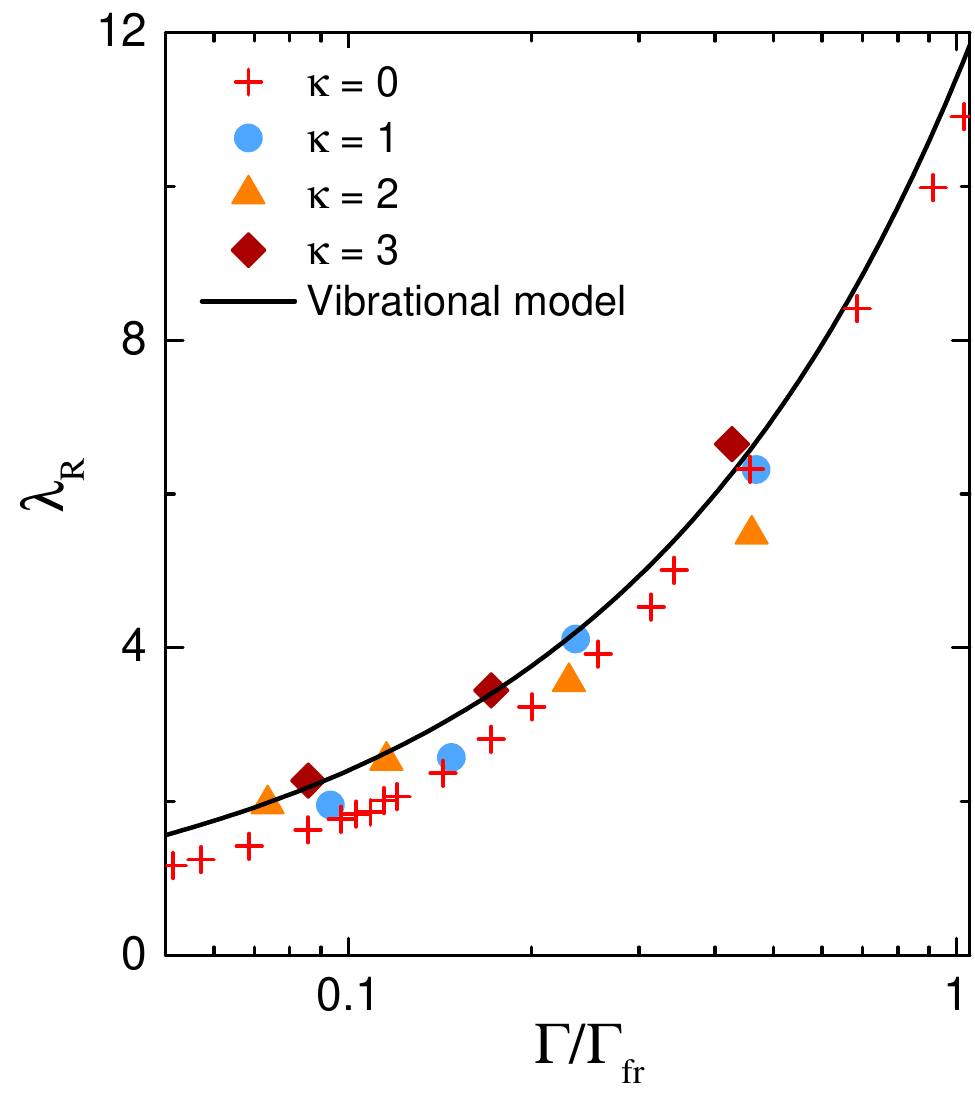}
\caption{(Color online) Reduced thermal conductivity coefficient $\lambda_{\rm R}$ vs the reduced coupling parameter $\Gamma/\Gamma_{\rm fr}$ of strongly coupled Yukawa fluids with different screening parameters.
Symbols denote molecular dynamics results, the notation is the same as in Fig.~\ref{Fig5}. The solid curve corresponds to the quasi-universal scaling of Eq.~(\ref{lambdascaling}).
}
\label{Figlambdascaling}
\end{figure}

The dependence of $\lambda_{\rm R}$ on $\Gamma/\Gamma_{\rm fr}$ in strongly coupled Yukawa fluids with different parameters $\kappa$ is plotted in Fig.~\ref{Figlambdascaling}. The symbols denote results from molecular dynamics simulations and are the same as used in Fig.~\ref{Fig5}. The solid line corresponds to Eq.~(\ref{lambdascaling}). A quasi-universality is observed, although a minor systematic tendency in increasing $\lambda_{\rm R}$ with $\kappa$ seems also present on top of the data scattering. The vibrational model prediction of Eq.~(\ref{lambdascaling}) agrees reasonably well with the simulation data in the considered strongly coupled regime $\Gamma/\Gamma_{\rm fr}\gtrsim 0.05$. 

\subsection{Lennard-Jones fluids}\label{LJliq}     

Next let us consider the Lennard-Jones model. The LJ system is one of the most popular and extensively studied model systems in condensed matter research, because it combines relative simplicity with adequate approximation of interatomic interactions in real substances, such as liquefied and solidified noble gases.  
The LJ potential is 
\begin{equation}
\phi(r)=4\epsilon\left[\left(\frac{\sigma}{r}\right)^{12}-\left(\frac{\sigma}{r}\right)^{6}\right], 
\end{equation}
where  $\epsilon$ and $\sigma$ are the energy and length scales (or LJ units), respectively. The reduced density and temperature expressed in LJ units are therefore $\rho_*=\rho\sigma^3$, $T_*=T/\epsilon$.  

Transport properties of LJ systems have been extensively studied in the literature. For recent reviews of available simulation data see for instance Refs.~\cite{BellJPCB2019,HarrisJCP2020,AllersJCP2020}. Particularly extensive and useful datasets have been published by Meier {\it et al.}~\cite{Meier2002,MeierJCP_1,MeierJCP_2} and by Baidakov {\it et al.}~\cite{BaidakovFPE2011,BaidakovJCP2012,BaidakovJCP2014}.   
These authors tabulated the transport data along different isotherms in a wide regions of the LJ system phase diagram. Though simulation protocols were different, the two datasets are in good agreement where they overlap~\cite{HarrisJCP2020}. 

It has been recently demonstrated that properly reduced transport coefficients (self-diffusion, shear viscosity, and thermal conductivity) of dense LJ fluids along isotherms exhibit quasi-universal scaling on the density divided by its value at the freezing point, $\rho_{\rm fr}$~\cite{KhrapakPRE04_2021}. This freezing density scaling is similar to the freezing temperature scaling and can be related to the quasi-universal excess entropy scaling approach~\cite{KhrapakJPCL2022,KhrapakJCP2022_1} and the isomorph theory~\cite{HeyesJCP2023}. Compared to the freezing temperature scaling, the freezing density scaling has a considerably wider applicability domain on the phase diagram of
LJ and related systems. Thus, it represents a very
useful corresponding states principle for the transport properties
of simple fluids. 

\begin{figure}
\includegraphics[width=8.5cm]{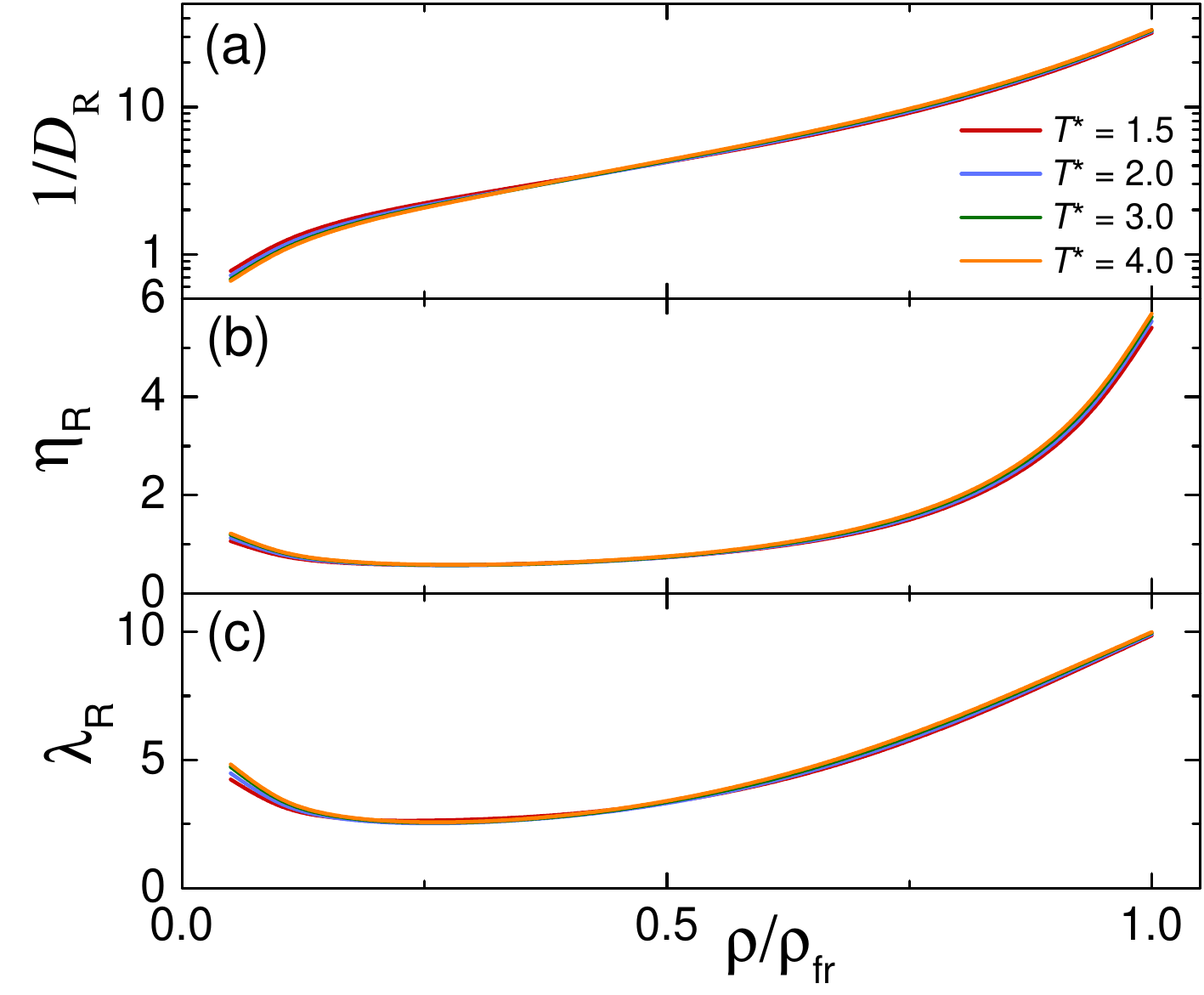}
\caption{(Color online) Inverse reduced self-diffusion (a), shear viscosity (b), and thermal conductivity (c) coefficients of Lennard-Jones fluids versus the density $\rho$ divided by the density at the freezing point $\rho_{\rm fr}$. The four curves correspond to supercritical isotherms: $T_*=1.5$, $T_*=2.0$, $T_*=3.0$, and $T_*=4.0$ (the color scheme is specified in panel (a)). The transport coefficients are evaluated using the approach from Ref.~\cite{BellJPCB2019}. The main observation is that the curves are overlapping to a very good accuracy, thus illustrating the quality of the freezing density scaling.}
\label{FigLJ_FDS}
\end{figure}

The quality of the freezing density scaling is illustrated in Fig.~\ref{FigLJ_FDS}, where the inverse diffusion, shear viscosity and thermal conductivity coefficients are plotted versus the the ratio $\rho/\rho_{\rm fr}$. The four curves for each transport property correspond to different supercritical isotherms (the color scheme is provided in Fig.~\ref{FigLJ_FDS}(a) and are calculated using the approach from Ref.~\cite{BellJPCB2019}. Note that the freezing density scaling applies to both supercritical and subcritical temperatures (not shown in Fig.~\ref{FigLJ_FDS}). The curves corresponding to different isotherms are overlapping to a good accuracy demonstrating the success of the freezing density scaling.  

\begin{figure}
\includegraphics[width=8.5cm]{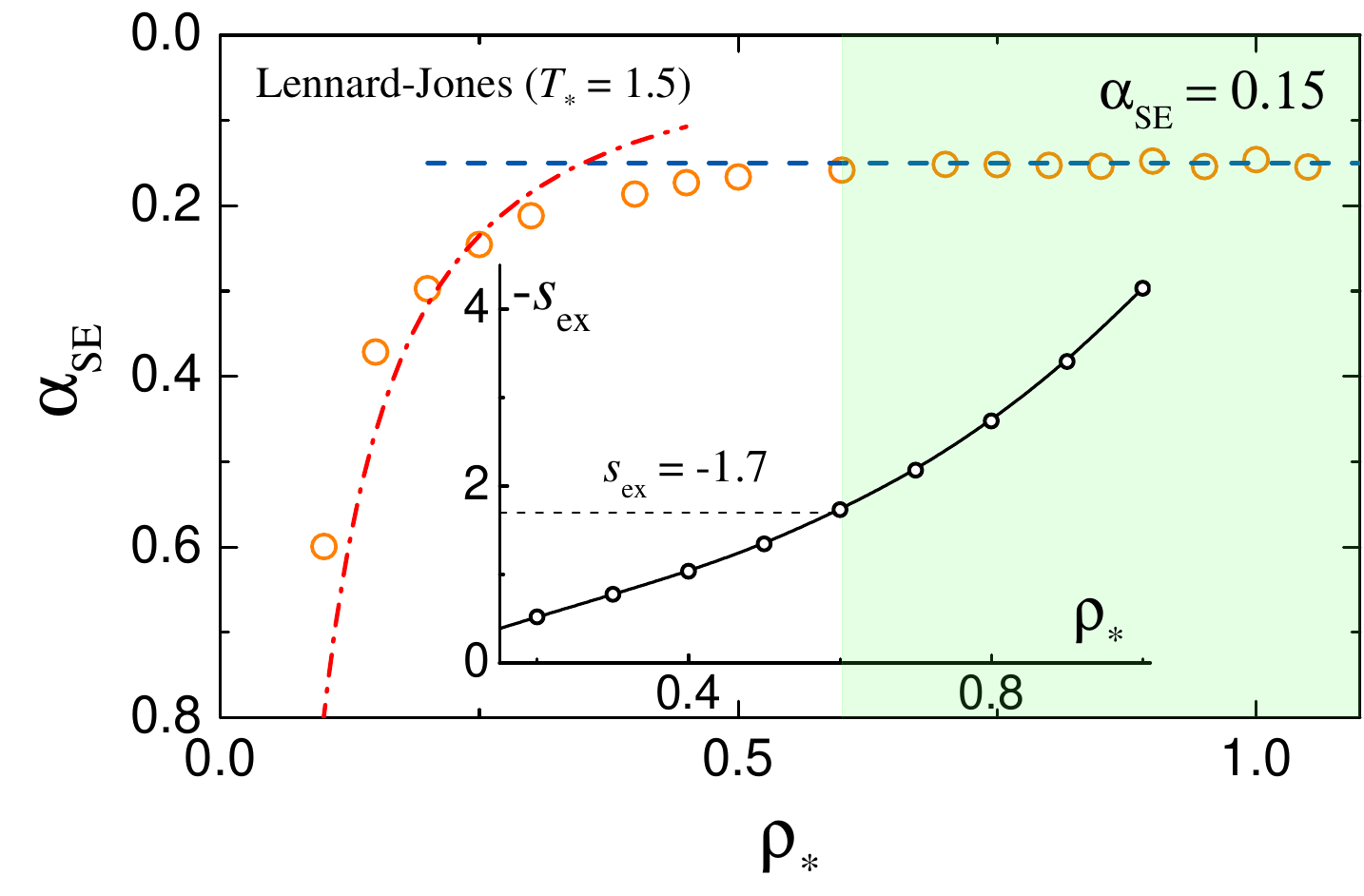}
\caption{(Color online) Stokes-Einstein coefficient $\alpha_{\rm SE}$ versus the reduced density $\rho_*$ for the LJ liquid. The symbols correspond to MD simulation results from Ref.~\cite{Meier2002}. The dashed line is the fluid asymptote $\alpha_{\rm SE}\simeq 0.15$. The dash-dotted curve corresponds to the dilute HS gas asymptote $\alpha_{\rm SE}\simeq 0.037/\rho_{*}^{4/3}$. In the shaded area at $\rho_{*}> 0.6$ the SE coefficient is constant (lies in the narrow range $\alpha_{\rm SE}\simeq 0.15\pm 0.01)$. The inset shows the dependence of the negative excess entropy on the reduced density~\cite{JaksePRE2003}. Onset of the validity of the SE relation corresponds to $s_{\rm ex}\lesssim -1.7$.  Reproduced from S. Khrapak and A. Khrapak, Phys. Rev. E {\bf 104}, 044110 (2021). }
\label{Fig6}
\end{figure}

Let us now focus on the applicability of SE relation to the LJ fluid. In view of the freezing density scaling, it is sufficient to consider a single isotherm. We have chosen a reference isotherm $T_*=1.5$ and employ the diffusion and viscosity coefficients tabulated in Ref.~\cite{Meier2002}. The resulting dependence of $\alpha_{\rm SE}$ on reduced density $\rho_*$ is plotted in Fig.~\ref{Fig6}. It is observed that as the density increases, the SE coefficient approaches the asymptotic value of $\alpha_{\rm SE}\simeq 0.15$. For $\rho_{*}\gtrsim 0.6$, the SE coefficient is located in a narrow range $\alpha_{\rm SE}\simeq 0.15\pm 0.01$. This is where, according to the pragmatic definition used here, the SE relation is satisfied (shaded area in Fig.~\ref{Fig6}). Note that Ohtori {\it et al}. reported slightly higher values for the SE coefficient in the LJ fluid~\cite{OhtoriPRE2015,OhtoriPRE2017}. They obtained $\alpha_{\rm SE}$ in a narrow range between $1/2\pi\simeq 0.16$ and $1/6\simeq 0.17$. The reason for this minor mismatch and the correct value of $\alpha_{\rm SE}$ in LJ fluids should be determined in the future.

At lower density, the dash-dotted curve shows the dilute hard-sphere (HS) gas asymptote. This asymptote has been obtained as follows. For a dilute HS gas the self-diffusion and viscosity coefficients within the Chapman-Enskog approach are given in the first approximation by~\cite{LifshitzKinetics}  
\begin{equation}
D = \frac{3}{8\rho\sigma^2}\left(\frac{T}{\pi m}\right)^{1/2}, \quad\quad \eta = \frac{5}{16\sigma^2}\left(\frac{mT}{\pi}\right)^{1/2}, \label{HStransport}
\end{equation}   
where $\sigma$ is now the HS diameter. In the first approximation we can associate the HS diameter with the LJ length scale. For the SE relation this gives 
\begin{equation}\label{HS2}
D\eta\left(\frac{\Delta}{T}\right)=\frac{15}{128\pi\rho_*^{4/3}}\simeq \frac{0.037}{\rho_*^{4/3}}.
 \end{equation} 
Similar scaling could be also obtained from Eqs.~(\ref{Rgas}), which leads to $D_{\rm R}\eta_{\rm R}\sim 1/\Sigma^2\rho^{4/3}$. Clearly, the density scaling of Eq.~(\ref{HS2}) is only approximate for dilute LJ gases. The effective hard-sphere diameter is not exactly equal to the LJ length scale $\sigma$. Moreover, the actual transport cross sections are different from the hard-sphere model and specifics of scattering in the LJ potential with long-range attraction has to be properly accounted for (see e.g. Refs.~\cite{HirschfelderBook,HirschfelderJCP1948,SmithJCP1964,
KhrapakPRE2014_scattering,KhrapakEPJD2014,KimJCompPhys2014,
Kristiansen2020} and references therein for some related works). Nevertheless, the simple Eq.~(\ref{HS2}) already provides a reasonable approximation for MD data as clearly observed in Fig.~\ref{Fig6}. 

The low-density asymptote $\alpha_{\rm SE}\simeq 0.037/\rho_*^{4/3}$ and the high-density asymptote $\alpha_{\rm SE}\simeq 0.15$ are intersecting at about $\rho_*\simeq 0.35$. This intersection can serve as a practical indicator of the crossover between the gas-like and liquid-like regions on the LJ system phase diagram~\cite{KhrapakPRE04_2021,KhrapakJCP2022}. We will elaborate on this further in Section~\ref{Crossover}.  

The inset in Fig.~\ref{Fig6} shows the dependence of the minus reduced excess entropy $-s_{\rm ex}$ on the density as tabulated in Ref.~\cite{JaksePRE2003} for the LJ liquid isotherm $T_*=1.5$. The onset of validity of the SE relation corresponds to $s_{\rm ex}\lesssim -1.7$, according to a pragmatic definition used above. This is slightly higher, but comparable with the onset of SE relation validity in plasma-related fluids considered previously.


\begin{figure}
\includegraphics[width=8.5cm]{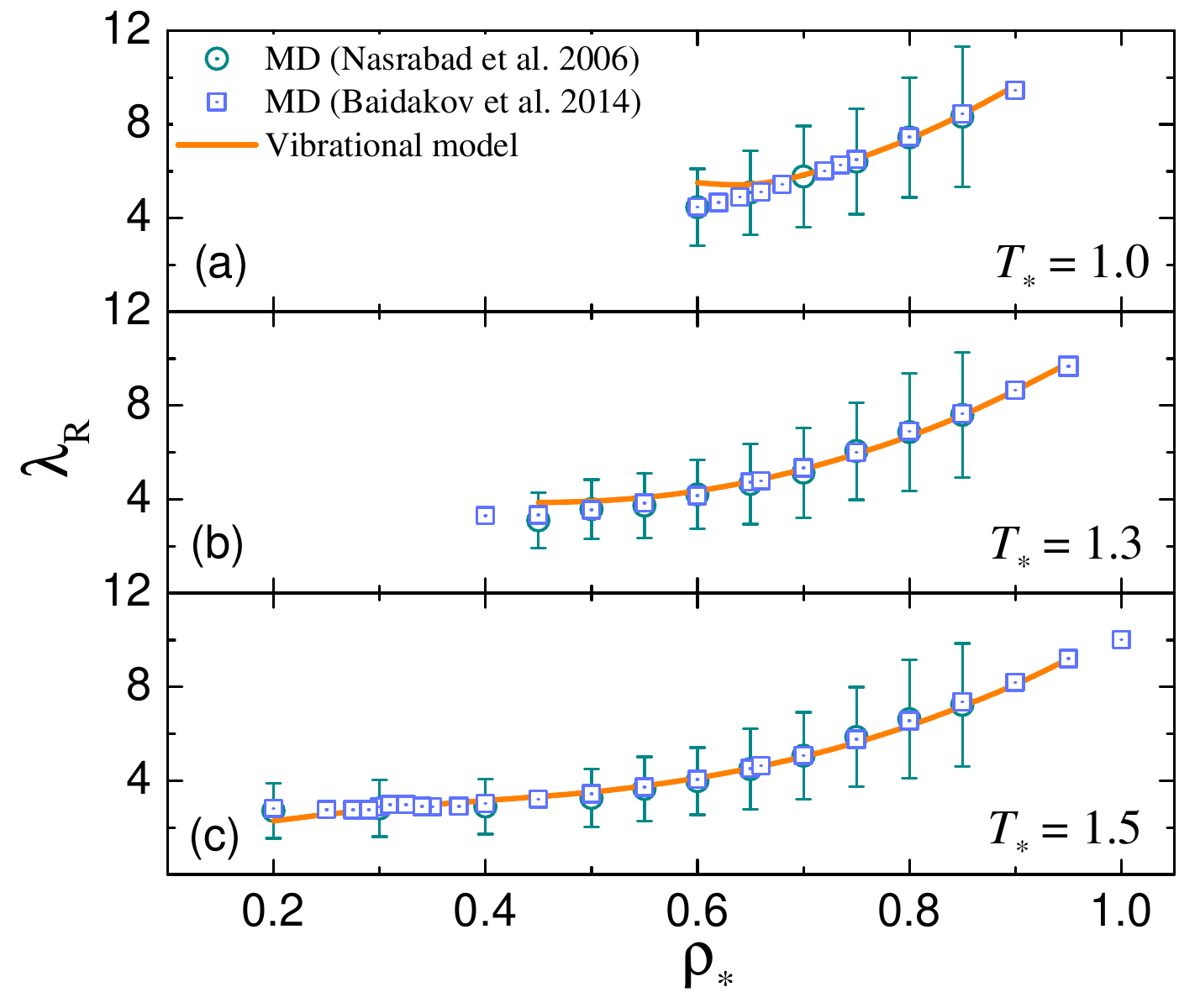}
\caption{(Color online) Reduced thermal conductivity coefficient $\lambda_{\rm R}$ vs the reduced density $\rho_*$ of the LJ fluid along the isotherms $T_*=1.0$ (a), $T_*=1.3$ (b) and $T_*=1.5$ (c).
Circles with error bars correspond to MD results from Nasrabad {\it et al}.~\cite{NasrabadJCP2006}. Squares are the MD results from Baidakov {\it et al}.~\cite{BaidakovJCP2014}. The solid curves are the theoretical calculations using Eq.~(\ref{Cahill1}).
}
\label{Fig7}
\end{figure}

The vibrational model of thermal conductivity has been first applied to the LJ fluid in Ref.~\cite{KhrapakPRE01_2021}. It has been  demonstrated that Eq.~(\ref{Cahill1}) describes well the numerical data from Ref.~\cite{Meier2002} along the near-critical isotherm $T_*=1.35$. Since for the thermal conductivity coefficient critical enhancement can be substantial it makes sense to consider other isotherms. Here we use the numerical data from Refs.~\cite{NasrabadJCP2006,BaidakovJCP2014} for the three isotherms: $T_*=1.0$, $1.3$, and $1.5$. To obtain theoretical curves, we use Eq.~(\ref{Cahill1}) complemented with thermodynamic data from Ref.~\cite{Meier2002}. In particular, we use the tabulated values for specific heat $c_{\rm v}$ and evaluate the longitudinal and transverse sound velocities using the existing relations with the excess pressure and energy for the LJ interaction potential~\cite{ZwanzigJCP1965,KhrapakMolecules2020}. For completeness, the explicit expressions are provided below:
\begin{equation}\label{cl}
\frac{c_l^2}{v_{\rm T}^2}=3-\frac{72}{5}u_{\rm ex}+11p_{\rm ex},
\end{equation} 
\begin{equation}\label{ct}
\frac{c_t^2}{v_{\rm T}^2}=1-\frac{24}{5}u_{\rm ex}+3p_{\rm ex}.
\end{equation}
Here $u_{\rm ex}=U/NT-3/2$ is the excess internal energy, $p_{\rm ex}=P/\rho T-1$ is the excess pressure, $U$, $P$, and $N$ being the internal energy,  pressure, and the number of atoms, respectively. Note that Cauchy relation is satisfied
\begin{equation}
\left(\frac{c_l}{v_{\rm T}}\right)^2-3\left(\frac{c_t}{v_{\rm T}}\right)^2=2p_{\rm ex}.
\end{equation} 
This is a generalization of the conventional Cauchy identity for isotropic solids made of molecules interacting with two-body central forces, which is now valid for fluids at any temperature and pressure~\cite{ZwanzigJCP1965}. Note that similar relations can be derived for the generalized $n$-$m$ LJ potential~\cite{KhrapakMolecules2021}. 

Thus, knowledge of thermodynamic parameters  $c_{\rm v}$, $u_{\rm ex}$, and $p_{\rm ex}$ is sufficient to evaluate the thermal conductivity coefficient within the vibrational paradigm. The comparison between theory and numerical results is shown in Fig.~\ref{Fig7}. Excellent agreement at high density is observed. Thus, the vibrational model of thermal conductivity works very well for dense LJ fluids, both in subcritical and supercritical regimes.

\subsection{Hard-Sphere fluids}

Another simple model system that should be considered is the fluid made of hard spheres. The HS interaction potential is extremely hard and short ranged. The interaction energy is infinite for $r<\sigma$ and is zero otherwise, where $\sigma$ is the sphere diameter. The HS system is a very important simple model for the behaviour of condensed matter in its various states~\cite{SmirnovUFN1982,MuleroBook, PuseyPhylTrans2009,ParisiRMP2010,BerthierRMP2011,
KlumovPRB2011,DyreJPCM2016}.  

From the beginning it is clear that neither the Zwanzig's derivation of the SE relation nor the vibrational model of heat transfer are consistent with the dynamical picture in HS fluids. In contrast to softer interactions, the velocity autocorrelation function $Z(t)$ rapidly vanishes after the first rebound against the initial cage and does not exhibit a pronounced oscillatory character~\cite{AlderPRL1967,WilliamsPRL2006,Daligault2020}. HS fluids are extremely anharmonic and this is for instance reflected by the divergence of the Einstein frequency. However, similar to other simple fluids with sufficiently steep interactions, dense HS fluids do support the acoustic-like longitudinal and transverse collective modes~\cite{BrykJCP2017} (with a forbidden long wavelength region for the transverse mode, the so-called  ''$k$-gap''~\cite{MurilloPRL2000,OhtaPRL2000,GoreePRE2012,BolmatovSciRep2016,
BrykJCP2017,YangPRL2017,KryuchkovSciRep2019,KhrapakJCP2019}). 
The elastic moduli and hence the longitudinal and transverse sound velocities remain finite and well defined~\cite{Miller1969,KhrapakPRE09_2019,KhrapakPRE05_2021}. This implies that Eqs.~(\ref{alpha}) and (\ref{Cahill1}) are formally applicable. It makes sense to compare their prediction with the existing results from numerical simulations.        

In HS systems the thermodynamic and transport properties  depend on a single reduced density parameter $\rho_*=\rho\sigma^3$ (the packing fraction, $\pi\rho\sigma^3/6$, is also often used). Transport properties of HS fluids have been extensively studied (see e.g. Ref.~\cite{MuleroBook} for a review). Here we use recent MD simulation results by Pieprzyk {\it et al}.~\cite{Pieprzyk2019,Pieprzyk2020}. The use of large simulation systems and long simulation times allowed accurate prediction of the transport coefficients in the thermodynamic limit. 

\begin{figure}
\includegraphics[width=8.5cm]{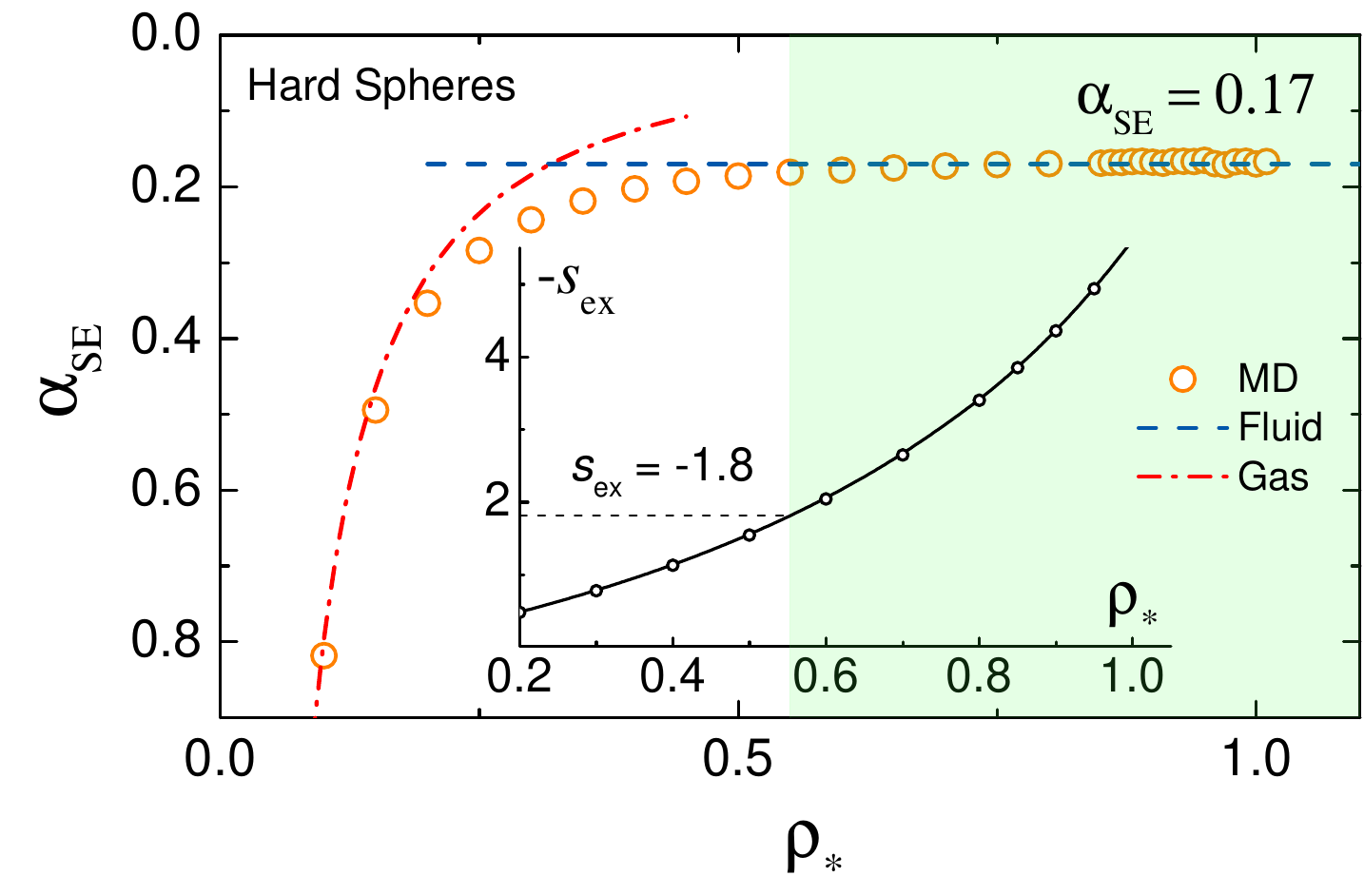}
\caption{(Color online) Stokes-Einstein parameter $\alpha_{\rm SE}$ versus the reduced density $\rho_*$ for the HS fluid. The symbols correspond to MD simulation results from Refs.~\cite{Pieprzyk2019,Pieprzyk2020}. The dashed line is the dense fluid asymptote, $\alpha_{\rm SE}\simeq 0.17$. The dash-dotted curve corresponds to the dilute HS gas asymptote $\alpha_{\rm SE}\simeq 0.037/\rho_{*}^{4/3}$. In the shaded area at $\rho_{*}> 0.55$ the SE coefficient is constant (lies in the narrow range $\alpha_{\rm SE}\simeq 0.17\pm 0.01)$. The inset shows the dependence of the negative excess entropy on the reduced density. Onset of the validity of the SE relation corresponds to $s_{\rm ex}\lesssim -1.8$. Reproduced from S. Khrapak and A. Khrapak, Phys. Rev. E {\bf 104}, 044110 (2021).}
\label{Fig8}
\end{figure}

Based on the tabulated data for the self-diffusion and viscosity coefficients the SE coefficient has been evaluated and plotted as a function of the reduced density in Fig.~\ref{Fig8}. It is observed that the data points stick to the two asymptotes: the gaseous Eq.~(\ref{HS2}) at low density and the liquid-like $\alpha_{\rm SE}\simeq 0.17$ at sufficiently high density. This numerical value correlates well with the results from other studies~\cite{OhtoriJCP2018}. The inset shows the dependence of the negative excess entropy on the reduced density. These low-density and high-density asymptotes are intersecting at $\rho_*\simeq 0.32$, which corresponds to $s_{\rm ex}\simeq -0.8$. The shaded region in Fig.~\ref{Fig4} is where the SE coefficient lies in a narrow range $\alpha_{\rm SE}\simeq 0.17\pm 0.01$. This is approximately the regime of SE relation validity according to the present pragmatic definition. Numerically, the onset of SE relation validity  occurs at $\rho_*\gtrsim 0.55$, which corresponds to $s_{\rm ex}\simeq -1.8$. Thus, we have to conclude that the SE relation is still satisfied for HS fluids, even though it is not expected to be. The magnitude of the SE coefficient is somewhat higher than for other fluids considered and slightly higher than the Zwanzig's model predicts [from the ratio of the transverse to longitudinal velocities $c_t/c_l\simeq 0.5$ in the HS limit~\cite{KhrapakPRE05_2021} one would expect $\alpha_{\rm SE}\simeq 0.15$ according to Eq.~(\ref{alpha})]. This difference can possibly arise due to obvious inconsistencies between theoretical assumptions and actual dynamical picture in HS fluids. 
 
\begin{figure}
\includegraphics[width=8.5cm]{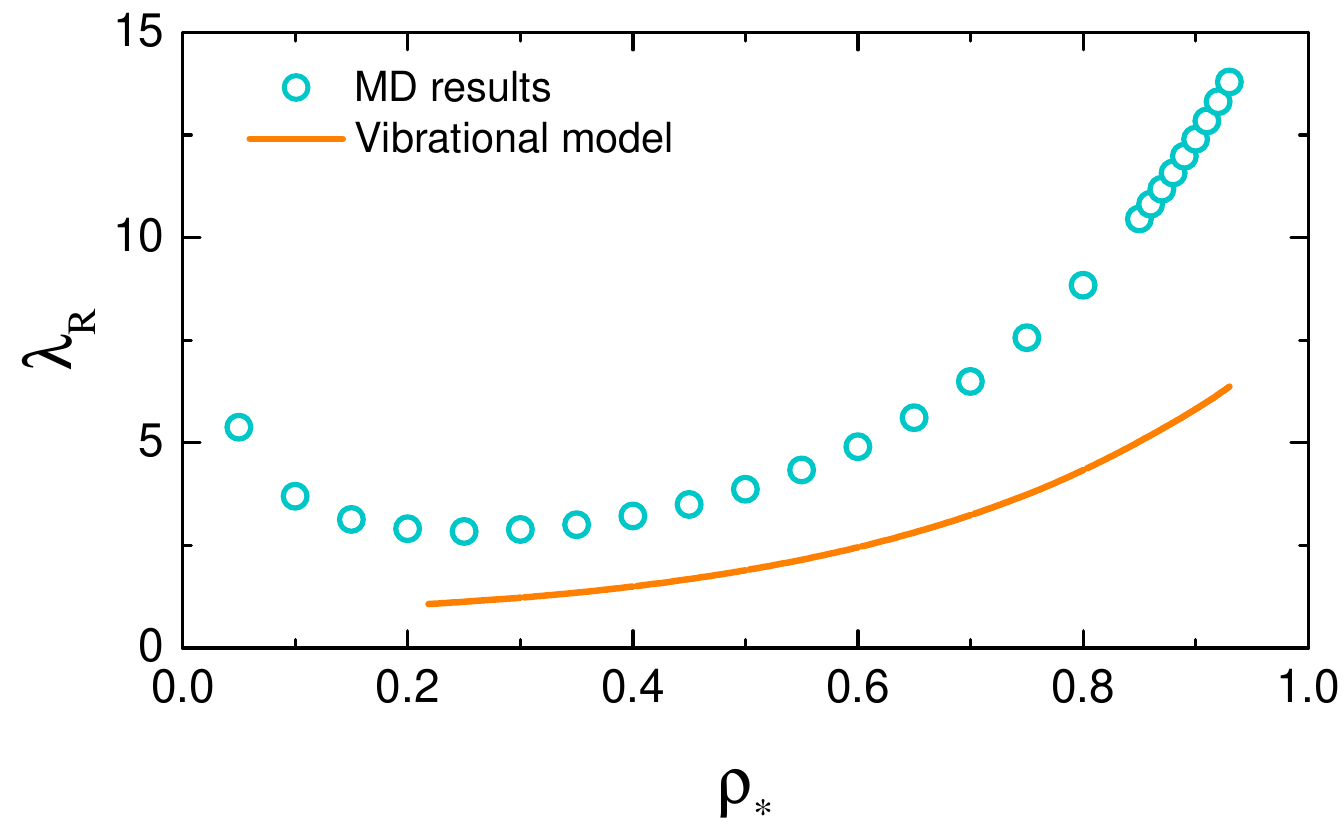}
\caption{(Color online) Reduced thermal conductivity coefficient of the HS fluid, $\lambda_{\rm R}$,  versus the reduced density $\rho_*$. The circles correspond to the recent MD results from Ref.~\cite{Pieprzyk2020}. The solid curve represents a variant of vibrational model of heat transfer corresponding to Eq.~(\ref{Cahill1}).  }
\label{Fig9}
\end{figure}  

To evaluate the thermal conductivity coefficient of the HS fluid from Eq.~(\ref{Cahill1}) we need to know the longitudinal and transverse sound velocities. These can be obtained from the instantaneous elastic moduli of HS fluids derived by Miller~\cite{Miller1969} and discussed in detail in Ref.~\cite{KhrapakPRE09_2019}. Note that since hard spheres possess only kinetic energy, we have $(\partial U/\partial V)_T = 0$ and therefore $(\partial U/\partial T)_P = c_{\rm v}$.  
Furthermore, since the excess internal energy of the HS system is identically zero, we have $c_{\rm v}=3/2$. The results of calculating $\lambda_{\rm R}$ by means of the vibrational model are shown in Fig.~\ref{Fig9} along with the recent MD data from Ref.~\cite{Pieprzyk2020}. The vibrational model underestimates the data from numerical simulations by a practically constant factor $\simeq 2$ in the density range considered. Much better agreement between the vibrational model and MD simulations was reported in Ref.~\cite{KhrapakApplSci2022}. 
However, this appears to be an incorrect result: $(\partial U/\partial T)_P$ was erroneously approximated by $c_{\rm p}$ in this work, leading to better match between theory and simulations. Actually, we see that the vibrational model cannot provide more than an order of magnitude estimate in this case. This is not surprising, taking into account the extreme anharmonicity of the HS model. Note that the thermal conductivity of the HS fluid is rather good described by the Enskog theory (deviations are within $\sim 5\%$ across the entire fluid regime~\cite{Pieprzyk2020}), which is thus clearly superior to the vibrational model in this case.   
   
\section{Link to the excess entropy scaling}\label{Link} 

Among different thermodynamic properties of liquids, the entropy is one of the hardest quantities to estimate. Therefore, development of models allowing accurate estimations of the entropy for different mechanisms of interatomic interactions represents an important problem. Recently, a method for estimating the excess entropy of simple liquids not too far from the liquid-solid phase transition, based on the vibrational picture of atomic dynamics has been proposed~\cite{KhrapakJCP2021}. The method represents a variant of cell theory, which particularly emphasises relations between liquid state thermodynamics and collective modes properties. The method works very well for inverse-power-law (IPL) fluids in the entire range of the potential softness from the very soft Coulomb limit to the opposite hard sphere interaction limit. Although it is less successful for the Lennard-Jones potential, it represents an appropriate tool to reveal the link between the vibrational model and excess entropy scaling that was pointed out previously~\cite{GroverJCP1985,HooverBook}.

Within the vibrational paradigm, the excess entropy is obtained as an appropriate averaging over liquid excitation frequencies~\cite{KhrapakJCP2021}
\begin{equation}\label{sex}
s_{\rm ex} = \frac{3}{2}-\frac{3}{2}\left\langle\ln \frac{m\Delta^2\omega^2}{2\pi T}\right\rangle.
\end{equation} 
On the other hand, the heat conductivity coefficient is related to the average oscillation frequency by virtue of Eq.~(\ref{Cond1}). 
Thus, $\lambda$ and $s_{\rm ex}$ are interrelated through the liquid collective excitation spectrum. Within the simplest Einstein model, a closed form relation between $\lambda_{\rm R}$ and $s_{\rm ex}$ can be obtained. Using $\langle \omega\rangle = \Omega_{\rm E}$ and $\langle \ln\omega^2\rangle = 2\ln\Omega_{\rm E}$ in Eqs.~(\ref{Cond1}) and (\ref{sex}) and assuming $c_{\rm p}\simeq c_{\rm v}$ we arrive at
\begin{equation}
\lambda_{\rm R}\simeq \frac{c_{\rm v}}{\sqrt{2\pi}}\exp\left(\frac{1}{2}-\frac{s_{\rm ex}}{3}\right)\simeq 0.66c_{\rm v}\exp\left(-\frac{s_{\rm ex}}{3}\right).
\end{equation} 
This predicts exponential scaling of the reduced thermal conductivity coefficient with the excess entropy with a universal slope of $1/3$. The dependence of $c_{\rm v}$ on $s_{\rm ex}$ is, however, not accounted for. However, the latter is weak and close to the freezing transition we can simply set $c_{\rm v}\sim 3$ according to the Dulong-Petit law. In this case the relation between $\lambda_{\rm R}$ and $s_{\rm ex}$ takes the form $\lambda_{\rm R}\simeq 2e^{-s_{\rm ex}/3}$, comparable to the scaling $\lambda_{\rm R}\simeq 1.5 e^{-0.5s_{\rm ex}}$ quoted originally by Rosenfeld~\cite{RosenfeldJPCM1999}.

Thus, the vibrational model of dense liquid transport is consistent with the exponential scaling of the heat conductivity coefficient with the excess entropy. The exponential scaling of the self-diffusion and viscosity coefficients is not explained since the vibrational model does not allow to evaluate them separately, but only regulates their product.

\section{Gas-liquid crossover}\label{Crossover}

An important question is what are the quantitative applicability conditions of the vibrational model of atomic transport in dense fluids. First of all, the vibrational picture is more appropriate for sufficiently soft interactions, for which caging phenomena and a pronounced oscillatory character of atomic motion are representative, like for instance the OCP fluid, considered above. We have seen, however, that although the vibrational picture is clearly inappropriate for the HS fluid, the SE relation is nevertheless satisfied. The Debye-like averaging of the excitation frequency provides reasonable results for the thermal conductivity coefficient of the HS fluid. Thus, the softness of the interatomic interaction does not represent a harsh limitation on the applicability of the model.

The vibrational picture requires the characteristic solid-like oscillation frequency to be much higher than the inverse relaxation time, i.e. the condition $\omega\tau >1$ should be satisfied. A characteristic vibrational frequency $\omega$ can be approximated by the Einstein frequency $\Omega_{\rm E}$. Still, we would need to know the behaviour of the shear viscosity $\eta$ and the instantaneous shear modulus $G_{\infty}$ in order to evaluate $\tau$ from Eq.~(\ref{tau}) and estimate $\tau\Omega_{\rm E}$. This is not very realistic in general (although for some special systems this program is feasible). Other options should be considered.

The vibrational picture is clearly not applicable in the gaseous regime, where the atoms mostly move freely between pair collisions. Here the vibrational model would result in wrong quantitative and qualitative predictions. Thus, it makes sense first to look into the crossover between the gas-like and liquid-like dynamical regimes. The possibility to define a demarcation line between liquid-like and gas-like behaviors of supercritical fluids has been a topic of major recent interest~\cite{GorelliPRL2006,Simeoni2010,McMillanNatPhys2010,BrazhkinJPCB2011,BrazhkinPRE2012,BrazhkinUFN2012,
BrazhkinPRL2013,GorelliSciRep2013,YangPRE2015,
BrykJPCL2017,BrazhkinJPCB2018,BrykJPCB2018,BellJCP2020,
ProctorJPCL2019,PloetzJPCB2019,BanutiJSupFluids2020,HaJPCL2020,
MaximNatCom2019,SunPRL2020,BellJPCL2021,
CockrellPRE2021,CockrellPhysRep2021}. This problem is not just a matter of curiosity since the structural and dynamical properties are very different in these regimes and quite different approaches are required for their description. It is of great fundamental and practical interest to understand where the crossover between the gas-like and liquid-like regimes takes place. This problem is also related to a long-standing debate about the nature of the supercritical fluid and a more general question ``What is liquid?''~\cite{BarkerRMP1976,Sengers1979,Woodcock2017}.                    
Multiple definitions for the gas-to-liquid dynamical crossover have been proposed and discussed in recent years~\cite{BrazhkinPRE2012,BrazhkinUFN2012,BellJCP2020,
KhrapakPRE04_2021,KhrapakJCP2022}. Not all of these definitions are consistent or universal, which generated a significant amount of debate. From the isomorph theory perspective~\cite{DyreJPCB2014}, it seems not unreasonable to assume that a demarcation line between the gas-like and fluid-like regimes should itself be an approximate isomorph. As such, it should be characterized by a quasi-universal value of the excess entropy. Properly reduced structural and dynamical properties should also be quasi-invariant along the demarcation line. This point of view has been elaborated in recent works~\cite{BellJCP2020,BellJPCL2021}. One of the possible definitions is based on the location of the minimum of the macroscopically scaled shear viscosity coefficient. This minimum corresponds to the crossover between the gas-like and liquid-like mechanisms of momentum transfer, which have quite different nature and hence different asymptotes. It has been observed that for several different model systems the location of the minima in the macroscopically reduced shear viscosity coefficients occurs at approximately the same value of the excess entropy per particle ($s_{\rm ex}\simeq -2/3$)~\cite{BellJCP2020}, 
and the minimum values themselves are also quasiuniversal~\cite{TrachenkoSciAdv2020,TrachenkoPRB2021,TrachenkoPhysToday2021,KhrapakPoF2022}. This is also where kinetic and potential contributions to the shear viscosity coefficients are equal to a good accuracy~\cite{BellJPCL2021}. 

Brazhkin {\it et al}. proposed to call the demarcation line between gas-like and liquid-like dynamics on the phase diagram as ``Frenkel line''~\cite{BrazhkinPRE2012}. A simple and popular definition of the demarcation line is the condition of constant specific heat $c_{\rm v}=2$~\cite{BrazhkinPRE2012,BrazhkinPRL2013,BrazhkinUFN2017}. It is argued that this condition corresponds to a qualitative change of the excitation spectrum in the liquid -- the loss of solid-like transverse modes~\cite{TrachenkoRPP2015,CockrellPRE2021}. Although this definition cannot be truly universal because for instance in HS fluids the specific heat is fixed, $c_{\rm v}\equiv 3/2$, it nevertheless remains quite useful for fluids with soft interactions. Another definition proposed by this group is related to the loss of oscillatory component of particle dynamics at the Frenkel line. Disappearance of the minima of the velocity autocorrelation function was proposed as a dynamical criterion of the gas-liquid crossover. For soft spheres and the LJ system the two criteria give lines that coincide to a good accuracy~\cite{BrazhkinPRL2013,BrazhkinUFN2017}.    

Independently, the SE relation has been recently analyzed in the context of gas-liquid crossover~\cite{KhrapakPRE10_2021}.
As we have already discussed above (and have seen in Figs.~\ref{Fig2}, \ref{Fig4}, \ref{Fig6}, and \ref{Fig8}), for various simple fluids (LJ, Coulomb, Yukawa, and HS) there exist two clear asymptotes for the product $D\eta(\Delta/T)\equiv \alpha_{\rm SE}$. In dense fluids near freezing transition this product approaches a slightly system-dependent constant value. This is where the SE relation without the hydrodynamic diameter holds. Far away from the freezing point, $\alpha_{\rm SE}$ decreases with increasing density. The intersection of these two asymptotes has been suggested as a convenient practical indicator for the crossover between the gas-like and liquid-like regions on the phase diagram. We have already observed that for the systems considered, intersection is characterised by very close values of the reduced excess entropy, $s_{\rm ex}=-0.9 \pm 0.1$. A similar value was also obtained in Ref.~\cite{BellJPCL2021}, where it was also recognized that this is nearly the critical point entropy for simple fluids exhibiting a critical point. 

\begin{figure}
\includegraphics[width=8.5cm]{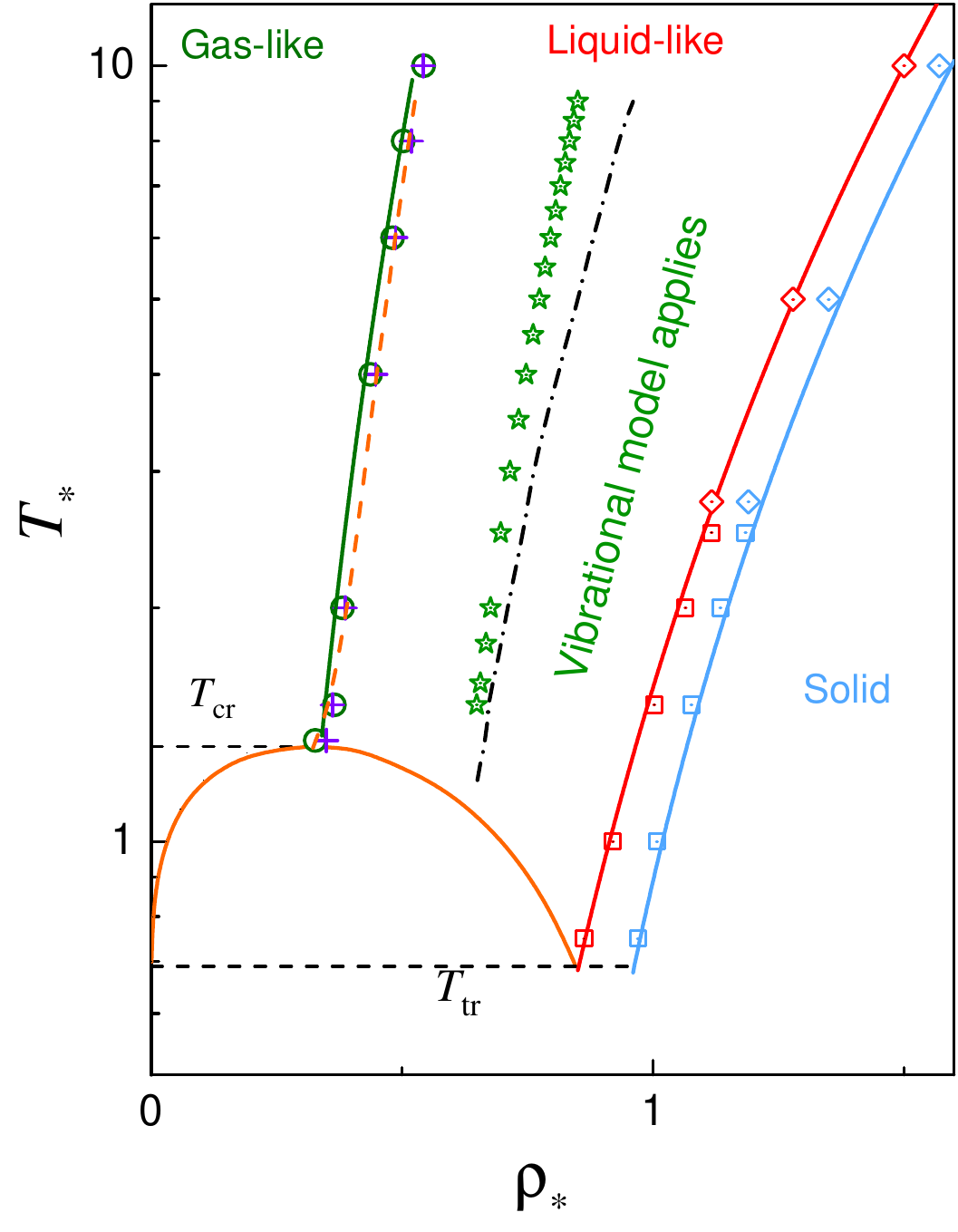}
\caption{(Color online) Phase diagram of the LJ system on the density-temperature plane. The squares and rhombs correspond to the fluid-solid coexistence boundaries as tabulated in Refs.~\cite{SousaJCP2012} and \cite{HansenPRA1970}, respectively; the corresponding curves are simple fits from Ref.~\cite{KhrapakJCP2011_2}. The liquid-vapour boundary is plotted using the formulas provided in Ref.~\cite{HeyesJCP2019}. 
The reduced triple point and critical temperatures are $T_{\rm tr}\simeq 0.69$~\cite{SousaJCP2012} and $T_{\rm c}\simeq 1.33$~\cite{HeyesJCP2019}, respectively. 
The circles correspond to the location of minima of kinematic viscosity~\cite{BellJPCL2021}. The crosses denote the points where the contribution to viscosity due to atomic translation is exactly equal to that due to interatomic interactions~\cite{BellJPCL2021}. The dashed curve emanating from the critical point corresponds to constant reduced excess entropy of $s_{\rm ex}=-0.9$, as evaluated from Thol's {\it et al}. equation of state~\cite{Thol2016}. The solid curve is plotted using Eq.~(\ref{Demarcation}) obtained from the constant density ratio $\rho/\rho_{\rm fr}=0.35$.  All considered definitions of the crossover between the gas-like and liquid-like regimes of atomic dynamics are in good agreement. The stars correspond to the thermodynamics criterion $c_{\rm v}=2$ and the dash-dotted curve to the condition $s_{\rm ex}=-2$, both evaluated from Thol {\it et al}.'s equation of state~\cite{Thol2016}. These are approximately coinciding. The vibrational model applies for $s_{\rm ex}\lesssim -2$ and up to the freezing curve, characterized roughly by $s_{\rm ex}\sim -4$.}
\label{Fig10}
\end{figure}

Let us now focus on the LJ system. The phase diagram of the LJ system in ($\rho_*$, $T_*$) plane is sketched in Fig.~\ref{Fig10}. The fluid-solid coexistence data points are taken from Refs.~\cite{SousaJCP2012,HansenPRA1970}. The curves are the fits of the form  $T_*^{\rm L,S}= {\mathcal A}^{\rm L,S}\rho_*^4-{\mathcal B}^{\rm L,S}\rho_*^2$ (superscripts L and S correspond to liquid and solid, respectively). This shape of the fluid-solid coexistence of LJ system with constant (or very weakly $\rho_*$-dependent) parameters ${\mathcal A}$ and ${\mathcal B}$ is a very robust result reproduced in a number of various theories and approximations~\cite{PedersenNatCom2016,RosenfeldMolPhys1976,KhrapakJCP2011_2,
HeyesPSS2015,KhrapakAIPAdv2016,CostigliolaPCCP2016,KhrapakPRR2020,
HeyesPRE2021}. Here we take constant values of ${\mathcal A}^{\rm L}=2.29$ and ${\mathcal B}^{\rm L}=0.71$ at freezing (liquidus) and ${\mathcal A}^{\rm S}=1.97$ and ${\mathcal B}^{\rm S}=1.08$ at melting (solidus) proposed in Ref.~\cite{KhrapakJCP2011_2}. The liquid-vapour boundary is plotted using the formulas provided in Ref.~\cite{HeyesJCP2019}. The reduced triple point and critical temperatures are $T_*^{\rm tr}\simeq 0.694$~\cite{SousaJCP2012} and $T_*^{\rm cr}\simeq 1.326$~\cite{HeyesJCP2019}, respectively. 
Additional symbols appearing in the supercritical region are: The circles correspond to the location of minima of kinematic viscosity~\cite{BellJPCL2021}; the crosses denote the points where the contribution to viscosity due to kinetic and potential contributions are equal~\cite{BellJPCL2021}; the solid curve emanating from very nearly the critical point and terminating at $T_*=9$ corresponds to constant excess entropy of $s_{\rm ex}=-0.9$, where gas-like and liquid-like asymptotes of the Stokes-Einstein product intersect; it has been calculated from the EoS by Thol {\it et al}.~\cite{Thol2016} in the domain of its applicability. The dashed curve corresponds to the condition $\rho/\rho_{\rm fr}=0.35$. The latter two are close, illustrating that in the LJ fluid fixed value of excess entropy corresponds to approximately fixed value of the reduced density ratio $\rho/\rho_{\rm fr}$~\cite{KhrapakJPCL2022,KhrapakJCP2022_1}. While different definitions agree to a good accuracy, the definition based on the freezing density is generally more practical, because freezing density is usually relatively well known (as compared to the exact location of the minima in reduced transport coefficients, or lines of constant excess entropy). In particular, in view of the relation between the freezing temperature and density of the LJ fluid, a very simple expression for the crossover line emerges~\cite{KhrapakJCP2022}:
\begin{equation}\label{Demarcation}
T_*=2.29 (\rho_*/0.35)^4-0.71(\rho_*/0.35)^2.
\end{equation}          

The stars in Fig.~\ref{Fig10} correspond to the thermodynamic condition $c_{\rm v}=2$ (for near-critical temperatures the condition $c_{\rm v}=2$ has multiple roots; we only consider temperatures for which a single root exists). The dash-dotted curve corresponds to the the condition $s_{\rm ex}=-2$. Both conditions have been evaluated with the help of Thol's EoS~\cite{Thol2016} and there is a reasonable agreement between them. The vibrational model operates to the right from these conditions, where pronounced solid-like oscillations dominate the dynamical picture. 

\begin{figure}
\includegraphics[width=8.5cm]{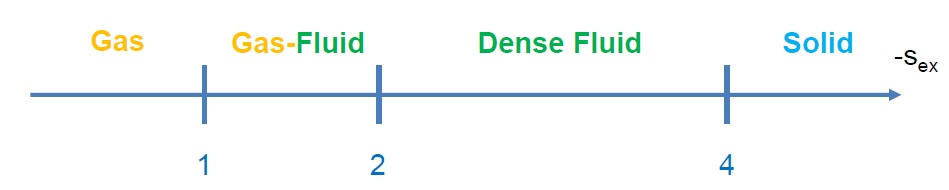}
\caption{(Color online)  Sketch of the one-dimensional phase diagram of simple systems in terms of the excess entropy $s_{\rm ex}$. For $s_{\rm ex}\gtrsim -1$ the system is in gas-like state, and the gas-liquid crossover occurs at around $s_{\rm ex}\simeq -1$. The condition  $s_{\rm ex}\simeq -2$ marks the point where vibrational dynamics starts to dominate. The vibrational model of transport is applicable from here up to the freezing point, which is roughly located at $s_{\rm ex}\sim -4$.}
\label{Fig11}
\end{figure}

Thus, the two useful lines on the phase diagram of simple fluids can be identified. The first corresponds to the  intersecton of gas-like and liquid-like asymptotes of reduced dynamical characteristics (such as the macroscopically reduced shear viscosity coefficient or dimensionless Stokes-Einstein product). This occurs at the excess entropy $s_{\rm ex}\simeq -0.9$ (and a fixed density ratio $\rho/\rho_{\rm fr}\simeq 0.35$ for the LJ fluid). The second line corresponds to $s_{\rm ex}\simeq -2$ (and $c_{\rm v}\simeq 2$ for the LJ fluid). This marks the onset of applicability for the  vibrational model of transport and thermodynamics of simple fluids. The phase diagram of simple systems becomes essentially one-dimensional in terms of the excess entropy. A sketch is shown in Fig.~\ref{Fig11}.         

So far the gas-liquid crossover (Frenkel line) has been illustrated using the LJ system. Note, however, that the long-range attraction is not a prerequisite of this crossover and it occurs in other systems as well, including soft and hard purely repulsive potentials. For example, the location of Frenkel line in HS systems was considered in Ref.~\cite{BellJPCL2021}, hard spheres and square-well potentials were investigated in Ref.~\cite{PruteanuACSOmega2023}, Yukawa systems in the context of dusty plasma were considered in Ref.~\cite{HuangPRR2023}.    

\section{Real liquids}\label{Realliq}

In this paper most attention was focused on model fluids consisting of particles interacting via several popular pairwise interaction potentials. All necessary information is available for these systems, including structural, dynamical, and thermodynamic properties. This allowed us to perform very detailed comparison between the actual transport properties and predictions based on the vibrational paradigm of atomic dynamics and to demonstrate its adequacy. The purpose of this Section is to provide a related brief overview of the transport properties of real liquids. The main point is to show that in many cases qualitative to semi-quantitative agreement between real and model systems is present and thus the vibrational model represents a useful tool
for better understanding and predicting transport properties of the liquid state.        

\begin{figure}
\includegraphics[width=8.5cm]{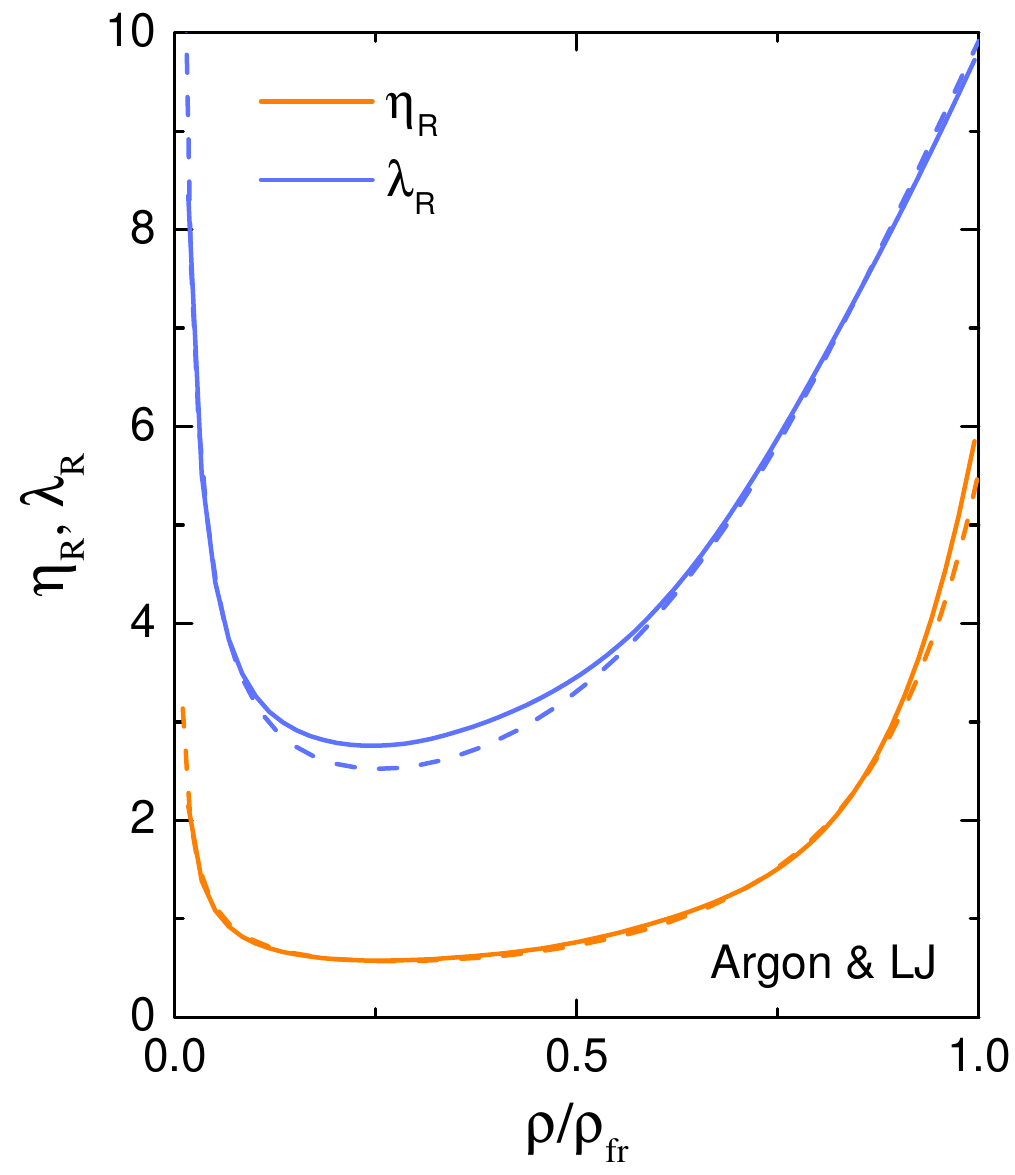}
\caption{(Color online) The reduced shear viscosity and thermal conductivity coefficients of liquefied argon (solid curves) and LJ fluid (dashed curves). Transport properties of LJ fluid are calculated using the approach from Ref.~\cite{BellJPCB2019}; transport properties of liquefied argon are taken from the REFPROP 10 database~\cite{Refprop}.  }
\label{FigLJvsArgon}
\end{figure}    

Figure~\ref{FigLJvsArgon} provides comparison between the shear viscosity and thermal conductivity coefficients of liquefied argon along $T=226$ K isotherm and LJ fluid with $T_*=2$. LJ data are shown by dashed curves and are calculated using the approach from Ref.~\cite{BellJPCB2019}. Argon data are shown by the solid curves and are taken from the Institute of Standards and Technology (NIST) Reference Fluid Thermodynamic and Transport Properties Database (REFPROP 10)~\cite{Refprop}. These are based on the original model described in Ref.~\cite{LemmonIJT2004}. It is observed that the solid and dashed curves nearly coincide implying that the transport properties of LJ and argon fluid are very similar. The situation is essentially the same for liquefied krypton and xenon~\cite{KhrapakJCP2022_1}. Note that the transport coefficients are plotted as functions of the reduced density $\rho/\rho_{\rm fr}$, demonstrating the success of the freezing density scaling. Remarkably, no information about LJ potential parameters $\sigma$ and $\epsilon$ was required to produce Fig.~\ref{FigLJvsArgon}. Taking further into account that the the functional forms of the freezing density scaling for the LJ and HS fluids are relatively close, it might be expected that the similarity observed in Fig.~\ref{FigLJvsArgon} would arise for a wide class of systems with relatively steep interactions and apply beyond the trivial case of liquefied noble gases and LJ potential.

The density dependence of the shear viscosity and thermal conductivity coefficients is qualitatively similar to other model systems considered above. They exhibit minima at an intermediate density $\rho/\rho_{\rm fr}\sim 0.3$ and then increase towards the freezing point. The values of $\eta_{\rm R}$ and $\lambda_{\rm R}$ at their respective minima can be estimated from simple classical arguments. For monatomic liquids these minimal value appear quasi-universal: for typical model and real fluids considered in Ref.~\cite{KhrapakPoF2022} it is observed that $\eta_{\rm R}^{\rm min}\simeq 0.6\pm 0.1$ and $\lambda_{\rm R}^{\rm min}\simeq 2.8\pm 0.2$. The only exception identified corresponds to the Coulomb one-component plasma fluid, where these minimal values are considerably lower ($\eta_{\rm R}^{\rm min}\sim 0.3$ and $\lambda_{\rm R}^{\rm min}\sim 1.2$). This difference was attributed to the extremely soft and long-ranged character of the Coulomb potential~\cite{KhrapakPoF2022}. It has been also observed that for molecular liquids, the minimal value of $\eta_{\rm R}$ is close to that in monatomic liquids. In contrast, the minimal value of $\lambda_{\rm R}$ somewhat increases, presumably due to the contribution from additional degrees of freedom to the energy transfer.    

As discussed previously, minima in $\eta_{\rm R}$ and $\lambda_{\rm R}$ signal the transition between different mechanisms of momentum and energy transfer and can be used as indicators of the gas-to-liquid dynamical crossover (Frenkel line). Figure~\ref{FigLJvsArgon} suggests that the minimum in $\lambda_{\rm R}$ is perhaps more convenient because the minimum in $\eta_{\rm R}$ seems rather shallow.  The vibrational model  starts to operate at densities above the minima and this is where the SE relation is expected to work (see Fig.~\ref{Fig6}). Interestingly, the applicability of SE relation is not limited to simplest monatomic model and real liquids.  
Several important non-spherical molecular liquids have been examined using numerical simulations in Ref.~\cite{OhtoriChemLett2020} and the applicability  of the SE relation has been confirmed. A few recent examples confirming the applicability of SE relation to real liquids include liquid iron at
conditions of planetary cores~\cite{LiJCP2021}, dense supercritical methane (at least for the most state points investigated)~\cite{Ranieri2021,KhrapakJMolLiq2022}, silicon melt at high temperatures~\cite{Luo2022}, and liquid water as modelled by the TIP4P/Ice model~\cite{BaranJCP2023,KhrapakJCP2023} (the TIP4P/Ice model was
specifically designed to cope with water near the fluid-solid phase transition and solid-phase properties~\cite{AbascalJCP2005}).    

The vibrational model does not allow us to estimate the values of diffusion and viscosity coefficients individually, but only their product by means of the SE relation. Nevertheless, it appears that $D_{\rm R}$ and $\eta_{\rm R}$ at freezing are indeed quasi-universal for many model and real fluids. As an example, in Table~\ref{Tab2} the values of $D_{\rm R}$, $\eta_{\rm R}$ and their product    
$D_{\rm R}\eta_{\rm R}=\alpha_{\rm SE}$ are presented for several liquid metals at their respective melting temperatures (this corresponds to the freezing point in our notation). The data are taken from Refs.~\cite{KhrapakAIPAdv2018,KhrapakJPCO2018}, where they were calculated using the original data from Ref.~\cite{MarchBook}. In a later paper~\cite{KhrapakMolPhys2019} the coefficient $\alpha_{\rm SE}$ was re-evaluated using the data from Ref.~\cite{Battezzati1989} and the results for Na, K, Rb, and Cu decreased to meet the theoretical expectation. However, there are still examples of unexpectedly high SE coefficients. For example, $\alpha_{\rm SE}\simeq 0.21$ for Zn, $\alpha_{\rm SE}\simeq 0.19$ for Hg, $\alpha_{\rm SE}\simeq 0.22$
for Ga, $\alpha_{\rm SE}\simeq 0.19$ for Sn, and $\alpha_{\rm SE}\simeq 0.22$ for Pb, according to the
data tabulated in Ref.~\cite{Battezzati1989}. Whether there are some
physical explanations behind these significant deviations, or
this merely reflects the quality of the available data needs to be clarified. 

We remind that the thermal conductivity of liquid metals is usually determined by the electron component and thus is not considered here. The same is true for a conventional multicomponent plasma~\cite{KumarPRE2021}. Electrical and thermal conductivity are mainly regulated by the electrons. The results discussed in this work could be relevant to the viscosity of strongly coupled plasmas, but this requires further analysis.      

\begin{table}
\caption{\label{Tab2} Reduced diffusion ($D_{\rm R}$), shear viscosity ($\eta_{\rm R}$)  coefficients and their product (SE coefficient $\alpha_{\rm SE}$) for several liquid metals at their corresponding melting temperatures as calculated from the data summarized in Ref.~\onlinecite{MarchBook} ($D_{\rm R}$ and $\eta_{\rm R}$ are taken from Refs.~\cite{KhrapakJPCO2018} and \cite{KhrapakAIPAdv2018}, respectively).  }
\begin{ruledtabular}
\begin{tabular}{lrrrrrrr}
Metal & Li & Na & K & Rb & Cu & Ag &  In   \\ \hline
$D_{\rm R}$ & 0.029 & 0.033 & 0.032 & 0.034 &0.04 & 0.031 & 0.032 \\
$\eta_{\rm R}$ & 5.6 & 5.9 & 5.7 & 5.7 & 5.1 & 5.1 & 5.1  \\
$\alpha_{\rm SE}$ & 0.16 & 0.19 & 0.18 & 0.19 & 0.20 & 0.16 & 0.16 \\
\end{tabular}
\end{ruledtabular}
\end{table}  

Let us provide some more reference numbers for completeness. The viscosity coefficient of liquefied noble gases on the freezing line is usually contained in a relatively narrow range $\eta_{\rm R}\simeq 5\pm 1$~\cite{KhrapakPoF2022,KhrapakAIPAdv2018,CostigliolaJCP2018}. Deviations from this value can occur for molecular liquids (for example, $\eta_{\rm R}\simeq 6.9$ at freezing of N$_2$ at 10 MPa and $\eta_{\rm R}\simeq 2.92$ at freezing of CO$_2$ at 30 MPa~\cite{KhrapakPoF2022}). For Yukawa fluids $D_{\rm R}\simeq 0.03$ and $\eta_{\rm R}\simeq 4.8$~\cite{KhrapakJPCO2018,KhrapakAIPAdv2018}, resulting in $\alpha_{\rm SE}\simeq 0.14$. For HS fluids at freezing we have $D_{\rm R}\simeq 0.025$ and $\eta_{\rm R}\simeq 6.5$~\cite{Pieprzyk2019,Pieprzyk2020}, resulting in $\alpha_{\rm SE}\simeq 0.16$. 

The vibrational model of heat
transfer combined with the simplest version of Lindemann’s melting
criterion allows for a rough but reasonable estimate  of the reduced heat conductivity coefficient at the freezing point [Eq.~(\ref{lambda8})]. As a reference, the thermal conductivity coefficient of liquefied noble gases at freezing is usually within the range $\lambda_{\rm R}\simeq 10\pm 1$~\cite{KhrapakPoF2022,KhrapakJCP2022_1}. It can increase for molecular liquids (for example, $\lambda_{\rm R}\simeq 13.15$ at freezing of N$_2$ at 10 MPa)~\cite{KhrapakPoF2022}. Regarding the density dependence of $\lambda_{\rm R}$, the detailed comparison is complicated by the fact that neither the Einstein frequency [to use in Eq.~(\ref{Horrocks})], nor the longitudinal and transverse sound velocity [to use in Eq.~(\ref{Cahill1})] of real fluids are directly available. However, there is a simple and useful alternative, known as Bridgman formula. 
Bridgman proposed this formula about a century ago. Using elementary consideration he expressed the coefficient of thermal conductivity via the sound velocity and the density as~\cite{Bridgman1923}:
\begin{equation}\label{tc1}
\lambda = \frac{2c_s}{\Delta^{2}},
\end{equation}
where $c_s$ is the sound velocity. Basically, Bridgman postulated that the energy is transferred at the speed of sound, which leads to a linear correlation between the thermal conductivity coefficient and the sound velocity as in Eq.~(\ref{tc1}). Similar correlation can be obtained within a vibrational model of heat transfer assuming that fluid supports only a sound-like longitudinal mode~\cite{KhrapakPRE01_2021}. A fixed numerical coefficient has remained somewhat contradictory, values between 2 and 3 were used in the literature~\cite{ZhaoJAP2021,XiCPL2020,KhrapakPRE01_2021,BirdBook}.

In a recent paper a systematic analysis of correlations between thermal conductivity  and sound velocity in simple model systems (hard sphere and Lennard-Jones fluids), monatomic liquids (argon and krypton), diatomic liquids (nitrogen and oxygen), and several polyatomic liquids (water, carbon dioxide, methane, and ethane) has been performed~\cite{KhrapakJMolLiq2023}. Correlation between the thermal conductivity and the sound velocity at freezing conditions have also been investigated~\cite{KhrapakPoF2023}.   The main result is that the linear correlations are observed for model fluids as well as real monatomic and diatomic liquids, but seem less convincing in polyatomic molecular liquids. The fixed coefficient of proportionality is not a good choice. In fact, this coefficient is about unity for monatomic liquids and generally increases with molecular complexity. Although, Bridgman formula cannot provide more than a very rough order of magnitude estimate, it can still be useful as a first approximation, because the information about the sound velocities is usually readily available. 

Another potentially useful correlation is the Leibfried-Schl\"omann equation for the pressure dependence of thermal conductivity of solids~\cite{Leibfried1954,ChenPRB2011}. To the best of my knowledge a possibility to apply it in the fluid regime (possibly with some modifications) has not yet been discussed.   

\section{Conclusion}\label{Concl}

A coherent and internally consistent picture of transport processes in dense fluids has been discussed. According to the vibrational dynamical paradigm, on short time scales atoms exhibit solid-like  oscillations around their temporary equilibrium positions, while the latter diffuse on much longer time scales. This approach allows us to predict the coefficients of self-diffusion, shear viscosity, and thermal conductivity, which are related to the properties of collective excitations supported in fluids and can be expressed using certain averages over the excitation frequency spectrum.

For example, self diffusion can be considered as random walk process. Its characteristic length scale is given by short-time mean-square displacement of an atom from its local equilibrium position due to solid-like vibrations. The characteristic time scale is given by the Maxwellian relaxation time. As a result, the product of the self-diffusion and viscosity coefficients can be expressed as $D\eta\propto \langle \omega^{-2}\rangle$. This leads to the Stokes-Einstein relation without the hydrodynamic diameter (hydrodynamic diameter is essentially replaced by the average interatomic separation) of the form $D\eta(\Delta/T)\simeq\alpha_{\rm SE}$, with a weakly system-dependent coefficient $\alpha_{\rm SE}$ determined by the properties of collective modes.

Within the vibrational model paradigm, heat transport occurs at an average excitation frequency $\langle \omega \rangle$, which again depends on the properties of the excitation spectrum. Interestingly, the excess entropy associated with solid-like oscillations can be related to $\langle \ln \omega^2\rangle$ and this provides a direct link for the excess entropy scaling of the thermal conductivity coefficient.

Application of the vibrational model to various simple systems such as plasma-related Coulomb and screened Coulomb (Yukawa) fluids, the Lennard-Jones fluid, and the hard-sphere fluid has been considered in detail. Comparison with available numerical data on the transport coefficients of different soft-interacting fluids demonstrates good to excellent agreement. The vibrational model is not designed for absolutely anharmonic HS interactions, where the assumed vibrational picture of atomic dynamics is clearly violated. Interestingly enough, however, the SE relation and an expression for the thermal conductivity coefficient that relates the average frequency to the instantaneous elastic moduli remain approximately valid even in this extreme case. Many reference properties following from vibrational picture of atomic dynamics are shared by real atomic and molecular liquids.   

Conditions of applicability of the vibrational model as well as the related question regarding the location of the gas-liquid crossover have been discussed. The excess entropy appears as a convenient parameter, regulating the dynamical behavior. Two important transition points emerge. The first, at $s_{\rm ex}\simeq -1$, corresponds to gas-liquid dynamical crossover, where gas-like and liquid-like asymptotes of some transport properties (such as macroscopically reduced shear viscosity coefficient or reduced SE product) intersect. The second, at $s_{\rm ex}\simeq -2.0$ corresponds to the onset of validity of the vibrational model. The model remains valid up to the liquid-solid phase transitions, which is characterised by $s_{\rm ex}\simeq -4$ for the systems considered here.                   

In future studies it would be interesting to ascertain to which extent the simple picture discussed above applies to a wider range of atomic and molecular liquids, mixtures, multicomponent plasmas, and other related systems. This can be of interest in the context of various disciplines including in particular physics of fluids, physics of plasmas, soft condensed matter, and materials science.    

\acknowledgments

I would like to thank Lenaic Couedel, Alexey Khrapak, Boris Klumov, Stanislav Yurchenko, with whom I collaborated on some of the topics discussed in this paper and my wife Natalia Khrapak for continued support. 


\bibliography{PhysRep}

\providecommand{\noopsort}[1]{}\providecommand{\singleletter}[1]{#1}%
\begin{thebibliography}{257}%
\makeatletter
\providecommand \@ifxundefined [1]{%
 \@ifx{#1\undefined}
}%
\providecommand \@ifnum [1]{%
 \ifnum #1\expandafter \@firstoftwo
 \else \expandafter \@secondoftwo
 \fi
}%
\providecommand \@ifx [1]{%
 \ifx #1\expandafter \@firstoftwo
 \else \expandafter \@secondoftwo
 \fi
}%
\providecommand \natexlab [1]{#1}%
\providecommand \enquote  [1]{``#1''}%
\providecommand \bibnamefont  [1]{#1}%
\providecommand \bibfnamefont [1]{#1}%
\providecommand \citenamefont [1]{#1}%
\providecommand \href@noop [0]{\@secondoftwo}%
\providecommand \href [0]{\begingroup \@sanitize@url \@href}%
\providecommand \@href[1]{\@@startlink{#1}\@@href}%
\providecommand \@@href[1]{\endgroup#1\@@endlink}%
\providecommand \@sanitize@url [0]{\catcode `\\12\catcode `\$12\catcode
  `\&12\catcode `\#12\catcode `\^12\catcode `\_12\catcode `\%12\relax}%
\providecommand \@@startlink[1]{}%
\providecommand \@@endlink[0]{}%
\providecommand \url  [0]{\begingroup\@sanitize@url \@url }%
\providecommand \@url [1]{\endgroup\@href {#1}{\urlprefix }}%
\providecommand \urlprefix  [0]{URL }%
\providecommand \Eprint [0]{\href }%
\providecommand \doibase [0]{http://dx.doi.org/}%
\providecommand \selectlanguage [0]{\@gobble}%
\providecommand \bibinfo  [0]{\@secondoftwo}%
\providecommand \bibfield  [0]{\@secondoftwo}%
\providecommand \translation [1]{[#1]}%
\providecommand \BibitemOpen [0]{}%
\providecommand \bibitemStop [0]{}%
\providecommand \bibitemNoStop [0]{.\EOS\space}%
\providecommand \EOS [0]{\spacefactor3000\relax}%
\providecommand \BibitemShut  [1]{\csname bibitem#1\endcsname}%
\let\auto@bib@innerbib\@empty
\bibitem [{\citenamefont {Frenkel}(1955)}]{FrenkelBook}%
  \BibitemOpen
  \bibfield  {author} {\bibinfo {author} {\bibfnamefont {Y.}~\bibnamefont
  {Frenkel}},\ }\href {https://cds.cern.ch/record/106808} {\emph {\bibinfo
  {title} {{Kinetic theory of liquids}}}}\ (\bibinfo  {publisher} {Dover},\
  \bibinfo {address} {New York, NY},\ \bibinfo {year} {1955})\BibitemShut
  {NoStop}%
\bibitem [{\citenamefont {Groot}\ and\ \citenamefont
  {Mazur}(1984)}]{GrootBook}%
  \BibitemOpen
  \bibfield  {author} {\bibinfo {author} {\bibfnamefont {S.~R.}\ \bibnamefont
  {Groot}}\ and\ \bibinfo {author} {\bibfnamefont {P.}~\bibnamefont {Mazur}},\
  }\href@noop {} {\emph {\bibinfo {title} {Non-equilibrium Thermodynamics}}}\
  (\bibinfo  {publisher} {Courier Corporation},\ \bibinfo {address} {New
  York},\ \bibinfo {year} {1984})\BibitemShut {NoStop}%
\bibitem [{\citenamefont {Balucani}\ and\ \citenamefont
  {Zoppi}(1994)}]{BalucaniBook}%
  \BibitemOpen
  \bibfield  {author} {\bibinfo {author} {\bibfnamefont {U.}~\bibnamefont
  {Balucani}}\ and\ \bibinfo {author} {\bibfnamefont {M.}~\bibnamefont
  {Zoppi}},\ }\href@noop {} {\emph {\bibinfo {title} {Dynamics of the Liquid
  State}}}\ (\bibinfo  {publisher} {Clarendon Press},\ \bibinfo {address}
  {Oxford},\ \bibinfo {year} {1994})\BibitemShut {NoStop}%
\bibitem [{\citenamefont {March}\ and\ \citenamefont {Tosi}(2002)}]{MarchBook}%
  \BibitemOpen
  \bibfield  {author} {\bibinfo {author} {\bibfnamefont {N.~H.}\ \bibnamefont
  {March}}\ and\ \bibinfo {author} {\bibfnamefont {M.~P.}\ \bibnamefont
  {Tosi}},\ }\href@noop {} {\emph {\bibinfo {title} {Introduction to Liquid
  State Physics}}}\ (\bibinfo  {publisher} {World Scientific Pub Co Inc},\
  \bibinfo {year} {2002})\BibitemShut {NoStop}%
\bibitem [{\citenamefont {Hansen}\ and\ \citenamefont
  {McDonald}(2006)}]{HansenBook}%
  \BibitemOpen
  \bibfield  {author} {\bibinfo {author} {\bibfnamefont {J.-P.}\ \bibnamefont
  {Hansen}}\ and\ \bibinfo {author} {\bibfnamefont {I.~R.}\ \bibnamefont
  {McDonald}},\ }\href@noop {} {\emph {\bibinfo {title} {Theory of Simple
  Liquids -}}}\ (\bibinfo  {publisher} {Elsevier},\ \bibinfo {address}
  {Amsterdam},\ \bibinfo {year} {2006})\BibitemShut {NoStop}%
\bibitem [{\citenamefont {Lifshitz}\ and\ \citenamefont
  {Pitaevskii}(1995)}]{LifshitzKinetics}%
  \BibitemOpen
  \bibfield  {author} {\bibinfo {author} {\bibfnamefont {E.M.}\ \bibnamefont
  {Lifshitz}}\ and\ \bibinfo {author} {\bibfnamefont {L.~P.}\ \bibnamefont
  {Pitaevskii}},\ }\href@noop {} {\emph {\bibinfo {title} {Physical
  Kinetics}}}\ (\bibinfo  {publisher} {Elsevier Science},\ \bibinfo {address}
  {Stanford},\ \bibinfo {year} {1995})\BibitemShut {NoStop}%
\bibitem [{\citenamefont {Brazhkin}(2017)}]{BrazhkinUFN2017}%
  \BibitemOpen
  \bibfield  {author} {\bibinfo {author} {\bibfnamefont {V~V}\ \bibnamefont
  {Brazhkin}},\ }\bibfield  {title} {\enquote {\bibinfo {title} {Phase
  transformations in liquids and the liquid{\textendash}gas transition in
  fluids at supercritical pressures},}\ }\href {\doibase
  10.3367/ufne.2016.12.038118} {\bibfield  {journal} {\bibinfo  {journal}
  {Phys.-Usp.}\ }\textbf {\bibinfo {volume} {60}},\ \bibinfo {pages} {954--957}
  (\bibinfo {year} {2017})}\BibitemShut {NoStop}%
\bibitem [{\citenamefont {Ziman}(2001)}]{ZimanBook}%
  \BibitemOpen
  \bibfield  {author} {\bibinfo {author} {\bibfnamefont {J~M}\ \bibnamefont
  {Ziman}},\ }\href@noop {} {\emph {\bibinfo {title} {Electrons and phonons:
  {T}he Theory of Transport Phenomena in Solids}}},\ Oxford Classic Texts in
  the Physical Sciences\ (\bibinfo  {publisher} {Oxford University Press},\
  \bibinfo {address} {London, England},\ \bibinfo {year} {2001})\BibitemShut
  {NoStop}%
\bibitem [{\citenamefont {Klemens}(1993)}]{Klemens1993}%
  \BibitemOpen
  \bibfield  {author} {\bibinfo {author} {\bibfnamefont {P.G.}\ \bibnamefont
  {Klemens}},\ }\bibfield  {title} {\enquote {\bibinfo {title} {Heat conduction
  in solids by phonons},}\ }\href@noop {} {\bibfield  {journal} {\bibinfo
  {journal} {Thermochimica Acta}\ }\textbf {\bibinfo {volume} {218}},\ \bibinfo
  {pages} {247–255} (\bibinfo {year} {1993})}\BibitemShut {NoStop}%
\bibitem [{\citenamefont {Chapman}\ and\ \citenamefont
  {Cowling}(1990)}]{ChapmanBook}%
  \BibitemOpen
  \bibfield  {author} {\bibinfo {author} {\bibfnamefont {S.}~\bibnamefont
  {Chapman}}\ and\ \bibinfo {author} {\bibfnamefont {T.~G.}\ \bibnamefont
  {Cowling}},\ }\href@noop {} {\emph {\bibinfo {title} {The Mathematical Theory
  of Non-uniform Gases - An Account of the Kinetic Theory of Viscosity, Thermal
  Conduction and Diffusion in Gases}}}\ (\bibinfo  {publisher} {Cambridge
  University Press},\ \bibinfo {address} {Cambridge},\ \bibinfo {year}
  {1990})\BibitemShut {NoStop}%
\bibitem [{\citenamefont {Lemmon}\ \emph {et~al.}(2018)\citenamefont {Lemmon},
  \citenamefont {Bell}, \citenamefont {Huber},\ and\ \citenamefont
  {McLinden}}]{Refprop}%
  \BibitemOpen
  \bibfield  {author} {\bibinfo {author} {\bibfnamefont {E.~W.}\ \bibnamefont
  {Lemmon}}, \bibinfo {author} {\bibfnamefont {I.H.}\ \bibnamefont {Bell}},
  \bibinfo {author} {\bibfnamefont {M.~L.}\ \bibnamefont {Huber}}, \ and\
  \bibinfo {author} {\bibfnamefont {M.~O.}\ \bibnamefont {McLinden}},\ }\href
  {\doibase https://doi.org/10.18434/T4/1502528} {\enquote {\bibinfo {title}
  {{NIST Standard Reference Database 23: Reference Fluid Thermodynamic and
  Transport Properties-REFPROP, Version 10.0, National Institute of Standards
  and Technology}},}\ } (\bibinfo {year} {2018})\BibitemShut {NoStop}%
\bibitem [{\citenamefont {Rosenfeld}(1977)}]{RosenfeldPRA1977}%
  \BibitemOpen
  \bibfield  {author} {\bibinfo {author} {\bibfnamefont {Y.}~\bibnamefont
  {Rosenfeld}},\ }\bibfield  {title} {\enquote {\bibinfo {title} {Relation
  between the transport coefficients and the internal entropy of simple
  systems},}\ }\href {\doibase 10.1103/physreva.15.2545} {\bibfield  {journal}
  {\bibinfo  {journal} {Phys. Rev. A}\ }\textbf {\bibinfo {volume} {15}},\
  \bibinfo {pages} {2545--2549} (\bibinfo {year} {1977})}\BibitemShut {NoStop}%
\bibitem [{\citenamefont {Rosenfeld}(1999)}]{RosenfeldJPCM1999}%
  \BibitemOpen
  \bibfield  {author} {\bibinfo {author} {\bibfnamefont {Y.}~\bibnamefont
  {Rosenfeld}},\ }\bibfield  {title} {\enquote {\bibinfo {title} {A
  quasi-universal scaling law for atomic transport in simple fluids},}\ }\href
  {\doibase 10.1088/0953-8984/11/28/303} {\bibfield  {journal} {\bibinfo
  {journal} {J. Phys.: Condens. Matter}\ }\textbf {\bibinfo {volume} {11}},\
  \bibinfo {pages} {5415--5427} (\bibinfo {year} {1999})}\BibitemShut {NoStop}%
\bibitem [{\citenamefont {Andrade}(1931)}]{Andrade1931}%
  \BibitemOpen
  \bibfield  {author} {\bibinfo {author} {\bibfnamefont {E.~N. DA~C.}\
  \bibnamefont {Andrade}},\ }\bibfield  {title} {\enquote {\bibinfo {title}
  {Viscosity of liquids},}\ }\href {\doibase 10.1038/128835a0} {\bibfield
  {journal} {\bibinfo  {journal} {Nature}\ }\textbf {\bibinfo {volume} {128}},\
  \bibinfo {pages} {835--835} (\bibinfo {year} {1931})}\BibitemShut {NoStop}%
\bibitem [{\citenamefont {da~C.~Andrade}(1952)}]{Andrade1952}%
  \BibitemOpen
  \bibfield  {author} {\bibinfo {author} {\bibfnamefont {E.~N.}\ \bibnamefont
  {da~C.~Andrade}},\ }\bibfield  {title} {\enquote {\bibinfo {title} {Viscosity
  and thermal conductivity of liquid argon},}\ }\href {\doibase
  10.1038/170794b0} {\bibfield  {journal} {\bibinfo  {journal} {Nature}\
  }\textbf {\bibinfo {volume} {170}},\ \bibinfo {pages} {794--794} (\bibinfo
  {year} {1952})}\BibitemShut {NoStop}%
\bibitem [{\citenamefont {Bell}(2019)}]{BellPNAS2019}%
  \BibitemOpen
  \bibfield  {author} {\bibinfo {author} {\bibfnamefont {I.~H.}\ \bibnamefont
  {Bell}},\ }\bibfield  {title} {\enquote {\bibinfo {title} {Probing the link
  between residual entropy and viscosity of molecular fluids and model
  potentials},}\ }\href {\doibase 10.1073/pnas.1815943116} {\bibfield
  {journal} {\bibinfo  {journal} {PNAS}\ }\textbf {\bibinfo {volume} {116}},\
  \bibinfo {pages} {4070--4079} (\bibinfo {year} {2019})}\BibitemShut {NoStop}%
\bibitem [{\citenamefont {Bell}\ \emph {et~al.}(2019)\citenamefont {Bell},
  \citenamefont {Messerly}, \citenamefont {Thol}, \citenamefont {Costigliola},\
  and\ \citenamefont {Dyre}}]{BellJPCB2019}%
  \BibitemOpen
  \bibfield  {author} {\bibinfo {author} {\bibfnamefont {I.~H.}\ \bibnamefont
  {Bell}}, \bibinfo {author} {\bibfnamefont {R.}~\bibnamefont {Messerly}},
  \bibinfo {author} {\bibfnamefont {M.}~\bibnamefont {Thol}}, \bibinfo {author}
  {\bibfnamefont {L.}~\bibnamefont {Costigliola}}, \ and\ \bibinfo {author}
  {\bibfnamefont {J.~C.}\ \bibnamefont {Dyre}},\ }\bibfield  {title} {\enquote
  {\bibinfo {title} {Modified entropy scaling of the transport properties of
  the {L}ennard-{J}ones fluid},}\ }\href {\doibase 10.1021/acs.jpcb.9b05808}
  {\bibfield  {journal} {\bibinfo  {journal} {J. Phys. Chem. B}\ }\textbf
  {\bibinfo {volume} {123}},\ \bibinfo {pages} {6345--6363} (\bibinfo {year}
  {2019})}\BibitemShut {NoStop}%
\bibitem [{\citenamefont {Krekelberg}\ \emph
  {et~al.}(2009{\natexlab{a}})\citenamefont {Krekelberg}, \citenamefont
  {Kumar}, \citenamefont {Mittal}, \citenamefont {Errington},\ and\
  \citenamefont {Truskett}}]{KrekelbergPRE03_2009}%
  \BibitemOpen
  \bibfield  {author} {\bibinfo {author} {\bibfnamefont {W.~P.}\ \bibnamefont
  {Krekelberg}}, \bibinfo {author} {\bibfnamefont {T.}~\bibnamefont {Kumar}},
  \bibinfo {author} {\bibfnamefont {J.}~\bibnamefont {Mittal}}, \bibinfo
  {author} {\bibfnamefont {J.~R.}\ \bibnamefont {Errington}}, \ and\ \bibinfo
  {author} {\bibfnamefont {T.~M.}\ \bibnamefont {Truskett}},\ }\bibfield
  {title} {\enquote {\bibinfo {title} {Anomalous structure and dynamics of the
  {G}aussian-core fluid},}\ }\href {\doibase 10.1103/physreve.79.031203}
  {\bibfield  {journal} {\bibinfo  {journal} {Phys. Rev. E}\ }\textbf {\bibinfo
  {volume} {79}},\ \bibinfo {pages} {031203} (\bibinfo {year}
  {2009}{\natexlab{a}})}\BibitemShut {NoStop}%
\bibitem [{\citenamefont {Krekelberg}\ \emph
  {et~al.}(2009{\natexlab{b}})\citenamefont {Krekelberg}, \citenamefont {Pond},
  \citenamefont {Goel}, \citenamefont {Shen}, \citenamefont {Errington},\ and\
  \citenamefont {Truskett}}]{KrekelbergPRE12_2009}%
  \BibitemOpen
  \bibfield  {author} {\bibinfo {author} {\bibfnamefont {W.~P.}\ \bibnamefont
  {Krekelberg}}, \bibinfo {author} {\bibfnamefont {M.~J.}\ \bibnamefont
  {Pond}}, \bibinfo {author} {\bibfnamefont {G.}~\bibnamefont {Goel}}, \bibinfo
  {author} {\bibfnamefont {V.~K.}\ \bibnamefont {Shen}}, \bibinfo {author}
  {\bibfnamefont {J.~R.}\ \bibnamefont {Errington}}, \ and\ \bibinfo {author}
  {\bibfnamefont {T.~M.}\ \bibnamefont {Truskett}},\ }\bibfield  {title}
  {\enquote {\bibinfo {title} {Generalized {R}osenfeld scalings for tracer
  diffusivities in not-so-simple fluids: Mixtures and soft particles},}\ }\href
  {\doibase 10.1103/physreve.80.061205} {\bibfield  {journal} {\bibinfo
  {journal} {Phys. Rev. E}\ }\textbf {\bibinfo {volume} {80}},\ \bibinfo
  {pages} {061205} (\bibinfo {year} {2009}{\natexlab{b}})}\BibitemShut
  {NoStop}%
\bibitem [{\citenamefont {Fomin}\ \emph {et~al.}(2010)\citenamefont {Fomin},
  \citenamefont {Ryzhov},\ and\ \citenamefont {Gribova}}]{FominPRE2010}%
  \BibitemOpen
  \bibfield  {author} {\bibinfo {author} {\bibfnamefont {Yu.~D.}\ \bibnamefont
  {Fomin}}, \bibinfo {author} {\bibfnamefont {V.~N.}\ \bibnamefont {Ryzhov}}, \
  and\ \bibinfo {author} {\bibfnamefont {N.~V.}\ \bibnamefont {Gribova}},\
  }\bibfield  {title} {\enquote {\bibinfo {title} {Breakdown of excess entropy
  scaling for systems with thermodynamic anomalies},}\ }\href {\doibase
  10.1103/physreve.81.061201} {\bibfield  {journal} {\bibinfo  {journal} {Phys.
  Rev. E}\ }\textbf {\bibinfo {volume} {81}},\ \bibinfo {pages} {061201}
  (\bibinfo {year} {2010})}\BibitemShut {NoStop}%
\bibitem [{\citenamefont {Dyre}(2014)}]{DyreJPCB2014}%
  \BibitemOpen
  \bibfield  {author} {\bibinfo {author} {\bibfnamefont {J.~C.}\ \bibnamefont
  {Dyre}},\ }\bibfield  {title} {\enquote {\bibinfo {title} {Hidden scale
  invariance in condensed matter},}\ }\href {\doibase 10.1021/jp501852b}
  {\bibfield  {journal} {\bibinfo  {journal} {J. Phys. Chem. B}\ }\textbf
  {\bibinfo {volume} {118}},\ \bibinfo {pages} {10007--10024} (\bibinfo {year}
  {2014})}\BibitemShut {NoStop}%
\bibitem [{\citenamefont {Dyre}(2018)}]{DyreJCP2018}%
  \BibitemOpen
  \bibfield  {author} {\bibinfo {author} {\bibfnamefont {J.~C.}\ \bibnamefont
  {Dyre}},\ }\bibfield  {title} {\enquote {\bibinfo {title} {Perspective:
  {E}xcess-entropy scaling},}\ }\href {\doibase 10.1063/1.5055064} {\bibfield
  {journal} {\bibinfo  {journal} {J. Chem. Phys.}\ }\textbf {\bibinfo {volume}
  {149}},\ \bibinfo {pages} {210901} (\bibinfo {year} {2018})}\BibitemShut
  {NoStop}%
\bibitem [{\citenamefont {Gnan}\ \emph {et~al.}(2009)\citenamefont {Gnan},
  \citenamefont {Schroder}, \citenamefont {Pedersen}, \citenamefont {Bailey},\
  and\ \citenamefont {Dyre}}]{GnanJCP2009}%
  \BibitemOpen
  \bibfield  {author} {\bibinfo {author} {\bibfnamefont {N.}~\bibnamefont
  {Gnan}}, \bibinfo {author} {\bibfnamefont {T.~B.}\ \bibnamefont {Schroder}},
  \bibinfo {author} {\bibfnamefont {U.~R.}\ \bibnamefont {Pedersen}}, \bibinfo
  {author} {\bibfnamefont {N.~P.}\ \bibnamefont {Bailey}}, \ and\ \bibinfo
  {author} {\bibfnamefont {J.~C.}\ \bibnamefont {Dyre}},\ }\bibfield  {title}
  {\enquote {\bibinfo {title} {Pressure-energy correlations in liquids. {IV}.
  {I}somorphs in liquid phase diagrams},}\ }\href {\doibase 10.1063/1.3265957}
  {\bibfield  {journal} {\bibinfo  {journal} {J. Chem. Phys.}\ }\textbf
  {\bibinfo {volume} {131}},\ \bibinfo {pages} {234504} (\bibinfo {year}
  {2009})}\BibitemShut {NoStop}%
\bibitem [{\citenamefont {Castello}\ \emph {et~al.}(2021)\citenamefont
  {Castello}, \citenamefont {Tolias},\ and\ \citenamefont
  {Dyre}}]{CastelloJCP2021}%
  \BibitemOpen
  \bibfield  {author} {\bibinfo {author} {\bibfnamefont {F.~Lucco}\
  \bibnamefont {Castello}}, \bibinfo {author} {\bibfnamefont {P.}~\bibnamefont
  {Tolias}}, \ and\ \bibinfo {author} {\bibfnamefont {J.~C.}\ \bibnamefont
  {Dyre}},\ }\bibfield  {title} {\enquote {\bibinfo {title} {Testing the
  isomorph invariance of the bridge functions of {Y}ukawa one-component
  plasmas},}\ }\href {\doibase 10.1063/5.0036226} {\bibfield  {journal}
  {\bibinfo  {journal} {J. Chem. Phys.}\ }\textbf {\bibinfo {volume} {154}},\
  \bibinfo {pages} {034501} (\bibinfo {year} {2021})}\BibitemShut {NoStop}%
\bibitem [{\citenamefont {Rahman}\ \emph {et~al.}(2021)\citenamefont {Rahman},
  \citenamefont {Carter}, \citenamefont {Saw}, \citenamefont {Douglass},
  \citenamefont {Costigliola}, \citenamefont {Ingebrigtsen}, \citenamefont
  {Schr{\o}der}, \citenamefont {Pedersen},\ and\ \citenamefont
  {Dyre}}]{RahmanMolecules2021}%
  \BibitemOpen
  \bibfield  {author} {\bibinfo {author} {\bibfnamefont {M.}~\bibnamefont
  {Rahman}}, \bibinfo {author} {\bibfnamefont {B.~M. G.~D.}\ \bibnamefont
  {Carter}}, \bibinfo {author} {\bibfnamefont {S.}~\bibnamefont {Saw}},
  \bibinfo {author} {\bibfnamefont {I.~M.}\ \bibnamefont {Douglass}}, \bibinfo
  {author} {\bibfnamefont {L.}~\bibnamefont {Costigliola}}, \bibinfo {author}
  {\bibfnamefont {T.~S.}\ \bibnamefont {Ingebrigtsen}}, \bibinfo {author}
  {\bibfnamefont {T.~B.}\ \bibnamefont {Schr{\o}der}}, \bibinfo {author}
  {\bibfnamefont {U.~R.}\ \bibnamefont {Pedersen}}, \ and\ \bibinfo {author}
  {\bibfnamefont {J.~C.}\ \bibnamefont {Dyre}},\ }\bibfield  {title} {\enquote
  {\bibinfo {title} {Isomorph invariance of higher-order structural measures in
  four {L}ennard-{J}ones systems},}\ }\href {\doibase
  10.3390/molecules26061746} {\bibfield  {journal} {\bibinfo  {journal}
  {Molecules}\ }\textbf {\bibinfo {volume} {26}},\ \bibinfo {pages} {1746}
  (\bibinfo {year} {2021})}\BibitemShut {NoStop}%
\bibitem [{\citenamefont {Khrapak}\ and\ \citenamefont
  {Khrapak}(2021{\natexlab{a}})}]{KhrapakJETPLett2021}%
  \BibitemOpen
  \bibfield  {author} {\bibinfo {author} {\bibfnamefont {S.~A.}\ \bibnamefont
  {Khrapak}}\ and\ \bibinfo {author} {\bibfnamefont {A.~G.}\ \bibnamefont
  {Khrapak}},\ }\bibfield  {title} {\enquote {\bibinfo {title} {Correlations
  between the shear viscosity and thermal conductivity coefficients of dense
  simple liquids},}\ }\href {\doibase 10.1134/s0021364021210037} {\bibfield
  {journal} {\bibinfo  {journal} {{JETP} Lett.}\ }\textbf {\bibinfo {volume}
  {114}},\ \bibinfo {pages} {540} (\bibinfo {year}
  {2021}{\natexlab{a}})}\BibitemShut {NoStop}%
\bibitem [{\citenamefont {Dzugutov}(1996)}]{DzugutovNature1996}%
  \BibitemOpen
  \bibfield  {author} {\bibinfo {author} {\bibfnamefont {M.}~\bibnamefont
  {Dzugutov}},\ }\bibfield  {title} {\enquote {\bibinfo {title} {A universal
  scaling law for atomic diffusion in condensed matter},}\ }\href {\doibase
  10.1038/381137a0} {\bibfield  {journal} {\bibinfo  {journal} {Nature}\
  }\textbf {\bibinfo {volume} {381}},\ \bibinfo {pages} {137--139} (\bibinfo
  {year} {1996})}\BibitemShut {NoStop}%
\bibitem [{\citenamefont {Laird}\ and\ \citenamefont
  {Haymet}(1992)}]{LairdPRA1992}%
  \BibitemOpen
  \bibfield  {author} {\bibinfo {author} {\bibfnamefont {B.~B.}\ \bibnamefont
  {Laird}}\ and\ \bibinfo {author} {\bibfnamefont {A.~D.~J.}\ \bibnamefont
  {Haymet}},\ }\bibfield  {title} {\enquote {\bibinfo {title} {Calculation of
  the entropy from multiparticle correlation functions},}\ }\href {\doibase
  10.1103/physreva.45.5680} {\bibfield  {journal} {\bibinfo  {journal} {Phys.
  Rev. A}\ }\textbf {\bibinfo {volume} {45}},\ \bibinfo {pages} {5680--5689}
  (\bibinfo {year} {1992})}\BibitemShut {NoStop}%
\bibitem [{\citenamefont {Giaquinta}\ and\ \citenamefont
  {Giunta}(1992)}]{GiaquintaPhysA1992}%
  \BibitemOpen
  \bibfield  {author} {\bibinfo {author} {\bibfnamefont {P.V.}\ \bibnamefont
  {Giaquinta}}\ and\ \bibinfo {author} {\bibfnamefont {G.}~\bibnamefont
  {Giunta}},\ }\bibfield  {title} {\enquote {\bibinfo {title} {About entropy
  and correlations in a fluid of hard spheres},}\ }\href {\doibase
  10.1016/0378-4371(92)90415-m} {\bibfield  {journal} {\bibinfo  {journal}
  {Phys. A}\ }\textbf {\bibinfo {volume} {187}},\ \bibinfo {pages} {145--158}
  (\bibinfo {year} {1992})}\BibitemShut {NoStop}%
\bibitem [{\citenamefont {Giaquinta}\ \emph {et~al.}(1992)\citenamefont
  {Giaquinta}, \citenamefont {Giunta},\ and\ \citenamefont
  {Giarritta}}]{GiaquintaPRA1992}%
  \BibitemOpen
  \bibfield  {author} {\bibinfo {author} {\bibfnamefont {P.~V.}\ \bibnamefont
  {Giaquinta}}, \bibinfo {author} {\bibfnamefont {G.}~\bibnamefont {Giunta}}, \
  and\ \bibinfo {author} {\bibfnamefont {S.~Prestipino}\ \bibnamefont
  {Giarritta}},\ }\bibfield  {title} {\enquote {\bibinfo {title} {Entropy and
  the freezing of simple liquids},}\ }\href {\doibase
  10.1103/physreva.45.r6966} {\bibfield  {journal} {\bibinfo  {journal} {Phys.
  Rev. A}\ }\textbf {\bibinfo {volume} {45}},\ \bibinfo {pages} {R6966--R6968}
  (\bibinfo {year} {1992})}\BibitemShut {NoStop}%
\bibitem [{\citenamefont {Saija}\ \emph {et~al.}(2006)\citenamefont {Saija},
  \citenamefont {Prestipino},\ and\ \citenamefont {Giaquinta}}]{SaijaJCP2006}%
  \BibitemOpen
  \bibfield  {author} {\bibinfo {author} {\bibfnamefont {F.}~\bibnamefont
  {Saija}}, \bibinfo {author} {\bibfnamefont {S.}~\bibnamefont {Prestipino}}, \
  and\ \bibinfo {author} {\bibfnamefont {P.~V.}\ \bibnamefont {Giaquinta}},\
  }\bibfield  {title} {\enquote {\bibinfo {title} {Evaluation of
  phenomenological one-phase criteria for the melting and freezing of softly
  repulsive particles},}\ }\href {\doibase 10.1063/1.2208357} {\bibfield
  {journal} {\bibinfo  {journal} {J. Chem. Phys.}\ }\textbf {\bibinfo {volume}
  {124}},\ \bibinfo {pages} {244504} (\bibinfo {year} {2006})}\BibitemShut
  {NoStop}%
\bibitem [{\citenamefont {Fomin}\ \emph {et~al.}(2014)\citenamefont {Fomin},
  \citenamefont {Ryzhov}, \citenamefont {Klumov},\ and\ \citenamefont
  {Tsiok}}]{FominJCP2014}%
  \BibitemOpen
  \bibfield  {author} {\bibinfo {author} {\bibfnamefont {Yu.~D.}\ \bibnamefont
  {Fomin}}, \bibinfo {author} {\bibfnamefont {V.~N.}\ \bibnamefont {Ryzhov}},
  \bibinfo {author} {\bibfnamefont {B.~A.}\ \bibnamefont {Klumov}}, \ and\
  \bibinfo {author} {\bibfnamefont {E.~N.}\ \bibnamefont {Tsiok}},\ }\bibfield
  {title} {\enquote {\bibinfo {title} {How to quantify structural anomalies in
  fluids?}}\ }\href {\doibase 10.1063/1.4890211} {\bibfield  {journal}
  {\bibinfo  {journal} {J. Chem. Phys.}\ }\textbf {\bibinfo {volume} {141}},\
  \bibinfo {pages} {034508} (\bibinfo {year} {2014})}\BibitemShut {NoStop}%
\bibitem [{\citenamefont {Klumov}\ and\ \citenamefont
  {Khrapak}(2020)}]{KlumovResPhys2020}%
  \BibitemOpen
  \bibfield  {author} {\bibinfo {author} {\bibfnamefont {B.~A.}\ \bibnamefont
  {Klumov}}\ and\ \bibinfo {author} {\bibfnamefont {S.~A.}\ \bibnamefont
  {Khrapak}},\ }\bibfield  {title} {\enquote {\bibinfo {title} {Two-body
  entropy of two-dimensional fluids},}\ }\href {\doibase
  10.1016/j.rinp.2020.103020} {\bibfield  {journal} {\bibinfo  {journal}
  {Results in Physics}\ }\textbf {\bibinfo {volume} {17}},\ \bibinfo {pages}
  {103020} (\bibinfo {year} {2020})}\BibitemShut {NoStop}%
\bibitem [{\citenamefont {Khrapak}\ and\ \citenamefont
  {Yurchenko}(2021)}]{KhrapakJCP2021}%
  \BibitemOpen
  \bibfield  {author} {\bibinfo {author} {\bibfnamefont {S.~A.}\ \bibnamefont
  {Khrapak}}\ and\ \bibinfo {author} {\bibfnamefont {S.~O.}\ \bibnamefont
  {Yurchenko}},\ }\bibfield  {title} {\enquote {\bibinfo {title} {Entropy of
  simple fluids with repulsive interactions near freezing},}\ }\href {\doibase
  10.1063/5.0063559} {\bibfield  {journal} {\bibinfo  {journal} {J. Chem.
  Phys.}\ }\textbf {\bibinfo {volume} {155}},\ \bibinfo {pages} {134501}
  (\bibinfo {year} {2021})}\BibitemShut {NoStop}%
\bibitem [{\citenamefont {Pond}\ \emph {et~al.}(2011)\citenamefont {Pond},
  \citenamefont {Errington},\ and\ \citenamefont {Truskett}}]{PondSM2011}%
  \BibitemOpen
  \bibfield  {author} {\bibinfo {author} {\bibfnamefont {M.~J.}\ \bibnamefont
  {Pond}}, \bibinfo {author} {\bibfnamefont {J.~R.}\ \bibnamefont {Errington}},
  \ and\ \bibinfo {author} {\bibfnamefont {T.~M.}\ \bibnamefont {Truskett}},\
  }\bibfield  {title} {\enquote {\bibinfo {title} {Mapping between long-time
  molecular and {B}rownian dynamics},}\ }\href {\doibase 10.1039/c1sm06493b}
  {\bibfield  {journal} {\bibinfo  {journal} {Soft Matter}\ }\textbf {\bibinfo
  {volume} {7}},\ \bibinfo {pages} {9859} (\bibinfo {year} {2011})}\BibitemShut
  {NoStop}%
\bibitem [{\citenamefont {Khrapak}(2018)}]{KhrapakAIPAdv2018}%
  \BibitemOpen
  \bibfield  {author} {\bibinfo {author} {\bibfnamefont {S.}~\bibnamefont
  {Khrapak}},\ }\bibfield  {title} {\enquote {\bibinfo {title} {Practical
  formula for the shear viscosity of {Y}ukawa fluids},}\ }\href {\doibase
  10.1063/1.5044703} {\bibfield  {journal} {\bibinfo  {journal} {{AIP} Adv.}\
  }\textbf {\bibinfo {volume} {8}},\ \bibinfo {pages} {105226} (\bibinfo {year}
  {2018})}\BibitemShut {NoStop}%
\bibitem [{\citenamefont {Khrapak}\ and\ \citenamefont
  {Khrapak}(2022{\natexlab{a}})}]{KhrapakPoF2022}%
  \BibitemOpen
  \bibfield  {author} {\bibinfo {author} {\bibfnamefont {S.~A.}\ \bibnamefont
  {Khrapak}}\ and\ \bibinfo {author} {\bibfnamefont {A.~G.}\ \bibnamefont
  {Khrapak}},\ }\bibfield  {title} {\enquote {\bibinfo {title} {Minima of shear
  viscosity and thermal conductivity coefficients of classical fluids},}\
  }\href {\doibase 10.1063/5.0082465} {\bibfield  {journal} {\bibinfo
  {journal} {Phys. Fluids}\ }\textbf {\bibinfo {volume} {34}},\ \bibinfo
  {pages} {027102} (\bibinfo {year} {2022}{\natexlab{a}})}\BibitemShut
  {NoStop}%
\bibitem [{\citenamefont {Trachenko}\ and\ \citenamefont
  {Brazhkin}(2020)}]{TrachenkoSciAdv2020}%
  \BibitemOpen
  \bibfield  {author} {\bibinfo {author} {\bibfnamefont {K.}~\bibnamefont
  {Trachenko}}\ and\ \bibinfo {author} {\bibfnamefont {V.~V.}\ \bibnamefont
  {Brazhkin}},\ }\bibfield  {title} {\enquote {\bibinfo {title} {Minimal
  quantum viscosity from fundamental physical constants},}\ }\href {\doibase
  10.1126/sciadv.aba3747} {\bibfield  {journal} {\bibinfo  {journal} {Sci.
  Adv.}\ }\textbf {\bibinfo {volume} {6}},\ \bibinfo {pages} {eaba3747}
  (\bibinfo {year} {2020})}\BibitemShut {NoStop}%
\bibitem [{\citenamefont {Trachenko}\ \emph {et~al.}(2021)\citenamefont
  {Trachenko}, \citenamefont {Baggioli}, \citenamefont {Behnia},\ and\
  \citenamefont {Brazhkin}}]{TrachenkoPRB2021}%
  \BibitemOpen
  \bibfield  {author} {\bibinfo {author} {\bibfnamefont {K.}~\bibnamefont
  {Trachenko}}, \bibinfo {author} {\bibfnamefont {M.}~\bibnamefont {Baggioli}},
  \bibinfo {author} {\bibfnamefont {K.}~\bibnamefont {Behnia}}, \ and\ \bibinfo
  {author} {\bibfnamefont {V.~V.}\ \bibnamefont {Brazhkin}},\ }\bibfield
  {title} {\enquote {\bibinfo {title} {Universal lower bounds on energy and
  momentum diffusion in liquids},}\ }\href {\doibase
  10.1103/physrevb.103.014311} {\bibfield  {journal} {\bibinfo  {journal}
  {Phys. Rev. B}\ }\textbf {\bibinfo {volume} {103}},\ \bibinfo {pages}
  {014311} (\bibinfo {year} {2021})}\BibitemShut {NoStop}%
\bibitem [{\citenamefont {Hubbard}\ and\ \citenamefont
  {Beeby}(1969)}]{Hubbard1969}%
  \BibitemOpen
  \bibfield  {author} {\bibinfo {author} {\bibfnamefont {J}~\bibnamefont
  {Hubbard}}\ and\ \bibinfo {author} {\bibfnamefont {J~L}\ \bibnamefont
  {Beeby}},\ }\bibfield  {title} {\enquote {\bibinfo {title} {Collective motion
  in liquids},}\ }\href {\doibase 10.1088/0022-3719/2/3/318} {\bibfield
  {journal} {\bibinfo  {journal} {J. Phys. C}\ }\textbf {\bibinfo {volume}
  {2}},\ \bibinfo {pages} {556--571} (\bibinfo {year} {1969})}\BibitemShut
  {NoStop}%
\bibitem [{\citenamefont {Stillinger}\ and\ \citenamefont
  {Weber}(1982)}]{Stillinger1982}%
  \BibitemOpen
  \bibfield  {author} {\bibinfo {author} {\bibfnamefont {F.~H.}\ \bibnamefont
  {Stillinger}}\ and\ \bibinfo {author} {\bibfnamefont {T.~A.}\ \bibnamefont
  {Weber}},\ }\bibfield  {title} {\enquote {\bibinfo {title} {Hidden structure
  in liquids},}\ }\href {\doibase 10.1103/physreva.25.978} {\bibfield
  {journal} {\bibinfo  {journal} {Phys. Rev. A}\ }\textbf {\bibinfo {volume}
  {25}},\ \bibinfo {pages} {978--989} (\bibinfo {year} {1982})}\BibitemShut
  {NoStop}%
\bibitem [{\citenamefont {Zwanzig}(1983)}]{ZwanzigJCP1983}%
  \BibitemOpen
  \bibfield  {author} {\bibinfo {author} {\bibfnamefont {R.}~\bibnamefont
  {Zwanzig}},\ }\bibfield  {title} {\enquote {\bibinfo {title} {On the relation
  between self-diffusion and viscosity of liquids},}\ }\href {\doibase
  10.1063/1.446338} {\bibfield  {journal} {\bibinfo  {journal} {J. Chem.
  Phys.}\ }\textbf {\bibinfo {volume} {79}},\ \bibinfo {pages} {4507--4508}
  (\bibinfo {year} {1983})}\BibitemShut {NoStop}%
\bibitem [{\citenamefont {Golden}\ and\ \citenamefont
  {Kalman}(2000)}]{GoldenPoP2000}%
  \BibitemOpen
  \bibfield  {author} {\bibinfo {author} {\bibfnamefont {K.~I.}\ \bibnamefont
  {Golden}}\ and\ \bibinfo {author} {\bibfnamefont {G.~J.}\ \bibnamefont
  {Kalman}},\ }\bibfield  {title} {\enquote {\bibinfo {title} {Quasilocalized
  charge approximation in strongly coupled plasma physics},}\ }\href {\doibase
  10.1063/1.873814} {\bibfield  {journal} {\bibinfo  {journal} {Phys. Plasmas}\
  }\textbf {\bibinfo {volume} {7}},\ \bibinfo {pages} {14--32} (\bibinfo {year}
  {2000})}\BibitemShut {NoStop}%
\bibitem [{\citenamefont
  {Khrapak}(2021{\natexlab{a}})}]{KhrapakMolecules12_2021}%
  \BibitemOpen
  \bibfield  {author} {\bibinfo {author} {\bibfnamefont {S.~A.}\ \bibnamefont
  {Khrapak}},\ }\bibfield  {title} {\enquote {\bibinfo {title} {Self-diffusion
  in simple liquids as a random walk process},}\ }\href {\doibase
  10.3390/molecules26247499} {\bibfield  {journal} {\bibinfo  {journal}
  {Molecules}\ }\textbf {\bibinfo {volume} {26}},\ \bibinfo {pages} {7499}
  (\bibinfo {year} {2021}{\natexlab{a}})}\BibitemShut {NoStop}%
\bibitem [{\citenamefont {Buchenau}\ \emph {et~al.}(2014)\citenamefont
  {Buchenau}, \citenamefont {Zorn},\ and\ \citenamefont
  {Ramos}}]{BuchenauPRE2014}%
  \BibitemOpen
  \bibfield  {author} {\bibinfo {author} {\bibfnamefont {U.}~\bibnamefont
  {Buchenau}}, \bibinfo {author} {\bibfnamefont {R.}~\bibnamefont {Zorn}}, \
  and\ \bibinfo {author} {\bibfnamefont {M.~A.}\ \bibnamefont {Ramos}},\
  }\bibfield  {title} {\enquote {\bibinfo {title} {Probing cooperative liquid
  dynamics with the mean square displacement},}\ }\href {\doibase
  10.1103/physreve.90.042312} {\bibfield  {journal} {\bibinfo  {journal} {Phys.
  Rev. E}\ }\textbf {\bibinfo {volume} {90}},\ \bibinfo {pages} {042312}
  (\bibinfo {year} {2014})}\BibitemShut {NoStop}%
\bibitem [{\citenamefont {Khrapak}(2020{\natexlab{a}})}]{KhrapakPRR2020}%
  \BibitemOpen
  \bibfield  {author} {\bibinfo {author} {\bibfnamefont {S.~A.}\ \bibnamefont
  {Khrapak}},\ }\bibfield  {title} {\enquote {\bibinfo {title} {Lindemann
  melting criterion in two dimensions},}\ }\href {\doibase
  10.1103/physrevresearch.2.012040} {\bibfield  {journal} {\bibinfo  {journal}
  {Phys. Rev. Research}\ }\textbf {\bibinfo {volume} {2}},\ \bibinfo {pages}
  {012040} (\bibinfo {year} {2020}{\natexlab{a}})}\BibitemShut {NoStop}%
\bibitem [{\citenamefont {Khrapak}(2019{\natexlab{a}})}]{KhrapakMolPhys2019}%
  \BibitemOpen
  \bibfield  {author} {\bibinfo {author} {\bibfnamefont {S.}~\bibnamefont
  {Khrapak}},\ }\bibfield  {title} {\enquote {\bibinfo {title}
  {Stokes{\textendash}{E}instein relation in simple fluids revisited},}\ }\href
  {\doibase 10.1080/00268976.2019.1643045} {\bibfield  {journal} {\bibinfo
  {journal} {Mol. Phys.}\ }\textbf {\bibinfo {volume} {118}},\ \bibinfo {pages}
  {e1643045} (\bibinfo {year} {2019}{\natexlab{a}})}\BibitemShut {NoStop}%
\bibitem [{\citenamefont {Berezhkovskii}\ and\ \citenamefont
  {Sutmann}(2002)}]{BerezhkovskiiPRE2002}%
  \BibitemOpen
  \bibfield  {author} {\bibinfo {author} {\bibfnamefont {A.~M.}\ \bibnamefont
  {Berezhkovskii}}\ and\ \bibinfo {author} {\bibfnamefont {G.}~\bibnamefont
  {Sutmann}},\ }\bibfield  {title} {\enquote {\bibinfo {title} {Time and length
  scales for diffusion in liquids},}\ }\href {\doibase
  10.1103/physreve.65.060201} {\bibfield  {journal} {\bibinfo  {journal} {Phys.
  Rev. E}\ }\textbf {\bibinfo {volume} {65}},\ \bibinfo {pages} {060201}
  (\bibinfo {year} {2002})}\BibitemShut {NoStop}%
\bibitem [{\citenamefont {Costigliola}\ \emph {et~al.}(2019)\citenamefont
  {Costigliola}, \citenamefont {Heyes}, \citenamefont {Schr{\o}der},\ and\
  \citenamefont {Dyre}}]{CostigliolaJCP2019}%
  \BibitemOpen
  \bibfield  {author} {\bibinfo {author} {\bibfnamefont {L.}~\bibnamefont
  {Costigliola}}, \bibinfo {author} {\bibfnamefont {D.~M.}\ \bibnamefont
  {Heyes}}, \bibinfo {author} {\bibfnamefont {T.~B.}\ \bibnamefont
  {Schr{\o}der}}, \ and\ \bibinfo {author} {\bibfnamefont {J.~C.}\ \bibnamefont
  {Dyre}},\ }\bibfield  {title} {\enquote {\bibinfo {title} {Revisiting the
  {S}tokes-{E}instein relation without a hydrodynamic diameter},}\ }\href
  {\doibase 10.1063/1.5080662} {\bibfield  {journal} {\bibinfo  {journal} {J.
  Chem. Phys.}\ }\textbf {\bibinfo {volume} {150}},\ \bibinfo {pages} {021101}
  (\bibinfo {year} {2019})}\BibitemShut {NoStop}%
\bibitem [{\citenamefont {Balucani}\ \emph {et~al.}(1990)\citenamefont
  {Balucani}, \citenamefont {Vallauri},\ and\ \citenamefont
  {Gaskell}}]{Balucani1990}%
  \BibitemOpen
  \bibfield  {author} {\bibinfo {author} {\bibfnamefont {U.}~\bibnamefont
  {Balucani}}, \bibinfo {author} {\bibfnamefont {R.}~\bibnamefont {Vallauri}},
  \ and\ \bibinfo {author} {\bibfnamefont {T.}~\bibnamefont {Gaskell}},\
  }\bibfield  {title} {\enquote {\bibinfo {title} {Generalized
  {S}tokes-{E}instein relation},}\ }\href {\doibase 10.1002/bbpc.19900940313}
  {\bibfield  {journal} {\bibinfo  {journal} {Berichte der Bunsengesellschaft
  f\"{u}r physikalische Chemie}\ }\textbf {\bibinfo {volume} {94}},\ \bibinfo
  {pages} {261--264} (\bibinfo {year} {1990})}\BibitemShut {NoStop}%
\bibitem [{\citenamefont {Hodgdon}\ and\ \citenamefont
  {Stillinger}(1993)}]{HodgdonPRE1993}%
  \BibitemOpen
  \bibfield  {author} {\bibinfo {author} {\bibfnamefont {J.~A.}\ \bibnamefont
  {Hodgdon}}\ and\ \bibinfo {author} {\bibfnamefont {F.~H.}\ \bibnamefont
  {Stillinger}},\ }\bibfield  {title} {\enquote {\bibinfo {title}
  {{S}tokes-{E}instein violation in glass-forming liquids},}\ }\href {\doibase
  10.1103/physreve.48.207} {\bibfield  {journal} {\bibinfo  {journal} {Phys.
  Rev. E}\ }\textbf {\bibinfo {volume} {48}},\ \bibinfo {pages} {207} (\bibinfo
  {year} {1993})}\BibitemShut {NoStop}%
\bibitem [{\citenamefont {Tarjus}\ and\ \citenamefont
  {Kivelson}(1995)}]{TarjusJCP1995}%
  \BibitemOpen
  \bibfield  {author} {\bibinfo {author} {\bibfnamefont {G.}~\bibnamefont
  {Tarjus}}\ and\ \bibinfo {author} {\bibfnamefont {D.}~\bibnamefont
  {Kivelson}},\ }\bibfield  {title} {\enquote {\bibinfo {title} {Breakdown of
  the {S}tokes{\textendash}{E}instein relation in supercooled liquids},}\
  }\href {\doibase 10.1063/1.470495} {\bibfield  {journal} {\bibinfo  {journal}
  {J. Chem. Phys.}\ }\textbf {\bibinfo {volume} {103}},\ \bibinfo {pages}
  {3071--3073} (\bibinfo {year} {1995})}\BibitemShut {NoStop}%
\bibitem [{\citenamefont {Bordat}\ \emph {et~al.}(2003)\citenamefont {Bordat},
  \citenamefont {Affouard}, \citenamefont {Descamps},\ and\ \citenamefont
  {Muller-Plathe}}]{BordatJPCM2003}%
  \BibitemOpen
  \bibfield  {author} {\bibinfo {author} {\bibfnamefont {P.}~\bibnamefont
  {Bordat}}, \bibinfo {author} {\bibfnamefont {F.}~\bibnamefont {Affouard}},
  \bibinfo {author} {\bibfnamefont {M.}~\bibnamefont {Descamps}}, \ and\
  \bibinfo {author} {\bibfnamefont {F.~M.}\ \bibnamefont {Muller-Plathe}},\
  }\bibfield  {title} {\enquote {\bibinfo {title} {The breakdown of the
  {S}tokes{\textendash}{E}instein relation in supercooled binary liquids},}\
  }\href {\doibase 10.1088/0953-8984/15/32/301} {\bibfield  {journal} {\bibinfo
   {journal} {J. Phys.: Condens. Matter}\ }\textbf {\bibinfo {volume} {15}},\
  \bibinfo {pages} {5397--5407} (\bibinfo {year} {2003})}\BibitemShut {NoStop}%
\bibitem [{\citenamefont {Chen}\ \emph {et~al.}(2006)\citenamefont {Chen},
  \citenamefont {Mallamace}, \citenamefont {Mou}, \citenamefont {Broccio},
  \citenamefont {Corsaro}, \citenamefont {Faraone},\ and\ \citenamefont
  {Liu}}]{ChenPNAS2006}%
  \BibitemOpen
  \bibfield  {author} {\bibinfo {author} {\bibfnamefont {S.-H.}\ \bibnamefont
  {Chen}}, \bibinfo {author} {\bibfnamefont {F.}~\bibnamefont {Mallamace}},
  \bibinfo {author} {\bibfnamefont {C.-Y.}\ \bibnamefont {Mou}}, \bibinfo
  {author} {\bibfnamefont {M.}~\bibnamefont {Broccio}}, \bibinfo {author}
  {\bibfnamefont {C.}~\bibnamefont {Corsaro}}, \bibinfo {author} {\bibfnamefont
  {A.}~\bibnamefont {Faraone}}, \ and\ \bibinfo {author} {\bibfnamefont
  {L.}~\bibnamefont {Liu}},\ }\bibfield  {title} {\enquote {\bibinfo {title}
  {The violation of the {S}tokes-{E}instein relation in supercooled water},}\
  }\href {\doibase 10.1073/pnas.0603253103} {\bibfield  {journal} {\bibinfo
  {journal} {PNAS}\ }\textbf {\bibinfo {volume} {103}},\ \bibinfo {pages}
  {12974--12978} (\bibinfo {year} {2006})}\BibitemShut {NoStop}%
\bibitem [{\citenamefont {Puosi}\ \emph {et~al.}(2018)\citenamefont {Puosi},
  \citenamefont {Pasturel}, \citenamefont {Jakse},\ and\ \citenamefont
  {Leporini}}]{PuosiJCP2018}%
  \BibitemOpen
  \bibfield  {author} {\bibinfo {author} {\bibfnamefont {F.}~\bibnamefont
  {Puosi}}, \bibinfo {author} {\bibfnamefont {A.}~\bibnamefont {Pasturel}},
  \bibinfo {author} {\bibfnamefont {N.}~\bibnamefont {Jakse}}, \ and\ \bibinfo
  {author} {\bibfnamefont {D.}~\bibnamefont {Leporini}},\ }\bibfield  {title}
  {\enquote {\bibinfo {title} {Communication: Fast dynamics perspective on the
  breakdown of the {S}tokes-{E}instein law in fragile glassformers},}\ }\href
  {\doibase 10.1063/1.5025614} {\bibfield  {journal} {\bibinfo  {journal} {J.
  Chem. Phys.}\ }\textbf {\bibinfo {volume} {148}},\ \bibinfo {pages} {131102}
  (\bibinfo {year} {2018})}\BibitemShut {NoStop}%
\bibitem [{\citenamefont {Brazhkin}\ \emph
  {et~al.}(2012{\natexlab{a}})\citenamefont {Brazhkin}, \citenamefont {Fomin},
  \citenamefont {Lyapin}, \citenamefont {Ryzhov},\ and\ \citenamefont
  {Trachenko}}]{BrazhkinPRE2012}%
  \BibitemOpen
  \bibfield  {author} {\bibinfo {author} {\bibfnamefont {V.~V.}\ \bibnamefont
  {Brazhkin}}, \bibinfo {author} {\bibfnamefont {Yu.~D.}\ \bibnamefont
  {Fomin}}, \bibinfo {author} {\bibfnamefont {A.~G.}\ \bibnamefont {Lyapin}},
  \bibinfo {author} {\bibfnamefont {V.~N.}\ \bibnamefont {Ryzhov}}, \ and\
  \bibinfo {author} {\bibfnamefont {K.}~\bibnamefont {Trachenko}},\ }\bibfield
  {title} {\enquote {\bibinfo {title} {Two liquid states of matter: A dynamic
  line on a phase diagram},}\ }\href {\doibase 10.1103/physreve.85.031203}
  {\bibfield  {journal} {\bibinfo  {journal} {Phys. Rev. E}\ }\textbf {\bibinfo
  {volume} {85}},\ \bibinfo {pages} {031203} (\bibinfo {year}
  {2012}{\natexlab{a}})}\BibitemShut {NoStop}%
\bibitem [{\citenamefont {Brazhkin}\ \emph
  {et~al.}(2012{\natexlab{b}})\citenamefont {Brazhkin}, \citenamefont {Lyapin},
  \citenamefont {Ryzhov}, \citenamefont {Trachenko}, \citenamefont {Fomin},\
  and\ \citenamefont {Tsiok}}]{BrazhkinUFN2012}%
  \BibitemOpen
  \bibfield  {author} {\bibinfo {author} {\bibfnamefont {V.~V.}\ \bibnamefont
  {Brazhkin}}, \bibinfo {author} {\bibfnamefont {A.G.}\ \bibnamefont {Lyapin}},
  \bibinfo {author} {\bibfnamefont {V.~N.}\ \bibnamefont {Ryzhov}}, \bibinfo
  {author} {\bibfnamefont {K.}~\bibnamefont {Trachenko}}, \bibinfo {author}
  {\bibfnamefont {Y.~D.}\ \bibnamefont {Fomin}}, \ and\ \bibinfo {author}
  {\bibfnamefont {E.~N.}\ \bibnamefont {Tsiok}},\ }\bibfield  {title} {\enquote
  {\bibinfo {title} {Where is the supercritical fluid on the phase diagram?}}\
  }\href {\doibase 10.3367/ufnr.0182.201211a.1137} {\bibfield  {journal}
  {\bibinfo  {journal} {Phys.-Usp.}\ }\textbf {\bibinfo {volume} {182}},\
  \bibinfo {pages} {1137--1156} (\bibinfo {year}
  {2012}{\natexlab{b}})}\BibitemShut {NoStop}%
\bibitem [{\citenamefont {Bryk}\ \emph {et~al.}(2018)\citenamefont {Bryk},
  \citenamefont {Gorelli}, \citenamefont {Mryglod}, \citenamefont {Ruocco},
  \citenamefont {Santoro},\ and\ \citenamefont {Scopigno}}]{BrykJPCB2018}%
  \BibitemOpen
  \bibfield  {author} {\bibinfo {author} {\bibfnamefont {T.}~\bibnamefont
  {Bryk}}, \bibinfo {author} {\bibfnamefont {F.~A.}\ \bibnamefont {Gorelli}},
  \bibinfo {author} {\bibfnamefont {I.}~\bibnamefont {Mryglod}}, \bibinfo
  {author} {\bibfnamefont {G.}~\bibnamefont {Ruocco}}, \bibinfo {author}
  {\bibfnamefont {M.}~\bibnamefont {Santoro}}, \ and\ \bibinfo {author}
  {\bibfnamefont {T.}~\bibnamefont {Scopigno}},\ }\bibfield  {title} {\enquote
  {\bibinfo {title} {Reply to comment on behavior of supercritical fluids
  across the {F}renkel line},}\ }\href {\doibase 10.1021/acs.jpcb.8b01900}
  {\bibfield  {journal} {\bibinfo  {journal} {J. Phys. Chem. B}\ }\textbf
  {\bibinfo {volume} {122}},\ \bibinfo {pages} {6120--6123} (\bibinfo {year}
  {2018})}\BibitemShut {NoStop}%
\bibitem [{\citenamefont {Lindemann}(1910)}]{Lindemann}%
  \BibitemOpen
  \bibfield  {author} {\bibinfo {author} {\bibfnamefont {F.}~\bibnamefont
  {Lindemann}},\ }\bibfield  {title} {\enquote {\bibinfo {title} {The
  calculation of molecular vibration frequencies},}\ }\href@noop {} {\bibfield
  {journal} {\bibinfo  {journal} {Z. Phys.}\ }\textbf {\bibinfo {volume}
  {11}},\ \bibinfo {pages} {609} (\bibinfo {year} {1910})}\BibitemShut
  {NoStop}%
\bibitem [{\citenamefont {Khrapak}(2021{\natexlab{b}})}]{KhrapakPRE01_2021}%
  \BibitemOpen
  \bibfield  {author} {\bibinfo {author} {\bibfnamefont {S.~A.}\ \bibnamefont
  {Khrapak}},\ }\bibfield  {title} {\enquote {\bibinfo {title} {Vibrational
  model of thermal conduction for fluids with soft interactions},}\ }\href
  {\doibase 10.1103/physreve.103.013207} {\bibfield  {journal} {\bibinfo
  {journal} {Phys. Rev. E}\ }\textbf {\bibinfo {volume} {103}},\ \bibinfo
  {pages} {013207} (\bibinfo {year} {2021}{\natexlab{b}})}\BibitemShut
  {NoStop}%
\bibitem [{\citenamefont {Horrocks}\ and\ \citenamefont
  {McLaughlin}(1960)}]{Horrocks1960}%
  \BibitemOpen
  \bibfield  {author} {\bibinfo {author} {\bibfnamefont {J.~K.}\ \bibnamefont
  {Horrocks}}\ and\ \bibinfo {author} {\bibfnamefont {E.}~\bibnamefont
  {McLaughlin}},\ }\bibfield  {title} {\enquote {\bibinfo {title} {Thermal
  conductivity of simple molecules in the condensed state},}\ }\href {\doibase
  10.1039/tf9605600206} {\bibfield  {journal} {\bibinfo  {journal} {Trans.
  Faraday Soc.}\ }\textbf {\bibinfo {volume} {56}},\ \bibinfo {pages} {206}
  (\bibinfo {year} {1960})}\BibitemShut {NoStop}%
\bibitem [{\citenamefont {Rao}(1941)}]{Rao1941}%
  \BibitemOpen
  \bibfield  {author} {\bibinfo {author} {\bibfnamefont {M.~R.}\ \bibnamefont
  {Rao}},\ }\bibfield  {title} {\enquote {\bibinfo {title} {Thermal
  conductivity of liquids},}\ }\href {\doibase 10.1103/physrev.59.212}
  {\bibfield  {journal} {\bibinfo  {journal} {Phys. Rev.}\ }\textbf {\bibinfo
  {volume} {59}},\ \bibinfo {pages} {212--212} (\bibinfo {year}
  {1941})}\BibitemShut {NoStop}%
\bibitem [{\citenamefont {Cahill}\ and\ \citenamefont
  {Pohl}(1989)}]{Cahill1989}%
  \BibitemOpen
  \bibfield  {author} {\bibinfo {author} {\bibfnamefont {D.~G.}\ \bibnamefont
  {Cahill}}\ and\ \bibinfo {author} {\bibfnamefont {R.O.}\ \bibnamefont
  {Pohl}},\ }\bibfield  {title} {\enquote {\bibinfo {title} {Heat flow and
  lattice vibrations in glasses},}\ }\href {\doibase
  10.1016/0038-1098(89)90630-3} {\bibfield  {journal} {\bibinfo  {journal}
  {Solid State Commun.}\ }\textbf {\bibinfo {volume} {70}},\ \bibinfo {pages}
  {927--930} (\bibinfo {year} {1989})}\BibitemShut {NoStop}%
\bibitem [{\citenamefont {Cahill}\ \emph {et~al.}(1992)\citenamefont {Cahill},
  \citenamefont {Watson},\ and\ \citenamefont {Pohl}}]{Cahill1992}%
  \BibitemOpen
  \bibfield  {author} {\bibinfo {author} {\bibfnamefont {D.~G.}\ \bibnamefont
  {Cahill}}, \bibinfo {author} {\bibfnamefont {S.~K.}\ \bibnamefont {Watson}},
  \ and\ \bibinfo {author} {\bibfnamefont {R.~O.}\ \bibnamefont {Pohl}},\
  }\bibfield  {title} {\enquote {\bibinfo {title} {Lower limit to the thermal
  conductivity of disordered crystals},}\ }\href {\doibase
  10.1103/physrevb.46.6131} {\bibfield  {journal} {\bibinfo  {journal} {Phys.
  Rev. B}\ }\textbf {\bibinfo {volume} {46}},\ \bibinfo {pages} {6131--6140}
  (\bibinfo {year} {1992})}\BibitemShut {NoStop}%
\bibitem [{\citenamefont {Xie}\ \emph {et~al.}(2017)\citenamefont {Xie},
  \citenamefont {Yang}, \citenamefont {Li}, \citenamefont {Tsai}, \citenamefont
  {Shin}, \citenamefont {Braun},\ and\ \citenamefont {Cahill}}]{XiePRB2017}%
  \BibitemOpen
  \bibfield  {author} {\bibinfo {author} {\bibfnamefont {X.}~\bibnamefont
  {Xie}}, \bibinfo {author} {\bibfnamefont {K.}~\bibnamefont {Yang}}, \bibinfo
  {author} {\bibfnamefont {D.}~\bibnamefont {Li}}, \bibinfo {author}
  {\bibfnamefont {T.-H.}\ \bibnamefont {Tsai}}, \bibinfo {author}
  {\bibfnamefont {J.}~\bibnamefont {Shin}}, \bibinfo {author} {\bibfnamefont
  {P.~V.}\ \bibnamefont {Braun}}, \ and\ \bibinfo {author} {\bibfnamefont
  {D.~G.}\ \bibnamefont {Cahill}},\ }\bibfield  {title} {\enquote {\bibinfo
  {title} {High and low thermal conductivity of amorphous macromolecules},}\
  }\href {\doibase 10.1103/physrevb.95.035406} {\bibfield  {journal} {\bibinfo
  {journal} {Phys. Rev. B}\ }\textbf {\bibinfo {volume} {95}},\ \bibinfo
  {pages} {035406} (\bibinfo {year} {2017})}\BibitemShut {NoStop}%
\bibitem [{\citenamefont {Berens}\ \emph {et~al.}(1983)\citenamefont {Berens},
  \citenamefont {Mackay}, \citenamefont {White},\ and\ \citenamefont
  {Wilson}}]{Berens1983}%
  \BibitemOpen
  \bibfield  {author} {\bibinfo {author} {\bibfnamefont {P.~H.}\ \bibnamefont
  {Berens}}, \bibinfo {author} {\bibfnamefont {D.~H.~J.}\ \bibnamefont
  {Mackay}}, \bibinfo {author} {\bibfnamefont {G.~M.}\ \bibnamefont {White}}, \
  and\ \bibinfo {author} {\bibfnamefont {K.~R.}\ \bibnamefont {Wilson}},\
  }\bibfield  {title} {\enquote {\bibinfo {title} {Thermodynamics and quantum
  corrections from molecular dynamics for liquid water},}\ }\href {\doibase
  10.1063/1.446044} {\bibfield  {journal} {\bibinfo  {journal} {J. Chem.
  Phys.}\ }\textbf {\bibinfo {volume} {79}},\ \bibinfo {pages} {2375–2389}
  (\bibinfo {year} {1983})}\BibitemShut {NoStop}%
\bibitem [{\citenamefont {Ohta}\ and\ \citenamefont
  {Hamaguchi}(2000{\natexlab{a}})}]{OhtaPoP2000}%
  \BibitemOpen
  \bibfield  {author} {\bibinfo {author} {\bibfnamefont {H.}~\bibnamefont
  {Ohta}}\ and\ \bibinfo {author} {\bibfnamefont {S.}~\bibnamefont
  {Hamaguchi}},\ }\bibfield  {title} {\enquote {\bibinfo {title} {Molecular
  dynamics evaluation of self-diffusion in {Y}ukawa systems},}\ }\href
  {\doibase 10.1063/1.1316084} {\bibfield  {journal} {\bibinfo  {journal}
  {Phys. Plasmas}\ }\textbf {\bibinfo {volume} {7}},\ \bibinfo {pages}
  {4506--4514} (\bibinfo {year} {2000}{\natexlab{a}})}\BibitemShut {NoStop}%
\bibitem [{\citenamefont {Zaccone}\ and\ \citenamefont
  {Baggioli}(2021)}]{ZacconePNAS2021}%
  \BibitemOpen
  \bibfield  {author} {\bibinfo {author} {\bibfnamefont {A.}~\bibnamefont
  {Zaccone}}\ and\ \bibinfo {author} {\bibfnamefont {M.}~\bibnamefont
  {Baggioli}},\ }\bibfield  {title} {\enquote {\bibinfo {title} {Universal law
  for the vibrational density of states of liquids},}\ }\href {\doibase
  10.1073/pnas.2022303118} {\bibfield  {journal} {\bibinfo  {journal} {Proc.
  Natl. Acad. Sci.}\ }\textbf {\bibinfo {volume} {118}},\ \bibinfo {pages}
  {e2022303118} (\bibinfo {year} {2021})}\BibitemShut {NoStop}%
\bibitem [{\citenamefont {Stamper}\ \emph {et~al.}(2022)\citenamefont
  {Stamper}, \citenamefont {Cortie}, \citenamefont {Yue}, \citenamefont
  {Wang},\ and\ \citenamefont {Yu}}]{StamperJPCL2022}%
  \BibitemOpen
  \bibfield  {author} {\bibinfo {author} {\bibfnamefont {C.}~\bibnamefont
  {Stamper}}, \bibinfo {author} {\bibfnamefont {D.}~\bibnamefont {Cortie}},
  \bibinfo {author} {\bibfnamefont {Z.}~\bibnamefont {Yue}}, \bibinfo {author}
  {\bibfnamefont {X.}~\bibnamefont {Wang}}, \ and\ \bibinfo {author}
  {\bibfnamefont {D.}~\bibnamefont {Yu}},\ }\bibfield  {title} {\enquote
  {\bibinfo {title} {Experimental confirmation of the universal law for the
  vibrational density of states of liquids},}\ }\href {\doibase
  10.1021/acs.jpclett.2c00297} {\bibfield  {journal} {\bibinfo  {journal} {J.
  Phys. Chem. Lett.}\ }\textbf {\bibinfo {volume} {13}},\ \bibinfo {pages}
  {3105--3111} (\bibinfo {year} {2022})}\BibitemShut {NoStop}%
\bibitem [{\citenamefont {Pieprzyk}\ \emph {et~al.}(2020)\citenamefont
  {Pieprzyk}, \citenamefont {Bra{\'{n}}ka}, \citenamefont {Heyes},\ and\
  \citenamefont {Bannerman}}]{Pieprzyk2020}%
  \BibitemOpen
  \bibfield  {author} {\bibinfo {author} {\bibfnamefont {S.}~\bibnamefont
  {Pieprzyk}}, \bibinfo {author} {\bibfnamefont {A.~C.}\ \bibnamefont
  {Bra{\'{n}}ka}}, \bibinfo {author} {\bibfnamefont {D.~M.}\ \bibnamefont
  {Heyes}}, \ and\ \bibinfo {author} {\bibfnamefont {M.~N.}\ \bibnamefont
  {Bannerman}},\ }\bibfield  {title} {\enquote {\bibinfo {title} {A
  comprehensive study of the thermal conductivity of the hard sphere fluid and
  solid by molecular dynamics simulation},}\ }\href {\doibase
  10.1039/d0cp00494d} {\bibfield  {journal} {\bibinfo  {journal} {Phys. Chem.
  Chem. Phys.}\ }\textbf {\bibinfo {volume} {22}},\ \bibinfo {pages}
  {8834--8845} (\bibinfo {year} {2020})}\BibitemShut {NoStop}%
\bibitem [{\citenamefont {Khrapak}(2022{\natexlab{a}})}]{KhrapakApplSci2022}%
  \BibitemOpen
  \bibfield  {author} {\bibinfo {author} {\bibfnamefont {S.}~\bibnamefont
  {Khrapak}},\ }\bibfield  {title} {\enquote {\bibinfo {title} {Vibrational
  model of heat conduction in a fluid of hard spheres},}\ }\href {\doibase
  10.3390/app12157939} {\bibfield  {journal} {\bibinfo  {journal} {Appl. Sci.}\
  }\textbf {\bibinfo {volume} {12}},\ \bibinfo {pages} {7939} (\bibinfo {year}
  {2022}{\natexlab{a}})}\BibitemShut {NoStop}%
\bibitem [{\citenamefont {Khrapak}\ and\ \citenamefont
  {Khrapak}(2022{\natexlab{b}})}]{KhrapakJCP2022_1}%
  \BibitemOpen
  \bibfield  {author} {\bibinfo {author} {\bibfnamefont {S.~A.}\ \bibnamefont
  {Khrapak}}\ and\ \bibinfo {author} {\bibfnamefont {A.~G.}\ \bibnamefont
  {Khrapak}},\ }\bibfield  {title} {\enquote {\bibinfo {title} {Freezing
  density scaling of fluid transport properties: {A}pplication to liquefied
  noble gases},}\ }\href {\doibase 10.1063/5.0096947} {\bibfield  {journal}
  {\bibinfo  {journal} {J. Chem. Phys.}\ }\textbf {\bibinfo {volume} {157}},\
  \bibinfo {pages} {014501} (\bibinfo {year} {2022}{\natexlab{b}})}\BibitemShut
  {NoStop}%
\bibitem [{\citenamefont {Donko}\ \emph {et~al.}(2002)\citenamefont {Donko},
  \citenamefont {Kalman},\ and\ \citenamefont {Golden}}]{DonkoPRL2002}%
  \BibitemOpen
  \bibfield  {author} {\bibinfo {author} {\bibfnamefont {Z.}~\bibnamefont
  {Donko}}, \bibinfo {author} {\bibfnamefont {G.~J.}\ \bibnamefont {Kalman}}, \
  and\ \bibinfo {author} {\bibfnamefont {K.~I.}\ \bibnamefont {Golden}},\
  }\bibfield  {title} {\enquote {\bibinfo {title} {Caging of particles in
  one-component plasmas},}\ }\href {\doibase 10.1103/physrevlett.88.225001}
  {\bibfield  {journal} {\bibinfo  {journal} {Phys. Rev. Lett.}\ }\textbf
  {\bibinfo {volume} {88}},\ \bibinfo {pages} {225001} (\bibinfo {year}
  {2002})}\BibitemShut {NoStop}%
\bibitem [{\citenamefont {Daligault}(2020)}]{Daligault2020}%
  \BibitemOpen
  \bibfield  {author} {\bibinfo {author} {\bibfnamefont {J.}~\bibnamefont
  {Daligault}},\ }\href@noop {} {\enquote {\bibinfo {title} {Universal
  character of atomic motions at the liquid-solid transition},}\ } (\bibinfo
  {year} {2020}),\ \Eprint {http://arxiv.org/abs/2009.14718} {arXiv:2009.14718
  [cond-mat.stat-mech]} \BibitemShut {NoStop}%
\bibitem [{\citenamefont {Brush}\ \emph {et~al.}(1966)\citenamefont {Brush},
  \citenamefont {Sahlin},\ and\ \citenamefont {Teller}}]{BrushJCP1966}%
  \BibitemOpen
  \bibfield  {author} {\bibinfo {author} {\bibfnamefont {S.~G.}\ \bibnamefont
  {Brush}}, \bibinfo {author} {\bibfnamefont {H.~L.}\ \bibnamefont {Sahlin}}, \
  and\ \bibinfo {author} {\bibfnamefont {E.}~\bibnamefont {Teller}},\
  }\bibfield  {title} {\enquote {\bibinfo {title} {Monte {C}arlo study of a
  one-component plasma},}\ }\href {\doibase 10.1063/1.1727895} {\bibfield
  {journal} {\bibinfo  {journal} {J. Chem. Phys.}\ }\textbf {\bibinfo {volume}
  {45}},\ \bibinfo {pages} {2102--2118} (\bibinfo {year} {1966})}\BibitemShut
  {NoStop}%
\bibitem [{\citenamefont {Hansen}(1973)}]{HansenPRA1973}%
  \BibitemOpen
  \bibfield  {author} {\bibinfo {author} {\bibfnamefont {J.~P.}\ \bibnamefont
  {Hansen}},\ }\bibfield  {title} {\enquote {\bibinfo {title} {Statistical
  mechanics of dense ionized matter. {I}. {E}quilibrium properties of the
  classical one-component plasma},}\ }\href {\doibase 10.1103/physreva.8.3096}
  {\bibfield  {journal} {\bibinfo  {journal} {Phys. Rev. A}\ }\textbf {\bibinfo
  {volume} {8}},\ \bibinfo {pages} {3096--3109} (\bibinfo {year}
  {1973})}\BibitemShut {NoStop}%
\bibitem [{\citenamefont {DeWitt}(1978)}]{deWitt1978}%
  \BibitemOpen
  \bibfield  {author} {\bibinfo {author} {\bibfnamefont {H.~E.}\ \bibnamefont
  {DeWitt}},\ }\bibfield  {title} {\enquote {\bibinfo {title} {Statistical
  mechnics of dense plasmas : Numerical simulation and theory},}\ }\href
  {\doibase 10.1051/jphyscol:1978132} {\bibfield  {journal} {\bibinfo
  {journal} {J. Phys. Colloques}\ }\textbf {\bibinfo {volume} {39}},\ \bibinfo
  {pages} {C1--173--C1--180} (\bibinfo {year} {1978})}\BibitemShut {NoStop}%
\bibitem [{\citenamefont {Baus}\ and\ \citenamefont
  {Hansen}(1980)}]{BausPR1980}%
  \BibitemOpen
  \bibfield  {author} {\bibinfo {author} {\bibfnamefont {M}~\bibnamefont
  {Baus}}\ and\ \bibinfo {author} {\bibfnamefont {J.~P.}\ \bibnamefont
  {Hansen}},\ }\bibfield  {title} {\enquote {\bibinfo {title} {Statistical
  mechanics of simple {C}oulomb systems},}\ }\href {\doibase
  10.1016/0370-1573(80)90022-8} {\bibfield  {journal} {\bibinfo  {journal}
  {Phys. Rep.}\ }\textbf {\bibinfo {volume} {59}},\ \bibinfo {pages} {1--94}
  (\bibinfo {year} {1980})}\BibitemShut {NoStop}%
\bibitem [{\citenamefont {Ichimaru}(1982)}]{IchimaruRMP1982}%
  \BibitemOpen
  \bibfield  {author} {\bibinfo {author} {\bibfnamefont {S.}~\bibnamefont
  {Ichimaru}},\ }\bibfield  {title} {\enquote {\bibinfo {title} {Strongly
  coupled plasmas: {H}igh-density classical plasmas and degenerate electron
  liquids},}\ }\href {\doibase 10.1103/revmodphys.54.1017} {\bibfield
  {journal} {\bibinfo  {journal} {Rev. Mod. Phys.}\ }\textbf {\bibinfo {volume}
  {54}},\ \bibinfo {pages} {1017--1059} (\bibinfo {year} {1982})}\BibitemShut
  {NoStop}%
\bibitem [{\citenamefont {Stringfellow}\ \emph {et~al.}(1990)\citenamefont
  {Stringfellow}, \citenamefont {DeWitt},\ and\ \citenamefont
  {Slattery}}]{StringfellowPRA1990}%
  \BibitemOpen
  \bibfield  {author} {\bibinfo {author} {\bibfnamefont {G.~S.}\ \bibnamefont
  {Stringfellow}}, \bibinfo {author} {\bibfnamefont {H.~E.}\ \bibnamefont
  {DeWitt}}, \ and\ \bibinfo {author} {\bibfnamefont {W.~L.}\ \bibnamefont
  {Slattery}},\ }\bibfield  {title} {\enquote {\bibinfo {title} {Equation of
  state of the one-component plasma derived from precision {M}onte {C}arlo
  calculations},}\ }\href {\doibase 10.1103/physreva.41.1105} {\bibfield
  {journal} {\bibinfo  {journal} {Phys. Rev. A}\ }\textbf {\bibinfo {volume}
  {41}},\ \bibinfo {pages} {1105--1111} (\bibinfo {year} {1990})}\BibitemShut
  {NoStop}%
\bibitem [{\citenamefont {Dubin}\ and\ \citenamefont
  {O'Neil}(1999)}]{DubinRMP1999}%
  \BibitemOpen
  \bibfield  {author} {\bibinfo {author} {\bibfnamefont {D.~H.~E.}\
  \bibnamefont {Dubin}}\ and\ \bibinfo {author} {\bibfnamefont {T.~M.}\
  \bibnamefont {O'Neil}},\ }\bibfield  {title} {\enquote {\bibinfo {title}
  {Trapped nonneutral plasmas, liquids, and crystals (the thermal equilibrium
  states)},}\ }\href {\doibase 10.1103/revmodphys.71.87} {\bibfield  {journal}
  {\bibinfo  {journal} {Rev. Mod. Phys.}\ }\textbf {\bibinfo {volume} {71}},\
  \bibinfo {pages} {87--172} (\bibinfo {year} {1999})}\BibitemShut {NoStop}%
\bibitem [{\citenamefont {Khrapak}\ and\ \citenamefont
  {Khrapak}(2016)}]{KhrapakCPP2016}%
  \BibitemOpen
  \bibfield  {author} {\bibinfo {author} {\bibfnamefont {S.~A.}\ \bibnamefont
  {Khrapak}}\ and\ \bibinfo {author} {\bibfnamefont {A.~G.}\ \bibnamefont
  {Khrapak}},\ }\bibfield  {title} {\enquote {\bibinfo {title} {Internal energy
  of the classical two- and three-dimensional one-component-plasma},}\ }\href
  {\doibase 10.1002/ctpp.201500104} {\bibfield  {journal} {\bibinfo  {journal}
  {Contrib. Plasma Phys.}\ }\textbf {\bibinfo {volume} {56}},\ \bibinfo {pages}
  {270--280} (\bibinfo {year} {2016})}\BibitemShut {NoStop}%
\bibitem [{\citenamefont {Khrapak}\ \emph {et~al.}(2016)\citenamefont
  {Khrapak}, \citenamefont {Klumov}, \citenamefont {Couedel},\ and\
  \citenamefont {Thomas}}]{KhrapakPoP2016}%
  \BibitemOpen
  \bibfield  {author} {\bibinfo {author} {\bibfnamefont {S.~A.}\ \bibnamefont
  {Khrapak}}, \bibinfo {author} {\bibfnamefont {B.}~\bibnamefont {Klumov}},
  \bibinfo {author} {\bibfnamefont {L.}~\bibnamefont {Couedel}}, \ and\
  \bibinfo {author} {\bibfnamefont {H.~M.}\ \bibnamefont {Thomas}},\ }\bibfield
   {title} {\enquote {\bibinfo {title} {On the long-waves dispersion in
  {Y}ukawa systems},}\ }\href {\doibase 10.1063/1.4942169} {\bibfield
  {journal} {\bibinfo  {journal} {Phys. Plasmas}\ }\textbf {\bibinfo {volume}
  {23}},\ \bibinfo {pages} {023702} (\bibinfo {year} {2016})}\BibitemShut
  {NoStop}%
\bibitem [{\citenamefont {Singwi}\ \emph {et~al.}(1970)\citenamefont {Singwi},
  \citenamefont {Sk\"{o}ld},\ and\ \citenamefont {Tosi}}]{SingwiPRA1970}%
  \BibitemOpen
  \bibfield  {author} {\bibinfo {author} {\bibfnamefont {K.~S.}\ \bibnamefont
  {Singwi}}, \bibinfo {author} {\bibfnamefont {K.}~\bibnamefont {Sk\"{o}ld}}, \
  and\ \bibinfo {author} {\bibfnamefont {M.~P.}\ \bibnamefont {Tosi}},\
  }\bibfield  {title} {\enquote {\bibinfo {title} {Collective motions in
  classical liquids},}\ }\href {\doibase 10.1103/physreva.1.454} {\bibfield
  {journal} {\bibinfo  {journal} {Phys. Rev. A}\ }\textbf {\bibinfo {volume}
  {1}},\ \bibinfo {pages} {454--463} (\bibinfo {year} {1970})}\BibitemShut
  {NoStop}%
\bibitem [{\citenamefont {Takeno}\ and\ \citenamefont
  {Goda}(1971)}]{Takeno1971}%
  \BibitemOpen
  \bibfield  {author} {\bibinfo {author} {\bibfnamefont {S.}~\bibnamefont
  {Takeno}}\ and\ \bibinfo {author} {\bibfnamefont {M.}~\bibnamefont {Goda}},\
  }\bibfield  {title} {\enquote {\bibinfo {title} {A theory of phonons in
  amorphous solids and its implications to collective motion in simple
  liquids},}\ }\href {\doibase 10.1143/ptp.45.331} {\bibfield  {journal}
  {\bibinfo  {journal} {Progr. Theor. Phys.}\ }\textbf {\bibinfo {volume}
  {45}},\ \bibinfo {pages} {331--352} (\bibinfo {year} {1971})}\BibitemShut
  {NoStop}%
\bibitem [{\citenamefont {Khrapak}\ \emph
  {et~al.}(2017{\natexlab{a}})\citenamefont {Khrapak}, \citenamefont {Klumov},\
  and\ \citenamefont {Couedel}}]{KhrapakSciRep2017}%
  \BibitemOpen
  \bibfield  {author} {\bibinfo {author} {\bibfnamefont {S.}~\bibnamefont
  {Khrapak}}, \bibinfo {author} {\bibfnamefont {B.}~\bibnamefont {Klumov}}, \
  and\ \bibinfo {author} {\bibfnamefont {L.}~\bibnamefont {Couedel}},\
  }\bibfield  {title} {\enquote {\bibinfo {title} {Collective modes in simple
  melts: Transition from soft spheres to the hard sphere limit},}\ }\href
  {\doibase 10.1038/s41598-017-08429-5} {\bibfield  {journal} {\bibinfo
  {journal} {Sci. Rep.}\ }\textbf {\bibinfo {volume} {7}},\ \bibinfo {pages}
  {7985} (\bibinfo {year} {2017}{\natexlab{a}})}\BibitemShut {NoStop}%
\bibitem [{\citenamefont {Golden}\ \emph {et~al.}(1992)\citenamefont {Golden},
  \citenamefont {Kalman},\ and\ \citenamefont {Wyns}}]{GoldenPRA1992}%
  \BibitemOpen
  \bibfield  {author} {\bibinfo {author} {\bibfnamefont {K.~I.}\ \bibnamefont
  {Golden}}, \bibinfo {author} {\bibfnamefont {G.}~\bibnamefont {Kalman}}, \
  and\ \bibinfo {author} {\bibfnamefont {P.}~\bibnamefont {Wyns}},\ }\bibfield
  {title} {\enquote {\bibinfo {title} {Dielectric tensor and shear-mode
  dispersion for strongly coupled {C}oulomb liquids: {T}hree-dimensional
  one-component plasmas},}\ }\href {\doibase 10.1103/physreva.46.3454}
  {\bibfield  {journal} {\bibinfo  {journal} {Phys. Rev. A}\ }\textbf {\bibinfo
  {volume} {46}},\ \bibinfo {pages} {3454--3462} (\bibinfo {year}
  {1992})}\BibitemShut {NoStop}%
\bibitem [{\citenamefont {Rosenberg}\ and\ \citenamefont
  {Kalman}(1997)}]{RosenbergPRE1997}%
  \BibitemOpen
  \bibfield  {author} {\bibinfo {author} {\bibfnamefont {M.}~\bibnamefont
  {Rosenberg}}\ and\ \bibinfo {author} {\bibfnamefont {G.}~\bibnamefont
  {Kalman}},\ }\bibfield  {title} {\enquote {\bibinfo {title} {Dust acoustic
  waves in strongly coupled dusty plasmas},}\ }\href {\doibase
  10.1103/physreve.56.7166} {\bibfield  {journal} {\bibinfo  {journal} {Phys.
  Rev. E}\ }\textbf {\bibinfo {volume} {56}},\ \bibinfo {pages} {7166--7173}
  (\bibinfo {year} {1997})}\BibitemShut {NoStop}%
\bibitem [{\citenamefont {Kalman}\ \emph {et~al.}(2000)\citenamefont {Kalman},
  \citenamefont {Rosenberg},\ and\ \citenamefont {DeWitt}}]{KalmanPRL2000}%
  \BibitemOpen
  \bibfield  {author} {\bibinfo {author} {\bibfnamefont {G.}~\bibnamefont
  {Kalman}}, \bibinfo {author} {\bibfnamefont {M.}~\bibnamefont {Rosenberg}}, \
  and\ \bibinfo {author} {\bibfnamefont {H.~E.}\ \bibnamefont {DeWitt}},\
  }\bibfield  {title} {\enquote {\bibinfo {title} {Collective modes in strongly
  correlated {Y}ukawa liquids: {W}aves in dusty plasmas},}\ }\href {\doibase
  10.1103/physrevlett.84.6030} {\bibfield  {journal} {\bibinfo  {journal}
  {Phys. Rev. Lett.}\ }\textbf {\bibinfo {volume} {84}},\ \bibinfo {pages}
  {6030--6033} (\bibinfo {year} {2000})}\BibitemShut {NoStop}%
\bibitem [{\citenamefont {Schmidt}\ \emph {et~al.}(1997)\citenamefont
  {Schmidt}, \citenamefont {Zwicknagel}, \citenamefont {Reinhard},\ and\
  \citenamefont {Toepffer}}]{SchmidtPRE1997}%
  \BibitemOpen
  \bibfield  {author} {\bibinfo {author} {\bibfnamefont {P.}~\bibnamefont
  {Schmidt}}, \bibinfo {author} {\bibfnamefont {G.}~\bibnamefont {Zwicknagel}},
  \bibinfo {author} {\bibfnamefont {P.~G.}\ \bibnamefont {Reinhard}}, \ and\
  \bibinfo {author} {\bibfnamefont {C.}~\bibnamefont {Toepffer}},\ }\bibfield
  {title} {\enquote {\bibinfo {title} {Longitudinal and transversal collective
  modes in strongly correlated plasmas},}\ }\href {\doibase
  10.1103/physreve.56.7310} {\bibfield  {journal} {\bibinfo  {journal} {Phys.
  Rev. E}\ }\textbf {\bibinfo {volume} {56}},\ \bibinfo {pages} {7310--7313}
  (\bibinfo {year} {1997})}\BibitemShut {NoStop}%
\bibitem [{\citenamefont {Khrapak}\ \emph
  {et~al.}(2017{\natexlab{b}})\citenamefont {Khrapak}, \citenamefont {Klumov},\
  and\ \citenamefont {Thomas}}]{KhrapakPoP02_2017}%
  \BibitemOpen
  \bibfield  {author} {\bibinfo {author} {\bibfnamefont {S.~A.}\ \bibnamefont
  {Khrapak}}, \bibinfo {author} {\bibfnamefont {B.~A.}\ \bibnamefont {Klumov}},
  \ and\ \bibinfo {author} {\bibfnamefont {H.~M.}\ \bibnamefont {Thomas}},\
  }\bibfield  {title} {\enquote {\bibinfo {title} {Fingerprints of different
  interaction mechanisms on the collective modes in complex (dusty) plasmas},}\
  }\href {\doibase 10.1063/1.4976124} {\bibfield  {journal} {\bibinfo
  {journal} {Phys. Plasmas}\ }\textbf {\bibinfo {volume} {24}},\ \bibinfo
  {pages} {023702} (\bibinfo {year} {2017}{\natexlab{b}})}\BibitemShut
  {NoStop}%
\bibitem [{\citenamefont {Khrapak}\ and\ \citenamefont
  {Khrapak}(2018)}]{KhrapakIEEE2018}%
  \BibitemOpen
  \bibfield  {author} {\bibinfo {author} {\bibfnamefont {S.}~\bibnamefont
  {Khrapak}}\ and\ \bibinfo {author} {\bibfnamefont {A.}~\bibnamefont
  {Khrapak}},\ }\bibfield  {title} {\enquote {\bibinfo {title} {Simple
  dispersion relations for {C}oulomb and {Y}ukawa fluids},}\ }\href {\doibase
  10.1109/tps.2017.2763741} {\bibfield  {journal} {\bibinfo  {journal} {{IEEE}
  Trans. Plasma Sci.}\ }\textbf {\bibinfo {volume} {46}},\ \bibinfo {pages}
  {737--742} (\bibinfo {year} {2018})}\BibitemShut {NoStop}%
\bibitem [{\citenamefont {Khrapak}(2016)}]{KhrapakPoP10_2016}%
  \BibitemOpen
  \bibfield  {author} {\bibinfo {author} {\bibfnamefont {Sergey~A.}\
  \bibnamefont {Khrapak}},\ }\bibfield  {title} {\enquote {\bibinfo {title}
  {Onset of negative dispersion in one-component-plasma revisited},}\ }\href
  {\doibase 10.1063/1.4965903} {\bibfield  {journal} {\bibinfo  {journal}
  {Phys. Plasmas}\ }\textbf {\bibinfo {volume} {23}},\ \bibinfo {pages}
  {104506} (\bibinfo {year} {2016})}\BibitemShut {NoStop}%
\bibitem [{\citenamefont {Khrapak}(2017)}]{KhrapakAIPAdv2017}%
  \BibitemOpen
  \bibfield  {author} {\bibinfo {author} {\bibfnamefont {S.~A.}\ \bibnamefont
  {Khrapak}},\ }\bibfield  {title} {\enquote {\bibinfo {title} {Practical
  dispersion relations for strongly coupled plasma fluids},}\ }\href {\doibase
  10.1063/1.5002130} {\bibfield  {journal} {\bibinfo  {journal} {{AIP} Adv.}\
  }\textbf {\bibinfo {volume} {7}},\ \bibinfo {pages} {125026} (\bibinfo {year}
  {2017})}\BibitemShut {NoStop}%
\bibitem [{\citenamefont {Fairushin}\ \emph {et~al.}(2020)\citenamefont
  {Fairushin}, \citenamefont {Khrapak},\ and\ \citenamefont
  {Mokshin}}]{FairushinResPhys2020}%
  \BibitemOpen
  \bibfield  {author} {\bibinfo {author} {\bibfnamefont {I.I.}\ \bibnamefont
  {Fairushin}}, \bibinfo {author} {\bibfnamefont {S.A.}\ \bibnamefont
  {Khrapak}}, \ and\ \bibinfo {author} {\bibfnamefont {A.V.}\ \bibnamefont
  {Mokshin}},\ }\bibfield  {title} {\enquote {\bibinfo {title} {Direct
  evaluation of the physical characteristics of {Y}ukawa fluids based on a
  simple approximation for the radial distribution function},}\ }\href
  {\doibase 10.1016/j.rinp.2020.103359} {\bibfield  {journal} {\bibinfo
  {journal} {Res. Phys.}\ }\textbf {\bibinfo {volume} {19}},\ \bibinfo {pages}
  {103359} (\bibinfo {year} {2020})}\BibitemShut {NoStop}%
\bibitem [{\citenamefont {Khrapak}\ \emph {et~al.}(2019)\citenamefont
  {Khrapak}, \citenamefont {Khrapak}, \citenamefont {Kryuchkov},\ and\
  \citenamefont {Yurchenko}}]{KhrapakJCP2019}%
  \BibitemOpen
  \bibfield  {author} {\bibinfo {author} {\bibfnamefont {S.~A.}\ \bibnamefont
  {Khrapak}}, \bibinfo {author} {\bibfnamefont {A.~G.}\ \bibnamefont
  {Khrapak}}, \bibinfo {author} {\bibfnamefont {N.~P.}\ \bibnamefont
  {Kryuchkov}}, \ and\ \bibinfo {author} {\bibfnamefont {S.~O.}\ \bibnamefont
  {Yurchenko}},\ }\bibfield  {title} {\enquote {\bibinfo {title} {Onset of
  transverse (shear) waves in strongly-coupled {Y}ukawa fluids},}\ }\href
  {\doibase 10.1063/1.5088141} {\bibfield  {journal} {\bibinfo  {journal} {J.
  Chem. Phys.}\ }\textbf {\bibinfo {volume} {150}},\ \bibinfo {pages} {104503}
  (\bibinfo {year} {2019})}\BibitemShut {NoStop}%
\bibitem [{\citenamefont {Hansen}\ \emph {et~al.}(1975)\citenamefont {Hansen},
  \citenamefont {McDonald},\ and\ \citenamefont {Pollock}}]{HansenPRA1975}%
  \BibitemOpen
  \bibfield  {author} {\bibinfo {author} {\bibfnamefont {J.~P.}\ \bibnamefont
  {Hansen}}, \bibinfo {author} {\bibfnamefont {I.~R.}\ \bibnamefont
  {McDonald}}, \ and\ \bibinfo {author} {\bibfnamefont {E.~L.}\ \bibnamefont
  {Pollock}},\ }\bibfield  {title} {\enquote {\bibinfo {title} {Statistical
  mechanics of dense ionized matter. {III}. dynamical properties of the
  classical one-component plasma},}\ }\href {\doibase 10.1103/physreva.11.1025}
  {\bibfield  {journal} {\bibinfo  {journal} {Phys. Rev. A}\ }\textbf {\bibinfo
  {volume} {11}},\ \bibinfo {pages} {1025--1039} (\bibinfo {year}
  {1975})}\BibitemShut {NoStop}%
\bibitem [{\citenamefont {Donk{\'{o}}}\ \emph {et~al.}(1998)\citenamefont
  {Donk{\'{o}}}, \citenamefont {Ny{\'{\i}}ri}, \citenamefont {Szalai},\ and\
  \citenamefont {Holl{\'{o}}}}]{DonkoPRL1998}%
  \BibitemOpen
  \bibfield  {author} {\bibinfo {author} {\bibfnamefont {Z.}~\bibnamefont
  {Donk{\'{o}}}}, \bibinfo {author} {\bibfnamefont {B.}~\bibnamefont
  {Ny{\'{\i}}ri}}, \bibinfo {author} {\bibfnamefont {L.}~\bibnamefont
  {Szalai}}, \ and\ \bibinfo {author} {\bibfnamefont {S.}~\bibnamefont
  {Holl{\'{o}}}},\ }\bibfield  {title} {\enquote {\bibinfo {title} {Thermal
  conductivity of the classical electron one-component plasma},}\ }\href
  {\doibase 10.1103/physrevlett.81.1622} {\bibfield  {journal} {\bibinfo
  {journal} {Phys. Rev. Lett.}\ }\textbf {\bibinfo {volume} {81}},\ \bibinfo
  {pages} {1622--1625} (\bibinfo {year} {1998})}\BibitemShut {NoStop}%
\bibitem [{\citenamefont {Donko}\ and\ \citenamefont
  {Nyiri}(2000)}]{DonkoPoP2000}%
  \BibitemOpen
  \bibfield  {author} {\bibinfo {author} {\bibfnamefont {Z.}~\bibnamefont
  {Donko}}\ and\ \bibinfo {author} {\bibfnamefont {B.}~\bibnamefont {Nyiri}},\
  }\bibfield  {title} {\enquote {\bibinfo {title} {Molecular dynamics
  calculation of the thermal conductivity and shear viscosity of the classical
  one-component plasma},}\ }\href {\doibase 10.1063/1.873824} {\bibfield
  {journal} {\bibinfo  {journal} {Phys. Plasmas}\ }\textbf {\bibinfo {volume}
  {7}},\ \bibinfo {pages} {45--50} (\bibinfo {year} {2000})}\BibitemShut
  {NoStop}%
\bibitem [{\citenamefont {Salin}\ and\ \citenamefont
  {Caillol}(2002)}]{SalinPRL2002}%
  \BibitemOpen
  \bibfield  {author} {\bibinfo {author} {\bibfnamefont {G.}~\bibnamefont
  {Salin}}\ and\ \bibinfo {author} {\bibfnamefont {J.-M.}\ \bibnamefont
  {Caillol}},\ }\bibfield  {title} {\enquote {\bibinfo {title} {Transport
  coefficients of the {Y}ukawa one-component plasma},}\ }\href {\doibase
  10.1103/physrevlett.88.065002} {\bibfield  {journal} {\bibinfo  {journal}
  {Phys. Rev. Lett.}\ }\textbf {\bibinfo {volume} {88}},\ \bibinfo {pages}
  {065002} (\bibinfo {year} {2002})}\BibitemShut {NoStop}%
\bibitem [{\citenamefont {Vaulina}\ \emph {et~al.}(2002)\citenamefont
  {Vaulina}, \citenamefont {Khrapak},\ and\ \citenamefont
  {Morfill}}]{VaulinaPRE2002}%
  \BibitemOpen
  \bibfield  {author} {\bibinfo {author} {\bibfnamefont {O.}~\bibnamefont
  {Vaulina}}, \bibinfo {author} {\bibfnamefont {S.}~\bibnamefont {Khrapak}}, \
  and\ \bibinfo {author} {\bibfnamefont {G.}~\bibnamefont {Morfill}},\
  }\bibfield  {title} {\enquote {\bibinfo {title} {Universal scaling in complex
  (dusty) plasmas},}\ }\href {\doibase 10.1103/physreve.66.016404} {\bibfield
  {journal} {\bibinfo  {journal} {Phys. Rev. E}\ }\textbf {\bibinfo {volume}
  {66}},\ \bibinfo {pages} {016404} (\bibinfo {year} {2002})}\BibitemShut
  {NoStop}%
\bibitem [{\citenamefont {Bastea}(2005)}]{BasteaPRE2005}%
  \BibitemOpen
  \bibfield  {author} {\bibinfo {author} {\bibfnamefont {S.}~\bibnamefont
  {Bastea}},\ }\bibfield  {title} {\enquote {\bibinfo {title} {Viscosity and
  mutual diffusion in strongly asymmetric binary ionic mixtures},}\ }\href
  {\doibase 10.1103/physreve.71.056405} {\bibfield  {journal} {\bibinfo
  {journal} {Phys. Rev. E}\ }\textbf {\bibinfo {volume} {71}},\ \bibinfo
  {pages} {056405} (\bibinfo {year} {2005})}\BibitemShut {NoStop}%
\bibitem [{\citenamefont {Daligault}(2006)}]{DaligaultPRL2006}%
  \BibitemOpen
  \bibfield  {author} {\bibinfo {author} {\bibfnamefont {J.}~\bibnamefont
  {Daligault}},\ }\bibfield  {title} {\enquote {\bibinfo {title} {Liquid-state
  properties of a one-component plasma},}\ }\href {\doibase
  10.1103/physrevlett.96.065003} {\bibfield  {journal} {\bibinfo  {journal}
  {Phys. Rev. Lett.}\ }\textbf {\bibinfo {volume} {96}},\ \bibinfo {pages}
  {065003} (\bibinfo {year} {2006})}\BibitemShut {NoStop}%
\bibitem [{\citenamefont {Daligault}(2012{\natexlab{a}})}]{DaligaultPRL2012}%
  \BibitemOpen
  \bibfield  {author} {\bibinfo {author} {\bibfnamefont {J.}~\bibnamefont
  {Daligault}},\ }\bibfield  {title} {\enquote {\bibinfo {title} {Diffusion in
  ionic mixtures across coupling regimes},}\ }\href {\doibase
  10.1103/physrevlett.108.225004} {\bibfield  {journal} {\bibinfo  {journal}
  {Phys. Rev. Lett.}\ }\textbf {\bibinfo {volume} {108}},\ \bibinfo {pages}
  {225004} (\bibinfo {year} {2012}{\natexlab{a}})}\BibitemShut {NoStop}%
\bibitem [{\citenamefont {Daligault}(2012{\natexlab{b}})}]{DaligaultPRE2012}%
  \BibitemOpen
  \bibfield  {author} {\bibinfo {author} {\bibfnamefont {J.}~\bibnamefont
  {Daligault}},\ }\bibfield  {title} {\enquote {\bibinfo {title} {Practical
  model for the self-diffusion coefficient in {Y}ukawa one-component
  plasmas},}\ }\href {\doibase 10.1103/physreve.86.047401} {\bibfield
  {journal} {\bibinfo  {journal} {Phys. Rev. E}\ }\textbf {\bibinfo {volume}
  {86}},\ \bibinfo {pages} {047401} (\bibinfo {year}
  {2012}{\natexlab{b}})}\BibitemShut {NoStop}%
\bibitem [{\citenamefont {Khrapak}(2013)}]{KhrapakPoP2013}%
  \BibitemOpen
  \bibfield  {author} {\bibinfo {author} {\bibfnamefont {S.~A.}\ \bibnamefont
  {Khrapak}},\ }\bibfield  {title} {\enquote {\bibinfo {title} {Effective
  {C}oulomb logarithm for one component plasma},}\ }\href {\doibase
  10.1063/1.4804341} {\bibfield  {journal} {\bibinfo  {journal} {Phys.
  Plasmas}\ }\textbf {\bibinfo {volume} {20}},\ \bibinfo {pages} {054501}
  (\bibinfo {year} {2013})}\BibitemShut {NoStop}%
\bibitem [{\citenamefont {Daligault}\ \emph {et~al.}(2014)\citenamefont
  {Daligault}, \citenamefont {Rasmussen},\ and\ \citenamefont
  {Baalrud}}]{DaligaultPRE2014}%
  \BibitemOpen
  \bibfield  {author} {\bibinfo {author} {\bibfnamefont {J.}~\bibnamefont
  {Daligault}}, \bibinfo {author} {\bibfnamefont {K.}~\bibnamefont
  {Rasmussen}}, \ and\ \bibinfo {author} {\bibfnamefont {S.~D.}\ \bibnamefont
  {Baalrud}},\ }\bibfield  {title} {\enquote {\bibinfo {title} {Determination
  of the shear viscosity of the one-component plasma},}\ }\href {\doibase
  10.1103/physreve.90.033105} {\bibfield  {journal} {\bibinfo  {journal} {Phys.
  Rev. E}\ }\textbf {\bibinfo {volume} {90}},\ \bibinfo {pages} {033105}
  (\bibinfo {year} {2014})}\BibitemShut {NoStop}%
\bibitem [{\citenamefont {Scheiner}\ and\ \citenamefont
  {Baalrud}(2019)}]{ScheinerPRE2019}%
  \BibitemOpen
  \bibfield  {author} {\bibinfo {author} {\bibfnamefont {B.}~\bibnamefont
  {Scheiner}}\ and\ \bibinfo {author} {\bibfnamefont {S.~D.}\ \bibnamefont
  {Baalrud}},\ }\bibfield  {title} {\enquote {\bibinfo {title} {Testing thermal
  conductivity models with equilibrium molecular dynamics simulations of the
  one-component plasma},}\ }\href {\doibase 10.1103/physreve.100.043206}
  {\bibfield  {journal} {\bibinfo  {journal} {Phys. Rev. E}\ }\textbf {\bibinfo
  {volume} {100}},\ \bibinfo {pages} {043206} (\bibinfo {year}
  {2019})}\BibitemShut {NoStop}%
\bibitem [{\citenamefont {Khrapak}\ and\ \citenamefont
  {Khrapak}(2021{\natexlab{b}})}]{KhrapakPRE10_2021}%
  \BibitemOpen
  \bibfield  {author} {\bibinfo {author} {\bibfnamefont {S.~A.}\ \bibnamefont
  {Khrapak}}\ and\ \bibinfo {author} {\bibfnamefont {A.~G.}\ \bibnamefont
  {Khrapak}},\ }\bibfield  {title} {\enquote {\bibinfo {title} {Excess entropy
  and {S}tokes-{E}instein relation in simple fluids},}\ }\href {\doibase
  10.1103/physreve.104.044110} {\bibfield  {journal} {\bibinfo  {journal}
  {Phys. Rev. E}\ }\textbf {\bibinfo {volume} {104}},\ \bibinfo {pages}
  {044110} (\bibinfo {year} {2021}{\natexlab{b}})}\BibitemShut {NoStop}%
\bibitem [{\citenamefont {Khrapak}\ and\ \citenamefont
  {Khrapak}(2014)}]{KhrapakPoP10_2014}%
  \BibitemOpen
  \bibfield  {author} {\bibinfo {author} {\bibfnamefont {S.~A.}\ \bibnamefont
  {Khrapak}}\ and\ \bibinfo {author} {\bibfnamefont {A.~G.}\ \bibnamefont
  {Khrapak}},\ }\bibfield  {title} {\enquote {\bibinfo {title} {Simple
  thermodynamics of strongly coupled one-component-plasma in two and three
  dimensions},}\ }\href {\doibase 10.1063/1.4897386} {\bibfield  {journal}
  {\bibinfo  {journal} {Phys. Plasmas}\ }\textbf {\bibinfo {volume} {21}},\
  \bibinfo {pages} {104505} (\bibinfo {year} {2014})}\BibitemShut {NoStop}%
\bibitem [{\citenamefont {Caillol}(1999)}]{CaillolJCP1999}%
  \BibitemOpen
  \bibfield  {author} {\bibinfo {author} {\bibfnamefont {J.~M.}\ \bibnamefont
  {Caillol}},\ }\bibfield  {title} {\enquote {\bibinfo {title} {Thermodynamic
  limit of the excess internal energy of the fluid phase of a one-component
  plasma: A {M}onte {C}arlo study},}\ }\href {\doibase 10.1063/1.479965}
  {\bibfield  {journal} {\bibinfo  {journal} {J. Chem. Phys.}\ }\textbf
  {\bibinfo {volume} {111}},\ \bibinfo {pages} {6538--6547} (\bibinfo {year}
  {1999})}\BibitemShut {NoStop}%
\bibitem [{\citenamefont {Khrapak}\ and\ \citenamefont
  {Morfill}(2009)}]{KhrapakCPP2009}%
  \BibitemOpen
  \bibfield  {author} {\bibinfo {author} {\bibfnamefont {S.}~\bibnamefont
  {Khrapak}}\ and\ \bibinfo {author} {\bibfnamefont {G.}~\bibnamefont
  {Morfill}},\ }\bibfield  {title} {\enquote {\bibinfo {title} {Basic processes
  in complex (dusty) plasmas: Charging, interactions, and ion drag force},}\
  }\href {\doibase 10.1002/ctpp.200910018} {\bibfield  {journal} {\bibinfo
  {journal} {Contrib. Plasma Phys.}\ }\textbf {\bibinfo {volume} {49}},\
  \bibinfo {pages} {148--168} (\bibinfo {year} {2009})}\BibitemShut {NoStop}%
\bibitem [{\citenamefont {Semenov}\ \emph {et~al.}(2015)\citenamefont
  {Semenov}, \citenamefont {Khrapak},\ and\ \citenamefont
  {Thomas}}]{SemenovPoP2015}%
  \BibitemOpen
  \bibfield  {author} {\bibinfo {author} {\bibfnamefont {I.~L.}\ \bibnamefont
  {Semenov}}, \bibinfo {author} {\bibfnamefont {S.~A.}\ \bibnamefont
  {Khrapak}}, \ and\ \bibinfo {author} {\bibfnamefont {H.~M.}\ \bibnamefont
  {Thomas}},\ }\bibfield  {title} {\enquote {\bibinfo {title} {Approximate
  expression for the electric potential around an absorbing particle in
  isotropic collisionless plasma},}\ }\href {\doibase 10.1063/1.4921249}
  {\bibfield  {journal} {\bibinfo  {journal} {Phys. Plasmas}\ }\textbf
  {\bibinfo {volume} {22}},\ \bibinfo {pages} {053704} (\bibinfo {year}
  {2015})}\BibitemShut {NoStop}%
\bibitem [{\citenamefont {{Tsytovich}}(1997)}]{TsytovichUFN1997}%
  \BibitemOpen
  \bibfield  {author} {\bibinfo {author} {\bibfnamefont {V.}~\bibnamefont
  {{Tsytovich}}},\ }\bibfield  {title} {\enquote {\bibinfo {title} {{Dust
  plasma crystals, drops, and clouds.}}}\ }\href {\doibase
  10.1070/PU1997v040n01ABEH000201} {\bibfield  {journal} {\bibinfo  {journal}
  {Phys.-Usp.}\ }\textbf {\bibinfo {volume} {40}},\ \bibinfo {pages} {53--94}
  (\bibinfo {year} {1997})}\BibitemShut {NoStop}%
\bibitem [{\citenamefont {Fortov}\ \emph {et~al.}(2004)\citenamefont {Fortov},
  \citenamefont {Khrapak}, \citenamefont {Khrapak}, \citenamefont {Molotkov},\
  and\ \citenamefont {Petrov}}]{FortovUFN}%
  \BibitemOpen
  \bibfield  {author} {\bibinfo {author} {\bibfnamefont {V.~E.}\ \bibnamefont
  {Fortov}}, \bibinfo {author} {\bibfnamefont {A.~G.}\ \bibnamefont {Khrapak}},
  \bibinfo {author} {\bibfnamefont {S.~A.}\ \bibnamefont {Khrapak}}, \bibinfo
  {author} {\bibfnamefont {V.~I.}\ \bibnamefont {Molotkov}}, \ and\ \bibinfo
  {author} {\bibfnamefont {O.~F.}\ \bibnamefont {Petrov}},\ }\bibfield  {title}
  {\enquote {\bibinfo {title} {Dusty plasmas},}\ }\href {\doibase
  10.3367/ufnr.0174.200405b.0495} {\bibfield  {journal} {\bibinfo  {journal}
  {Phys.-Usp.}\ }\textbf {\bibinfo {volume} {47}},\ \bibinfo {pages} {447 --
  492} (\bibinfo {year} {2004})}\BibitemShut {NoStop}%
\bibitem [{\citenamefont {Fortov}\ \emph {et~al.}(2005)\citenamefont {Fortov},
  \citenamefont {Ivlev}, \citenamefont {Khrapak}, \citenamefont {Khrapak},\
  and\ \citenamefont {Morfill}}]{FortovPR}%
  \BibitemOpen
  \bibfield  {author} {\bibinfo {author} {\bibfnamefont {V.~E.}\ \bibnamefont
  {Fortov}}, \bibinfo {author} {\bibfnamefont {A.~V.}\ \bibnamefont {Ivlev}},
  \bibinfo {author} {\bibfnamefont {S.~A.}\ \bibnamefont {Khrapak}}, \bibinfo
  {author} {\bibfnamefont {A.~G.}\ \bibnamefont {Khrapak}}, \ and\ \bibinfo
  {author} {\bibfnamefont {G.~E.}\ \bibnamefont {Morfill}},\ }\bibfield
  {title} {\enquote {\bibinfo {title} {Complex (dusty) plasmas: Current status,
  open issues, perspectives},}\ }\href@noop {} {\bibfield  {journal} {\bibinfo
  {journal} {Phys. Rep.}\ }\textbf {\bibinfo {volume} {421}},\ \bibinfo {pages}
  {1--103} (\bibinfo {year} {2005})}\BibitemShut {NoStop}%
\bibitem [{\citenamefont {Ivlev}\ \emph {et~al.}(2012)\citenamefont {Ivlev},
  \citenamefont {L\"owen}, \citenamefont {Morfill},\ and\ \citenamefont
  {Royall}}]{IvlevBook}%
  \BibitemOpen
  \bibfield  {author} {\bibinfo {author} {\bibfnamefont {A.}~\bibnamefont
  {Ivlev}}, \bibinfo {author} {\bibfnamefont {H.}~\bibnamefont {L\"owen}},
  \bibinfo {author} {\bibfnamefont {G.}~\bibnamefont {Morfill}}, \ and\
  \bibinfo {author} {\bibfnamefont {C.~P.}\ \bibnamefont {Royall}},\
  }\href@noop {} {\emph {\bibinfo {title} {Complex Plasmas and Colloidal
  Dispersions: Particle-Resolved Studies of Classical Liquids and Solids}}}\
  (\bibinfo  {publisher} {World Scientific},\ \bibinfo {year}
  {2012})\BibitemShut {NoStop}%
\bibitem [{\citenamefont {Khrapak}\ \emph {et~al.}(2008)\citenamefont
  {Khrapak}, \citenamefont {Klumov},\ and\ \citenamefont
  {Morfill}}]{KhrapakPRL2008}%
  \BibitemOpen
  \bibfield  {author} {\bibinfo {author} {\bibfnamefont {S.~A.}\ \bibnamefont
  {Khrapak}}, \bibinfo {author} {\bibfnamefont {B.~A.}\ \bibnamefont {Klumov}},
  \ and\ \bibinfo {author} {\bibfnamefont {G.~E.}\ \bibnamefont {Morfill}},\
  }\bibfield  {title} {\enquote {\bibinfo {title} {Electric potential around an
  absorbing body in plasmas: Effect of ion-neutral collisions},}\ }\href
  {\doibase 10.1103/physrevlett.100.225003} {\bibfield  {journal} {\bibinfo
  {journal} {Phys. Rev. Lett.}\ }\textbf {\bibinfo {volume} {100}},\ \bibinfo
  {pages} {225003} (\bibinfo {year} {2008})}\BibitemShut {NoStop}%
\bibitem [{\citenamefont {Klumov}(2010)}]{KlumovUFN2010}%
  \BibitemOpen
  \bibfield  {author} {\bibinfo {author} {\bibfnamefont {B.~A}\ \bibnamefont
  {Klumov}},\ }\bibfield  {title} {\enquote {\bibinfo {title} {On melting
  criteria for complex plasma},}\ }\href {\doibase
  10.3367/ufne.0180.201010e.1095} {\bibfield  {journal} {\bibinfo  {journal}
  {Phys.-Usp.}\ }\textbf {\bibinfo {volume} {53}},\ \bibinfo {pages}
  {1053--1065} (\bibinfo {year} {2010})}\BibitemShut {NoStop}%
\bibitem [{\citenamefont {Chaudhuri}\ \emph {et~al.}(2011)\citenamefont
  {Chaudhuri}, \citenamefont {Ivlev}, \citenamefont {Khrapak}, \citenamefont
  {Thomas},\ and\ \citenamefont {Morfill}}]{ChaudhuriSM2011}%
  \BibitemOpen
  \bibfield  {author} {\bibinfo {author} {\bibfnamefont {M.}~\bibnamefont
  {Chaudhuri}}, \bibinfo {author} {\bibfnamefont {A.~V.}\ \bibnamefont
  {Ivlev}}, \bibinfo {author} {\bibfnamefont {S.~A.}\ \bibnamefont {Khrapak}},
  \bibinfo {author} {\bibfnamefont {H.~M.}\ \bibnamefont {Thomas}}, \ and\
  \bibinfo {author} {\bibfnamefont {G.~E.}\ \bibnamefont {Morfill}},\
  }\bibfield  {title} {\enquote {\bibinfo {title} {Complex
  plasma{\textemdash}the plasma state of soft matter},}\ }\href {\doibase
  10.1039/c0sm00813c} {\bibfield  {journal} {\bibinfo  {journal} {Soft Matter}\
  }\textbf {\bibinfo {volume} {7}},\ \bibinfo {pages} {1287--1298} (\bibinfo
  {year} {2011})}\BibitemShut {NoStop}%
\bibitem [{\citenamefont {Lampe}\ and\ \citenamefont
  {Joyce}(2015)}]{LampePoP2015}%
  \BibitemOpen
  \bibfield  {author} {\bibinfo {author} {\bibfnamefont {M.}~\bibnamefont
  {Lampe}}\ and\ \bibinfo {author} {\bibfnamefont {G.}~\bibnamefont {Joyce}},\
  }\bibfield  {title} {\enquote {\bibinfo {title} {Grain-grain interaction in
  stationary dusty plasma},}\ }\href {\doibase 10.1063/1.4907649} {\bibfield
  {journal} {\bibinfo  {journal} {Phys. Plasmas}\ }\textbf {\bibinfo {volume}
  {22}},\ \bibinfo {pages} {023704} (\bibinfo {year} {2015})}\BibitemShut
  {NoStop}%
\bibitem [{\citenamefont {Robbins}\ \emph {et~al.}(1988)\citenamefont
  {Robbins}, \citenamefont {Kremer},\ and\ \citenamefont
  {Grest}}]{RobbinsJCP1988}%
  \BibitemOpen
  \bibfield  {author} {\bibinfo {author} {\bibfnamefont {M.~O.}\ \bibnamefont
  {Robbins}}, \bibinfo {author} {\bibfnamefont {K.}~\bibnamefont {Kremer}}, \
  and\ \bibinfo {author} {\bibfnamefont {G.~S.}\ \bibnamefont {Grest}},\
  }\bibfield  {title} {\enquote {\bibinfo {title} {Phase diagram and dynamics
  of {Y}ukawa systems},}\ }\href {\doibase 10.1063/1.453924} {\bibfield
  {journal} {\bibinfo  {journal} {J. Chem. Phys.}\ }\textbf {\bibinfo {volume}
  {88}},\ \bibinfo {pages} {3286--3312} (\bibinfo {year} {1988})}\BibitemShut
  {NoStop}%
\bibitem [{\citenamefont {{Hamaguchi}}\ \emph {et~al.}(1996)\citenamefont
  {{Hamaguchi}}, \citenamefont {{Farouki}},\ and\ \citenamefont
  {{Dubin}}}]{HamaguchiJCP1996}%
  \BibitemOpen
  \bibfield  {author} {\bibinfo {author} {\bibfnamefont {S.}~\bibnamefont
  {{Hamaguchi}}}, \bibinfo {author} {\bibfnamefont {R.~T.}\ \bibnamefont
  {{Farouki}}}, \ and\ \bibinfo {author} {\bibfnamefont {D.~H.~E.}\
  \bibnamefont {{Dubin}}},\ }\bibfield  {title} {\enquote {\bibinfo {title}
  {{Phase diagram of {Y}ukawa systems near the one-component-plasma limit
  revisited}},}\ }\href {\doibase 10.1063/1.472802} {\bibfield  {journal}
  {\bibinfo  {journal} {J. Chem. Phys.}\ }\textbf {\bibinfo {volume} {105}},\
  \bibinfo {pages} {7641--7647} (\bibinfo {year} {1996})}\BibitemShut {NoStop}%
\bibitem [{\citenamefont {Hamaguchi}\ \emph {et~al.}(1997)\citenamefont
  {Hamaguchi}, \citenamefont {Farouki},\ and\ \citenamefont
  {Dubin}}]{HamaguchiPRE1997}%
  \BibitemOpen
  \bibfield  {author} {\bibinfo {author} {\bibfnamefont {S.}~\bibnamefont
  {Hamaguchi}}, \bibinfo {author} {\bibfnamefont {R.~T.}\ \bibnamefont
  {Farouki}}, \ and\ \bibinfo {author} {\bibfnamefont {D.~H.~E.}\ \bibnamefont
  {Dubin}},\ }\bibfield  {title} {\enquote {\bibinfo {title} {Triple point of
  {Y}ukawa systems},}\ }\href {\doibase 10.1103/physreve.56.4671} {\bibfield
  {journal} {\bibinfo  {journal} {Phys. Rev. E}\ }\textbf {\bibinfo {volume}
  {56}},\ \bibinfo {pages} {4671--4682} (\bibinfo {year} {1997})}\BibitemShut
  {NoStop}%
\bibitem [{\citenamefont {Vaulina}\ and\ \citenamefont
  {Khrapak}(2000)}]{VaulinaJETP2000}%
  \BibitemOpen
  \bibfield  {author} {\bibinfo {author} {\bibfnamefont {O.~S.}\ \bibnamefont
  {Vaulina}}\ and\ \bibinfo {author} {\bibfnamefont {S.~A.}\ \bibnamefont
  {Khrapak}},\ }\bibfield  {title} {\enquote {\bibinfo {title} {Scaling law for
  the fluid-solid phase transition in {Y}ukawa systems (dusty plasmas)},}\
  }\href {\doibase 10.1134/1.559102} {\bibfield  {journal} {\bibinfo  {journal}
  {JETP}\ }\textbf {\bibinfo {volume} {90}},\ \bibinfo {pages} {287--289}
  (\bibinfo {year} {2000})}\BibitemShut {NoStop}%
\bibitem [{\citenamefont {Khrapak}\ \emph {et~al.}(2010)\citenamefont
  {Khrapak}, \citenamefont {Thomas},\ and\ \citenamefont
  {Morfill}}]{KhrapakEPL2010}%
  \BibitemOpen
  \bibfield  {author} {\bibinfo {author} {\bibfnamefont {S.~A.}\ \bibnamefont
  {Khrapak}}, \bibinfo {author} {\bibfnamefont {H.~M.}\ \bibnamefont {Thomas}},
  \ and\ \bibinfo {author} {\bibfnamefont {G.~E.}\ \bibnamefont {Morfill}},\
  }\bibfield  {title} {\enquote {\bibinfo {title} {Multiple phase transitions
  associated with charge cannibalism effect in complex (dusty) plasmas},}\
  }\href {\doibase 10.1209/0295-5075/91/25001} {\bibfield  {journal} {\bibinfo
  {journal} {{EPL} (Europhys. Lett.)}\ }\textbf {\bibinfo {volume} {91}},\
  \bibinfo {pages} {25001} (\bibinfo {year} {2010})}\BibitemShut {NoStop}%
\bibitem [{\citenamefont {Yazdi}\ \emph {et~al.}(2014)\citenamefont {Yazdi},
  \citenamefont {Ivlev}, \citenamefont {Khrapak}, \citenamefont {Thomas},
  \citenamefont {Morfill}, \citenamefont {L\"{o}wen}, \citenamefont {Wysocki},\
  and\ \citenamefont {Sperl}}]{YazdiPRE2014}%
  \BibitemOpen
  \bibfield  {author} {\bibinfo {author} {\bibfnamefont {A.}~\bibnamefont
  {Yazdi}}, \bibinfo {author} {\bibfnamefont {A.}~\bibnamefont {Ivlev}},
  \bibinfo {author} {\bibfnamefont {S.}~\bibnamefont {Khrapak}}, \bibinfo
  {author} {\bibfnamefont {H.}~\bibnamefont {Thomas}}, \bibinfo {author}
  {\bibfnamefont {G.~E.}\ \bibnamefont {Morfill}}, \bibinfo {author}
  {\bibfnamefont {H.}~\bibnamefont {L\"{o}wen}}, \bibinfo {author}
  {\bibfnamefont {A.}~\bibnamefont {Wysocki}}, \ and\ \bibinfo {author}
  {\bibfnamefont {M.}~\bibnamefont {Sperl}},\ }\bibfield  {title} {\enquote
  {\bibinfo {title} {Glass-transition properties of {Y}ukawa potentials: From
  charged point particles to hard spheres},}\ }\href {\doibase
  10.1103/PhysRevE.89.063105} {\bibfield  {journal} {\bibinfo  {journal} {Phys.
  Rev. E}\ }\textbf {\bibinfo {volume} {89}},\ \bibinfo {pages} {063105}
  (\bibinfo {year} {2014})}\BibitemShut {NoStop}%
\bibitem [{\citenamefont {Sanbonmatsu}\ and\ \citenamefont
  {Murillo}(2001)}]{SanbonmatsuPRL2001}%
  \BibitemOpen
  \bibfield  {author} {\bibinfo {author} {\bibfnamefont {K.~Y.}\ \bibnamefont
  {Sanbonmatsu}}\ and\ \bibinfo {author} {\bibfnamefont {M.~S.}\ \bibnamefont
  {Murillo}},\ }\bibfield  {title} {\enquote {\bibinfo {title} {Shear viscosity
  of strongly coupled {Y}ukawa systems on finite length scales},}\ }\href
  {\doibase 10.1103/physrevlett.86.1215} {\bibfield  {journal} {\bibinfo
  {journal} {Phys. Rev. Lett.}\ }\textbf {\bibinfo {volume} {86}},\ \bibinfo
  {pages} {1215--1218} (\bibinfo {year} {2001})}\BibitemShut {NoStop}%
\bibitem [{\citenamefont {Saigo}\ and\ \citenamefont
  {Hamaguchi}(2002)}]{SaigoPoP2002}%
  \BibitemOpen
  \bibfield  {author} {\bibinfo {author} {\bibfnamefont {T.}~\bibnamefont
  {Saigo}}\ and\ \bibinfo {author} {\bibfnamefont {S.}~\bibnamefont
  {Hamaguchi}},\ }\bibfield  {title} {\enquote {\bibinfo {title} {Shear
  viscosity of strongly coupled {Y}ukawa systems},}\ }\href {\doibase
  10.1063/1.1459708} {\bibfield  {journal} {\bibinfo  {journal} {Phys.
  Plasmas}\ }\textbf {\bibinfo {volume} {9}},\ \bibinfo {pages} {1210--1216}
  (\bibinfo {year} {2002})}\BibitemShut {NoStop}%
\bibitem [{\citenamefont {Salin}\ and\ \citenamefont
  {Caillol}(2003)}]{SalinPoP2003}%
  \BibitemOpen
  \bibfield  {author} {\bibinfo {author} {\bibfnamefont {G.}~\bibnamefont
  {Salin}}\ and\ \bibinfo {author} {\bibfnamefont {J.-M.}\ \bibnamefont
  {Caillol}},\ }\bibfield  {title} {\enquote {\bibinfo {title} {Equilibrium
  molecular dynamics simulations of the transport coefficients of the {Y}ukawa
  one component plasma},}\ }\href {\doibase 10.1063/1.1566749} {\bibfield
  {journal} {\bibinfo  {journal} {Phys. Plasmas}\ }\textbf {\bibinfo {volume}
  {10}},\ \bibinfo {pages} {1220--1230} (\bibinfo {year} {2003})}\BibitemShut
  {NoStop}%
\bibitem [{\citenamefont {Faussurier}\ and\ \citenamefont
  {Murillo}(2003)}]{FaussurierPRE2003}%
  \BibitemOpen
  \bibfield  {author} {\bibinfo {author} {\bibfnamefont {G.}~\bibnamefont
  {Faussurier}}\ and\ \bibinfo {author} {\bibfnamefont {M.~S.}\ \bibnamefont
  {Murillo}},\ }\bibfield  {title} {\enquote {\bibinfo {title}
  {Gibbs-{B}ogolyubov inequality and transport properties for strongly coupled
  {Y}ukawa fluids},}\ }\href {\doibase 10.1103/physreve.67.046404} {\bibfield
  {journal} {\bibinfo  {journal} {Phys. Rev. E}\ }\textbf {\bibinfo {volume}
  {67}},\ \bibinfo {pages} {046404} (\bibinfo {year} {2003})}\BibitemShut
  {NoStop}%
\bibitem [{\citenamefont {Donk{\'{o}}}\ and\ \citenamefont
  {Hartmann}(2004)}]{DonkoPRE2004}%
  \BibitemOpen
  \bibfield  {author} {\bibinfo {author} {\bibfnamefont {Z.}~\bibnamefont
  {Donk{\'{o}}}}\ and\ \bibinfo {author} {\bibfnamefont {P.}~\bibnamefont
  {Hartmann}},\ }\bibfield  {title} {\enquote {\bibinfo {title} {Thermal
  conductivity of strongly coupled {Y}ukawa liquids},}\ }\href {\doibase
  10.1103/physreve.69.016405} {\bibfield  {journal} {\bibinfo  {journal} {Phys.
  Rev. E}\ }\textbf {\bibinfo {volume} {69}},\ \bibinfo {pages} {016405}
  (\bibinfo {year} {2004})}\BibitemShut {NoStop}%
\bibitem [{\citenamefont {Donko}\ and\ \citenamefont
  {Hartmann}(2008)}]{DonkoPRE2008}%
  \BibitemOpen
  \bibfield  {author} {\bibinfo {author} {\bibfnamefont {Z.}~\bibnamefont
  {Donko}}\ and\ \bibinfo {author} {\bibfnamefont {P.}~\bibnamefont
  {Hartmann}},\ }\bibfield  {title} {\enquote {\bibinfo {title} {Shear
  viscosity of strongly coupled {Y}ukawa liquids},}\ }\href {\doibase
  10.1103/physreve.78.026408} {\bibfield  {journal} {\bibinfo  {journal} {Phys.
  Rev. E}\ }\textbf {\bibinfo {volume} {78}},\ \bibinfo {pages} {026408}
  (\bibinfo {year} {2008})}\BibitemShut {NoStop}%
\bibitem [{\citenamefont {Khrapak}\ \emph {et~al.}(2012)\citenamefont
  {Khrapak}, \citenamefont {Vaulina},\ and\ \citenamefont
  {Morfill}}]{KhrapakPoP2012}%
  \BibitemOpen
  \bibfield  {author} {\bibinfo {author} {\bibfnamefont {S.~A.}\ \bibnamefont
  {Khrapak}}, \bibinfo {author} {\bibfnamefont {O.~S.}\ \bibnamefont
  {Vaulina}}, \ and\ \bibinfo {author} {\bibfnamefont {G.~E.}\ \bibnamefont
  {Morfill}},\ }\bibfield  {title} {\enquote {\bibinfo {title} {Self-diffusion
  in strongly coupled {Y}ukawa systems (complex plasmas)},}\ }\href {\doibase
  10.1063/1.3691960} {\bibfield  {journal} {\bibinfo  {journal} {Phys.
  Plasmas}\ }\textbf {\bibinfo {volume} {19}},\ \bibinfo {pages} {034503}
  (\bibinfo {year} {2012})}\BibitemShut {NoStop}%
\bibitem [{\citenamefont {Khrapak}\ \emph {et~al.}(2018)\citenamefont
  {Khrapak}, \citenamefont {Klumov},\ and\ \citenamefont
  {Couedel}}]{KhrapakJPCO2018}%
  \BibitemOpen
  \bibfield  {author} {\bibinfo {author} {\bibfnamefont {S.}~\bibnamefont
  {Khrapak}}, \bibinfo {author} {\bibfnamefont {B.}~\bibnamefont {Klumov}}, \
  and\ \bibinfo {author} {\bibfnamefont {L.}~\bibnamefont {Couedel}},\
  }\bibfield  {title} {\enquote {\bibinfo {title} {Self-diffusion in
  single-component {Y}ukawa fluids},}\ }\href {\doibase
  10.1088/2399-6528/aaba23} {\bibfield  {journal} {\bibinfo  {journal} {J.
  Phys. Commun.}\ }\textbf {\bibinfo {volume} {2}},\ \bibinfo {pages} {045013}
  (\bibinfo {year} {2018})}\BibitemShut {NoStop}%
\bibitem [{\citenamefont {K\"{a}hlert}(2020)}]{KahlertPPR2020}%
  \BibitemOpen
  \bibfield  {author} {\bibinfo {author} {\bibfnamefont {H.}~\bibnamefont
  {K\"{a}hlert}},\ }\bibfield  {title} {\enquote {\bibinfo {title}
  {Thermodynamic and transport coefficients from the dynamic structure factor
  of {Y}ukawa liquids},}\ }\href {\doibase 10.1103/physrevresearch.2.033287}
  {\bibfield  {journal} {\bibinfo  {journal} {Phys. Rev. Research}\ }\textbf
  {\bibinfo {volume} {2}},\ \bibinfo {pages} {033287} (\bibinfo {year}
  {2020})}\BibitemShut {NoStop}%
\bibitem [{\citenamefont {Tolias}\ \emph {et~al.}(2014)\citenamefont {Tolias},
  \citenamefont {Ratynskaia},\ and\ \citenamefont
  {de~Angelis}}]{ToliasPRE2014}%
  \BibitemOpen
  \bibfield  {author} {\bibinfo {author} {\bibfnamefont {P.}~\bibnamefont
  {Tolias}}, \bibinfo {author} {\bibfnamefont {S.}~\bibnamefont {Ratynskaia}},
  \ and\ \bibinfo {author} {\bibfnamefont {U.}~\bibnamefont {de~Angelis}},\
  }\bibfield  {title} {\enquote {\bibinfo {title} {Soft mean spherical
  approximation for dusty plasma liquids: One-component {Y}ukawa systems with
  plasma shielding},}\ }\href {\doibase 10.1103/physreve.90.053101} {\bibfield
  {journal} {\bibinfo  {journal} {Phys. Rev. E}\ }\textbf {\bibinfo {volume}
  {90}},\ \bibinfo {pages} {053101} (\bibinfo {year} {2014})}\BibitemShut
  {NoStop}%
\bibitem [{\citenamefont {Tolias}\ \emph {et~al.}(2015)\citenamefont {Tolias},
  \citenamefont {Ratynskaia},\ and\ \citenamefont
  {de~Angelis}}]{ToliasPoP2015}%
  \BibitemOpen
  \bibfield  {author} {\bibinfo {author} {\bibfnamefont {P.}~\bibnamefont
  {Tolias}}, \bibinfo {author} {\bibfnamefont {S.}~\bibnamefont {Ratynskaia}},
  \ and\ \bibinfo {author} {\bibfnamefont {U.}~\bibnamefont {de~Angelis}},\
  }\bibfield  {title} {\enquote {\bibinfo {title} {Soft mean spherical
  approximation for dusty plasma liquids: Level of accuracy and analytic
  expressions},}\ }\href {\doibase 10.1063/1.4928113} {\bibfield  {journal}
  {\bibinfo  {journal} {Phys. Plasmas}\ }\textbf {\bibinfo {volume} {22}},\
  \bibinfo {pages} {083703} (\bibinfo {year} {2015})}\BibitemShut {NoStop}%
\bibitem [{\citenamefont {Khrapak}\ and\ \citenamefont
  {Thomas}(2015)}]{KhrapakPRE02_2015}%
  \BibitemOpen
  \bibfield  {author} {\bibinfo {author} {\bibfnamefont {S.~A.}\ \bibnamefont
  {Khrapak}}\ and\ \bibinfo {author} {\bibfnamefont {H.~M.}\ \bibnamefont
  {Thomas}},\ }\bibfield  {title} {\enquote {\bibinfo {title} {Practical
  expressions for the internal energy and pressure of {Y}ukawa fluids},}\
  }\href {\doibase 10.1103/physreve.91.023108} {\bibfield  {journal} {\bibinfo
  {journal} {Phys. Rev. E}\ }\textbf {\bibinfo {volume} {91}},\ \bibinfo
  {pages} {023108} (\bibinfo {year} {2015})}\BibitemShut {NoStop}%
\bibitem [{\citenamefont {Khrapak}\ \emph {et~al.}(2015)\citenamefont
  {Khrapak}, \citenamefont {Kryuchkov}, \citenamefont {Yurchenko},\ and\
  \citenamefont {Thomas}}]{KhrapakJCP2015}%
  \BibitemOpen
  \bibfield  {author} {\bibinfo {author} {\bibfnamefont {S.~A.}\ \bibnamefont
  {Khrapak}}, \bibinfo {author} {\bibfnamefont {N.~P.}\ \bibnamefont
  {Kryuchkov}}, \bibinfo {author} {\bibfnamefont {S.~O.}\ \bibnamefont
  {Yurchenko}}, \ and\ \bibinfo {author} {\bibfnamefont {H.~M.}\ \bibnamefont
  {Thomas}},\ }\bibfield  {title} {\enquote {\bibinfo {title} {Practical
  thermodynamics of {Y}ukawa systems at strong coupling},}\ }\href {\doibase
  10.1063/1.4921223} {\bibfield  {journal} {\bibinfo  {journal} {J. Chem.
  Phys.}\ }\textbf {\bibinfo {volume} {142}},\ \bibinfo {pages} {194903}
  (\bibinfo {year} {2015})}\BibitemShut {NoStop}%
\bibitem [{\citenamefont {Khrapak}(2015)}]{KhrapakPPCF2015}%
  \BibitemOpen
  \bibfield  {author} {\bibinfo {author} {\bibfnamefont {S.~A.}\ \bibnamefont
  {Khrapak}},\ }\bibfield  {title} {\enquote {\bibinfo {title} {Thermodynamics
  of {Y}ukawa systems and sound velocity in dusty plasmas},}\ }\href {\doibase
  10.1088/0741-3335/58/1/014022} {\bibfield  {journal} {\bibinfo  {journal}
  {Plasma Phys. Control. Fusion}\ }\textbf {\bibinfo {volume} {58}},\ \bibinfo
  {pages} {014022} (\bibinfo {year} {2015})}\BibitemShut {NoStop}%
\bibitem [{\citenamefont {Veldhorst}\ \emph {et~al.}(2015)\citenamefont
  {Veldhorst}, \citenamefont {Schroder},\ and\ \citenamefont
  {Dyre}}]{VeldhorstPoP2015}%
  \BibitemOpen
  \bibfield  {author} {\bibinfo {author} {\bibfnamefont {A.~A.}\ \bibnamefont
  {Veldhorst}}, \bibinfo {author} {\bibfnamefont {T.~B.}\ \bibnamefont
  {Schroder}}, \ and\ \bibinfo {author} {\bibfnamefont {J.~C.}\ \bibnamefont
  {Dyre}},\ }\bibfield  {title} {\enquote {\bibinfo {title} {Invariants in the
  {Y}ukawa system thermodynamic phase diagram},}\ }\href {\doibase
  10.1063/1.4926822} {\bibfield  {journal} {\bibinfo  {journal} {Phys.
  Plasmas}\ }\textbf {\bibinfo {volume} {22}},\ \bibinfo {pages} {073705}
  (\bibinfo {year} {2015})}\BibitemShut {NoStop}%
\bibitem [{\citenamefont {Castello}\ and\ \citenamefont
  {Tolias}(2020)}]{CastelloCPP2020}%
  \BibitemOpen
  \bibfield  {author} {\bibinfo {author} {\bibfnamefont {F.~L.}\ \bibnamefont
  {Castello}}\ and\ \bibinfo {author} {\bibfnamefont {P.}~\bibnamefont
  {Tolias}},\ }\bibfield  {title} {\enquote {\bibinfo {title} {On the advanced
  integral equation theory description of dense {Y}ukawa one-component plasma
  liquids},}\ }\href {\doibase 10.1002/ctpp.202000105} {\bibfield  {journal}
  {\bibinfo  {journal} {Contrib. Plasma Phys.}\ }\textbf {\bibinfo {volume}
  {61}},\ \bibinfo {pages} {e202000105} (\bibinfo {year} {2020})}\BibitemShut
  {NoStop}%
\bibitem [{\citenamefont {Rosenfeld}\ and\ \citenamefont
  {Tarazona}(1998)}]{RosenfeldMolPhys1998}%
  \BibitemOpen
  \bibfield  {author} {\bibinfo {author} {\bibfnamefont {Y.}~\bibnamefont
  {Rosenfeld}}\ and\ \bibinfo {author} {\bibfnamefont {P.}~\bibnamefont
  {Tarazona}},\ }\bibfield  {title} {\enquote {\bibinfo {title} {Density
  functional theory and the asymptotic high density expansion of the free
  energy of classical solids and fluids},}\ }\href {\doibase
  10.1080/00268979809483145} {\bibfield  {journal} {\bibinfo  {journal} {Mol.
  Phys.}\ }\textbf {\bibinfo {volume} {95}},\ \bibinfo {pages} {141--150}
  (\bibinfo {year} {1998})}\BibitemShut {NoStop}%
\bibitem [{\citenamefont {Rosenfeld}(2000)}]{RosenfeldPRE2000}%
  \BibitemOpen
  \bibfield  {author} {\bibinfo {author} {\bibfnamefont {Y.}~\bibnamefont
  {Rosenfeld}},\ }\bibfield  {title} {\enquote {\bibinfo {title}
  {Excess-entropy and freezing-temperature scalings for transport coefficients:
  Self-diffusion in {Y}ukawa systems},}\ }\href {\doibase
  10.1103/physreve.62.7524} {\bibfield  {journal} {\bibinfo  {journal} {Phys.
  Rev. E}\ }\textbf {\bibinfo {volume} {62}},\ \bibinfo {pages} {7524--7527}
  (\bibinfo {year} {2000})}\BibitemShut {NoStop}%
\bibitem [{\citenamefont {Ingebrigtsen}\ \emph {et~al.}(2013)\citenamefont
  {Ingebrigtsen}, \citenamefont {Veldhorst}, \citenamefont {Schr{\o}der},\ and\
  \citenamefont {Dyre}}]{IngebrigtsenJCP2013}%
  \BibitemOpen
  \bibfield  {author} {\bibinfo {author} {\bibfnamefont {T.~S.}\ \bibnamefont
  {Ingebrigtsen}}, \bibinfo {author} {\bibfnamefont {A.~A.}\ \bibnamefont
  {Veldhorst}}, \bibinfo {author} {\bibfnamefont {T.~B.}\ \bibnamefont
  {Schr{\o}der}}, \ and\ \bibinfo {author} {\bibfnamefont {J.~C.}\ \bibnamefont
  {Dyre}},\ }\bibfield  {title} {\enquote {\bibinfo {title} {Communication: The
  {R}osenfeld-{T}arazona expression for liquids' specific heat: A numerical
  investigation of eighteen systems},}\ }\href {\doibase 10.1063/1.4827865}
  {\bibfield  {journal} {\bibinfo  {journal} {J. Chem. Phys.}\ }\textbf
  {\bibinfo {volume} {139}},\ \bibinfo {pages} {171101} (\bibinfo {year}
  {2013})}\BibitemShut {NoStop}%
\bibitem [{\citenamefont {Tolias}\ and\ \citenamefont
  {Castello}(2019)}]{ToliasPoP2019}%
  \BibitemOpen
  \bibfield  {author} {\bibinfo {author} {\bibfnamefont {P.}~\bibnamefont
  {Tolias}}\ and\ \bibinfo {author} {\bibfnamefont {F.~Lucco}\ \bibnamefont
  {Castello}},\ }\bibfield  {title} {\enquote {\bibinfo {title} {Isomorph-based
  empirically modified hypernetted-chain approach for strongly coupled {Y}ukawa
  one-component plasmas},}\ }\href {\doibase 10.1063/1.5089663} {\bibfield
  {journal} {\bibinfo  {journal} {Phys. of Plasmas}\ }\textbf {\bibinfo
  {volume} {26}},\ \bibinfo {pages} {043703} (\bibinfo {year}
  {2019})}\BibitemShut {NoStop}%
\bibitem [{\citenamefont {Castello}\ \emph {et~al.}(2019)\citenamefont
  {Castello}, \citenamefont {Tolias}, \citenamefont {Hansen},\ and\
  \citenamefont {Dyre}}]{CastelloPoP2019}%
  \BibitemOpen
  \bibfield  {author} {\bibinfo {author} {\bibfnamefont {F.~Lucco}\
  \bibnamefont {Castello}}, \bibinfo {author} {\bibfnamefont {P.}~\bibnamefont
  {Tolias}}, \bibinfo {author} {\bibfnamefont {J.~S.}\ \bibnamefont {Hansen}},
  \ and\ \bibinfo {author} {\bibfnamefont {J.~C.}\ \bibnamefont {Dyre}},\
  }\bibfield  {title} {\enquote {\bibinfo {title} {Isomorph invariance and
  thermodynamics of repulsive dense bi-{Y}ukawa one-component plasmas},}\
  }\href {\doibase 10.1063/1.5100150} {\bibfield  {journal} {\bibinfo
  {journal} {Phys. Plasmas}\ }\textbf {\bibinfo {volume} {26}},\ \bibinfo
  {pages} {053705} (\bibinfo {year} {2019})}\BibitemShut {NoStop}%
\bibitem [{\citenamefont {Khrapak}(2021{\natexlab{c}})}]{KhrapakPoP08_2021}%
  \BibitemOpen
  \bibfield  {author} {\bibinfo {author} {\bibfnamefont {S.~A.}\ \bibnamefont
  {Khrapak}},\ }\bibfield  {title} {\enquote {\bibinfo {title} {Thermal
  conductivity of strongly coupled {Y}ukawa fluids},}\ }\href {\doibase
  10.1063/5.0056763} {\bibfield  {journal} {\bibinfo  {journal} {Phys.
  Plasmas}\ }\textbf {\bibinfo {volume} {28}},\ \bibinfo {pages} {084501}
  (\bibinfo {year} {2021}{\natexlab{c}})}\BibitemShut {NoStop}%
\bibitem [{\citenamefont {Rosenfeld}(2001)}]{RosenfeldJPCM2001}%
  \BibitemOpen
  \bibfield  {author} {\bibinfo {author} {\bibfnamefont {Y.}~\bibnamefont
  {Rosenfeld}},\ }\bibfield  {title} {\enquote {\bibinfo {title}
  {Quasi-universal melting-temperature scaling of transport coefficients in
  {Y}ukawa systems},}\ }\href {\doibase 10.1088/0953-8984/13/2/101} {\bibfield
  {journal} {\bibinfo  {journal} {J. Phys.: Condens. Matter}\ }\textbf
  {\bibinfo {volume} {13}},\ \bibinfo {pages} {L39--L43} (\bibinfo {year}
  {2001})}\BibitemShut {NoStop}%
\bibitem [{\citenamefont {Costigliola}\ \emph {et~al.}(2018)\citenamefont
  {Costigliola}, \citenamefont {Pedersen}, \citenamefont {Heyes}, \citenamefont
  {Schr{\o}der},\ and\ \citenamefont {Dyre}}]{CostigliolaJCP2018}%
  \BibitemOpen
  \bibfield  {author} {\bibinfo {author} {\bibfnamefont {L.}~\bibnamefont
  {Costigliola}}, \bibinfo {author} {\bibfnamefont {U.~R.}\ \bibnamefont
  {Pedersen}}, \bibinfo {author} {\bibfnamefont {D.~M.}\ \bibnamefont {Heyes}},
  \bibinfo {author} {\bibfnamefont {T.~B.}\ \bibnamefont {Schr{\o}der}}, \ and\
  \bibinfo {author} {\bibfnamefont {J.~C.}\ \bibnamefont {Dyre}},\ }\bibfield
  {title} {\enquote {\bibinfo {title} {Communication: Simple liquids'
  high-density viscosity},}\ }\href {\doibase 10.1063/1.5022058} {\bibfield
  {journal} {\bibinfo  {journal} {J. Chem. Phys.}\ }\textbf {\bibinfo {volume}
  {148}},\ \bibinfo {pages} {081101} (\bibinfo {year} {2018})}\BibitemShut
  {NoStop}%
\bibitem [{\citenamefont {Khrapak}(2023{\natexlab{a}})}]{KhrapakPPR2023}%
  \BibitemOpen
  \bibfield  {author} {\bibinfo {author} {\bibfnamefont {S.~A.}\ \bibnamefont
  {Khrapak}},\ }\bibfield  {title} {\enquote {\bibinfo {title} {Vibrational
  model of heat transfer in strongly coupled {Y}ukawa fluids (dusty plasma
  liquids)},}\ }\href {\doibase 10.1134/s1063780x22600876} {\bibfield
  {journal} {\bibinfo  {journal} {Plasma Phys. Rep.}\ }\textbf {\bibinfo
  {volume} {49}},\ \bibinfo {pages} {15--22} (\bibinfo {year}
  {2023}{\natexlab{a}})}\BibitemShut {NoStop}%
\bibitem [{\citenamefont {Harris}(2020)}]{HarrisJCP2020}%
  \BibitemOpen
  \bibfield  {author} {\bibinfo {author} {\bibfnamefont {K.~R.}\ \bibnamefont
  {Harris}},\ }\bibfield  {title} {\enquote {\bibinfo {title} {Thermodynamic or
  density scaling of the thermal conductivity of liquids},}\ }\href {\doibase
  10.1063/5.0016389} {\bibfield  {journal} {\bibinfo  {journal} {J. Chem.
  Phys.}\ }\textbf {\bibinfo {volume} {153}},\ \bibinfo {pages} {104504}
  (\bibinfo {year} {2020})}\BibitemShut {NoStop}%
\bibitem [{\citenamefont {Allers}\ \emph {et~al.}(2020)\citenamefont {Allers},
  \citenamefont {Harvey}, \citenamefont {Garzon},\ and\ \citenamefont
  {Alam}}]{AllersJCP2020}%
  \BibitemOpen
  \bibfield  {author} {\bibinfo {author} {\bibfnamefont {J.~P.}\ \bibnamefont
  {Allers}}, \bibinfo {author} {\bibfnamefont {J.~A.}\ \bibnamefont {Harvey}},
  \bibinfo {author} {\bibfnamefont {F.~H.}\ \bibnamefont {Garzon}}, \ and\
  \bibinfo {author} {\bibfnamefont {T.~M.}\ \bibnamefont {Alam}},\ }\bibfield
  {title} {\enquote {\bibinfo {title} {Machine learning prediction of
  self-diffusion in {L}ennard-{J}ones fluids},}\ }\href {\doibase
  10.1063/5.0011512} {\bibfield  {journal} {\bibinfo  {journal} {J. Chem.
  Phys.}\ }\textbf {\bibinfo {volume} {153}},\ \bibinfo {pages} {034102}
  (\bibinfo {year} {2020})}\BibitemShut {NoStop}%
\bibitem [{\citenamefont {Meier}(2002)}]{Meier2002}%
  \BibitemOpen
  \bibfield  {author} {\bibinfo {author} {\bibfnamefont {K.}~\bibnamefont
  {Meier}},\ }\href@noop {} {\emph {\bibinfo {title} {Computer Simulation and
  Interpretation of the Transport Coefficients of the {L}ennard-{J}ones Model
  Fluid (PhD Thesis)}}}\ (\bibinfo  {publisher} {Shaker},\ \bibinfo {address}
  {Aachen},\ \bibinfo {year} {2002})\BibitemShut {NoStop}%
\bibitem [{\citenamefont {Meier}\ \emph
  {et~al.}(2004{\natexlab{a}})\citenamefont {Meier}, \citenamefont {Laesecke},\
  and\ \citenamefont {Kabelac}}]{MeierJCP_1}%
  \BibitemOpen
  \bibfield  {author} {\bibinfo {author} {\bibfnamefont {K.}~\bibnamefont
  {Meier}}, \bibinfo {author} {\bibfnamefont {A.}~\bibnamefont {Laesecke}}, \
  and\ \bibinfo {author} {\bibfnamefont {S.}~\bibnamefont {Kabelac}},\
  }\bibfield  {title} {\enquote {\bibinfo {title} {Transport coefficients of
  the {L}ennard-{J}ones model fluid. {I}. {V}iscosity},}\ }\href {\doibase
  10.1063/1.1770695} {\bibfield  {journal} {\bibinfo  {journal} {J. Chem.
  Phys.}\ }\textbf {\bibinfo {volume} {121}},\ \bibinfo {pages} {3671--3687}
  (\bibinfo {year} {2004}{\natexlab{a}})}\BibitemShut {NoStop}%
\bibitem [{\citenamefont {Meier}\ \emph
  {et~al.}(2004{\natexlab{b}})\citenamefont {Meier}, \citenamefont {Laesecke},\
  and\ \citenamefont {Kabelac}}]{MeierJCP_2}%
  \BibitemOpen
  \bibfield  {author} {\bibinfo {author} {\bibfnamefont {K.}~\bibnamefont
  {Meier}}, \bibinfo {author} {\bibfnamefont {A.}~\bibnamefont {Laesecke}}, \
  and\ \bibinfo {author} {\bibfnamefont {S.}~\bibnamefont {Kabelac}},\
  }\bibfield  {title} {\enquote {\bibinfo {title} {Transport coefficients of
  the {L}ennard-{J}ones model fluid. {II}. {S}elf-diffusion},}\ }\href
  {\doibase 10.1063/1.1786579} {\bibfield  {journal} {\bibinfo  {journal} {J.
  Chem. Phys.}\ }\textbf {\bibinfo {volume} {121}},\ \bibinfo {pages}
  {9526--9535} (\bibinfo {year} {2004}{\natexlab{b}})}\BibitemShut {NoStop}%
\bibitem [{\citenamefont {Baidakov}\ \emph {et~al.}(2011)\citenamefont
  {Baidakov}, \citenamefont {Protsenko},\ and\ \citenamefont
  {Kozlova}}]{BaidakovFPE2011}%
  \BibitemOpen
  \bibfield  {author} {\bibinfo {author} {\bibfnamefont {V.G.}\ \bibnamefont
  {Baidakov}}, \bibinfo {author} {\bibfnamefont {S.P.}\ \bibnamefont
  {Protsenko}}, \ and\ \bibinfo {author} {\bibfnamefont {Z.R.}\ \bibnamefont
  {Kozlova}},\ }\bibfield  {title} {\enquote {\bibinfo {title} {The
  self-diffusion coefficient in stable and metastable states of the
  {L}ennard{\textendash}{J}ones fluid},}\ }\href {\doibase
  10.1016/j.fluid.2011.03.002} {\bibfield  {journal} {\bibinfo  {journal}
  {Fluid Phase Equilibria}\ }\textbf {\bibinfo {volume} {305}},\ \bibinfo
  {pages} {106--113} (\bibinfo {year} {2011})}\BibitemShut {NoStop}%
\bibitem [{\citenamefont {Baidakov}\ \emph {et~al.}(2012)\citenamefont
  {Baidakov}, \citenamefont {Protsenko},\ and\ \citenamefont
  {Kozlova}}]{BaidakovJCP2012}%
  \BibitemOpen
  \bibfield  {author} {\bibinfo {author} {\bibfnamefont {V.~G.}\ \bibnamefont
  {Baidakov}}, \bibinfo {author} {\bibfnamefont {S.~P.}\ \bibnamefont
  {Protsenko}}, \ and\ \bibinfo {author} {\bibfnamefont {Z.~R.}\ \bibnamefont
  {Kozlova}},\ }\bibfield  {title} {\enquote {\bibinfo {title} {Metastable
  {L}ennard-{J}ones fluids. {I}. {S}hear viscosity},}\ }\href {\doibase
  10.1063/1.4758806} {\bibfield  {journal} {\bibinfo  {journal} {J. Chem.
  Phys.}\ }\textbf {\bibinfo {volume} {137}},\ \bibinfo {pages} {164507}
  (\bibinfo {year} {2012})}\BibitemShut {NoStop}%
\bibitem [{\citenamefont {Baidakov}\ and\ \citenamefont
  {Protsenko}(2014)}]{BaidakovJCP2014}%
  \BibitemOpen
  \bibfield  {author} {\bibinfo {author} {\bibfnamefont {V.~G.}\ \bibnamefont
  {Baidakov}}\ and\ \bibinfo {author} {\bibfnamefont {S.~P.}\ \bibnamefont
  {Protsenko}},\ }\bibfield  {title} {\enquote {\bibinfo {title} {Metastable
  {L}ennard-{J}ones fluids. {II}. {T}hermal conductivity},}\ }\href {\doibase
  10.1063/1.4880958} {\bibfield  {journal} {\bibinfo  {journal} {J. Chem.
  Phys.}\ }\textbf {\bibinfo {volume} {140}},\ \bibinfo {pages} {214506}
  (\bibinfo {year} {2014})}\BibitemShut {NoStop}%
\bibitem [{\citenamefont {Khrapak}\ and\ \citenamefont
  {Khrapak}(2021{\natexlab{c}})}]{KhrapakPRE04_2021}%
  \BibitemOpen
  \bibfield  {author} {\bibinfo {author} {\bibfnamefont {S.~A.}\ \bibnamefont
  {Khrapak}}\ and\ \bibinfo {author} {\bibfnamefont {A.~G.}\ \bibnamefont
  {Khrapak}},\ }\bibfield  {title} {\enquote {\bibinfo {title} {Transport
  properties of {L}ennard-{J}ones fluids: Freezing density scaling along
  isotherms},}\ }\href {\doibase 10.1103/physreve.103.042122} {\bibfield
  {journal} {\bibinfo  {journal} {Phys. Rev. E}\ }\textbf {\bibinfo {volume}
  {103}},\ \bibinfo {pages} {042122} (\bibinfo {year}
  {2021}{\natexlab{c}})}\BibitemShut {NoStop}%
\bibitem [{\citenamefont {Khrapak}\ and\ \citenamefont
  {Khrapak}(2022{\natexlab{c}})}]{KhrapakJPCL2022}%
  \BibitemOpen
  \bibfield  {author} {\bibinfo {author} {\bibfnamefont {S.~A.}\ \bibnamefont
  {Khrapak}}\ and\ \bibinfo {author} {\bibfnamefont {A.~G.}\ \bibnamefont
  {Khrapak}},\ }\bibfield  {title} {\enquote {\bibinfo {title} {Freezing
  temperature and density scaling of transport coefficients},}\ }\href
  {\doibase 10.1021/acs.jpclett.2c00408} {\bibfield  {journal} {\bibinfo
  {journal} {J. Phys. Chem. Lett.}\ ,\ \bibinfo {pages} {2674--2678}} (\bibinfo
  {year} {2022}{\natexlab{c}})}\BibitemShut {NoStop}%
\bibitem [{\citenamefont {Heyes}\ \emph {et~al.}(2023)\citenamefont {Heyes},
  \citenamefont {Dini}, \citenamefont {Pieprzyk},\ and\ \citenamefont
  {Bra{\'{n}}ka}}]{HeyesJCP2023}%
  \BibitemOpen
  \bibfield  {author} {\bibinfo {author} {\bibfnamefont {D.~M.}\ \bibnamefont
  {Heyes}}, \bibinfo {author} {\bibfnamefont {D.}~\bibnamefont {Dini}},
  \bibinfo {author} {\bibfnamefont {S.}~\bibnamefont {Pieprzyk}}, \ and\
  \bibinfo {author} {\bibfnamefont {A.~C.}\ \bibnamefont {Bra{\'{n}}ka}},\
  }\bibfield  {title} {\enquote {\bibinfo {title} {Departures from perfect
  isomorph behavior in {L}ennard-{J}ones fluids and solids},}\ }\href {\doibase
  10.1063/5.0143651} {\bibfield  {journal} {\bibinfo  {journal} {J. Chem.
  Phys.}\ }\textbf {\bibinfo {volume} {158}},\ \bibinfo {pages} {134502}
  (\bibinfo {year} {2023})}\BibitemShut {NoStop}%
\bibitem [{\citenamefont {Ohtori}\ and\ \citenamefont
  {Ishii}(2015)}]{OhtoriPRE2015}%
  \BibitemOpen
  \bibfield  {author} {\bibinfo {author} {\bibfnamefont {N.}~\bibnamefont
  {Ohtori}}\ and\ \bibinfo {author} {\bibfnamefont {Y.}~\bibnamefont {Ishii}},\
  }\bibfield  {title} {\enquote {\bibinfo {title} {Explicit expression for the
  {S}tokes-{E}instein relation for pure {L}ennard-{J}ones liquids},}\ }\href
  {\doibase 10.1103/physreve.91.012111} {\bibfield  {journal} {\bibinfo
  {journal} {Phys. Rev. E}\ }\textbf {\bibinfo {volume} {91}},\ \bibinfo
  {pages} {012111} (\bibinfo {year} {2015})}\BibitemShut {NoStop}%
\bibitem [{\citenamefont {Ohtori}\ \emph {et~al.}(2017)\citenamefont {Ohtori},
  \citenamefont {Miyamoto},\ and\ \citenamefont {Ishii}}]{OhtoriPRE2017}%
  \BibitemOpen
  \bibfield  {author} {\bibinfo {author} {\bibfnamefont {N.}~\bibnamefont
  {Ohtori}}, \bibinfo {author} {\bibfnamefont {S.}~\bibnamefont {Miyamoto}}, \
  and\ \bibinfo {author} {\bibfnamefont {Y.}~\bibnamefont {Ishii}},\ }\bibfield
   {title} {\enquote {\bibinfo {title} {Breakdown of the {S}tokes-{E}instein
  relation in pure {L}ennard-{J}ones fluids: From gas to liquid via
  supercritical states},}\ }\href {\doibase 10.1103/physreve.95.052122}
  {\bibfield  {journal} {\bibinfo  {journal} {Phys. Rev. E}\ }\textbf {\bibinfo
  {volume} {95}},\ \bibinfo {pages} {052122} (\bibinfo {year}
  {2017})}\BibitemShut {NoStop}%
\bibitem [{\citenamefont {Jakse}\ and\ \citenamefont
  {Charpentier}(2003)}]{JaksePRE2003}%
  \BibitemOpen
  \bibfield  {author} {\bibinfo {author} {\bibfnamefont {N.}~\bibnamefont
  {Jakse}}\ and\ \bibinfo {author} {\bibfnamefont {I.}~\bibnamefont
  {Charpentier}},\ }\bibfield  {title} {\enquote {\bibinfo {title} {Direct
  excess entropy calculation for a {L}ennard-{J}ones fluid by the integral
  equation method},}\ }\href {\doibase 10.1103/physreve.67.061203} {\bibfield
  {journal} {\bibinfo  {journal} {Phys. Rev. E}\ }\textbf {\bibinfo {volume}
  {67}},\ \bibinfo {pages} {061203} (\bibinfo {year} {2003})}\BibitemShut
  {NoStop}%
\bibitem [{\citenamefont {Hirschfelder}\ \emph {et~al.}(1954)\citenamefont
  {Hirschfelder}, \citenamefont {Curtiss},\ and\ \citenamefont
  {Bird}}]{HirschfelderBook}%
  \BibitemOpen
  \bibfield  {author} {\bibinfo {author} {\bibfnamefont {J.~O.}\ \bibnamefont
  {Hirschfelder}}, \bibinfo {author} {\bibfnamefont {C.~F.}\ \bibnamefont
  {Curtiss}}, \ and\ \bibinfo {author} {\bibfnamefont {R.~B.}\ \bibnamefont
  {Bird}},\ }\href@noop {} {\emph {\bibinfo {title} {The Molecular Theory of
  Gases and Liquids -}}}\ (\bibinfo  {publisher} {Wiley},\ \bibinfo {address}
  {New York},\ \bibinfo {year} {1954})\BibitemShut {NoStop}%
\bibitem [{\citenamefont {Hirschfelder}\ \emph {et~al.}(1948)\citenamefont
  {Hirschfelder}, \citenamefont {Bird},\ and\ \citenamefont
  {Spotz}}]{HirschfelderJCP1948}%
  \BibitemOpen
  \bibfield  {author} {\bibinfo {author} {\bibfnamefont {J.~O.}\ \bibnamefont
  {Hirschfelder}}, \bibinfo {author} {\bibfnamefont {R.~B.}\ \bibnamefont
  {Bird}}, \ and\ \bibinfo {author} {\bibfnamefont {E.~L.}\ \bibnamefont
  {Spotz}},\ }\bibfield  {title} {\enquote {\bibinfo {title} {The transport
  properties for non-polar gases},}\ }\href {\doibase 10.1063/1.1746696}
  {\bibfield  {journal} {\bibinfo  {journal} {J. Chem. Phys.}\ }\textbf
  {\bibinfo {volume} {16}},\ \bibinfo {pages} {968--981} (\bibinfo {year}
  {1948})}\BibitemShut {NoStop}%
\bibitem [{\citenamefont {Smith}\ and\ \citenamefont
  {Munn}(1964)}]{SmithJCP1964}%
  \BibitemOpen
  \bibfield  {author} {\bibinfo {author} {\bibfnamefont {F.~J.}\ \bibnamefont
  {Smith}}\ and\ \bibinfo {author} {\bibfnamefont {R.~J.}\ \bibnamefont
  {Munn}},\ }\bibfield  {title} {\enquote {\bibinfo {title} {Automatic
  calculation of the transport collision integrals with tables for the {M}orse
  potential},}\ }\href {\doibase 10.1063/1.1725768} {\bibfield  {journal}
  {\bibinfo  {journal} {J. Chem. Phys.}\ }\textbf {\bibinfo {volume} {41}},\
  \bibinfo {pages} {3560--3568} (\bibinfo {year} {1964})}\BibitemShut {NoStop}%
\bibitem [{\citenamefont
  {Khrapak}(2014{\natexlab{a}})}]{KhrapakPRE2014_scattering}%
  \BibitemOpen
  \bibfield  {author} {\bibinfo {author} {\bibfnamefont {S.~A.}\ \bibnamefont
  {Khrapak}},\ }\bibfield  {title} {\enquote {\bibinfo {title} {Classical
  scattering in strongly attractive potentials},}\ }\href {\doibase
  10.1103/physreve.89.032145} {\bibfield  {journal} {\bibinfo  {journal} {Phys.
  Rev. E}\ }\textbf {\bibinfo {volume} {89}},\ \bibinfo {pages} {032145}
  (\bibinfo {year} {2014}{\natexlab{a}})}\BibitemShut {NoStop}%
\bibitem [{\citenamefont {Khrapak}(2014{\natexlab{b}})}]{KhrapakEPJD2014}%
  \BibitemOpen
  \bibfield  {author} {\bibinfo {author} {\bibfnamefont {S.~A.}\ \bibnamefont
  {Khrapak}},\ }\bibfield  {title} {\enquote {\bibinfo {title} {Accurate
  transport cross sections for the {L}ennard-{J}ones potential},}\ }\href
  {\doibase 10.1140/epjd/e2014-50449-y} {\bibfield  {journal} {\bibinfo
  {journal} {Eur. Phys. J. D}\ }\textbf {\bibinfo {volume} {68}},\ \bibinfo
  {pages} {276} (\bibinfo {year} {2014}{\natexlab{b}})}\BibitemShut {NoStop}%
\bibitem [{\citenamefont {Kim}\ and\ \citenamefont
  {Monroe}(2014)}]{KimJCompPhys2014}%
  \BibitemOpen
  \bibfield  {author} {\bibinfo {author} {\bibfnamefont {S.~U.}\ \bibnamefont
  {Kim}}\ and\ \bibinfo {author} {\bibfnamefont {C.~W.}\ \bibnamefont
  {Monroe}},\ }\bibfield  {title} {\enquote {\bibinfo {title} {High-accuracy
  calculations of sixteen collision integrals for {L}ennard-{J}ones 12-6 gases
  and their interpolation to parameterize neon, argon, and krypton},}\ }\href
  {\doibase 10.1016/j.jcp.2014.05.018} {\bibfield  {journal} {\bibinfo
  {journal} {J. Comp. Phys.}\ }\textbf {\bibinfo {volume} {273}},\ \bibinfo
  {pages} {358--373} (\bibinfo {year} {2014})}\BibitemShut {NoStop}%
\bibitem [{\citenamefont {Kristiansen}(2020)}]{Kristiansen2020}%
  \BibitemOpen
  \bibfield  {author} {\bibinfo {author} {\bibfnamefont {K.~R.}\ \bibnamefont
  {Kristiansen}},\ }\bibfield  {title} {\enquote {\bibinfo {title} {Transport
  properties of the simple {L}ennard-{J}ones/spline fluid {I}: Binary
  scattering and high-accuracy low-density transport coefficients},}\ }\href
  {\doibase 10.3389/fphy.2020.00271} {\bibfield  {journal} {\bibinfo  {journal}
  {Frontiers Phys.}\ }\textbf {\bibinfo {volume} {8}},\ \bibinfo {pages} {271}
  (\bibinfo {year} {2020})}\BibitemShut {NoStop}%
\bibitem [{\citenamefont {Khrapak}(2022{\natexlab{b}})}]{KhrapakJCP2022}%
  \BibitemOpen
  \bibfield  {author} {\bibinfo {author} {\bibfnamefont {S.~A.}\ \bibnamefont
  {Khrapak}},\ }\bibfield  {title} {\enquote {\bibinfo {title} {Gas-liquid
  crossover in the {L}ennard-{J}ones system},}\ }\href {\doibase
  10.1063/5.0085181} {\bibfield  {journal} {\bibinfo  {journal} {J. Chem.
  Phys.}\ }\textbf {\bibinfo {volume} {156}},\ \bibinfo {pages} {116101}
  (\bibinfo {year} {2022}{\natexlab{b}})}\BibitemShut {NoStop}%
\bibitem [{\citenamefont {Nasrabad}\ \emph {et~al.}(2006)\citenamefont
  {Nasrabad}, \citenamefont {Laghaei},\ and\ \citenamefont
  {Eu}}]{NasrabadJCP2006}%
  \BibitemOpen
  \bibfield  {author} {\bibinfo {author} {\bibfnamefont {A.~E.}\ \bibnamefont
  {Nasrabad}}, \bibinfo {author} {\bibfnamefont {R.}~\bibnamefont {Laghaei}}, \
  and\ \bibinfo {author} {\bibfnamefont {B.~C.}\ \bibnamefont {Eu}},\
  }\bibfield  {title} {\enquote {\bibinfo {title} {Molecular theory of thermal
  conductivity of the {L}ennard-{J}ones fluid},}\ }\href {\doibase
  10.1063/1.2166394} {\bibfield  {journal} {\bibinfo  {journal} {J. Chem.
  Phys.}\ }\textbf {\bibinfo {volume} {124}},\ \bibinfo {pages} {084506}
  (\bibinfo {year} {2006})}\BibitemShut {NoStop}%
\bibitem [{\citenamefont {Zwanzig}\ and\ \citenamefont
  {Mountain}(1965)}]{ZwanzigJCP1965}%
  \BibitemOpen
  \bibfield  {author} {\bibinfo {author} {\bibfnamefont {R.}~\bibnamefont
  {Zwanzig}}\ and\ \bibinfo {author} {\bibfnamefont {R.~D.}\ \bibnamefont
  {Mountain}},\ }\bibfield  {title} {\enquote {\bibinfo {title} {High-frequency
  elastic moduli of simple fluids},}\ }\href {\doibase 10.1063/1.1696718}
  {\bibfield  {journal} {\bibinfo  {journal} {J. Chem. Phys.}\ }\textbf
  {\bibinfo {volume} {43}},\ \bibinfo {pages} {4464--4471} (\bibinfo {year}
  {1965})}\BibitemShut {NoStop}%
\bibitem [{\citenamefont {Khrapak}(2020{\natexlab{b}})}]{KhrapakMolecules2020}%
  \BibitemOpen
  \bibfield  {author} {\bibinfo {author} {\bibfnamefont {S.~A.}\ \bibnamefont
  {Khrapak}},\ }\bibfield  {title} {\enquote {\bibinfo {title} {Sound
  velocities of {L}ennard-{J}ones systems near the liquid-solid phase
  transition},}\ }\href {\doibase 10.3390/molecules25153498} {\bibfield
  {journal} {\bibinfo  {journal} {Molecules}\ }\textbf {\bibinfo {volume}
  {25}},\ \bibinfo {pages} {3498} (\bibinfo {year}
  {2020}{\natexlab{b}})}\BibitemShut {NoStop}%
\bibitem [{\citenamefont {Khrapak}(2021{\natexlab{d}})}]{KhrapakMolecules2021}%
  \BibitemOpen
  \bibfield  {author} {\bibinfo {author} {\bibfnamefont {S.}~\bibnamefont
  {Khrapak}},\ }\bibfield  {title} {\enquote {\bibinfo {title} {Sound
  velocities of generalized {L}ennard-{J}ones (n - 6) fluids near freezing},}\
  }\href {\doibase 10.3390/molecules26061660} {\bibfield  {journal} {\bibinfo
  {journal} {Molecules}\ }\textbf {\bibinfo {volume} {26}},\ \bibinfo {pages}
  {1660} (\bibinfo {year} {2021}{\natexlab{d}})}\BibitemShut {NoStop}%
\bibitem [{\citenamefont {Smirnov}(1982)}]{SmirnovUFN1982}%
  \BibitemOpen
  \bibfield  {author} {\bibinfo {author} {\bibfnamefont {B.~M.}\ \bibnamefont
  {Smirnov}},\ }\bibfield  {title} {\enquote {\bibinfo {title} {The hard-sphere
  model in plasma and gas physics},}\ }\href {\doibase
  10.1070/pu1982v025n11abeh004663} {\bibfield  {journal} {\bibinfo  {journal}
  {Sov. Phys.-Usp.}\ }\textbf {\bibinfo {volume} {25}},\ \bibinfo {pages}
  {854--862} (\bibinfo {year} {1982})}\BibitemShut {NoStop}%
\bibitem [{\citenamefont {Mulero}(2008)}]{MuleroBook}%
  \BibitemOpen
  \bibinfo {editor} {\bibfnamefont {A.}~\bibnamefont {Mulero}},\ ed.,\ \href
  {\doibase 10.1007/978-3-540-78767-9} {\emph {\bibinfo {title} {Theory and
  Simulation of Hard-Sphere Fluids and Related Systems}}}\ (\bibinfo
  {publisher} {Springer Berlin Heidelberg},\ \bibinfo {year}
  {2008})\BibitemShut {NoStop}%
\bibitem [{\citenamefont {Pusey}\ \emph {et~al.}(2009)\citenamefont {Pusey},
  \citenamefont {Zaccarelli}, \citenamefont {Valeriani}, \citenamefont {Sanz},
  \citenamefont {Poon},\ and\ \citenamefont {Cates}}]{PuseyPhylTrans2009}%
  \BibitemOpen
  \bibfield  {author} {\bibinfo {author} {\bibfnamefont {P.~N.}\ \bibnamefont
  {Pusey}}, \bibinfo {author} {\bibfnamefont {E.}~\bibnamefont {Zaccarelli}},
  \bibinfo {author} {\bibfnamefont {C.}~\bibnamefont {Valeriani}}, \bibinfo
  {author} {\bibfnamefont {E.}~\bibnamefont {Sanz}}, \bibinfo {author}
  {\bibfnamefont {Wilson C.~K.}\ \bibnamefont {Poon}}, \ and\ \bibinfo {author}
  {\bibfnamefont {Michael~E.}\ \bibnamefont {Cates}},\ }\bibfield  {title}
  {\enquote {\bibinfo {title} {Hard spheres: {C}rystallization and glass
  formation},}\ }\href {\doibase 10.1098/rsta.2009.0181} {\bibfield  {journal}
  {\bibinfo  {journal} {Phil. Trans. Royal Soc. A: Math. Phys. and Engineering
  Sci.}\ }\textbf {\bibinfo {volume} {367}},\ \bibinfo {pages} {4993--5011}
  (\bibinfo {year} {2009})}\BibitemShut {NoStop}%
\bibitem [{\citenamefont {Parisi}\ and\ \citenamefont
  {Zamponi}(2010)}]{ParisiRMP2010}%
  \BibitemOpen
  \bibfield  {author} {\bibinfo {author} {\bibfnamefont {G.}~\bibnamefont
  {Parisi}}\ and\ \bibinfo {author} {\bibfnamefont {F.}~\bibnamefont
  {Zamponi}},\ }\bibfield  {title} {\enquote {\bibinfo {title} {Mean-field
  theory of hard sphere glasses and jamming},}\ }\href {\doibase
  10.1103/revmodphys.82.789} {\bibfield  {journal} {\bibinfo  {journal} {Rev.
  Mod. Phys.}\ }\textbf {\bibinfo {volume} {82}},\ \bibinfo {pages} {789--845}
  (\bibinfo {year} {2010})}\BibitemShut {NoStop}%
\bibitem [{\citenamefont {Berthier}\ and\ \citenamefont
  {Biroli}(2011)}]{BerthierRMP2011}%
  \BibitemOpen
  \bibfield  {author} {\bibinfo {author} {\bibfnamefont {L.}~\bibnamefont
  {Berthier}}\ and\ \bibinfo {author} {\bibfnamefont {G.}~\bibnamefont
  {Biroli}},\ }\bibfield  {title} {\enquote {\bibinfo {title} {Theoretical
  perspective on the glass transition and amorphous materials},}\ }\href
  {\doibase 10.1103/revmodphys.83.587} {\bibfield  {journal} {\bibinfo
  {journal} {Rev. Mod. Phys.}\ }\textbf {\bibinfo {volume} {83}},\ \bibinfo
  {pages} {587--645} (\bibinfo {year} {2011})}\BibitemShut {NoStop}%
\bibitem [{\citenamefont {Klumov}\ \emph {et~al.}(2011)\citenamefont {Klumov},
  \citenamefont {Khrapak},\ and\ \citenamefont {Morfill}}]{KlumovPRB2011}%
  \BibitemOpen
  \bibfield  {author} {\bibinfo {author} {\bibfnamefont {B.~A.}\ \bibnamefont
  {Klumov}}, \bibinfo {author} {\bibfnamefont {S.~A.}\ \bibnamefont {Khrapak}},
  \ and\ \bibinfo {author} {\bibfnamefont {G.~E.}\ \bibnamefont {Morfill}},\
  }\bibfield  {title} {\enquote {\bibinfo {title} {Structural properties of
  dense hard sphere packings},}\ }\href {\doibase 10.1103/physrevb.83.184105}
  {\bibfield  {journal} {\bibinfo  {journal} {Phys. Rev. B}\ }\textbf {\bibinfo
  {volume} {83}},\ \bibinfo {pages} {184105} (\bibinfo {year}
  {2011})}\BibitemShut {NoStop}%
\bibitem [{\citenamefont {Dyre}(2016)}]{DyreJPCM2016}%
  \BibitemOpen
  \bibfield  {author} {\bibinfo {author} {\bibfnamefont {J~C}\ \bibnamefont
  {Dyre}},\ }\bibfield  {title} {\enquote {\bibinfo {title} {Simple liquids'
  quasiuniversality and the hard-sphere paradigm},}\ }\href {\doibase
  10.1088/0953-8984/28/32/323001} {\bibfield  {journal} {\bibinfo  {journal}
  {J. Phys.: Condens. Matter}\ }\textbf {\bibinfo {volume} {28}},\ \bibinfo
  {pages} {323001} (\bibinfo {year} {2016})}\BibitemShut {NoStop}%
\bibitem [{\citenamefont {Alder}\ and\ \citenamefont
  {Wainwright}(1967)}]{AlderPRL1967}%
  \BibitemOpen
  \bibfield  {author} {\bibinfo {author} {\bibfnamefont {B.~J.}\ \bibnamefont
  {Alder}}\ and\ \bibinfo {author} {\bibfnamefont {T.~E.}\ \bibnamefont
  {Wainwright}},\ }\bibfield  {title} {\enquote {\bibinfo {title} {Velocity
  autocorrelations for hard spheres},}\ }\href {\doibase
  10.1103/physrevlett.18.988} {\bibfield  {journal} {\bibinfo  {journal} {Phys.
  Rev. Lett.}\ }\textbf {\bibinfo {volume} {18}},\ \bibinfo {pages} {988--990}
  (\bibinfo {year} {1967})}\BibitemShut {NoStop}%
\bibitem [{\citenamefont {Williams}\ \emph {et~al.}(2006)\citenamefont
  {Williams}, \citenamefont {Bryant}, \citenamefont {Snook},\ and\
  \citenamefont {van Megen}}]{WilliamsPRL2006}%
  \BibitemOpen
  \bibfield  {author} {\bibinfo {author} {\bibfnamefont {S.~R.}\ \bibnamefont
  {Williams}}, \bibinfo {author} {\bibfnamefont {G.}~\bibnamefont {Bryant}},
  \bibinfo {author} {\bibfnamefont {I.~K.}\ \bibnamefont {Snook}}, \ and\
  \bibinfo {author} {\bibfnamefont {W.}~\bibnamefont {van Megen}},\ }\bibfield
  {title} {\enquote {\bibinfo {title} {Velocity autocorrelation functions of
  hard-sphere fluids: Long-time tails upon undercooling},}\ }\href {\doibase
  10.1103/physrevlett.96.087801} {\bibfield  {journal} {\bibinfo  {journal}
  {Phys. Rev. Lett.}\ }\textbf {\bibinfo {volume} {96}},\ \bibinfo {pages}
  {087801} (\bibinfo {year} {2006})}\BibitemShut {NoStop}%
\bibitem [{\citenamefont {Bryk}\ \emph
  {et~al.}(2017{\natexlab{a}})\citenamefont {Bryk}, \citenamefont {Huerta},
  \citenamefont {Hordiichuk},\ and\ \citenamefont {Trokhymchuk}}]{BrykJCP2017}%
  \BibitemOpen
  \bibfield  {author} {\bibinfo {author} {\bibfnamefont {T.}~\bibnamefont
  {Bryk}}, \bibinfo {author} {\bibfnamefont {A.}~\bibnamefont {Huerta}},
  \bibinfo {author} {\bibfnamefont {V.}~\bibnamefont {Hordiichuk}}, \ and\
  \bibinfo {author} {\bibfnamefont {A.~D.}\ \bibnamefont {Trokhymchuk}},\
  }\bibfield  {title} {\enquote {\bibinfo {title} {Non-hydrodynamic transverse
  collective excitations in hard-sphere fluids},}\ }\href {\doibase
  10.1063/1.4997640} {\bibfield  {journal} {\bibinfo  {journal} {J. Chem.
  Phys.}\ }\textbf {\bibinfo {volume} {147}},\ \bibinfo {pages} {064509}
  (\bibinfo {year} {2017}{\natexlab{a}})}\BibitemShut {NoStop}%
\bibitem [{\citenamefont {Murillo}(2000)}]{MurilloPRL2000}%
  \BibitemOpen
  \bibfield  {author} {\bibinfo {author} {\bibfnamefont {M.~S.}\ \bibnamefont
  {Murillo}},\ }\bibfield  {title} {\enquote {\bibinfo {title} {Critical wave
  vectors for transverse modes in strongly coupled dusty plasmas},}\ }\href
  {\doibase 10.1103/physrevlett.85.2514} {\bibfield  {journal} {\bibinfo
  {journal} {Phys. Rev. Lett.}\ }\textbf {\bibinfo {volume} {85}},\ \bibinfo
  {pages} {2514--2517} (\bibinfo {year} {2000})}\BibitemShut {NoStop}%
\bibitem [{\citenamefont {Ohta}\ and\ \citenamefont
  {Hamaguchi}(2000{\natexlab{b}})}]{OhtaPRL2000}%
  \BibitemOpen
  \bibfield  {author} {\bibinfo {author} {\bibfnamefont {H.}~\bibnamefont
  {Ohta}}\ and\ \bibinfo {author} {\bibfnamefont {S.}~\bibnamefont
  {Hamaguchi}},\ }\bibfield  {title} {\enquote {\bibinfo {title} {Wave
  dispersion relations in {Y}ukawa fluids},}\ }\href {\doibase
  10.1103/physrevlett.84.6026} {\bibfield  {journal} {\bibinfo  {journal}
  {Phys. Rev. Lett.}\ }\textbf {\bibinfo {volume} {84}},\ \bibinfo {pages}
  {6026--6029} (\bibinfo {year} {2000}{\natexlab{b}})}\BibitemShut {NoStop}%
\bibitem [{\citenamefont {Goree}\ \emph {et~al.}(2012)\citenamefont {Goree},
  \citenamefont {Donk{\'{o}}},\ and\ \citenamefont {Hartmann}}]{GoreePRE2012}%
  \BibitemOpen
  \bibfield  {author} {\bibinfo {author} {\bibfnamefont {J.}~\bibnamefont
  {Goree}}, \bibinfo {author} {\bibfnamefont {Z.}~\bibnamefont {Donk{\'{o}}}},
  \ and\ \bibinfo {author} {\bibfnamefont {P.}~\bibnamefont {Hartmann}},\
  }\bibfield  {title} {\enquote {\bibinfo {title} {Cutoff wave number for shear
  waves and {M}axwell relaxation time in {Y}ukawa liquids},}\ }\href {\doibase
  10.1103/physreve.85.066401} {\bibfield  {journal} {\bibinfo  {journal} {Phys.
  Rev. E}\ }\textbf {\bibinfo {volume} {85}},\ \bibinfo {pages} {066401}
  (\bibinfo {year} {2012})}\BibitemShut {NoStop}%
\bibitem [{\citenamefont {Bolmatov}\ \emph {et~al.}(2016)\citenamefont
  {Bolmatov}, \citenamefont {Zhernenkov}, \citenamefont {Zav'yalov},
  \citenamefont {Stoupin}, \citenamefont {Cunsolo},\ and\ \citenamefont
  {Cai}}]{BolmatovSciRep2016}%
  \BibitemOpen
  \bibfield  {author} {\bibinfo {author} {\bibfnamefont {D.}~\bibnamefont
  {Bolmatov}}, \bibinfo {author} {\bibfnamefont {M.}~\bibnamefont
  {Zhernenkov}}, \bibinfo {author} {\bibfnamefont {D.}~\bibnamefont
  {Zav'yalov}}, \bibinfo {author} {\bibfnamefont {S.}~\bibnamefont {Stoupin}},
  \bibinfo {author} {\bibfnamefont {A.}~\bibnamefont {Cunsolo}}, \ and\
  \bibinfo {author} {\bibfnamefont {Y.~Q.}\ \bibnamefont {Cai}},\ }\bibfield
  {title} {\enquote {\bibinfo {title} {Thermally triggered phononic gaps in
  liquids at {THz} scale},}\ }\href {\doibase 10.1038/srep19469} {\bibfield
  {journal} {\bibinfo  {journal} {Sci. Rep.}\ }\textbf {\bibinfo {volume} {6}}
  (\bibinfo {year} {2016}),\ 10.1038/srep19469}\BibitemShut {NoStop}%
\bibitem [{\citenamefont {Yang}\ \emph {et~al.}(2017)\citenamefont {Yang},
  \citenamefont {Dove}, \citenamefont {Brazhkin},\ and\ \citenamefont
  {Trachenko}}]{YangPRL2017}%
  \BibitemOpen
  \bibfield  {author} {\bibinfo {author} {\bibfnamefont {C.}~\bibnamefont
  {Yang}}, \bibinfo {author} {\bibfnamefont {M.~T.}\ \bibnamefont {Dove}},
  \bibinfo {author} {\bibfnamefont {V.~V.}\ \bibnamefont {Brazhkin}}, \ and\
  \bibinfo {author} {\bibfnamefont {K.}~\bibnamefont {Trachenko}},\ }\bibfield
  {title} {\enquote {\bibinfo {title} {Emergence and evolution of the k-gap in
  spectra of liquid and supercritical states},}\ }\href {\doibase
  10.1103/physrevlett.118.215502} {\bibfield  {journal} {\bibinfo  {journal}
  {Phys. Rev. Lett.}\ }\textbf {\bibinfo {volume} {118}},\ \bibinfo {pages}
  {215502} (\bibinfo {year} {2017})}\BibitemShut {NoStop}%
\bibitem [{\citenamefont {Kryuchkov}\ \emph {et~al.}(2019)\citenamefont
  {Kryuchkov}, \citenamefont {Mistryukova}, \citenamefont {Brazhkin},\ and\
  \citenamefont {Yurchenko}}]{KryuchkovSciRep2019}%
  \BibitemOpen
  \bibfield  {author} {\bibinfo {author} {\bibfnamefont {N.~P.}\ \bibnamefont
  {Kryuchkov}}, \bibinfo {author} {\bibfnamefont {L.~A.}\ \bibnamefont
  {Mistryukova}}, \bibinfo {author} {\bibfnamefont {V.~V.}\ \bibnamefont
  {Brazhkin}}, \ and\ \bibinfo {author} {\bibfnamefont {S.~O.}\ \bibnamefont
  {Yurchenko}},\ }\bibfield  {title} {\enquote {\bibinfo {title} {Excitation
  spectra in fluids: How to analyze them properly},}\ }\href {\doibase
  10.1038/s41598-019-46979-y} {\bibfield  {journal} {\bibinfo  {journal} {Sci.
  Rep.}\ }\textbf {\bibinfo {volume} {9}},\ \bibinfo {pages} {10483} (\bibinfo
  {year} {2019})}\BibitemShut {NoStop}%
\bibitem [{\citenamefont {Miller}(1969)}]{Miller1969}%
  \BibitemOpen
  \bibfield  {author} {\bibinfo {author} {\bibfnamefont {B.~N.}\ \bibnamefont
  {Miller}},\ }\bibfield  {title} {\enquote {\bibinfo {title} {Elastic moduli
  of a fluid of rigid spheres},}\ }\href {\doibase 10.1063/1.1671437}
  {\bibfield  {journal} {\bibinfo  {journal} {J. Chem. Phys.}\ }\textbf
  {\bibinfo {volume} {50}},\ \bibinfo {pages} {2733--2740} (\bibinfo {year}
  {1969})}\BibitemShut {NoStop}%
\bibitem [{\citenamefont {Khrapak}(2019{\natexlab{b}})}]{KhrapakPRE09_2019}%
  \BibitemOpen
  \bibfield  {author} {\bibinfo {author} {\bibfnamefont {S.}~\bibnamefont
  {Khrapak}},\ }\bibfield  {title} {\enquote {\bibinfo {title} {Elastic
  properties of dense hard-sphere fluids},}\ }\href {\doibase
  10.1103/physreve.100.032138} {\bibfield  {journal} {\bibinfo  {journal}
  {Phys. Rev. E}\ }\textbf {\bibinfo {volume} {100}},\ \bibinfo {pages}
  {032138} (\bibinfo {year} {2019}{\natexlab{b}})}\BibitemShut {NoStop}%
\bibitem [{\citenamefont {Khrapak}\ \emph {et~al.}(2021)\citenamefont
  {Khrapak}, \citenamefont {Kryuchkov}, \citenamefont {Mistryukova},\ and\
  \citenamefont {Yurchenko}}]{KhrapakPRE05_2021}%
  \BibitemOpen
  \bibfield  {author} {\bibinfo {author} {\bibfnamefont {S.}~\bibnamefont
  {Khrapak}}, \bibinfo {author} {\bibfnamefont {N.~P.}\ \bibnamefont
  {Kryuchkov}}, \bibinfo {author} {\bibfnamefont {L.~A.}\ \bibnamefont
  {Mistryukova}}, \ and\ \bibinfo {author} {\bibfnamefont {S.~O.}\ \bibnamefont
  {Yurchenko}},\ }\bibfield  {title} {\enquote {\bibinfo {title} {From soft- to
  hard-sphere fluids: Crossover evidenced by high-frequency elastic moduli},}\
  }\href {\doibase 10.1103/physreve.103.052117} {\bibfield  {journal} {\bibinfo
   {journal} {Phys. Rev. E}\ }\textbf {\bibinfo {volume} {103}},\ \bibinfo
  {pages} {052117} (\bibinfo {year} {2021})}\BibitemShut {NoStop}%
\bibitem [{\citenamefont {Pieprzyk}\ \emph {et~al.}(2019)\citenamefont
  {Pieprzyk}, \citenamefont {Bannerman}, \citenamefont {Bra{\'{n}}ka},
  \citenamefont {Chudak},\ and\ \citenamefont {Heyes}}]{Pieprzyk2019}%
  \BibitemOpen
  \bibfield  {author} {\bibinfo {author} {\bibfnamefont {S.}~\bibnamefont
  {Pieprzyk}}, \bibinfo {author} {\bibfnamefont {M.~N.}\ \bibnamefont
  {Bannerman}}, \bibinfo {author} {\bibfnamefont {A.~C.}\ \bibnamefont
  {Bra{\'{n}}ka}}, \bibinfo {author} {\bibfnamefont {M.}~\bibnamefont
  {Chudak}}, \ and\ \bibinfo {author} {\bibfnamefont {D.~M.}\ \bibnamefont
  {Heyes}},\ }\bibfield  {title} {\enquote {\bibinfo {title} {Thermodynamic and
  dynamical properties of the hard sphere system revisited by molecular
  dynamics simulation},}\ }\href {\doibase 10.1039/c9cp00903e} {\bibfield
  {journal} {\bibinfo  {journal} {Phys. Chem. Chem. Phys.}\ }\textbf {\bibinfo
  {volume} {21}},\ \bibinfo {pages} {6886--6899} (\bibinfo {year}
  {2019})}\BibitemShut {NoStop}%
\bibitem [{\citenamefont {Ohtori}\ \emph {et~al.}(2018)\citenamefont {Ohtori},
  \citenamefont {Uchiyama},\ and\ \citenamefont {Ishii}}]{OhtoriJCP2018}%
  \BibitemOpen
  \bibfield  {author} {\bibinfo {author} {\bibfnamefont {N.}~\bibnamefont
  {Ohtori}}, \bibinfo {author} {\bibfnamefont {H.}~\bibnamefont {Uchiyama}}, \
  and\ \bibinfo {author} {\bibfnamefont {Y.}~\bibnamefont {Ishii}},\ }\bibfield
   {title} {\enquote {\bibinfo {title} {The {S}tokes-{E}instein relation for
  simple fluids: From hard-sphere to {L}ennard-{J}ones via {WCA} potentials},}\
  }\href {\doibase 10.1063/1.5054577} {\bibfield  {journal} {\bibinfo
  {journal} {J. Chem. Phys.}\ }\textbf {\bibinfo {volume} {149}},\ \bibinfo
  {pages} {214501} (\bibinfo {year} {2018})}\BibitemShut {NoStop}%
\bibitem [{\citenamefont {Grover}\ \emph {et~al.}(1985)\citenamefont {Grover},
  \citenamefont {Hoover},\ and\ \citenamefont {Moran}}]{GroverJCP1985}%
  \BibitemOpen
  \bibfield  {author} {\bibinfo {author} {\bibfnamefont {R.}~\bibnamefont
  {Grover}}, \bibinfo {author} {\bibfnamefont {W.~G.}\ \bibnamefont {Hoover}},
  \ and\ \bibinfo {author} {\bibfnamefont {B.}~\bibnamefont {Moran}},\
  }\bibfield  {title} {\enquote {\bibinfo {title} {Corresponding states for
  thermal conductivities via nonequilibrium molecular dynamics},}\ }\href
  {\doibase 10.1063/1.449441} {\bibfield  {journal} {\bibinfo  {journal} {J.
  Chem. Phys.}\ }\textbf {\bibinfo {volume} {83}},\ \bibinfo {pages}
  {1255--1259} (\bibinfo {year} {1985})}\BibitemShut {NoStop}%
\bibitem [{\citenamefont {Hoover}(1986)}]{HooverBook}%
  \BibitemOpen
  \bibfield  {author} {\bibinfo {author} {\bibfnamefont {W.~G.}\ \bibnamefont
  {Hoover}},\ }\href@noop {} {\emph {\bibinfo {title} {Molecular Dynamics}}}\
  (\bibinfo  {publisher} {Springer},\ \bibinfo {year} {1986})\BibitemShut
  {NoStop}%
\bibitem [{\citenamefont {Gorelli}\ \emph {et~al.}(2006)\citenamefont
  {Gorelli}, \citenamefont {Santoro}, \citenamefont {Scopigno}, \citenamefont
  {Krisch},\ and\ \citenamefont {Ruocco}}]{GorelliPRL2006}%
  \BibitemOpen
  \bibfield  {author} {\bibinfo {author} {\bibfnamefont {F.}~\bibnamefont
  {Gorelli}}, \bibinfo {author} {\bibfnamefont {M.}~\bibnamefont {Santoro}},
  \bibinfo {author} {\bibfnamefont {T.}~\bibnamefont {Scopigno}}, \bibinfo
  {author} {\bibfnamefont {M.}~\bibnamefont {Krisch}}, \ and\ \bibinfo {author}
  {\bibfnamefont {G.}~\bibnamefont {Ruocco}},\ }\bibfield  {title} {\enquote
  {\bibinfo {title} {Liquidlike behavior of supercritical fluids},}\ }\href
  {\doibase 10.1103/physrevlett.97.245702} {\bibfield  {journal} {\bibinfo
  {journal} {Phys. Rev. Lett.}\ }\textbf {\bibinfo {volume} {97}},\ \bibinfo
  {pages} {245702} (\bibinfo {year} {2006})}\BibitemShut {NoStop}%
\bibitem [{\citenamefont {Simeoni}\ \emph {et~al.}(2010)\citenamefont
  {Simeoni}, \citenamefont {Bryk}, \citenamefont {Gorelli}, \citenamefont
  {Krisch}, \citenamefont {Ruocco}, \citenamefont {Santoro},\ and\
  \citenamefont {Scopigno}}]{Simeoni2010}%
  \BibitemOpen
  \bibfield  {author} {\bibinfo {author} {\bibfnamefont {G.~G.}\ \bibnamefont
  {Simeoni}}, \bibinfo {author} {\bibfnamefont {T.}~\bibnamefont {Bryk}},
  \bibinfo {author} {\bibfnamefont {F.~A.}\ \bibnamefont {Gorelli}}, \bibinfo
  {author} {\bibfnamefont {M.}~\bibnamefont {Krisch}}, \bibinfo {author}
  {\bibfnamefont {G.}~\bibnamefont {Ruocco}}, \bibinfo {author} {\bibfnamefont
  {M.}~\bibnamefont {Santoro}}, \ and\ \bibinfo {author} {\bibfnamefont
  {T.}~\bibnamefont {Scopigno}},\ }\bibfield  {title} {\enquote {\bibinfo
  {title} {The {W}idom line as the crossover between liquid-like and gas-like
  behaviour in supercritical~fluids},}\ }\href {\doibase 10.1038/nphys1683}
  {\bibfield  {journal} {\bibinfo  {journal} {Nature Phys.}\ }\textbf {\bibinfo
  {volume} {6}},\ \bibinfo {pages} {503--507} (\bibinfo {year}
  {2010})}\BibitemShut {NoStop}%
\bibitem [{\citenamefont {McMillan}\ and\ \citenamefont
  {Stanley}(2010)}]{McMillanNatPhys2010}%
  \BibitemOpen
  \bibfield  {author} {\bibinfo {author} {\bibfnamefont {P.~F.}\ \bibnamefont
  {McMillan}}\ and\ \bibinfo {author} {\bibfnamefont {H.~E.}\ \bibnamefont
  {Stanley}},\ }\bibfield  {title} {\enquote {\bibinfo {title} {Going
  supercritical},}\ }\href {\doibase 10.1038/nphys1711} {\bibfield  {journal}
  {\bibinfo  {journal} {Nature Phys.}\ }\textbf {\bibinfo {volume} {6}},\
  \bibinfo {pages} {479--480} (\bibinfo {year} {2010})}\BibitemShut {NoStop}%
\bibitem [{\citenamefont {Brazhkin}\ \emph {et~al.}(2011)\citenamefont
  {Brazhkin}, \citenamefont {Fomin}, \citenamefont {Lyapin}, \citenamefont
  {Ryzhov},\ and\ \citenamefont {Tsiok}}]{BrazhkinJPCB2011}%
  \BibitemOpen
  \bibfield  {author} {\bibinfo {author} {\bibfnamefont {V.~V.}\ \bibnamefont
  {Brazhkin}}, \bibinfo {author} {\bibfnamefont {Yu.~D.}\ \bibnamefont
  {Fomin}}, \bibinfo {author} {\bibfnamefont {A.~G.}\ \bibnamefont {Lyapin}},
  \bibinfo {author} {\bibfnamefont {V.~N.}\ \bibnamefont {Ryzhov}}, \ and\
  \bibinfo {author} {\bibfnamefont {E.~N.}\ \bibnamefont {Tsiok}},\ }\bibfield
  {title} {\enquote {\bibinfo {title} {Widom line for the
  liquid{\textendash}gas transition in {L}ennard-{J}ones system},}\ }\href
  {\doibase 10.1021/jp2039898} {\bibfield  {journal} {\bibinfo  {journal} {J.
  Phys. Chem. B}\ }\textbf {\bibinfo {volume} {115}},\ \bibinfo {pages}
  {14112--14115} (\bibinfo {year} {2011})}\BibitemShut {NoStop}%
\bibitem [{\citenamefont {Brazhkin}\ \emph {et~al.}(2013)\citenamefont
  {Brazhkin}, \citenamefont {Fomin}, \citenamefont {Lyapin}, \citenamefont
  {Ryzhov}, \citenamefont {Tsiok},\ and\ \citenamefont
  {Trachenko}}]{BrazhkinPRL2013}%
  \BibitemOpen
  \bibfield  {author} {\bibinfo {author} {\bibfnamefont {V.~V.}\ \bibnamefont
  {Brazhkin}}, \bibinfo {author} {\bibfnamefont {Yu.~D.}\ \bibnamefont
  {Fomin}}, \bibinfo {author} {\bibfnamefont {A.~G.}\ \bibnamefont {Lyapin}},
  \bibinfo {author} {\bibfnamefont {V.~N.}\ \bibnamefont {Ryzhov}}, \bibinfo
  {author} {\bibfnamefont {E.~N.}\ \bibnamefont {Tsiok}}, \ and\ \bibinfo
  {author} {\bibfnamefont {K.}~\bibnamefont {Trachenko}},\ }\bibfield  {title}
  {\enquote {\bibinfo {title} {Liquid-gas'' transition in the supercritical
  region: fundamental changes in the particle dynamics},}\ }\href@noop {}
  {\bibfield  {journal} {\bibinfo  {journal} {Phys. Rev. Lett.}\ }\textbf
  {\bibinfo {volume} {111}},\ \bibinfo {pages} {145901} (\bibinfo {year}
  {2013})}\BibitemShut {NoStop}%
\bibitem [{\citenamefont {Gorelli}\ \emph {et~al.}(2013)\citenamefont
  {Gorelli}, \citenamefont {Bryk}, \citenamefont {Krisch}, \citenamefont
  {Ruocco}, \citenamefont {Santoro},\ and\ \citenamefont
  {Scopigno}}]{GorelliSciRep2013}%
  \BibitemOpen
  \bibfield  {author} {\bibinfo {author} {\bibfnamefont {F.~A.}\ \bibnamefont
  {Gorelli}}, \bibinfo {author} {\bibfnamefont {T.}~\bibnamefont {Bryk}},
  \bibinfo {author} {\bibfnamefont {M.}~\bibnamefont {Krisch}}, \bibinfo
  {author} {\bibfnamefont {G.}~\bibnamefont {Ruocco}}, \bibinfo {author}
  {\bibfnamefont {M.}~\bibnamefont {Santoro}}, \ and\ \bibinfo {author}
  {\bibfnamefont {T.}~\bibnamefont {Scopigno}},\ }\bibfield  {title} {\enquote
  {\bibinfo {title} {Dynamics and thermodynamics beyond the critical point},}\
  }\href {\doibase 10.1038/srep01203} {\bibfield  {journal} {\bibinfo
  {journal} {Sci. Rep.}\ }\textbf {\bibinfo {volume} {3}},\ \bibinfo {pages}
  {1203} (\bibinfo {year} {2013})}\BibitemShut {NoStop}%
\bibitem [{\citenamefont {Yang}\ \emph {et~al.}(2015)\citenamefont {Yang},
  \citenamefont {Brazhkin}, \citenamefont {Dove},\ and\ \citenamefont
  {Trachenko}}]{YangPRE2015}%
  \BibitemOpen
  \bibfield  {author} {\bibinfo {author} {\bibfnamefont {C.}~\bibnamefont
  {Yang}}, \bibinfo {author} {\bibfnamefont {V.~V.}\ \bibnamefont {Brazhkin}},
  \bibinfo {author} {\bibfnamefont {M.~T.}\ \bibnamefont {Dove}}, \ and\
  \bibinfo {author} {\bibfnamefont {K.}~\bibnamefont {Trachenko}},\ }\bibfield
  {title} {\enquote {\bibinfo {title} {Frenkel line and solubility maximum in
  supercritical fluids},}\ }\href {\doibase 10.1103/physreve.91.012112}
  {\bibfield  {journal} {\bibinfo  {journal} {Phys. Rev. E}\ }\textbf {\bibinfo
  {volume} {91}},\ \bibinfo {pages} {012112} (\bibinfo {year}
  {2015})}\BibitemShut {NoStop}%
\bibitem [{\citenamefont {Bryk}\ \emph
  {et~al.}(2017{\natexlab{b}})\citenamefont {Bryk}, \citenamefont {Gorelli},
  \citenamefont {Mryglod}, \citenamefont {Ruocco}, \citenamefont {Santoro},\
  and\ \citenamefont {Scopigno}}]{BrykJPCL2017}%
  \BibitemOpen
  \bibfield  {author} {\bibinfo {author} {\bibfnamefont {T.}~\bibnamefont
  {Bryk}}, \bibinfo {author} {\bibfnamefont {F.~A.}\ \bibnamefont {Gorelli}},
  \bibinfo {author} {\bibfnamefont {I.}~\bibnamefont {Mryglod}}, \bibinfo
  {author} {\bibfnamefont {G.}~\bibnamefont {Ruocco}}, \bibinfo {author}
  {\bibfnamefont {M.}~\bibnamefont {Santoro}}, \ and\ \bibinfo {author}
  {\bibfnamefont {T.}~\bibnamefont {Scopigno}},\ }\bibfield  {title} {\enquote
  {\bibinfo {title} {Behavior of supercritical fluids across the {F}renkel
  line},}\ }\href {\doibase 10.1021/acs.jpclett.7b02176} {\bibfield  {journal}
  {\bibinfo  {journal} {J. Phys. Chem. Lett.}\ }\textbf {\bibinfo {volume}
  {8}},\ \bibinfo {pages} {4995--5001} (\bibinfo {year}
  {2017}{\natexlab{b}})}\BibitemShut {NoStop}%
\bibitem [{\citenamefont {Brazhkin}\ \emph {et~al.}(2018)\citenamefont
  {Brazhkin}, \citenamefont {Prescher}, \citenamefont {Fomin}, \citenamefont
  {Tsiok}, \citenamefont {Lyapin}, \citenamefont {Ryzhov}, \citenamefont
  {Prakapenka}, \citenamefont {Stefanski}, \citenamefont {Trachenko},\ and\
  \citenamefont {Sapelkin}}]{BrazhkinJPCB2018}%
  \BibitemOpen
  \bibfield  {author} {\bibinfo {author} {\bibfnamefont {V.~V.}\ \bibnamefont
  {Brazhkin}}, \bibinfo {author} {\bibfnamefont {C.}~\bibnamefont {Prescher}},
  \bibinfo {author} {\bibfnamefont {Yu.~D.}\ \bibnamefont {Fomin}}, \bibinfo
  {author} {\bibfnamefont {E.~N.}\ \bibnamefont {Tsiok}}, \bibinfo {author}
  {\bibfnamefont {A.~G.}\ \bibnamefont {Lyapin}}, \bibinfo {author}
  {\bibfnamefont {V.~N.}\ \bibnamefont {Ryzhov}}, \bibinfo {author}
  {\bibfnamefont {V.~B.}\ \bibnamefont {Prakapenka}}, \bibinfo {author}
  {\bibfnamefont {J.}~\bibnamefont {Stefanski}}, \bibinfo {author}
  {\bibfnamefont {K.}~\bibnamefont {Trachenko}}, \ and\ \bibinfo {author}
  {\bibfnamefont {A.}~\bibnamefont {Sapelkin}},\ }\bibfield  {title} {\enquote
  {\bibinfo {title} {Comment on {B}ehavior of supercritical fluids across the
  {F}renkel line},}\ }\href {\doibase 10.1021/acs.jpcb.7b11359} {\bibfield
  {journal} {\bibinfo  {journal} {J. Phys. Chem. B}\ }\textbf {\bibinfo
  {volume} {122}},\ \bibinfo {pages} {6124--6128} (\bibinfo {year}
  {2018})}\BibitemShut {NoStop}%
\bibitem [{\citenamefont {Bell}\ \emph {et~al.}(2020)\citenamefont {Bell},
  \citenamefont {Galliero}, \citenamefont {Delage-Santacreu},\ and\
  \citenamefont {Costigliola}}]{BellJCP2020}%
  \BibitemOpen
  \bibfield  {author} {\bibinfo {author} {\bibfnamefont {I.~H.}\ \bibnamefont
  {Bell}}, \bibinfo {author} {\bibfnamefont {G.}~\bibnamefont {Galliero}},
  \bibinfo {author} {\bibfnamefont {S.}~\bibnamefont {Delage-Santacreu}}, \
  and\ \bibinfo {author} {\bibfnamefont {L.}~\bibnamefont {Costigliola}},\
  }\bibfield  {title} {\enquote {\bibinfo {title} {An entropy scaling
  demarcation of gas- and liquid-like fluid behaviors},}\ }\href {\doibase
  10.1063/1.5143854} {\bibfield  {journal} {\bibinfo  {journal} {J. Chem.
  Phys.}\ }\textbf {\bibinfo {volume} {152}},\ \bibinfo {pages} {191102}
  (\bibinfo {year} {2020})}\BibitemShut {NoStop}%
\bibitem [{\citenamefont {Proctor}\ \emph {et~al.}(2019)\citenamefont
  {Proctor}, \citenamefont {Pruteanu}, \citenamefont {Morrison}, \citenamefont
  {Crowe},\ and\ \citenamefont {Loveday}}]{ProctorJPCL2019}%
  \BibitemOpen
  \bibfield  {author} {\bibinfo {author} {\bibfnamefont {J.~E.}\ \bibnamefont
  {Proctor}}, \bibinfo {author} {\bibfnamefont {C.~G.}\ \bibnamefont
  {Pruteanu}}, \bibinfo {author} {\bibfnamefont {I.}~\bibnamefont {Morrison}},
  \bibinfo {author} {\bibfnamefont {I.~F.}\ \bibnamefont {Crowe}}, \ and\
  \bibinfo {author} {\bibfnamefont {J.~S.}\ \bibnamefont {Loveday}},\
  }\bibfield  {title} {\enquote {\bibinfo {title} {Transition from gas-like to
  liquid-like behavior in supercritical {N}$_2$},}\ }\href {\doibase
  10.1021/acs.jpclett.9b02358} {\bibfield  {journal} {\bibinfo  {journal} {J.
  Phys. Chem. Lett.}\ }\textbf {\bibinfo {volume} {10}},\ \bibinfo {pages}
  {6584--6589} (\bibinfo {year} {2019})}\BibitemShut {NoStop}%
\bibitem [{\citenamefont {Ploetz}\ and\ \citenamefont
  {Smith}(2019)}]{PloetzJPCB2019}%
  \BibitemOpen
  \bibfield  {author} {\bibinfo {author} {\bibfnamefont {E.~A.}\ \bibnamefont
  {Ploetz}}\ and\ \bibinfo {author} {\bibfnamefont {P.~E.}\ \bibnamefont
  {Smith}},\ }\bibfield  {title} {\enquote {\bibinfo {title} {Gas or liquid?
  {T}he supercritical behavior of pure fluids},}\ }\href {\doibase
  10.1021/acs.jpcb.9b04058} {\bibfield  {journal} {\bibinfo  {journal} {J.
  Phys. Chem. B}\ }\textbf {\bibinfo {volume} {123}},\ \bibinfo {pages}
  {6554--6563} (\bibinfo {year} {2019})}\BibitemShut {NoStop}%
\bibitem [{\citenamefont {Banuti}\ \emph {et~al.}(2020)\citenamefont {Banuti},
  \citenamefont {Raju},\ and\ \citenamefont {Ihme}}]{BanutiJSupFluids2020}%
  \BibitemOpen
  \bibfield  {author} {\bibinfo {author} {\bibfnamefont {D.T.}\ \bibnamefont
  {Banuti}}, \bibinfo {author} {\bibfnamefont {M.}~\bibnamefont {Raju}}, \ and\
  \bibinfo {author} {\bibfnamefont {M.}~\bibnamefont {Ihme}},\ }\bibfield
  {title} {\enquote {\bibinfo {title} {Between supercritical liquids and gases
  {\textendash} reconciling dynamic and thermodynamic state transitions},}\
  }\href {\doibase 10.1016/j.supflu.2020.104895} {\bibfield  {journal}
  {\bibinfo  {journal} {J. Supercrit. Fluids}\ }\textbf {\bibinfo {volume}
  {165}},\ \bibinfo {pages} {104895} (\bibinfo {year} {2020})}\BibitemShut
  {NoStop}%
\bibitem [{\citenamefont {Ha}\ \emph {et~al.}(2019)\citenamefont {Ha},
  \citenamefont {Yoon}, \citenamefont {Tlusty}, \citenamefont {Jho},\ and\
  \citenamefont {Lee}}]{HaJPCL2020}%
  \BibitemOpen
  \bibfield  {author} {\bibinfo {author} {\bibfnamefont {M.Y.}\ \bibnamefont
  {Ha}}, \bibinfo {author} {\bibfnamefont {T.~J.}\ \bibnamefont {Yoon}},
  \bibinfo {author} {\bibfnamefont {T.}~\bibnamefont {Tlusty}}, \bibinfo
  {author} {\bibfnamefont {Y.}~\bibnamefont {Jho}}, \ and\ \bibinfo {author}
  {\bibfnamefont {W.~B.}\ \bibnamefont {Lee}},\ }\bibfield  {title} {\enquote
  {\bibinfo {title} {Universality, scaling, and collapse in supercritical
  fluids},}\ }\href {\doibase 10.1021/acs.jpclett.9b03360} {\bibfield
  {journal} {\bibinfo  {journal} {J. Phys. Chem. Lett.}\ }\textbf {\bibinfo
  {volume} {11}},\ \bibinfo {pages} {451--455} (\bibinfo {year}
  {2019})}\BibitemShut {NoStop}%
\bibitem [{\citenamefont {Maxim}\ \emph {et~al.}(2019)\citenamefont {Maxim},
  \citenamefont {Contescu}, \citenamefont {Boillat}, \citenamefont {Niceno},
  \citenamefont {Karalis}, \citenamefont {Testino},\ and\ \citenamefont
  {Ludwig}}]{MaximNatCom2019}%
  \BibitemOpen
  \bibfield  {author} {\bibinfo {author} {\bibfnamefont {F.}~\bibnamefont
  {Maxim}}, \bibinfo {author} {\bibfnamefont {C.}~\bibnamefont {Contescu}},
  \bibinfo {author} {\bibfnamefont {P.}~\bibnamefont {Boillat}}, \bibinfo
  {author} {\bibfnamefont {B.}~\bibnamefont {Niceno}}, \bibinfo {author}
  {\bibfnamefont {K.}~\bibnamefont {Karalis}}, \bibinfo {author} {\bibfnamefont
  {A.}~\bibnamefont {Testino}}, \ and\ \bibinfo {author} {\bibfnamefont
  {C.}~\bibnamefont {Ludwig}},\ }\bibfield  {title} {\enquote {\bibinfo {title}
  {Visualization of supercritical water pseudo-boiling at {W}idom line
  crossover},}\ }\href {\doibase 10.1038/s41467-019-12117-5} {\bibfield
  {journal} {\bibinfo  {journal} {Nature Commun.}\ }\textbf {\bibinfo {volume}
  {10}},\ \bibinfo {pages} {4114} (\bibinfo {year} {2019})}\BibitemShut
  {NoStop}%
\bibitem [{\citenamefont {Sun}\ \emph {et~al.}(2020)\citenamefont {Sun},
  \citenamefont {Hastings}, \citenamefont {Ishikawa}, \citenamefont {Baron},\
  and\ \citenamefont {Monaco}}]{SunPRL2020}%
  \BibitemOpen
  \bibfield  {author} {\bibinfo {author} {\bibfnamefont {P.}~\bibnamefont
  {Sun}}, \bibinfo {author} {\bibfnamefont {J.~B.}\ \bibnamefont {Hastings}},
  \bibinfo {author} {\bibfnamefont {D.}~\bibnamefont {Ishikawa}}, \bibinfo
  {author} {\bibfnamefont {A.~Q.~R.}\ \bibnamefont {Baron}}, \ and\ \bibinfo
  {author} {\bibfnamefont {G.}~\bibnamefont {Monaco}},\ }\bibfield  {title}
  {\enquote {\bibinfo {title} {Two-component dynamics and the liquidlike to
  gaslike crossover in supercritical water},}\ }\href {\doibase
  10.1103/physrevlett.125.256001} {\bibfield  {journal} {\bibinfo  {journal}
  {Phys. Rev. Lett.}\ }\textbf {\bibinfo {volume} {125}},\ \bibinfo {pages}
  {256001} (\bibinfo {year} {2020})}\BibitemShut {NoStop}%
\bibitem [{\citenamefont {Bell}\ \emph {et~al.}(2021)\citenamefont {Bell},
  \citenamefont {Delage-Santacreu}, \citenamefont {Hoang},\ and\ \citenamefont
  {Galliero}}]{BellJPCL2021}%
  \BibitemOpen
  \bibfield  {author} {\bibinfo {author} {\bibfnamefont {I.~H.}\ \bibnamefont
  {Bell}}, \bibinfo {author} {\bibfnamefont {S.}~\bibnamefont
  {Delage-Santacreu}}, \bibinfo {author} {\bibfnamefont {H.}~\bibnamefont
  {Hoang}}, \ and\ \bibinfo {author} {\bibfnamefont {G.}~\bibnamefont
  {Galliero}},\ }\bibfield  {title} {\enquote {\bibinfo {title} {Dynamic
  crossover in fluids: From hard spheres to molecules},}\ }\href {\doibase
  10.1021/acs.jpclett.1c01594} {\bibfield  {journal} {\bibinfo  {journal} {J.
  Phys. Chem. Lett.}\ }\textbf {\bibinfo {volume} {12}},\ \bibinfo {pages}
  {6411--6417} (\bibinfo {year} {2021})}\BibitemShut {NoStop}%
\bibitem [{\citenamefont {Cockrell}\ \emph
  {et~al.}(2021{\natexlab{a}})\citenamefont {Cockrell}, \citenamefont
  {Brazhkin},\ and\ \citenamefont {Trachenko}}]{CockrellPRE2021}%
  \BibitemOpen
  \bibfield  {author} {\bibinfo {author} {\bibfnamefont {C.}~\bibnamefont
  {Cockrell}}, \bibinfo {author} {\bibfnamefont {V.~V.}\ \bibnamefont
  {Brazhkin}}, \ and\ \bibinfo {author} {\bibfnamefont {K.}~\bibnamefont
  {Trachenko}},\ }\bibfield  {title} {\enquote {\bibinfo {title} {Universal
  interrelation between dynamics and thermodynamics and a dynamically driven
  {\textquotedblleft}c{\textquotedblright} transition in fluids},}\ }\href
  {\doibase 10.1103/physreve.104.034108} {\bibfield  {journal} {\bibinfo
  {journal} {Phys. Rev. E}\ }\textbf {\bibinfo {volume} {104}},\ \bibinfo
  {pages} {034108} (\bibinfo {year} {2021}{\natexlab{a}})}\BibitemShut
  {NoStop}%
\bibitem [{\citenamefont {Cockrell}\ \emph
  {et~al.}(2021{\natexlab{b}})\citenamefont {Cockrell}, \citenamefont
  {Brazhkin},\ and\ \citenamefont {Trachenko}}]{CockrellPhysRep2021}%
  \BibitemOpen
  \bibfield  {author} {\bibinfo {author} {\bibfnamefont {C.}~\bibnamefont
  {Cockrell}}, \bibinfo {author} {\bibfnamefont {V.V.}\ \bibnamefont
  {Brazhkin}}, \ and\ \bibinfo {author} {\bibfnamefont {K.}~\bibnamefont
  {Trachenko}},\ }\bibfield  {title} {\enquote {\bibinfo {title} {Transition in
  the supercritical state of matter: Review of experimental evidence},}\ }\href
  {\doibase 10.1016/j.physrep.2021.10.002} {\bibfield  {journal} {\bibinfo
  {journal} {Phys. Rep.}\ }\textbf {\bibinfo {volume} {941}},\ \bibinfo {pages}
  {1--27} (\bibinfo {year} {2021}{\natexlab{b}})}\BibitemShut {NoStop}%
\bibitem [{\citenamefont {Barker}\ and\ \citenamefont
  {Henderson}(1976)}]{BarkerRMP1976}%
  \BibitemOpen
  \bibfield  {author} {\bibinfo {author} {\bibfnamefont {J.~A.}\ \bibnamefont
  {Barker}}\ and\ \bibinfo {author} {\bibfnamefont {D.}~\bibnamefont
  {Henderson}},\ }\bibfield  {title} {\enquote {\bibinfo {title} {What is
  "liquid"? {U}nderstanding the states of matter},}\ }\href {\doibase
  10.1103/revmodphys.48.587} {\bibfield  {journal} {\bibinfo  {journal} {Rev.
  Mod. Phys.}\ }\textbf {\bibinfo {volume} {48}},\ \bibinfo {pages} {587--671}
  (\bibinfo {year} {1976})}\BibitemShut {NoStop}%
\bibitem [{\citenamefont {Sengers}(1979)}]{Sengers1979}%
  \BibitemOpen
  \bibfield  {author} {\bibinfo {author} {\bibfnamefont {J.M.H.~Levelt}\
  \bibnamefont {Sengers}},\ }\bibfield  {title} {\enquote {\bibinfo {title}
  {Liquidons and gasons: controversies about the continuity of states},}\
  }\href {\doibase 10.1016/0378-4371(79)90145-6} {\bibfield  {journal}
  {\bibinfo  {journal} {Phys. A}\ }\textbf {\bibinfo {volume} {98}},\ \bibinfo
  {pages} {363--402} (\bibinfo {year} {1979})}\BibitemShut {NoStop}%
\bibitem [{\citenamefont {{L. Woodcock}}(2017)}]{Woodcock2017}%
  \BibitemOpen
  \bibfield  {author} {\bibinfo {author} {\bibnamefont {{L. Woodcock}}},\
  }\bibfield  {title} {\enquote {\bibinfo {title} {Percolation transitions and
  fluid state boundaries},}\ }\href {\doibase 10.12921/CMST.2016.0000070}
  {\bibfield  {journal} {\bibinfo  {journal} {CMST}\ }\textbf {\bibinfo
  {volume} {23}},\ \bibinfo {pages} {281} (\bibinfo {year} {2017})}\BibitemShut
  {NoStop}%
\bibitem [{\citenamefont {Trachenko}\ and\ \citenamefont
  {Brazhkin}(2021)}]{TrachenkoPhysToday2021}%
  \BibitemOpen
  \bibfield  {author} {\bibinfo {author} {\bibfnamefont {K.}~\bibnamefont
  {Trachenko}}\ and\ \bibinfo {author} {\bibfnamefont {V.~V.}\ \bibnamefont
  {Brazhkin}},\ }\bibfield  {title} {\enquote {\bibinfo {title} {The quantum
  mechanics of viscosity},}\ }\href {\doibase 10.1063/pt.3.4908} {\bibfield
  {journal} {\bibinfo  {journal} {Phys. Today}\ }\textbf {\bibinfo {volume}
  {74}},\ \bibinfo {pages} {66--67} (\bibinfo {year} {2021})}\BibitemShut
  {NoStop}%
\bibitem [{\citenamefont {Trachenko}\ and\ \citenamefont
  {Brazhkin}(2015)}]{TrachenkoRPP2015}%
  \BibitemOpen
  \bibfield  {author} {\bibinfo {author} {\bibfnamefont {K.}~\bibnamefont
  {Trachenko}}\ and\ \bibinfo {author} {\bibfnamefont {V.~V.}\ \bibnamefont
  {Brazhkin}},\ }\bibfield  {title} {\enquote {\bibinfo {title} {Collective
  modes and thermodynamics of the liquid state},}\ }\href {\doibase
  10.1088/0034-4885/79/1/016502} {\bibfield  {journal} {\bibinfo  {journal}
  {Rep. Progr. Phys.}\ }\textbf {\bibinfo {volume} {79}},\ \bibinfo {pages}
  {016502} (\bibinfo {year} {2015})}\BibitemShut {NoStop}%
\bibitem [{\citenamefont {Sousa}\ \emph {et~al.}(2012)\citenamefont {Sousa},
  \citenamefont {Ferreira},\ and\ \citenamefont {Barroso}}]{SousaJCP2012}%
  \BibitemOpen
  \bibfield  {author} {\bibinfo {author} {\bibfnamefont {J.~M.~G.}\
  \bibnamefont {Sousa}}, \bibinfo {author} {\bibfnamefont {A.~L.}\ \bibnamefont
  {Ferreira}}, \ and\ \bibinfo {author} {\bibfnamefont {M.~A.}\ \bibnamefont
  {Barroso}},\ }\bibfield  {title} {\enquote {\bibinfo {title} {Determination
  of the solid-fluid coexistence of the n - 6 {L}ennard-{J}ones system from
  free energy calculations},}\ }\href {\doibase 10.1063/1.4707746} {\bibfield
  {journal} {\bibinfo  {journal} {J. Chem. Phys.}\ }\textbf {\bibinfo {volume}
  {136}},\ \bibinfo {pages} {174502} (\bibinfo {year} {2012})}\BibitemShut
  {NoStop}%
\bibitem [{\citenamefont {Hansen}(1970)}]{HansenPRA1970}%
  \BibitemOpen
  \bibfield  {author} {\bibinfo {author} {\bibfnamefont {J.-P.}\ \bibnamefont
  {Hansen}},\ }\bibfield  {title} {\enquote {\bibinfo {title} {Phase transition
  of the {L}ennard-{J}ones system. {II}. {H}igh-temperature limit},}\ }\href
  {\doibase 10.1103/physreva.2.221} {\bibfield  {journal} {\bibinfo  {journal}
  {Phys. Rev. A}\ }\textbf {\bibinfo {volume} {2}},\ \bibinfo {pages}
  {221--230} (\bibinfo {year} {1970})}\BibitemShut {NoStop}%
\bibitem [{\citenamefont {Khrapak}\ and\ \citenamefont
  {Morfill}(2011)}]{KhrapakJCP2011_2}%
  \BibitemOpen
  \bibfield  {author} {\bibinfo {author} {\bibfnamefont {S.~A.}\ \bibnamefont
  {Khrapak}}\ and\ \bibinfo {author} {\bibfnamefont {G.~E.}\ \bibnamefont
  {Morfill}},\ }\bibfield  {title} {\enquote {\bibinfo {title} {Accurate
  freezing and melting equations for the {L}ennard-{J}ones system},}\ }\href
  {\doibase 10.1063/1.3561698} {\bibfield  {journal} {\bibinfo  {journal} {J.
  Chem. Phys.}\ }\textbf {\bibinfo {volume} {134}},\ \bibinfo {pages} {094108}
  (\bibinfo {year} {2011})}\BibitemShut {NoStop}%
\bibitem [{\citenamefont {Heyes}\ \emph {et~al.}(2019)\citenamefont {Heyes},
  \citenamefont {Dini}, \citenamefont {Costigliola},\ and\ \citenamefont
  {Dyre}}]{HeyesJCP2019}%
  \BibitemOpen
  \bibfield  {author} {\bibinfo {author} {\bibfnamefont {D.~M.}\ \bibnamefont
  {Heyes}}, \bibinfo {author} {\bibfnamefont {D.}~\bibnamefont {Dini}},
  \bibinfo {author} {\bibfnamefont {L.}~\bibnamefont {Costigliola}}, \ and\
  \bibinfo {author} {\bibfnamefont {J.~C.}\ \bibnamefont {Dyre}},\ }\bibfield
  {title} {\enquote {\bibinfo {title} {Transport coefficients of the
  {L}ennard-{J}ones fluid close to the freezing line},}\ }\href {\doibase
  10.1063/1.5128707} {\bibfield  {journal} {\bibinfo  {journal} {J. Chem.
  Phys.}\ }\textbf {\bibinfo {volume} {151}},\ \bibinfo {pages} {204502}
  (\bibinfo {year} {2019})}\BibitemShut {NoStop}%
\bibitem [{\citenamefont {Thol}\ \emph {et~al.}(2016)\citenamefont {Thol},
  \citenamefont {Rutkai}, \citenamefont {K\"{o}ster}, \citenamefont {Lustig},
  \citenamefont {Span},\ and\ \citenamefont {Vrabec}}]{Thol2016}%
  \BibitemOpen
  \bibfield  {author} {\bibinfo {author} {\bibfnamefont {M.}~\bibnamefont
  {Thol}}, \bibinfo {author} {\bibfnamefont {G.}~\bibnamefont {Rutkai}},
  \bibinfo {author} {\bibfnamefont {A.}~\bibnamefont {K\"{o}ster}}, \bibinfo
  {author} {\bibfnamefont {R.}~\bibnamefont {Lustig}}, \bibinfo {author}
  {\bibfnamefont {R.}~\bibnamefont {Span}}, \ and\ \bibinfo {author}
  {\bibfnamefont {J.}~\bibnamefont {Vrabec}},\ }\bibfield  {title} {\enquote
  {\bibinfo {title} {Equation of state for the {L}ennard-{J}ones fluid},}\
  }\href {\doibase 10.1063/1.4945000} {\bibfield  {journal} {\bibinfo
  {journal} {J. Phys. Chem. Ref. Data}\ }\textbf {\bibinfo {volume} {45}},\
  \bibinfo {pages} {023101} (\bibinfo {year} {2016})}\BibitemShut {NoStop}%
\bibitem [{\citenamefont {Pedersen}\ \emph {et~al.}(2016)\citenamefont
  {Pedersen}, \citenamefont {Costigliola}, \citenamefont {Bailey},
  \citenamefont {Schr{\o}der},\ and\ \citenamefont
  {Dyre}}]{PedersenNatCom2016}%
  \BibitemOpen
  \bibfield  {author} {\bibinfo {author} {\bibfnamefont {U.~R.}\ \bibnamefont
  {Pedersen}}, \bibinfo {author} {\bibfnamefont {L.}~\bibnamefont
  {Costigliola}}, \bibinfo {author} {\bibfnamefont {N.~P.}\ \bibnamefont
  {Bailey}}, \bibinfo {author} {\bibfnamefont {T.~B.}\ \bibnamefont
  {Schr{\o}der}}, \ and\ \bibinfo {author} {\bibfnamefont {J.~C.}\ \bibnamefont
  {Dyre}},\ }\bibfield  {title} {\enquote {\bibinfo {title} {Thermodynamics of
  freezing and melting},}\ }\href {\doibase 10.1038/ncomms12386} {\bibfield
  {journal} {\bibinfo  {journal} {Nature Commun.}\ }\textbf {\bibinfo {volume}
  {7}},\ \bibinfo {pages} {12386} (\bibinfo {year} {2016})}\BibitemShut
  {NoStop}%
\bibitem [{\citenamefont {Rosenfeld}(1976)}]{RosenfeldMolPhys1976}%
  \BibitemOpen
  \bibfield  {author} {\bibinfo {author} {\bibfnamefont {Y.}~\bibnamefont
  {Rosenfeld}},\ }\bibfield  {title} {\enquote {\bibinfo {title} {Universality
  of melting and freezing indicators and additivity of melting curves},}\
  }\href {\doibase 10.1080/00268977600102381} {\bibfield  {journal} {\bibinfo
  {journal} {Mol. Phys.}\ }\textbf {\bibinfo {volume} {32}},\ \bibinfo {pages}
  {963--977} (\bibinfo {year} {1976})}\BibitemShut {NoStop}%
\bibitem [{\citenamefont {Heyes}\ \emph {et~al.}(2015)\citenamefont {Heyes},
  \citenamefont {Dini},\ and\ \citenamefont {Bra{\'{n}}ka}}]{HeyesPSS2015}%
  \BibitemOpen
  \bibfield  {author} {\bibinfo {author} {\bibfnamefont {D.~M.}\ \bibnamefont
  {Heyes}}, \bibinfo {author} {\bibfnamefont {D.}~\bibnamefont {Dini}}, \ and\
  \bibinfo {author} {\bibfnamefont {A.~C.}\ \bibnamefont {Bra{\'{n}}ka}},\
  }\bibfield  {title} {\enquote {\bibinfo {title} {Scaling of {L}ennard-{J}ones
  liquid elastic moduli, viscoelasticity and other properties along fluid-solid
  coexistence},}\ }\href {\doibase 10.1002/pssb.201451695} {\bibfield
  {journal} {\bibinfo  {journal} {Phys. Status Solidi (b)}\ }\textbf {\bibinfo
  {volume} {252}},\ \bibinfo {pages} {1514--1525} (\bibinfo {year}
  {2015})}\BibitemShut {NoStop}%
\bibitem [{\citenamefont {Khrapak}\ and\ \citenamefont
  {Ning}(2016)}]{KhrapakAIPAdv2016}%
  \BibitemOpen
  \bibfield  {author} {\bibinfo {author} {\bibfnamefont {S.~A.}\ \bibnamefont
  {Khrapak}}\ and\ \bibinfo {author} {\bibfnamefont {N.}~\bibnamefont {Ning}},\
  }\bibfield  {title} {\enquote {\bibinfo {title} {Freezing and melting
  equations for the n-6 {L}ennard-{J}ones systems},}\ }\href {\doibase
  10.1063/1.4952587} {\bibfield  {journal} {\bibinfo  {journal} {{AIP} Adv.}\
  }\textbf {\bibinfo {volume} {6}},\ \bibinfo {pages} {055215} (\bibinfo {year}
  {2016})}\BibitemShut {NoStop}%
\bibitem [{\citenamefont {Costigliola}\ \emph {et~al.}(2016)\citenamefont
  {Costigliola}, \citenamefont {Schr{\o}der},\ and\ \citenamefont
  {Dyre}}]{CostigliolaPCCP2016}%
  \BibitemOpen
  \bibfield  {author} {\bibinfo {author} {\bibfnamefont {L.}~\bibnamefont
  {Costigliola}}, \bibinfo {author} {\bibfnamefont {T.~B.}\ \bibnamefont
  {Schr{\o}der}}, \ and\ \bibinfo {author} {\bibfnamefont {J.~C.}\ \bibnamefont
  {Dyre}},\ }\bibfield  {title} {\enquote {\bibinfo {title} {Freezing and
  melting line invariants of the {L}ennard-{J}ones system},}\ }\href {\doibase
  10.1039/c5cp06363a} {\bibfield  {journal} {\bibinfo  {journal} {Phys. Chem.
  Chem. Phys.}\ }\textbf {\bibinfo {volume} {18}},\ \bibinfo {pages}
  {14678--14690} (\bibinfo {year} {2016})}\BibitemShut {NoStop}%
\bibitem [{\citenamefont {Heyes}\ \emph {et~al.}(2021)\citenamefont {Heyes},
  \citenamefont {Pieprzyk},\ and\ \citenamefont {Branka}}]{HeyesPRE2021}%
  \BibitemOpen
  \bibfield  {author} {\bibinfo {author} {\bibfnamefont {D.~M.}\ \bibnamefont
  {Heyes}}, \bibinfo {author} {\bibfnamefont {S.}~\bibnamefont {Pieprzyk}}, \
  and\ \bibinfo {author} {\bibfnamefont {A.~C.}\ \bibnamefont {Branka}},\
  }\bibfield  {title} {\enquote {\bibinfo {title} {Application of cell models
  to the melting and sublimation lines of the {L}ennard-{J}ones and related
  potential systems},}\ }\href {\doibase 10.1103/physreve.104.044119}
  {\bibfield  {journal} {\bibinfo  {journal} {Phys. Rev. E}\ }\textbf {\bibinfo
  {volume} {104}},\ \bibinfo {pages} {044119} (\bibinfo {year}
  {2021})}\BibitemShut {NoStop}%
\bibitem [{\citenamefont {Pruteanu}\ \emph {et~al.}(2023)\citenamefont
  {Pruteanu}, \citenamefont {Bannerman}, \citenamefont {Kirsz}, \citenamefont
  {Lue},\ and\ \citenamefont {Ackland}}]{PruteanuACSOmega2023}%
  \BibitemOpen
  \bibfield  {author} {\bibinfo {author} {\bibfnamefont {C.~G.}\ \bibnamefont
  {Pruteanu}}, \bibinfo {author} {\bibfnamefont {M.~N.}\ \bibnamefont
  {Bannerman}}, \bibinfo {author} {\bibfnamefont {M.}~\bibnamefont {Kirsz}},
  \bibinfo {author} {\bibfnamefont {L.}~\bibnamefont {Lue}}, \ and\ \bibinfo
  {author} {\bibfnamefont {G.~J.}\ \bibnamefont {Ackland}},\ }\bibfield
  {title} {\enquote {\bibinfo {title} {From atoms to colloids: Does the
  {F}renkel line exist in discontinuous potentials?}}\ }\href {\doibase
  10.1021/acsomega.2c08056} {\bibfield  {journal} {\bibinfo  {journal} {{ACS}
  Omega}\ }\textbf {\bibinfo {volume} {8}},\ \bibinfo {pages} {12144--12153}
  (\bibinfo {year} {2023})}\BibitemShut {NoStop}%
\bibitem [{\citenamefont {Huang}\ \emph {et~al.}(2023)\citenamefont {Huang},
  \citenamefont {Baggioli}, \citenamefont {Lu}, \citenamefont {Ma},\ and\
  \citenamefont {Feng}}]{HuangPRR2023}%
  \BibitemOpen
  \bibfield  {author} {\bibinfo {author} {\bibfnamefont {D.}~\bibnamefont
  {Huang}}, \bibinfo {author} {\bibfnamefont {M.}~\bibnamefont {Baggioli}},
  \bibinfo {author} {\bibfnamefont {S.}~\bibnamefont {Lu}}, \bibinfo {author}
  {\bibfnamefont {Z.}~\bibnamefont {Ma}}, \ and\ \bibinfo {author}
  {\bibfnamefont {Y.}~\bibnamefont {Feng}},\ }\bibfield  {title} {\enquote
  {\bibinfo {title} {Revealing the supercritical dynamics of dusty plasmas and
  their liquidlike to gaslike dynamical crossover},}\ }\href {\doibase
  10.1103/physrevresearch.5.013149} {\bibfield  {journal} {\bibinfo  {journal}
  {Phys. Rev. Research}\ }\textbf {\bibinfo {volume} {5}},\ \bibinfo {pages}
  {013149} (\bibinfo {year} {2023})}\BibitemShut {NoStop}%
\bibitem [{\citenamefont {Lemmon}\ and\ \citenamefont
  {Jacobsen}(2004)}]{LemmonIJT2004}%
  \BibitemOpen
  \bibfield  {author} {\bibinfo {author} {\bibfnamefont {E.~W.}\ \bibnamefont
  {Lemmon}}\ and\ \bibinfo {author} {\bibfnamefont {R.~T}\ \bibnamefont
  {Jacobsen}},\ }\bibfield  {title} {\enquote {\bibinfo {title} {Viscosity and
  thermal conductivity equations for {N}itrogen, {O}xygen, {A}rgon, and
  {A}ir},}\ }\href {\doibase 10.1023/b:ijot.0000022327.04529.f3} {\bibfield
  {journal} {\bibinfo  {journal} {Int. J. Thermophys.}\ }\textbf {\bibinfo
  {volume} {25}},\ \bibinfo {pages} {21--69} (\bibinfo {year}
  {2004})}\BibitemShut {NoStop}%
\bibitem [{\citenamefont {Ohtori}\ \emph {et~al.}(2020)\citenamefont {Ohtori},
  \citenamefont {Kondo}, \citenamefont {Shintani}, \citenamefont {Murakami},
  \citenamefont {Nobuta},\ and\ \citenamefont {Ishii}}]{OhtoriChemLett2020}%
  \BibitemOpen
  \bibfield  {author} {\bibinfo {author} {\bibfnamefont {N.}~\bibnamefont
  {Ohtori}}, \bibinfo {author} {\bibfnamefont {Y.}~\bibnamefont {Kondo}},
  \bibinfo {author} {\bibfnamefont {K.}~\bibnamefont {Shintani}}, \bibinfo
  {author} {\bibfnamefont {T.}~\bibnamefont {Murakami}}, \bibinfo {author}
  {\bibfnamefont {T.}~\bibnamefont {Nobuta}}, \ and\ \bibinfo {author}
  {\bibfnamefont {Y.}~\bibnamefont {Ishii}},\ }\bibfield  {title} {\enquote
  {\bibinfo {title} {The {S}tokes-{E}instein relation for non-spherical
  molecular liquids},}\ }\href {\doibase 10.1246/cl.200021} {\bibfield
  {journal} {\bibinfo  {journal} {Chem. Lett.}\ }\textbf {\bibinfo {volume}
  {49}},\ \bibinfo {pages} {379--382} (\bibinfo {year} {2020})}\BibitemShut
  {NoStop}%
\bibitem [{\citenamefont {Li}\ \emph {et~al.}(2021)\citenamefont {Li},
  \citenamefont {Sun}, \citenamefont {Zhang}, \citenamefont {Xian},\ and\
  \citenamefont {Vocadlo}}]{LiJCP2021}%
  \BibitemOpen
  \bibfield  {author} {\bibinfo {author} {\bibfnamefont {Q.}~\bibnamefont
  {Li}}, \bibinfo {author} {\bibfnamefont {T.}~\bibnamefont {Sun}}, \bibinfo
  {author} {\bibfnamefont {Y.}~\bibnamefont {Zhang}}, \bibinfo {author}
  {\bibfnamefont {J.-W.}\ \bibnamefont {Xian}}, \ and\ \bibinfo {author}
  {\bibfnamefont {L.}~\bibnamefont {Vocadlo}},\ }\bibfield  {title} {\enquote
  {\bibinfo {title} {Atomic transport properties of liquid iron at conditions
  of planetary cores},}\ }\href {\doibase 10.1063/5.0062081} {\bibfield
  {journal} {\bibinfo  {journal} {J. Chem. Phys.}\ }\textbf {\bibinfo {volume}
  {155}},\ \bibinfo {pages} {194505} (\bibinfo {year} {2021})}\BibitemShut
  {NoStop}%
\bibitem [{\citenamefont {Ranieri}\ \emph {et~al.}(2021)\citenamefont
  {Ranieri}, \citenamefont {Klotz}, \citenamefont {Gaal}, \citenamefont
  {Koza},\ and\ \citenamefont {Bove}}]{Ranieri2021}%
  \BibitemOpen
  \bibfield  {author} {\bibinfo {author} {\bibfnamefont {U.}~\bibnamefont
  {Ranieri}}, \bibinfo {author} {\bibfnamefont {S.}~\bibnamefont {Klotz}},
  \bibinfo {author} {\bibfnamefont {R.}~\bibnamefont {Gaal}}, \bibinfo {author}
  {\bibfnamefont {M.~M.}\ \bibnamefont {Koza}}, \ and\ \bibinfo {author}
  {\bibfnamefont {L.~E.}\ \bibnamefont {Bove}},\ }\bibfield  {title} {\enquote
  {\bibinfo {title} {Diffusion in dense supercritical methane from
  quasi-elastic neutron scattering measurements},}\ }\href {\doibase
  10.1038/s41467-021-22182-4} {\bibfield  {journal} {\bibinfo  {journal}
  {Nature Commun.}\ }\textbf {\bibinfo {volume} {12}},\ \bibinfo {pages} {1958}
  (\bibinfo {year} {2021})}\BibitemShut {NoStop}%
\bibitem [{\citenamefont {Khrapak}(2022{\natexlab{c}})}]{KhrapakJMolLiq2022}%
  \BibitemOpen
  \bibfield  {author} {\bibinfo {author} {\bibfnamefont {S.A.}\ \bibnamefont
  {Khrapak}},\ }\bibfield  {title} {\enquote {\bibinfo {title} {Diffusion,
  viscosity, and {S}tokes-{E}instein relation in dense supercritical
  methane},}\ }\href {\doibase 10.1016/j.molliq.2022.118840} {\bibfield
  {journal} {\bibinfo  {journal} {J. Mol. Liq.}\ }\textbf {\bibinfo {volume}
  {354}},\ \bibinfo {pages} {118840} (\bibinfo {year}
  {2022}{\natexlab{c}})}\BibitemShut {NoStop}%
\bibitem [{\citenamefont {Luo}\ \emph {et~al.}(2022)\citenamefont {Luo},
  \citenamefont {Zhou}, \citenamefont {Li}, \citenamefont {Lin},\ and\
  \citenamefont {Liu}}]{Luo2022}%
  \BibitemOpen
  \bibfield  {author} {\bibinfo {author} {\bibfnamefont {J.}~\bibnamefont
  {Luo}}, \bibinfo {author} {\bibfnamefont {C.}~\bibnamefont {Zhou}}, \bibinfo
  {author} {\bibfnamefont {Q.}~\bibnamefont {Li}}, \bibinfo {author}
  {\bibfnamefont {Y.}~\bibnamefont {Lin}}, \ and\ \bibinfo {author}
  {\bibfnamefont {L.}~\bibnamefont {Liu}},\ }\bibfield  {title} {\enquote
  {\bibinfo {title} {Atomic transport properties of silicon melt at high
  temperature},}\ }\href {\doibase 10.1016/j.jcrysgro.2022.126701} {\bibfield
  {journal} {\bibinfo  {journal} {J. Crystal Growth}\ }\textbf {\bibinfo
  {volume} {590}},\ \bibinfo {pages} {126701} (\bibinfo {year}
  {2022})}\BibitemShut {NoStop}%
\bibitem [{\citenamefont {Baran}\ \emph {et~al.}(2023)\citenamefont {Baran},
  \citenamefont {Rzysko},\ and\ \citenamefont {MacDowell}}]{BaranJCP2023}%
  \BibitemOpen
  \bibfield  {author} {\bibinfo {author} {\bibfnamefont {L.}~\bibnamefont
  {Baran}}, \bibinfo {author} {\bibfnamefont {W.}~\bibnamefont {Rzysko}}, \
  and\ \bibinfo {author} {\bibfnamefont {L.~G.}\ \bibnamefont {MacDowell}},\
  }\bibfield  {title} {\enquote {\bibinfo {title} {Self-diffusion and shear
  viscosity for the {TIP}4{P}/{I}ce water model},}\ }\href {\doibase
  10.1063/5.0134932} {\bibfield  {journal} {\bibinfo  {journal} {J. Chem.
  Phys.}\ }\textbf {\bibinfo {volume} {158}},\ \bibinfo {pages} {064503}
  (\bibinfo {year} {2023})}\BibitemShut {NoStop}%
\bibitem [{\citenamefont {Khrapak}\ and\ \citenamefont
  {Khrapak}(2023{\natexlab{a}})}]{KhrapakJCP2023}%
  \BibitemOpen
  \bibfield  {author} {\bibinfo {author} {\bibfnamefont {S.~A.}\ \bibnamefont
  {Khrapak}}\ and\ \bibinfo {author} {\bibfnamefont {A.~G.}\ \bibnamefont
  {Khrapak}},\ }\bibfield  {title} {\enquote {\bibinfo {title}
  {Stokes-{E}instein relation without hydrodynamic diameter in the
  {TIP4P}/{I}ce water model},}\ }\href {\doibase 10.1063/5.0150871} {\bibfield
  {journal} {\bibinfo  {journal} {J. Chem. Phys.}\ }\textbf {\bibinfo {volume}
  {158}},\ \bibinfo {pages} {206101} (\bibinfo {year}
  {2023}{\natexlab{a}})}\BibitemShut {NoStop}%
\bibitem [{\citenamefont {Abascal}\ \emph {et~al.}(2005)\citenamefont
  {Abascal}, \citenamefont {Sanz}, \citenamefont {Fernandez},\ and\
  \citenamefont {Vega}}]{AbascalJCP2005}%
  \BibitemOpen
  \bibfield  {author} {\bibinfo {author} {\bibfnamefont {J.~L.~F.}\
  \bibnamefont {Abascal}}, \bibinfo {author} {\bibfnamefont {E.}~\bibnamefont
  {Sanz}}, \bibinfo {author} {\bibfnamefont {R.~G.}\ \bibnamefont {Fernandez}},
  \ and\ \bibinfo {author} {\bibfnamefont {C.}~\bibnamefont {Vega}},\
  }\bibfield  {title} {\enquote {\bibinfo {title} {A potential model for the
  study of ices and amorphous water: {TIP}4{P}/{I}ce},}\ }\href {\doibase
  10.1063/1.1931662} {\bibfield  {journal} {\bibinfo  {journal} {J. Chem.
  Phys.}\ }\textbf {\bibinfo {volume} {122}},\ \bibinfo {pages} {234511}
  (\bibinfo {year} {2005})}\BibitemShut {NoStop}%
\bibitem [{\citenamefont {Battezzati}\ and\ \citenamefont
  {Greer}(1989)}]{Battezzati1989}%
  \BibitemOpen
  \bibfield  {author} {\bibinfo {author} {\bibfnamefont {L.}~\bibnamefont
  {Battezzati}}\ and\ \bibinfo {author} {\bibfnamefont {A.L.}\ \bibnamefont
  {Greer}},\ }\bibfield  {title} {\enquote {\bibinfo {title} {The viscosity of
  liquid metals and alloys},}\ }\href {\doibase 10.1016/0001-6160(89)90064-3}
  {\bibfield  {journal} {\bibinfo  {journal} {Acta Metallurgica}\ }\textbf
  {\bibinfo {volume} {37}},\ \bibinfo {pages} {1791--1802} (\bibinfo {year}
  {1989})}\BibitemShut {NoStop}%
\bibitem [{\citenamefont {Kumar}\ \emph {et~al.}(2021)\citenamefont {Kumar},
  \citenamefont {Poser}, \citenamefont {Sch\"{o}ttler}, \citenamefont
  {Kleinschmidt}, \citenamefont {Dietrich}, \citenamefont {Wicht},
  \citenamefont {French},\ and\ \citenamefont {Redmer}}]{KumarPRE2021}%
  \BibitemOpen
  \bibfield  {author} {\bibinfo {author} {\bibfnamefont {S.}~\bibnamefont
  {Kumar}}, \bibinfo {author} {\bibfnamefont {A.~J.}\ \bibnamefont {Poser}},
  \bibinfo {author} {\bibfnamefont {M.}~\bibnamefont {Sch\"{o}ttler}}, \bibinfo
  {author} {\bibfnamefont {U.}~\bibnamefont {Kleinschmidt}}, \bibinfo {author}
  {\bibfnamefont {W.}~\bibnamefont {Dietrich}}, \bibinfo {author}
  {\bibfnamefont {J.}~\bibnamefont {Wicht}}, \bibinfo {author} {\bibfnamefont
  {M.}~\bibnamefont {French}}, \ and\ \bibinfo {author} {\bibfnamefont
  {R.}~\bibnamefont {Redmer}},\ }\bibfield  {title} {\enquote {\bibinfo {title}
  {Ionization and transport in partially ionized multicomponent plasmas:
  {A}pplication to atmospheres of hot {J}upiters},}\ }\href {\doibase
  10.1103/physreve.103.063203} {\bibfield  {journal} {\bibinfo  {journal}
  {Phys. Rev. E}\ }\textbf {\bibinfo {volume} {103}} (\bibinfo {year} {2021}),\
  10.1103/physreve.103.063203}\BibitemShut {NoStop}%
\bibitem [{\citenamefont {Bridgman}(1923)}]{Bridgman1923}%
  \BibitemOpen
  \bibfield  {author} {\bibinfo {author} {\bibfnamefont {P.~W.}\ \bibnamefont
  {Bridgman}},\ }\bibfield  {title} {\enquote {\bibinfo {title} {The thermal
  conductivity of liquids under pressure},}\ }\href {\doibase 10.2307/20026073}
  {\bibfield  {journal} {\bibinfo  {journal} {PNAAS}\ }\textbf {\bibinfo
  {volume} {59}},\ \bibinfo {pages} {141} (\bibinfo {year} {1923})}\BibitemShut
  {NoStop}%
\bibitem [{\citenamefont {Zhao}\ \emph {et~al.}(2021)\citenamefont {Zhao},
  \citenamefont {Wingert}, \citenamefont {Chen},\ and\ \citenamefont
  {Garay}}]{ZhaoJAP2021}%
  \BibitemOpen
  \bibfield  {author} {\bibinfo {author} {\bibfnamefont {A.~Z.}\ \bibnamefont
  {Zhao}}, \bibinfo {author} {\bibfnamefont {M.~C.}\ \bibnamefont {Wingert}},
  \bibinfo {author} {\bibfnamefont {R.}~\bibnamefont {Chen}}, \ and\ \bibinfo
  {author} {\bibfnamefont {J.~E.}\ \bibnamefont {Garay}},\ }\bibfield  {title}
  {\enquote {\bibinfo {title} {Phonon gas model for thermal conductivity of
  dense, strongly interacting liquids},}\ }\href {\doibase 10.1063/5.0040734}
  {\bibfield  {journal} {\bibinfo  {journal} {J. Appl. Phys.}\ }\textbf
  {\bibinfo {volume} {129}},\ \bibinfo {pages} {235101} (\bibinfo {year}
  {2021})}\BibitemShut {NoStop}%
\bibitem [{\citenamefont {Xi}\ \emph {et~al.}(2020)\citenamefont {Xi},
  \citenamefont {Zhong}, \citenamefont {He}, \citenamefont {Xu}, \citenamefont
  {Nakayama}, \citenamefont {Wang}, \citenamefont {Liu}, \citenamefont {Zhou},\
  and\ \citenamefont {Li}}]{XiCPL2020}%
  \BibitemOpen
  \bibfield  {author} {\bibinfo {author} {\bibfnamefont {Q.}~\bibnamefont
  {Xi}}, \bibinfo {author} {\bibfnamefont {J.}~\bibnamefont {Zhong}}, \bibinfo
  {author} {\bibfnamefont {J.}~\bibnamefont {He}}, \bibinfo {author}
  {\bibfnamefont {X.}~\bibnamefont {Xu}}, \bibinfo {author} {\bibfnamefont
  {T.}~\bibnamefont {Nakayama}}, \bibinfo {author} {\bibfnamefont
  {Y.}~\bibnamefont {Wang}}, \bibinfo {author} {\bibfnamefont {J.}~\bibnamefont
  {Liu}}, \bibinfo {author} {\bibfnamefont {J.}~\bibnamefont {Zhou}}, \ and\
  \bibinfo {author} {\bibfnamefont {B.}~\bibnamefont {Li}},\ }\bibfield
  {title} {\enquote {\bibinfo {title} {A ubiquitous thermal conductivity
  formula for liquids, polymer glass, and amorphous solids},}\ }\href {\doibase
  10.1088/0256-307x/37/10/104401} {\bibfield  {journal} {\bibinfo  {journal}
  {Chin. Phys. Lett.}\ }\textbf {\bibinfo {volume} {37}},\ \bibinfo {pages}
  {104401} (\bibinfo {year} {2020})}\BibitemShut {NoStop}%
\bibitem [{\citenamefont {Bird}\ \emph {et~al.}(2002)\citenamefont {Bird},
  \citenamefont {Lightfoot},\ and\ \citenamefont {Stewart}}]{BirdBook}%
  \BibitemOpen
  \bibfield  {author} {\bibinfo {author} {\bibfnamefont {R.~B.}\ \bibnamefont
  {Bird}}, \bibinfo {author} {\bibfnamefont {E.~N.}\ \bibnamefont {Lightfoot}},
  \ and\ \bibinfo {author} {\bibfnamefont {W.~E.}\ \bibnamefont {Stewart}},\
  }\href@noop {} {\emph {\bibinfo {title} {Transport Phenomena -}}}\ (\bibinfo
  {publisher} {J. Wiley},\ \bibinfo {address} {New York},\ \bibinfo {year}
  {2002})\BibitemShut {NoStop}%
\bibitem [{\citenamefont {Khrapak}(2023{\natexlab{b}})}]{KhrapakJMolLiq2023}%
  \BibitemOpen
  \bibfield  {author} {\bibinfo {author} {\bibfnamefont {S.A.}\ \bibnamefont
  {Khrapak}},\ }\bibfield  {title} {\enquote {\bibinfo {title} {Bridgman
  formula for the thermal conductivity of atomic and molecular liquids},}\
  }\href {\doibase 10.1016/j.molliq.2023.121786} {\bibfield  {journal}
  {\bibinfo  {journal} {J. Mol. Liq.}\ }\textbf {\bibinfo {volume} {381}},\
  \bibinfo {pages} {121786} (\bibinfo {year} {2023}{\natexlab{b}})}\BibitemShut
  {NoStop}%
\bibitem [{\citenamefont {Khrapak}\ and\ \citenamefont
  {Khrapak}(2023{\natexlab{b}})}]{KhrapakPoF2023}%
  \BibitemOpen
  \bibfield  {author} {\bibinfo {author} {\bibfnamefont {S.~A.}\ \bibnamefont
  {Khrapak}}\ and\ \bibinfo {author} {\bibfnamefont {A.~G.}\ \bibnamefont
  {Khrapak}},\ }\bibfield  {title} {\enquote {\bibinfo {title} {Sound
  velocities in liquids near freezing: {D}ependence on the interaction
  potential and correlations with thermal conductivity},}\ }\href {\doibase
  10.1063/5.0157945} {\bibfield  {journal} {\bibinfo  {journal} {Phys. Fluids}\
  }\textbf {\bibinfo {volume} {35}},\ \bibinfo {pages} {077129} (\bibinfo
  {year} {2023}{\natexlab{b}})}\BibitemShut {NoStop}%
\bibitem [{\citenamefont {Leibfried}\ and\ \citenamefont
  {Schl\"omann}(1954)}]{Leibfried1954}%
  \BibitemOpen
  \bibfield  {author} {\bibinfo {author} {\bibfnamefont {G.}~\bibnamefont
  {Leibfried}}\ and\ \bibinfo {author} {\bibfnamefont {E.}~\bibnamefont
  {Schl\"omann}},\ }\href@noop {} {\bibfield  {journal} {\bibinfo  {journal}
  {Nach. Akad. Wiss. G\"ottingen Math. Phyz. Klasse}\ }\textbf {\bibinfo
  {volume} {4}},\ \bibinfo {pages} {71} (\bibinfo {year} {1954})}\BibitemShut
  {NoStop}%
\bibitem [{\citenamefont {Chen}\ \emph {et~al.}(2011)\citenamefont {Chen},
  \citenamefont {Hsieh}, \citenamefont {Cahill}, \citenamefont {Trinkle},\ and\
  \citenamefont {Li}}]{ChenPRB2011}%
  \BibitemOpen
  \bibfield  {author} {\bibinfo {author} {\bibfnamefont {B.}~\bibnamefont
  {Chen}}, \bibinfo {author} {\bibfnamefont {W.-P.}\ \bibnamefont {Hsieh}},
  \bibinfo {author} {\bibfnamefont {D.~G.}\ \bibnamefont {Cahill}}, \bibinfo
  {author} {\bibfnamefont {D.~R.}\ \bibnamefont {Trinkle}}, \ and\ \bibinfo
  {author} {\bibfnamefont {J.}~\bibnamefont {Li}},\ }\bibfield  {title}
  {\enquote {\bibinfo {title} {Thermal conductivity of compressed {H$_2$O} to
  22 {GP}a: {A} test of the {L}eibfried-{S}chl\"{o}mann equation},}\ }\href
  {\doibase 10.1103/physrevb.83.132301} {\bibfield  {journal} {\bibinfo
  {journal} {Phys. Rev. B}\ }\textbf {\bibinfo {volume} {83}},\ \bibinfo
  {pages} {132301} (\bibinfo {year} {2011})}\BibitemShut {NoStop}%
\end{thebibliography}%

\end{document}